\RequirePackage{ifpdf}
\ifpdf 
\documentclass[pdftex]{sigma14}
\else
\documentclass{sigma14}
\fi

 \numberwithin{theorem}{section}
 \numberwithin{proposition}{section}
 \numberwithin{lemma}{section}
 \numberwithin{corollary}{section}
 \numberwithin{definition}{section}
 \numberwithin{example}{section}
 \numberwithin{remark}{section}
 \numberwithin{note}{section}

\definecolor{darksienna}{rgb}{0.24, 0.08, 0.08}
\definecolor{forestgreen}{rgb}{0.13, 0.55, 0.13}
 \definecolor{sapphire}{rgb}{0.03, 0.15, 0.4}
\definecolor{charcoal}{rgb}{0.21, 0.27, 0.31}
\definecolor{darkjunglegreen}{rgb}{0.1, 0.14, 0.13}
 \definecolor{slategray}{rgb}{0.44, 0.5, 0.56}
\definecolor{warmblack}{rgb}{0.0, 0.26, 0.26}

 \usepackage[margin=2.5cm]{geometry}

\usepackage{amssymb}
\usepackage[mathscr]{euscript}
\usepackage{graphicx}
\usepackage[all,knot]{xy}
\usepackage{multicol} 
\usepackage{cmap-cyr-vf}
\usepackage{skt}
\usepackage{verbatim}

\usepackage[T1]{fontenc} 
 \usepackage{yfonts}
\usepackage{sectsty}

\newtheorem{prop}{Proposition}[section]

\newtheorem{defin}{Definition}[subsection]
\newtheorem{lemm}{Lemma}[section]

\newcommand{\iN}{\hbox{ {\leaders\hrule\hskip.2cm}{\vrule height .22cm} }}

\newcommand{\R}[1][]{\ensuremath{{\mathbb{R}^{#1}} }}

 \newcommand{\BL}{{\Big{[}}}
\newcommand{\BR}{{\Big{]}}}
\newcommand{\bl}{{\big{[}}}
\newcommand{\br}{{\big{]}}}
\newcommand{\bee}{\begin{equation}}
\newcommand{\eee}{\end{equation}}
\newcommand{\dd}{{\hbox{d}}}
\newcommand{\vol}{\beta}

\def\<{\langle} \def\>{\rangle}

\newcommand\dt{E}
\newcommand\dtt{{E}}

\newcommand\dttD{{D}}

\newcommand\dttG{G}

\newcommand\dttP{\hbox{\sffamily\bfseries P}}
\newcommand\dttQ{\hbox{\sffamily\bfseries Q}}
\newcommand\dttJ{J}

\newcommand{\PB}{{{\cal{P}}}}

\begin{document}

\allowdisplaybreaks

\renewcommand{\PaperNumber}{}
\thispagestyle{empty}
   
   \begin{center}
\ArticleName{{{Multisymplectic formulation of vielbein gravity}}}
{{{{{I.}}} De Donder-Weyl  formulation, Hamiltonian $(n-1)$-forms}}

\end{center}

\ShortArticleName{{{{{\sffamily{I.}}} De Donder-Weyl  formulation, Hamiltonian $(n-1)$-forms } }}

    \begin{center}

\Author{{{\em Dimitri VEY}}\footnote{SPHERE Laboratory - UMR 7219,  D. Diderot, Paris 7 University - Paris, France  ---  dim.vey@gmail.com.}}
\end{center}

\AuthorNameForHeading{Multisymplectic formulation of vielbein gravity}



 \thispagestyle{empty}

  \begin{center}

\begin{minipage}{0.95\textwidth}

   {{ {{
   
    {\textsf{Abstract.}}  
    We consider   the  De Donder-Weyl ({DW}) Hamiltonian formulation of  the Palatini action  of vielbein gravity formulated  in terms of   the solder form and  spin connection, which are treated as independent variables. The basic geometrical constructions   necessary for the   {DW} Hamiltonian theory    of  vielbein gravity are presented. We reproduce  the   {DW} Hamilton equations in the multisymplectic and pre-multisymplectic formulations. We also give   basic examples of Hamiltonian   $(n-1)$-forms and   related Poisson   brackets.     }}}}  \end{minipage}
 \end{center}


 \section{{{Introduction}}}\label{sec:introd}
 
The canonical Hamiltonian theory of the Palatini action of vierbein (tetrad) gravity has been studied  by  Deser and Isham   \cite{Deser} and Heanneaux {\em et al.} \cite{Henneaux}.  In the canonical  formulation,  {\em space} and {\em time} are  treated asymmetrically and the canonical variables are defined on spacelike hypersurfaces. Therefore, the dynamics  implies a global spacelike  foliation of the space-time manifold.  The     canonical commutation  relations are defined   on the  equal time hypersurfaces. Accordingly, the Dirac  canonical   quantization   is related to   the     instantaneous Hamiltonian formalism,  which adds an additional structure  of global hyperbolicity on the relativistic space-time. In this paper,  we consider  the  De Donder-Weyl ({DW})  Hamiltonian formulation of vielbein gravity in the broader   context  of  Multisymplectic Geometry {{(MG)}}.   The finite dimensional  {DW}  theory  is a   covariant Hamiltonian-like formulation for field theory,  where  the {\em space} and {\em time} coordinates are treated  symmetrically.   Hence,   {MG}  may give    a  profound   geometrical road  to   field quantization 
(see {\em e.g.} \cite{H-02,kann-geom}).   The {DW} Hamiltonian formulation of vielbein gravity based on the first order Palatini action is already  found in  some papers.   A constraints analysis of the Ashtekar theory based  on the    multisymplectic formalism  is   found in the paper by    Esposito   {\em et al.} \cite{Espo}. For   a glimpse of the {DW}   formulation of vierbein gravity,   see   also Rovelli \cite{Rovelli002,Carlo1}.   The work of  Bruno,  Cianci, and  Vignolo \cite{bruno,bruno1} gives a more detailed development  at the crossroad of 
  the 
  {natural bundles theory} and the jet bundle formalism. Finally,   the   papers of Kanatchikov \cite{kannnnnnnaa,kannnnnnnaa1}     focus  on the problem  of constraints    and   precanonical quantization \cite{kann-geom2} of vielbein gravity   in the {DW} formulation.    

 In this paper, we first outline    in section \ref{sec:introd}  the basic  ingredients needed for the subsequent   study   such as the {MG}, Palatini formulation and the configuration space  of vielbein gravity.  Then, in section \ref{sec:section2}, we present the  {DW} Hamiltonian formulation of the first order Palatini action of vielbein gravity. More precisely,  in section \ref{subsec:2.2}  we describe the Legendre correspondence in the   {DW}   setting. We define the constraint  hypersurface  $ {{\cal C}} \subset  {{\cal M}}_{\hbox{\tiny{\sffamily DW}}}$   in section \ref{subsec:2.3}. In section  \ref{subsec:Ham093U984} we give  the expression of the  {DW}  Hamiltonian density  related to the Palatini action {\em i.e.}    ${\cal{H}}^{\hbox{\tiny{\sffamily Palatini}}} := \iota^{\star} {\cal{H}}^{\hbox{\tiny{\sffamily DW}}} $, where    $\iota $ is the canonical inclusion $ \iota : {{\cal C}} \hookrightarrow  {{\cal M}}_{\hbox{\tiny{\sffamily DW}}}  $.  In section \ref{subsec:ext73645} we calculate  its exterior derivative $\dd {\cal{H}}^{\hbox{\tiny{\sffamily Palatini}}} := \iota^{\star} \dd {\cal{H}}^{\hbox{\tiny{\sffamily DW}}}$. Then, in section \ref{subsec:saapsjpj89},  
   we present a   brief comment on   the   primary constraints set and the extended {DW} Hamiltonian. Finally,  in section \ref{subsec:2.4} we derive the {DW} Hamilton equations in   three  and four dimensional cases.  In section \ref{sec:section17} we discuss   the pre-multisymplectic formulation of vielbein gravity, {\em i.e.} we work on the level set     $  {{\cal C}}_\circ  := ({\cal{H}}^{{\tiny\hbox{\sffamily DW}}})^{-1} (0)  \subset    {{\cal M}}_{\hbox{\tiny{\sffamily DW}}} $. Thus, the pre-multisymplectic formulation      of dreibein and vierbein    gravity   is presented in sections  \ref{subsec:3Dpremulti} and    \ref{subsec:4Dpremulti},  respectively.    In   section \ref{sec:section18} we focus on the notion of Hamiltonian   $(n-1)$-forms. In particular, we explore   its   relation to   homotopy Lie algebra   and   to the graded Poisson bracket    in sections   \ref{subsec:saison0010}  and \ref{subsec:saison001}, respectively. We also  present  some simple examples of    Hamiltonian $(n-1)$-forms in  sections \ref{subsec:SAOFex} and  \ref{subsec:saison002}. Finally, in   section \ref{subsec:saison003} we give   succinct  comments on   canonically conjugate  forms for vielbein gravity.   
   
\subsection{{{{Multisymplectic geometry}}}}  

Let us recall that {MG} is a generalization of  symplectic  geometry  to field theory.   It allows us to construct a general framework for the calculus of variations with several independent variables.  The origins of {{MG}} are connected with the names of  Carath\'eodory \cite{cara02},  Weyl \cite{Weyl}  on one hand and   De Donder  \cite{Donder01,Donder02}   on the other. We make this distinction since the motivations involved were different. Carath\'eodory and   Weyl were interested in the
generalization of the Hamilton-Jacobi equation to   the case of several independent  variables and the line of development
stemming from their work is concerned with the solutions of variational problems given by  an  action functional.  On the other hand,   Cartan \cite{cartan}  recognized   the crucial importance of developing an invariant language not dependent on local coordinates.  De Donder carried this development further  by exploring, in the context of field theory,  the relation between Hamilton equations and the theory of integral invariants.  
The {DW} system of   Hamiltonian  equations, as noted in   \cite{Donder01,H-02},   has been  discovered  
 already   by  Volterra   \cite{Volterra1,Volterra2}  at the end of the ninetieth century. Hence, the Hamilton-Volterra system of equations is today termed  the {DW} Hamilton equations   with the  reference to the work by De Donder \cite{Donder01,Donder02}  and Weyl   \cite{Weyl}.    As was first noted by   Lepage \cite{Lepage,Lepage01,Lepage02},    the  {DW} theory is a special case of  a more general   theory. The geometrical constructions permitting a fully general treatment  were provided by   Dedecker \cite{KS12,KS199,KS1994}.   Note also that the   line of research focusing  on 
 the 
 related  Lepagean equivalents was developed in particular by Krupka \cite{Kru01,Kru02,Kru03}, Krupkov\'{a} and Smetanov\'{a} \cite{Kru04,Kru05,Kru06}.  Finally, we refer   to  the review paper by   Kastrup \cite{Kastrup}, the book by  Rund \cite{Rund}, 
  Gotay  \cite{Gotay223,Gotayext},  
  and  Olver \cite{Olver,Olver1} for more details about 
 the Lepagean equivalents.     The {Legendre correspondence}, {\em i.e.} the generalization  of the Legendre transform  in the  context of the   Lepage-Dedecker  theory,   the  description of {observables} and   the construction of the   {Poisson brackets} are 
 the cornerstones of the  covariant Hamiltonian formalism for field theories. For example,  in the context of the Lepage-Dedecker theory, the papers by    H\'elein and Kouneiher     \cite{HK-02,HK-03}    develop  an insightful classification of 
 observable  forms   in terms of {\em algebraic observable forms} and {\em observable forms}.

A fruitful  step in the development of {MG} and  its relation to classical field theories was taken in the seventies  of the past century. In particular,  the Polish school  formulated important ideas and   developed the    {\guillemotleft {\em multisymplectic}\guillemotright}, or   {\guillemotleft {\em multiphase-space}\guillemotright},  formalism in the work of  Tulczyjew \cite{Tutu,Tutu1},  Kijowski \cite{JK-01,KS0}, Kijowski and Tulczyjew \cite{JKWMT}, Kijowski    and  Szczyrba \cite{ KS1,KS2}, and Gawedski \cite{GAGA}. 
   We find the notion of   an  observable form   already in their work.  A formulation of the notion of  a    dynamical observable  used in  \cite{HK-02,HK-03} already emerges in the work  of   Kijowski \cite{JK-01}.  Parallel to this development, the paper  by  Goldschmidt and Sternberg   \cite{HGSS},  gave a formulation of the {DW} Hamilton equations in terms of  the Poincar\'e-Cartan form and  the underlying jet bundles geometry,   and a related approach was also developed by   the Spanish school:  
   Garc\'{i}a \cite{Garcia00,Garcia01} and García and P\'{e}rez-Rend\'{u}n  \cite{Garcia02}.

  In this paper, we use the  multisymplectic formulation  based on the {DW}   {\guillemotleft{\em multimomentum phase space}\guillemotright}. Let us consider a theory with a covariant configuration space  given by a fiber bundle $({\cal Y},{\cal X}, \pi )$, where  $\pi : {{\cal{{\cal{Y}}}}}  {\longrightarrow} {{\cal{X}}}$ is the bundle projection. Let us denote  by $\{ x^{\mu} \}_{1\leq\mu\leq n} $ local coordinates on ${{\cal{X}}}$ the base space. The dimension of the space-time manifold is   $\hbox{dim}({{\cal{X}}}) = n$. We denote also   by $\{ y^{i} \}_{1\leq i\leq k} $   local coordinates on ${{\cal{{\cal{Y}}}}}_{x}$, where ${{\cal{{\cal{Y}}}}}_{x} := \pi^{-1}(x)$ is the fiber over a point of the space-time manifold. The dimension of the fiber is $\hbox{dim}({{\cal{{\cal{Y}}}}}_{x} ) = k$. Local coordinates on the total space ${{\cal{{\cal{Y}}}}}$ are denoted by $(x^{\mu} , y^{i})$. {We denote   $\Lambda^{n}_{\mathfrak{1}}T^{\star} {{{{\cal{{\cal{Y}}}}}}}$ the vector subbundle of $\Lambda^{n}T^{\star} {{{{\cal{{\cal{Y}}}}}}}$ whose fiber at $y \in {{{{\cal{{\cal{Y}}}}}}}$ consists   of all ${\pmb{\varphi}} \in \Lambda^{n}_{y}T^{\star} {{{{\cal{{\cal{Y}}}}}}}$ such that for any  vertical vector fields $ \zeta^{\hbox{\sffamily{v}}} , \chi^{\hbox{\sffamily{v}}} \in \hbox{\sffamily V}  {{{\cal{{\cal{Y}}}}}}  $ {\em i.e.} 
$ \Lambda^{n}_{\mathfrak{1}}T^{\star} {{{{\cal{{\cal{Y}}}}}}} =
{\big{\{}}  {\pmb{\varphi}}  \in \Lambda^{n}_{y}T^{\star} {{{{\cal{{\cal{Y}}}}}}} \ / \  \zeta^{\hbox{\sffamily{v}}} \iN \chi^{\hbox{\sffamily{v}}} \iN {\pmb{\varphi}} = 0 
{\big{\}}} 
$. We also denote     $\Lambda^{n}_{\mathfrak{0}} T^{\star} {{\cal{{\cal{Y}}}}}$ the space of horizontal $n$-forms on ${{\cal{{\cal{Y}}}}} $.  
 Thus, we denote  by ${{\cal M}}_{\hbox{\tiny{\sffamily DW}}}  :=  {{\cal M}}_{\hbox{\tiny{\sffamily DW}}}   ({\cal Y}) := \Lambda^{n}_{\mathfrak{1}} T^{\star}{{\cal{{\cal{Y}}}}} $ the {DW} multimomentum phase space.  The bundle $  \Lambda^{n}_{\mathfrak{1}} T^{\star}{{\cal{{\cal{Y}}}}}  \rightarrow {{\cal{{\cal{Y}}}}}$   carries a canonical structure  ${{\theta}}^{\hbox{\tiny{\sffamily DW}}} = \varkappa \vol + {{p}}^{\mu}_{i} \dd y^{i} \wedge \vol_\mu $ and leads to the multisymplectic structure: 
$ \displaystyle
{\pmb{\omega}}^{\hbox{\tiny{\sffamily DW}}} = \dd \varkappa \wedge \vol + \dd {{p}}^{\mu}_{i} \wedge \dd y^{i} \wedge \vol_\mu
$, {with $  {\beta}= \dd x^{1}  \wedge ... \wedge \dd x^{n}$ a volume $n$-form on ${{\cal{X}}}$ and $   {\beta}_{\mu}:= \partial_{\mu} \iN   {\beta}$ is a $(n-1)$-form}. 

To conclude this  overview we    mention examples of   more recent papers  in the field.   We refer  to Binz, Sniatycki and Fischer \cite{Binz},  G\"{u}nther  \cite{Gunt},  De Le\'{o}n, Cari\~{n}ena,  Crampin, Ibort \cite{cant,cari},  Forger, Paufler and R\"{o}mer \cite{Forger0010,Forger015,Forger011}, Gotay   {\em et al}.   \cite{Gotay,Gotay223,Gotay224,Gotayext},    H\'elein   \cite{FH-01,H-02},  H\'elein and Kouneiher       \cite{HK-01,HK-02,HK-03},   Kanatchikov   \cite{Kana-01,Kana-02,Kana-014,Kana-015}, and     Sardanashvily {\em et al.} \cite{Gaga01,Gaga02,Gaga03,Sarda01}.   Most of the literature on the subject      focuses on the  contact structure and  jet bundles formalism.  For a general presentation  of  multisymplectic, $k$-symplectic and $k$-cosymplectic geometries,  we refer to the review paper by Rom\'{a}n-Roy \cite{Roy} and   the book by De Le\'{o}n,  Salgado and Vilari\~{n}o  \cite{Leon}.       The multiplicity of formalisms  is  illustrated  by   the polysemy   of the term  {\guillemotleft {\em polysymplectic}\guillemotright}, first introduced by Günther \cite{Gunt}.  Thus,   Günther's  {\em polysymplectic} (or   $k$-symplectic, see  \cite{Leon}) formalism     is   different  from  the polysymplectic approaches  developed later by   Kanatchikov   \cite{Kana-01} and    Sardanashvily {\em et al.} \cite{Gaga02}, respectively.     In  the former,      the polysymplectic formulation  is based on the  {\em polymomentum phase space}    {\em i.e.} the quotient bundle  $  {\cal M}^{{\hbox{\tiny\sffamily Poly}}}_{{\hbox{\tiny\sffamily DW}}} ({{\cal{{\cal{Y}}}}}) =  \Lambda^{n}_{\mathfrak{1}} T^{\star} {{\cal{{\cal{Y}}}}}   /  \Lambda^{n}_{\mathfrak{0}} T^{\star} {{\cal{{\cal{Y}}}}}
$.  The polysymplectic structure on $ {\cal M}^{{\hbox{\tiny\sffamily Poly}}}_{{\hbox{\tiny\sffamily DW}}} ({{\cal{{\cal{Y}}}}})  $ is described as an equivalence class of canonical forms while the main object is ${{\pmb{\omega}}}^{\hbox{\sffamily v}} :=  \dd {{p}}^{\mu}_{i} \wedge \dd y^{i} \wedge \vol_\mu$,  the vertical part of the multisymplectic form $  {\pmb{\omega}}^{\hbox{\tiny{\sffamily DW}}}$. In the latter approach,   the {\em polymomentum phase space}   is   defined as $  {\cal M}^{{\hbox{\tiny\sffamily Poly}}}  ({{\cal{{\cal{Y}}}}})     = \pi^{\star} T{{\cal{X}}}  \otimes \hbox{\sffamily V}^{\star} ({{\cal{Y}}}) \otimes  \pi^{\star} \Lambda^{n} T^{\star} {{\cal{X}}}$  and   the canonical  polysymplectic form  is given by 
$ \displaystyle
{\pmb{\omega}}^{\tiny{\hbox{\sffamily Poly}}} = \dd {{p}}^{\mu}_{i} \wedge \dd y^{i} \wedge \vol \otimes \partial_{\mu}
$.

 \subsubsection{{{Poincar\'é-Cartan $n$-form,  multisymplectic $(n+1)$-form}}}

 In this section, we introduce the  multimomentum phase space in   {MG}, {\em i.e.} the bundle ${\cal M} := \Lambda^n T^\star {{\cal{Y}}} $ of $n$-forms over the configuration space ${{{\cal{{\cal{Y}}}}}}$. This   is a generalization  of the  phase space,   {\em i.e.} of the  cotangent bundle     introduced in  symplectic geometry.  We will follow the terminology found in \cite{H-02,HK-01,HK-02,HK-03}.
                      \begin{defin}\label{def03}
  A multisymplectic manifold  ${\big{(}}  {\cal M} , {\pmb{\omega}}{\big{)}}$    is a manifold ${\cal M}$  together with   $ {\pmb{\omega}} $,     a closed  and non degenerate differential $(n+1)$-form on $ {\cal M}$.
\end{defin}
   In field theory we are led to think of solutions of variational problems as  $n$-dimensional submanifolds  ${{\Gamma}}$ embedded in   the multimomentum phase space.  We introduce the notion of a Hamiltonian $n$-curve, see  \cite{HK-02,HK-03}:
  \begin{defin}\label{jfzajoo1002} Let ${\cal{H}} : {\cal M} \rightarrow \Bbb{R}$ be a smooth Hamiltonian function (such that {\em$\dd {\cal{H}} \neq 0$}).
A {Hamiltonian n-curve} is a $n$-dimensional oriented submanifold ${{\Gamma}} \subset {\cal M}$ such that   
{\em \begin{equation}
 \forall m \in {{\Gamma}}, \quad \quad \exists X \in \Lambda^{n}T_m {{\Gamma}}, \quad  \quad  X \iN {\pmb{\omega}}_m = (-1)^{n} \hbox{d} {\cal{H}}_m.
 \end{equation}}
\end{defin}
 A Hamiltonian $n$-curve is  parametrized by a map   $  x \mapsto (q(x), p(x))$ from the space-time manifold $  {\cal X} $ to the multimomentum phase space  $ {{{\cal M}}}$.  Actually,  in  definition \ref{jfzajoo1002}, the  generalized  Hamilton equations  are    written in geometric form as  $X \iN {\pmb{\omega}}_{m} = (-1)^n \hbox{d}{\cal{H}}_{m}$.

The Poincar\'e-Cartan    $n$-form ${{\theta}}$ 
on $\Lambda^nT^\star{{\cal{{\cal{Y}}}}}$ is defined as 
\begin{equation}\label{PC}
\forall q\in {{\cal{{\cal{Y}}}}}, \quad \forall p\in \Lambda^nT^\star_q{{\cal{{\cal{Y}}}}},
\quad 
 \quad
{{\theta}}_{(q,p)}(X_1, \cdots, X_{n}) =
p(  \Pi_{\star}(X_1),\cdots,  \Pi_{\star} (X_n)),
\end{equation}
  where $\Pi :  {\cal M}  := \Lambda^nT^\star{{\cal{{\cal{Y}}}}}   \overset{\Pi}{\longrightarrow} {{\cal{{\cal{Y}}}}} $ is the bundle projection on the configuration bundle and     $\Pi_{\star} := \dd \Pi :    T\Lambda^nT^\star{{\cal{{\cal{Y}}}}}   \overset{\Pi_{\star}}{\longrightarrow}  T{\cal Y}  $.  Note that the dimension of a fiber at  $q \in {{\cal{{\cal{Y}}}}}  $ is  $   \hbox{dim} \left( \Lambda^{n} T^{\star}_{q} {{{\cal{Y}}} } \right) =     {(n+k)!} /({n!k!})$, whereas the dimension of the total space of the fiber bundle is    $     \hbox{dim} \left( \Lambda^{n} T^{\star} {{{\cal{Y}}} } \right) = n+k+  {(n+k)!}/({n!k!}) $. 

Strictly speaking,   the object defined by \eqref{PC} is  the most general  Lepagean equivalent of the Poincaré-Cartan form. Nevertheless,  we term it the {\guillemotleft {\em Poincaré-Cartan}\guillemotright}  form,  according to  the terminology found in   \cite{HK-02,HK-03}. Let $(q^{{\pmb{\mu}}})_{{1 \leq {\pmb{\mu}} \leq n+k}}$ be  the local coordinates on ${{\cal{{\cal{Y}}}}}$, {\em i.e.} $ q^{{\pmb{\mu}}} := (x^{\mu} , y^{i})$.  Let  the family $(\dd q^{{\pmb{\mu}}_1} \wedge ... \wedge \dd q^{{\pmb{\mu}}_n})_{{1 \leq {{\pmb{\mu}}}_1  < ... {{\pmb{\mu}}}_n < n+k}}$     be a  basis of 
$\Lambda^nT^\star_q {{\cal{{{Y}}}}}$. We denote by  $p_{{{\pmb{\mu}}}_1 ... {{\pmb{\mu}}}_n}$  the  local coordinates of the Poincar\'e-Cartan form on  $ \Lambda^nT^\star_q {{\cal{{\cal{Y}}}}}$ in the basis $\dd q^{{\pmb{\mu}}_1} \wedge ... \wedge \dd q^{{\pmb{\mu}}_n}$. In particular, we denote $\varkappa := p_{1\cdots n}$, $p_{i}^{\mu} := p_{1\cdots (\mu-1)i(\mu+1)\cdots n }$, $p^{\mu_1   \mu_2}_{i_{1}   i_2} := p_{1 \cdots (\mu_1 - 1) i_1 (\mu_1 +1) \cdots  (\mu_2 - 1) i_2 (\mu_2 +1) \cdots n}$, ...  Finally, we use also the notations  ${\beta}_{\mu_1 \cdots \mu_p} :=   \partial_{\mu_1} \wedge \cdots \wedge \partial_{\mu_p} \iN {\beta} $, and   ${\beta}^{i_1 \cdots i_p}_{\mu_1 \cdots \mu_p} := \hbox{d} y^{i_1} \wedge \cdots \wedge \hbox{d} y^{i_p} \wedge {\big{(}} \partial_{\mu_1} \wedge \cdots \wedge \partial_{\mu_p} \iN {\beta} {\big{)}} $.   In local coordinates,    the   Poincar\'{e}-Cartan  $n$-form   ${{\theta}}$   is written as  
 \begin{equation}\label{PCmulti002}
  \left.
\begin{array}{ccl}
\displaystyle  {{\theta}}    & = & \displaystyle    \sum_{1 \leq {{\pmb{\mu}}}_1  < \cdots <  {{\pmb{\mu}}}_n < n+k} p_{{{\pmb{\mu}}}_1 \cdots  {{\pmb{\mu}}}_n} \dd q^{{\pmb{\mu}}_1} \wedge \cdots  \wedge \dd q^{{\pmb{\mu}}_n}, \\
\displaystyle   & = & \displaystyle 
    \varkappa   {\beta} +  \sum_{j=1}^{n} \sum_{\mu_1<\cdots<\mu_j} \sum_{i_1<\cdots<i_j}   p^{\mu_1 \cdots \mu_j }_{i_1 \cdots i_j}   {\beta}^{i_1 \cdots i_j}_{\mu_1 \cdots \mu_j} .
\\ 
 \end{array}
\right.
  \end{equation}
  The {multisymplectic} $(n+1)$-form    $ {\pmb{\omega}} := \hbox{d}{{\theta}}$ (called also the  {\guillemotleft {\em pataplectic form}\guillemotright} in  \cite{HK-01})  is  the exterior derivative of the Poincar\'e-Cartan form. Traditionaly the term   {\guillemotleft {\em multisymplectic form}\guillemotright}  refers   to Kijowski's multisymplectic form \cite{JK-01,KS0} {\em i.e.} in the {DW} formulation {\em only}.  Nonetheless, we will   follow   the terminology introduced in \cite{HK-02,HK-03}.   In local coordinates,   the multisymplectic  $(n+1)$-form  $ {\pmb{\omega}} := \dd  {{{\theta}}}  $ is written as
  \begin{equation}\label{PCmulti003}
  \left.
\begin{array}{ccl}
 \displaystyle   {\pmb{\omega}}   & = & \displaystyle    \sum_{1 \leq {{\pmb{\mu}}}_1  < \cdots < {{\pmb{\mu}}}_n < n+k} \dd p_{{{\pmb{\mu}}}_1 ...  {{\pmb{\mu}}}_n} \wedge \dd q^{{\pmb{\mu}}_1} \wedge \cdots  \wedge \dd q^{{\pmb{\mu}}_n} , 
\\ 
\displaystyle    & = & \displaystyle      
  \hbox{d} \varkappa \wedge {\beta} + \sum_{j=1}^{n} \sum_{\mu_1<\cdots<\mu_j} \sum_{i_1<\cdots<i_j} \hbox{d} p^{\mu_1 \cdots \mu_j }_{i_1 \cdots i_j} \wedge {\beta}^{i_1 \cdots i_j}_{\mu_1 \cdots \mu_j}.
\end{array}
\right.
  \end{equation}
 
 \subsubsection{{{Bundle of field derivatives}}}\label{field-derivative-bundle}

We  now describe the  Lagrangian side of the formulation of  a variational problem on fields $\varphi : {{\cal{X}}} \rightarrow {{\cal{{\cal{Y}}}}}$.   The Lagrangian density  ${{L}} (q,v) =   {{L}} (x^{\mu} , y^{i} , v^{i}_{\mu}) = {{L}} (x^{\mu} , y^{i} , \partial_{\mu} y^{i}) $   is defined on the  bundle  ${{\cal P}}$ of field derivatives.      We associate to $\varphi $ the bundle  $ \varphi^\star T  {{\cal{{\cal{Y}}}}} \otimes   T^\star {{\cal{X}}} $ over ${{\cal{X}}}$. A point  $(x,v) \in  \varphi^{\star}T  {{\cal{{\cal{Y}}}}} \otimes    T^\star {{\cal{X}}}    $
is given by 
$    v =  \sum_{1 \leq \mu \leq n} \sum_{1 \leq i \leq k} v^{i}_{\mu} \frac{\partial }{\partial {y}^{i}}   \otimes   \dd x^{\mu} $. On the bundle  $ {{\cal P}}^{\varphi} :=  \varphi^\star T  {{\cal{{\cal{Y}}}}} \otimes    T^\star {{\cal{X}}} $, which is included in the bundle   $  {{\cal P}}   ={{\{}} (x,y,v)   / (x,y) \in {{\cal{{\cal{Y}}}}}, \  v \in   T_{y}  {{\cal{{\cal{Y}}}}} \otimes    T^\star_{x} {{\cal{X}}}     {{\}}} $,   the local coordinates are  $(x^\mu,y^{i}, v^{i}_{\mu} )$.  Note that the dimension of the {fiber} is $ \hbox{dim}  (  {{\cal P}}_{(x,y)}  ) = nk$, whereas the dimension the bundle   is   $   \hbox{dim} \left(  {{\cal P}}  \right) = n+k+nk$. They     can be   equivalently thought of as the  local coordinates on the  first order jet bundle ${J}^{1} ({{\cal{{\cal{Y}}}}})$. We refer to  Saunders \cite{Saunders} for an introduction to the  jet bundle formalism,  and to  Cari\~{n}ena {\em et al.}   \cite{cari}, and Gotay  {\em et al.} \cite{Gotay} for the use  of it   in the multisymplectic context.
    
  Using the variational principle we obtain for the action    ${\cal S}[{\varphi}] = \int_{\cal{X}}L(x,{\varphi}(x),{\hbox{d}{\varphi}(x)}) {\beta} $    
 the related    Euler-Lagrange  \begin{equation}\label{eulerlagrange0}
\quad \frac{\partial}{\partial x^\mu}\left(\frac{\partial L}{\partial v^i_\mu}(x,{\varphi}(x),{\hbox{d}{\varphi} (x)})\right) = \frac{\partial L}{\partial y^i}(x,{\varphi}(x),{\hbox{d}{\varphi}(x)}).
\end{equation}
  We denote by $ \Lambda^{n}_{\textsc{n}} T  {{{\cal{Y}}} }  $ the normalized space of decomposable $n$-vector fields  on ${\cal Y}$:   for any $  {z}_1 , \cdots , {z}_n \in T_{q} {\cal{Y}}$,
$  \Lambda^{n}_{\textsc{n}}  T  {{{\cal{Y}}} } := {{\{}} (q, {z}) \in \Lambda^{n}T{{\cal{{\cal{Y}}}}}   / {z} = {z}_1 \wedge ... \wedge {z}_n       
   \ \hbox{and} \ \vol({z}_1 , ... , {z}_n) = 1 {{\}}} 
$.  We construct a  diffeomorphism between $  \Lambda^{n}_{\textsc{n}} T  {{{\cal{Y}}} }    $ and ${{\cal P}}$. More precisely, for any $(x^\mu,y^{i}) \in  {\cal{Y}}$  the fiber $  \Lambda^{n}_{\textsc{n}}   T_{(x,y)}   {{{\cal{{\cal{Y}}}}}} $ is   identified with $  {{\cal P}}_{(x,y)}   $ using the diffeomorphism  
$      \sum_{\mu} \sum_{i} v^{i}_{\mu} \frac{\partial }{\partial {y}^{i}}   \otimes   \dd x^{\mu}   \mapsto      {z} = {z}_1 \wedge ... \wedge {z}_n
  $, where for any $  1 \leq   \alpha   \leq n $, 
$  
{{z}}_\alpha =   \frac{\partial}{\partial x^\alpha} + \sum_{1 \leq i \leq k}  v^{i}_{\alpha} \frac{\partial}{\partial y^{i}}  $, see \cite{HK-01}.

\subsubsection{{{DW Multimomentum manifold}}}\label{subsec:DWmultimomentum}

  The DW multimomentum manifold   ${{\cal M}}_{\hbox{\tiny{\sffamily DW}}} $ is a submanifold of  ${\cal M} := \Lambda^nT^\star { {{\cal{{\cal{Y}}}}}} $. For any   $(q,p)\in \Lambda^nT^\star{{\cal{{\cal{Y}}}}}$  we restrict ourselves to the case where  the interior product   $\zeta^{\hbox{\sffamily{v}}} \iN \chi^{\hbox{\sffamily{v}}}  \iN p = 0 $ is identically vanishing, where   $ \zeta^{\hbox{\sffamily{v}}} , \chi^{\hbox{\sffamily{v}}}  \in   \hbox{\sffamily V}  {{{\cal{{\cal{Y}}}}}}  $ are any two  vertical vector fields. Let us recall that  a vector field is vertical if  any $\xi \in T_y{{\cal{Y}}}$ such that  $ \pi_{\star} (\xi) :=  \hbox{d}\pi (\xi ) = 0$, where $\pi$ is the bundle projection on the space-time manifold ${\cal Y}   \overset{\pi }{\longrightarrow} {{\cal{X}}}$.  Then, by definition 
\begin{equation}\label{DDWDDWDDW}
{{\cal M}}_{\hbox{\tiny{\sffamily DW}}}     :=    \Lambda^{n}_{\mathfrak{1}}T^{\star} {\cal{Y}} = \left\{ (q,p)\in \Lambda^nT^\star{{\cal{Y}}}    \ / \    \forall \zeta^{\hbox{\sffamily{v}}} , \chi^{\hbox{\sffamily{v}}} \in \hbox{\sffamily V}  {{{\cal{{\cal{Y}}}}}} ,  \quad \zeta^{\hbox{\sffamily{v}}}  \wedge \chi^{\hbox{\sffamily{v}}}  \iN p = 0 \right\} .
\end{equation}
Let ${\iota}_{\mathfrak{1}} : {{{\cal M}}}_{\tiny{\hbox{\sffamily DW}}}  \hookrightarrow {{{\cal M}}}$ be the canonical inclusion. Note that   $ {{\theta}}^{{\tiny{\hbox{\sffamily DW}}}} := \iota_{\mathfrak{1}}^{\star} {{\theta}}    \in   \Gamma ( {{\cal M}}_{\hbox{\tiny{\sffamily DW}}}  , \Lambda^{n} T^{\star} ( {{\cal M}}_{\hbox{\tiny{\sffamily DW}}} )  ) $ where ${{\theta}}   \in   \Gamma ( {\cal M}   , \Lambda^{n} T^{\star} ( {\cal M}  )  )  $. Since $ \dd {\big{(}}   \iota_{\mathfrak{1}}^{\star} {{\theta}}   {\big{)}}  =    \iota_{\mathfrak{1}}^{\star}   \dd {{\theta}}      =  \iota_{\mathfrak{1}}^{\star}   {\pmb{\omega}}  $,  we obtain $  {\pmb{\omega}}^{{\tiny{\hbox{\sffamily DW}}}} =  \dd  {{\theta}}^{{\tiny{\hbox{\sffamily DW}}}}  =  \iota_{\mathfrak{1}}^{\star}   {\pmb{\omega}}$. We denote by ${{\theta}}^{{\tiny{\hbox{\sffamily DW}}}}:= {{\theta}} |_{{{\cal M}}_{\hbox{\tiny{\sffamily DW}}}}$  the restriction of ${{\theta}}$ to  ${{\cal M}}_{\hbox{\tiny{\sffamily DW}}}$. Working on ${{\cal M}}_{\hbox{\tiny{\sffamily DW}}}$  is equivalent to setting  
 $p^{\mu_{1} \cdots \mu_{j}}_{{i}_{1} \cdots {i}_{j}} = 0 $ for all $j > 1$ in the expression of     ${{{\theta}}}$ given in \eqref{PCmulti002}. In local coordinates, the Poincaré-Cartan $n$-form is written as ${{{\theta}}}^{\tiny{\hbox{\sffamily DW}}} =  \varkappa   {\beta} +  p^{\mu}_i   \hbox{d}y^{i} \wedge {\beta}_{\mu}$. Then, following the terminology used by {\em e.g.} Kijowski  \cite{JK-01,KS0}, Cantrijn, Ibort and  Le\'{o}n  \cite{cant}, and Hélein \cite{H-02}, we introduce  also the   multisymplectic $(n+1)$-form  
\bee\label{ZZZ01}
{\pmb{\omega}}^{\tiny{\hbox{\sffamily DW}}} =  \hbox{d} \varkappa \wedge {\beta} + \sum_{\mu} \sum_{i} \hbox{d} p^{\mu}_i \wedge \hbox{d}y^{i} \wedge {\beta}_{\mu}.
\end{equation}

\subsubsection{{{Hamilton equation in DW formulation}}}\label{sec:GHE}

The {DW} Hamiltonian function 
$
H(x^\mu,y^i, p^\mu_i) = p^\mu_i v^i_\mu - L(x^\mu,y^i, v^i_\mu)
$ is defined  by introducing  the Legendre transform    $(x^\mu, y^{i} , {v^i_\mu})\mapsto (x^\mu, y^{i}, {p^{\mu}_i})$, with the multimomenta $   {p^{\mu}_i}:=  {\partial L}/ {\partial v^i_\mu}(x^\mu, y^i,{v^{i}_{\mu}}) $. If the Legendre transform is non singular, {\em i.e.} $ \hbox{det}  (   {\partial^{2} L }/{    \partial v^{i}_{\mu} \partial v^{j}_{\nu} }   )    \neq 0  $, the Euler-Lagrange equations \eqref{eulerlagrange0}  are equivalent to   the {DW} Hamilton equations:
\bee
   \left.
\begin{array}{ccl}
\displaystyle   \frac{\partial {\varphi}^i}{\partial x^\mu}(x)   &  = 
& \displaystyle       \frac{\partial H}{\partial p^\mu_i}   (x^\mu,{\varphi}^i(x), {p}^\mu_i(x)),
\\
  \end{array}
\right.
\quad \quad 
   \left.
\begin{array}{ccl}
 \displaystyle   \sum_\mu\frac{\partial  {p}^\mu_i}{\partial x^\mu}(x)    
   &
=    & \displaystyle   - \frac{\partial H}{\partial y^i}  (x^\mu,{\varphi}^i(x), {p}^\mu_i(x)) , 
\end{array}
\right.
\eee   
Following \cite{HK-01}, we introduce  the Legendre correspondence in the context of the most general Lepagean theory by the function $
{W} :  \Lambda^nT {{\cal{{\cal{Y}}}}}  \times \Lambda^nT^\star  {{\cal{{\cal{Y}}}}}  \rightarrow \Bbb{R}, 
(q,v,p)  \mapsto \langle p , v \rangle -L (q,v)$, where 
\bee\label{angle}
 \langle p , v \rangle \cong  \langle p , z \rangle = \langle   p , z_1 \wedge , \cdots , \wedge z_n \rangle
 = \sum_{{\pmb{\mu}}_1 < \cdots < {\pmb{\mu}}_n}  p_{{\pmb{\mu}}_1 , \cdots , {\pmb{\mu}}_n} z^{{\pmb{\mu}}_1}_{1} \cdots z^{{\pmb{\mu}}_n}_{n}.
 \eee
We   have denoted by $z^{{\pmb{\mu}}}_{\alpha}$   the coordinates of the vector fields ${z_{\alpha}} = \sum_{1 \leq {\pmb{\mu}} \leq n+k} z^{{\pmb{\mu}}}_{\alpha} \partial / \partial {q}^{\pmb{\mu}} \in T_q {\cal Y}$, which are  used  to construct the decomposable  $n$-vector field $z  = z_1 \wedge \cdots \wedge z_n \in \Lambda^{n}_{\textsc{n}} T  {{{\cal{Y}}} } \in \Lambda^{n}_{\textsc{n}} T  {{{\cal{Y}}} }  \cong {{\cal P}}$, see  section \ref{field-derivative-bundle}.
  The Legendre correspondence   is satisfied   if and only if, for any  $  (q,v,w)   \in     \Lambda^{n}_{\textsc{n}}  T  {{\cal{{\cal{Y}}}}}    \times \Bbb{R} $, and for any  $  (q,p)  \in \Lambda^{n} T^{\star}  {{\cal{{\cal{Y}}}}}$, we have
\bee
   \langle p,v \rangle -  L(q,v)   =   w
\quad \quad
\hbox{and}
\quad \quad
  \frac{\partial {W}}{\partial v} (q, v, p) =    0.
 \eee
 When the Legendre hypothesis is satisfied, {\em c.f.} \cite{HK-01,HK-02,HK-03}, we denote  $(q,v,w)  
\ {\pmb{\leftrightarrow}}   \      (q,p)$.
 To  obtain the     {DW}   Hamilton equations, we restrict ourselves to   the manifold ${{\cal M}}_{\hbox{\tiny{\sffamily DW}}}$ with a Hamiltonian function $\mathcal{H}:  {{\cal M}}_{\hbox{\tiny{\sffamily DW}}}     =   \Lambda^{n}_{\mathfrak{1}}T^{\star} {{{{\cal{{\cal{Y}}}}}}}    \subset    \Lambda^{n} T^{\star} {{{{\cal{{\cal{Y}}}}}}}     \rightarrow \mathbb{R}$.   Only  when
  the Legendre correspondence is non degenerate we have a unique correspondence $(q,v) {\pmb{\leftrightarrow}} (q,p)$, {\em i.e.} for any  $(q,p) \in {{\cal M}}_{\hbox{\tiny{\sffamily DW}}} $  there exists  a unique element   $ (q,v) \in T{\cal Y} \otimes T^{\star}{\cal X}$ such that $(p,v)  {\pmb{\leftrightarrow}}  (q,p) $.    The   {DW}   Hamilton equations (or the generalized Hamilton equations, as termed in  \cite{HK-01,HK-02})  are to be thought of as  necessary and sufficient conditions on the map $ x \mapsto (q(x), p(x)) := (x^{\mu} , \varphi^{i}(x) , \varkappa(x) ,  {p}^\mu_i(x))$   such that there exist fields $x \mapsto  {\varphi}  (x)$  for which: 
 \begin{itemize}
\item The Legendre condition is satisfied for any $x \in {\cal X}$, $ (x, {{\varphi}} (x),  \hbox{d} {{\varphi}} (x))  {\pmb{\leftrightarrow}}   (q(x),p(x))$.
\item  The fields $x \mapsto {\varphi} (x)$ are solutions of the  Euler-Lagrange equations \eqref{eulerlagrange0}, which are related to   the Lagrangian density $L(x,\varphi (x),\dd \varphi(x)) $.
\end{itemize}
  Note that we can always
write  ${\cal H} (q,p) = {\cal H} (x^\mu, y^{i} ,\varkappa, {p^{\mu}_i}  ) = \varkappa + H(x^\mu, y^{i} , {p^{\mu}_i} ) =  \varkappa + H (q,p)$ and then work on the level set ${\cal H}^{-1}(0)$. The variable  $\varkappa = p_{1\cdots n} $ is seen as the canonical variable conjugate to the volume form ${\beta}$, see \cite{HK-02,HK-03}. If we fix $ {\cal H} (q,p) = 0$, then $\varkappa = - H(q,p)$.        In this case, the pre-multisymplectic $(n+1)$-form $ {\pmb{\omega}}^\circ :=    {\pmb{\omega}}^{\tiny{\hbox{\sffamily DW}}}      |_{ {\cal H} = 0 }   $  is~
 \bee\label{premultiform1}
  {\pmb{\omega}}^\circ  =    \hbox{d}p^\mu_i\wedge \hbox{d}z^i\wedge {\beta}_\mu - \hbox{d}H\wedge {\beta},
\eee
the exterior derivative of the   Poincaré-Cartan $n$-form   $\theta^{\hbox{\sffamily\tiny PC}} :=  p^\mu_i  \hbox{d}z^i\wedge {\beta}_\mu - H  {\beta} $, see  Gotay \cite{Gotay223,Gotay224,Gotayext}, 
the    analogue   of the Poincaré-Cartan form of mechanics   
 in the multisymplectic context.  

We denote by  ${{\cal C}}_\circ$  the level set ${\cal H}^{-1} (0) = {\big{\{}}  (q, p) \in  {{{\cal M}}}  = \Lambda^{n}_{\mathfrak{1}} T^\star {\pmb{\mathfrak{Z}}}   /  {\cal H} (q,p) = 0  {\big{\}}}$. The triple   ${{(}} {{\cal C}}_\circ :=   {\cal H}^{-1} (0)  ,  {{ {\pmb{\omega}}}}     |_{ {{\cal C}}_\circ}   , \vol     |_{ {{\cal C}}_\circ}  {{)}}$  is a $n$-phase space, where $  \vol     |_{ {{\cal C}}_\circ} $  is a      nowhere vanishing volume $n$-form, and ${{ {\pmb{\omega}}}}     |_{ {{\cal C}}_\circ}    $ is a  closed $(n+1)$-form,      see Kijowski and Szczyrba    \cite{JK-01,KS0,KS1,KS2} and Hélein \cite{H-02}.    We consider the $n$-dimensional submanifold ${{\Gamma}} \subset {{\cal M}}_{\hbox{\tiny{\sffamily DW}}}$, {\em i.e.} the Hamiltonian $n$-curve defined by
$
{{\Gamma}} = {{\{}}  (x^\mu,y^i,{{{p}}^\mu_i}) \ /  \  y^i = {\varphi}^i(x) \ , \ {p^\mu_i}= {\partial L}/{\partial v^i_\mu}(x^\mu,{\varphi}^i(x),{\partial_\mu{\varphi}^i}(x))   {{\}}}
$.
Then, on the level set $ {{\cal C}}_\circ  \subset {{\cal M}}_{\hbox{\tiny{\sffamily DW}}}$,  the  DW  system   is written in   geometric form as   
    \begin{equation}\label{geo-pre-ham-equ}
 \left.
\begin{array}{cll}
\displaystyle  \forall m \in {{\Gamma}},  &      \forall X \in \Lambda^{n}T_m {{\Gamma}},         & \displaystyle 
  X \iN  {\pmb{\omega}}^\circ  = 0 \quad \hbox{and} \quad  \exists X \in \Lambda^{n}T_m {{\Gamma}},    \quad   
 X \iN \beta_m \neq 0.
   \end{array}
\right.
 \end{equation}}
We refer to   section \ref{sec:section17} for more details on the pre-multisymplectic scenario, where we reproduce  the DW Hamilton system of equations, which in turn is equivalent to the Einstein system.
 
\subsection{{{First order Palatini  formulation of vielbein gravity}}}
 
Dynamics  of General Relativity ({{GR}})  is described by the Einstein's equations. They are obtained from the  Einstein-Hilbert action functional   
 \begin{equation}\label{EH} 
{{{\cal S}}}_{\tiny{\hbox{\sffamily EH}}}[{\textsf{g}}_{\mu \nu}] =  {\kappa}  \int_{{\cal{X}}} { { {{L}}}}_{\tiny{\hbox{\sffamily EH}}} [{\textsf{g}}_{\mu\nu}] \vol =   {\kappa}   \int_{{\cal{X}}} {R} \sqrt{-{\textsf{g}}} \vol ,
\end{equation} 
where $\kappa := (16 \pi G)^{-1}$.   The Einstein-Hilbert Lagrangian density is ${{{L}}}_{\tiny{\hbox{\sffamily EH}}}[{\textsf{g}}_{\mu \nu}] $.
The functional  \eqref{EH}  depends on the metric ${\textsf{g}}_{\mu \nu}$ and its first and second derivatives. In this approach  the metric is the dynamical variable and it
satisfies the Euler-Lagrange equations. The fundamental objects:  the Levi-Civita connection $ \Gamma^\rho_{\mu \nu} $ and the curvature tensor ${{R}^\rho}_{\mu \nu \sigma}$, are
expressed via the metric ant its derivatives. In such a framework,   {{GR}} is described  as a {\em metric} theory. The variational principle is applied to the functional ${{{\cal S}}}_{\tiny{\hbox{\sffamily EH}}}[{\textsf{g}}_{\mu \nu}]$. Variations with respect to the metric ${\textsf{g}}_{\mu \nu}$ lead to  the vacuum Einstein field equations 
\begin{equation}\label{EE}
{\dttG}_{\mu\nu} = {R}_{\mu\nu} - {(1/2) } {\textsf{g}}_{\mu\nu} {R} = 0.
\end{equation}
Classical {{GR}} can be also formulated in terms of the {\em vierbein}  $  {e}_\mu^I   $,  or {\em vielbein} in the $n$-dimensional case,   and the spin connection $ \omega_\mu^{IJ} $, see section \ref{subsec:dynamicalfields} for details.  The passage from {{GR}} seen as a {metric theory} to  the   first order Palatini action of vielbein gravity  is built, as emphasized in \cite{Rovelli001}, in two steps. The first step is  the Palatini first order theory.  We consider the metric ${\textsf{g}}$ and the connection $\Gamma$ as independent variables. We write 
\begin{equation}\label{PAPA}
{{{\cal S}}}_{\hbox{\tiny{\sffamily Palatini}}} [{\textsf{g}} , \Gamma] =  \kappa  \int_{{\cal{X}}}  \sqrt{-{\textsf{g}}}{\textsf{g}}^{\mu \nu}{R}_{\mu \nu}[\Gamma] \vol ,
\end{equation}
and we perform the variations of $  \Gamma$ and ${\textsf{g}}$ independently. The variations with respect to  the connection coefficients    set  the connection $\Gamma$ to be  the Levi-Civita affine connection,  while variations with respect to the metric yield the Einstein vacuum equations \eqref{EE}. The second step   concerns the use of the vierbein (tetrad) field.  The   Einstein-Palatini first order theory is given by the  action  
\begin{equation}\label{qmamqma012}
{{{\cal S}}}_{\hbox{\tiny{\sffamily Palatini}}} [e , \omega] = \frac{\kappa}{2} \int  \epsilon_{IJKL} e^I \wedge e^J \wedge {F}^{KL}, 
\end{equation}
 which  uses of two independent dynamical fields: the {\em co-frame field} $e^I$, or the {\em solder form},  and the {\em spin connection} $\omega^{IJ} $. We refer to   appendix  \ref{app:lagrang}  for details on the action functional \eqref{qmamqma012}.  Using this formulation the    Einstein's equations  \eqref{EE} are  equivalent to the Euler-Lagrange   system of equations 
\begin{equation}\label{qmamqma01}
  \left.
\begin{array}{rcc}
\displaystyle  \dd_{\omega} e^I = \dd e^I + {\omega^I}_J \wedge  e^J & = & 0 ,
\\ 
\displaystyle  \epsilon_{IJKL} e^J \wedge {F}^{KL} & = & 0 ,
\end{array}
\right.
\end{equation}
 see \cite{Baezbook,H-01}. We call  the action functional $ {{{\cal S}}}_{\hbox{\tiny{\sffamily Palatini}}} [e , \omega]  $  given in  \eqref{qmamqma012}   the first order Palatini action of   vielbein gravity.

     \subsection{Vielbein gravity: dynamical fields}\label{subsec:dynamicalfields} 
  
As emphasized in many papers,  {\em e.g.} \cite{Fatibene,H-01,kobayaya01,Kol01},  the concept of {\em orthonormal moving frame}, or vielbein,   
 is distinct  from   the concept  of the {\em solder form}. A {\em moving frame}   ${\bf e}_{\mu}(x)$,  or {\em rep\`ere mobile} of  Cartan \cite{cartan01,cartan02},    is  thought of as a section  ${\bf e}_{\mu}(x) : {{\cal{X}}} \rightarrow {L}  ({{\cal{X}}})$ of the linear frame bundle $ {L}  ({{\cal{X}}}) $. In the same way,     an orthonormal moving frame     ${\bf e}_{I}(x)$  is   a section of the Lorentz frame bundle ${L}_{\tiny{{SO}}(1,3)} ({{\cal{X}}})$.  
We denote  a  local frame as $\{ {\bf e}_{\mu}^{(\alpha)} \}$ defined on an open subset ${\cal U}_{(\alpha)} \subset {\cal X}$, where the index $(\alpha)$ is related to a choice of trivialization.    If the space-time manifold is parallelizable, the local nature of the moving frame     extends to   a well-defined global object. The vielbein field is written as $ {\bf e}_{I} = e^{\mu}_{I} (x) \partial_{\mu}$ and is related to the metric by the formula ${\textsf{g}}_{\mu\nu} = e^{I}_{\mu} e^{J}_{\nu} {\textsf{h}}_{IJ}$. Note that  the dual object is ${\bf e}^{I} = e^{I}_{\mu} (x) \dd x^{\mu}$. In the next section,   the solder form is given as a global section of the bundle ${\cal V} \otimes  T^{\star} {{\cal{X}}}$ over ${{\cal{X}}}$, see the right side of figure \ref{fig:solder}. The solder form    is canonically represented by  a family of local frames   $\{ {\bf e}_{\mu}^{(\alpha)} \}$  on  the space-time manifold and is  termed alternatively    the  vielbein field or {\em co-frame field}. {In the subsequent section, we offer some basic remarks about the interplay between the concept of  vielbein, {\em i.e} a   section of the orthonormal frame bundle,  as opposed to the one of   {\em solder form}, or {\guillemotleft {\em forme de soudure}\guillemotright}  \cite{Ehresmann}, and the related description of the co-frame field as a bundle isomorphism.}

\subsubsection{{{Co-frame field: the solder form}}}\label{subsubsec:solderr}

\begin{figure}   [h!]
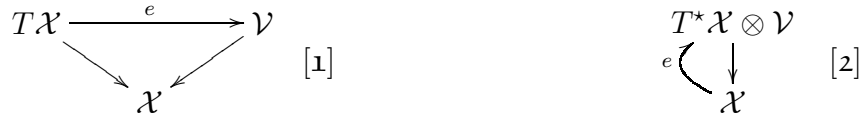
 
\centering
$$   \xygraph{
!{<0cm,0cm>;<1.5cm,0cm>:<0cm,0.7cm>::}
!{(-4,1.5) }*+{   T{{\cal{X}}}}="bioxx"
!{(-2,1.5) }*+{  {\cal V}    }="bioxx1"
!{(-3,0) }*+{  {{\cal{X}}} }="bioxx2"
!{(-1.5,0.75) }*+{ [\mathfrak{1}] }="oo"
  "bioxx":@{->}^{ e     } "bioxx1"
 "bioxx":@{->}_{ } "bioxx2"
  "bioxx1":@{->}_{   } "bioxx2"
}
\quad \quad 
\quad \quad \quad \quad 
\quad \quad \quad \quad   
 \xygraph{
!{<0cm,0cm>;<1.5cm,0cm>:<0cm,0.7cm>::}
!{(4,0) }*+{  {{\cal{X}}} }="bioxx2"
!{(4,1.5) }*+{    T^{\star} {{\cal{X}}} \otimes {\cal V} }="bioxx3"
!{(5.0,0.75) }*+{ [\mathfrak{2}] }="oo"
"bioxx3" :@{->}_{    } "bioxx2"
"bioxx2" :@/^{0.70cm}/^{ e } "bioxx3"
}
 $$
\caption{$[\mathfrak{1}]$ The solder form   the co-frame field as a bundle isomorphism.  $[\mathfrak{2}] $ Equivalently the co-frame field is pictured as a global section of  $ T^{\star} {{\cal{X}}} \otimes {\cal V} \longrightarrow {{\cal{X}}}$. Note that ${\cal V}$ does not need to be trivial.}
\label{fig:solder}
\end{figure}

 In the first order Palatini formulation of vielbein  gravity,  space-time is represented by an $n$-dimensional oriented manifold ${{\cal{X}}}$ which is not  equipped  
with a metric  {\em a priori}.   The metric is obtained via the pullback along the  co-frame field, or   solder form $e:T {{\cal{X}}} \rightarrow    {{\cal V}   }$.  Then, we work  in terms of   the bundle isomorphism $e:T {{\cal{X}}} \rightarrow    {\cal V}$ between the  not necessarily  trivial tangent bundle   $ T {{\cal{X}}}  \rightarrow  {{\cal{X}}}$ and   the  vector bundle 
$ {\cal V}  \rightarrow  {{\cal{X}}} $,  see  figure \ref{fig:solder}-$[\mathfrak{1}]$. The isomorphism $e$ is equivalently seen as a section of the vector bundle ${\cal V} \otimes  T^{\star} {{\cal{X}}} \rightarrow {{\cal{X}}}$ such that for any $x\in {{\cal{X}}}$, $e_{x} $ is an isomorphism,  see figure \ref{fig:solder}-$[\mathfrak{2}]$. Note that    $  {{\cal V}_{x}   }$ is   the {\em  internal} space.     The     notion of {solder form} was    introduced by Ehresmann in \cite{Ehresmann}, see also  \cite{kobayaya01,Kol01}.  
    As emphasized  in \cite{Baezbook,Wise}, the name {\em co-frame} is related  to  the case   the manifold is parallelizable, the tangent bundle is trivial, and the bundle isomorphism  $e:T {{\cal{X}}} \rightarrow    {{\cal V} = {{\cal{X}}} \times  \R^{1,3} }$   is equivalent to a choice of trivialization. In this context, the  solder  form   is identified locally, on any tangent space $T_x {{\cal{X}}}$, with the co-frame $e_{x} : T_{x} {{\cal{X}}} \rightarrow \Bbb{R}^{1,3} $.     
 
\subsubsection{The co-frame field: covariant exterior derivative}\label{ecddsd}
 
In this section, we consider the  solder form  $e \in \Omega^1({{{\cal{X}}}}, {\cal{V}}) = {\Gamma} ({\cal V}) \otimes \Omega^1 ({{\cal{X}}})$ previously introduced in section \ref{subsubsec:solderr}.  Let ${\textswab{e}}_{I}$ be a frame on the vector space ${\cal V}_{x} := \Bbb{R}^{1,3}$,   the Minkowski   space. Let    ${{\bf e}}^{\mu}$ be a moving co-frame, locally  defined  on  $\Omega^1({{\cal{X}}})$ (on an open subset ${\cal U}_{(\alpha)}  \subset {{\cal{X}}}  $).  Locally, for $x \in {\cal U}_{(\alpha)} $, we write  $e = e^{I}_{\mu} {{\bf e}}^{\mu}  \otimes  {{\textswab{e}}}_{I}  = {{\textswab{e}}}_{I} e^{I}_{\mu} {{{\bf e}}}^{\mu} =  {\textswab{e}}_{I} e^{I} $, {\em i.e.} $e$ is decomposed with respect to the basis $ {\textswab{e}}_{I} \otimes {\bf e}^{\mu}$  without  any reference to space-time indices.   We use the covariant derivative  ${\dttD}: \Gamma ({\cal V}) \longrightarrow \Omega^{1} ({{\cal{X}}} , {\cal V})  = \Gamma ({\cal V})  \otimes \Omega^{1}({{\cal{X}}})
$. Let $\sigma$ be a section of the vector bundle ${\cal V}\rightarrow {{{\cal{X}}}} $ so that ${\dttD} \sigma$ is a section $1$-form, ${\dttD} \sigma \in   \Gamma({\cal{V}}) \otimes \Omega^{n} ({{\cal{X}}}) = \Omega^{n} ({{\cal{X}}} , {\cal V})   $. By   means of the covariant exterior derivative defined for any $\lambda = (1/n!) \lambda^I_{\mu_1 \cdots \mu_n }   {{\bf e}}^{\mu_1}  \wedge \cdots \wedge {{\bf e}}^{\mu_n} \otimes  {{\textswab{e}}}_{I}   \in  \Omega^{n} ({{\cal{X}}} , {\cal V}) $ by 
 \bee\label{extcovariant}
\left.
\begin{array}{rcl}
 \displaystyle   \dd_{\omega}  :   \Omega^{n} ({{\cal{X}}} , {\cal V})   & \longrightarrow &  \Omega^{n+1} ({{\cal{X}}} , {\cal V})  ,
  \\
\displaystyle  \lambda   & \longmapsto &  \displaystyle   \dd_{\omega} \lambda.
\end{array}
\right.
\eee
We obtain the expression  of  $  \dd_{\omega}   e  \in \Gamma( {\cal V}) \otimes \Omega^2({{\cal{X}}}) $, {\em i.e.}
  \bee\label{ex-cov-deriv-2}
\left.
\begin{array}{rcl}
 \displaystyle    \dd_{\omega}   e   & = &  \displaystyle     \dd_{\omega}  {\big{(}}  {\textswab{e}}_{I} e^{I}_{\mu}   {\big{)}}  \wedge {{\bf e}}^{\mu} +{\textswab{e}}_{I}  e^{I}_{\mu} \dd {{\bf e}}^{\mu}   
=
    ({\dttD}{\textswab{e}}_{I} )e^{I}_{\mu} \wedge {{\bf e}}^{\mu}  + {\textswab{e}}_{I} \dd e^{I}_{\mu}  \wedge {{\bf e}}^{\mu}  + {\textswab{e}}_{I}  e^{I}_{\mu} \dd {{\bf e}}^{\mu}  ,
  \\
\displaystyle     & = &  \displaystyle    \omega_{\nu I}^{J} {\textswab{e}}_{J} {{\bf e}}^{\nu} e^{I}_{\mu} \wedge {{\bf e}}^{\mu}   + {\textswab{e}}_{I} \partial_\nu e^{I}_{\mu} {{\bf e}}^{\mu}  \wedge {{\bf e}}^{\nu} + {\textswab{e}}_{I}  e^{I}_{\mu} \dd {{\bf e}}^{\mu},
\end{array}
\right.
\eee
 where we have used in \eqref{ex-cov-deriv-2} the formula     $ {\dttD}{\textswab{e}}_{I}= \omega_{\nu I}^{J} {\textswab{e}}_{J} e^{\nu}$ as well as $  \dd e^{I}_{\mu} = \partial_\nu e^{I}_{\mu} e^{\mu} $. 
 We refer to the       section   \ref{sec:curvaexteriorcov} for  details on the connection $\omega_{\mu}^{IJ}$.  For a non  integrable moving  co-frame we obtain  $\dd {{\bf e}}^{\mu} = - {1}/{2} \mathfrak{c}^{\mu}_{\rho\nu} {{\bf e}}^{\rho} \wedge {{\bf e}}^{\nu}$.
Hence, in this case 
$
 \dd_{\omega}   e = {\textswab{e}}_{J}  {\big{(}}  \partial_\nu e^{J}_{\mu} +  \omega_{\nu I}^{J}  e^{I}_{\mu}  -  {1}/{2}  e^{J}_{\rho} \mathfrak{c}^{\rho}_{\nu\mu}    {\big{)}} {{\bf e}}^{\nu}  \wedge {{\bf e}}^{\mu} 
$.  For an  integrable moving co-frame  $ {{\bf e}}^{\mu} = \dd x^{\mu}$, we have  $\dd {{\bf e}}^{\mu} = 0$, and   we obtain  
\bee\label{kiuza02}
 \dd_{\omega}   e = {\textswab{e}}_{J} {\big{(}}  \partial_\nu e^{J}_{\mu} +  \omega_{\nu I}^{J}  e^{I}_{\mu}    {\big{)}} \dd x^{\nu}  \wedge \dd x^{\mu} \in\Gamma( {\cal V}) \otimes \Omega^1({{\cal{X}}}).
\eee
  Now we   write the object $  \dd_{\omega}  e$ decomposed  with respect to a basis of $ \Gamma( {\cal V}) \otimes \Omega^2({{\cal{X}}})$, {\em i.e.} ${\textswab{e}}_{I} \otimes {\bf e}^{\mu} \wedge   {\bf e}^{\nu}$.  Hence $ \dd_{\omega}  e$ is written as 
$ \dd_{\omega}  e  = (1/2) \left( \dd_{\omega}  e  \right)^{I}_{\mu\nu}  {\textswab{e}}_{I} \otimes {\bf e}^{\mu} \wedge   {\bf e}^{\nu} 
$.
The covariant exterior derivative $ \dd_{\omega}$ and  the gauge covariant derivative ${\cal D}$  are    related by $
 \dd_{\omega}   e = {\textswab{e}}_{I} {\cal D} e^{I}
$,  where $
 {\cal D} e^{I} = \hbox{d} e^{I} + \omega^{I}_{J} \wedge e^{J}$.  Since  ${\omega^{I}}_{J} = \omega^{I}_{\mu J} \dd x^{\mu}$ and $ \hbox{d}  e^{I} =  \hbox{d}  (e^{I}_{\mu} \dd x^{\mu}) =  \hbox{d}  e^{I}_{\mu} \wedge \dd x^{\mu}$ we obtain 
$     {\cal D} e^{I}      = \big{(} \partial_{\mu} e^{I}_{\nu}   +\omega^{I}_{\mu J}  e^{J}_\nu \big{)}   \dd x^{\mu} \wedge \dd x^{\nu}$.

\subsubsection{{{The Lorentz spin  connection}}}\label{Lorentzconnection}

Let $({{{\cal{P}}}} ,  {{\cal{X}}} , \pi , {{SO}}(1,3) )$ be a principal fiber bundle with a gauge group $ {{SO}}(1,3) $. We denote  by $\textswab{g} $ the $ \mathfrak{so}(1,3)$-Lie algebra. Equivalently, ${{{\cal{P}}}}$ is thought to be the total space of the ${\textsf{h}}$-orthonormal frame bundle over the space-time manifold.  Here, ${\textsf{h}}$ is the Minkowski metric.  We consider  an   Ehresmann   connection  on  $\PB$   {\em i.e.}  a smooth distribution of horizontal subspaces, see \cite{Ehresmann}, along with an equivariance property. In a given {trivialization}  we obtain  from the connection $1$-form  ${{{\pmb{\omega}}}} \in \Omega^1( {{{\cal{P}}}} ,  \textswab{g})$  on ${{{\cal{P}}}}$     the  {local   connection form}   a $1$-form $\omega \in \Omega^1( {{\cal{X}}} ,  \textswab{g})$ on   ${{\cal{X}}}$. Note that  the  local   connection form or     {\em gauge potential} is   the pull back of the connection form ${{{\pmb{\omega}}}}$ by a section $\sigma^{(\alpha)} : {\cal U}_{\alpha} \subset {{\cal{X}}} \rightarrow {{{\cal{P}}}}$  - and denoted as $ \omega = (\sigma^{(\alpha)})^\star( {{{\pmb{\omega}}}} ) \in T^\star{{\cal{X}}} \otimes {\textswab{g}} $.  The local connection form is only described {\em in} the local trivialization $\sigma^{(\alpha)}$ and therefore is a notion that depends on the choice of  trivialization.    In the context of vierbein gravity, the  Lorentz spin connection   is written as $\omega = \omega_{\mu} \dd x^{\mu} =   {{\textswab{b}}}_{i}   \omega_{\mu}^{{i}} \dd x^{\mu} = \omega_{\mu}^{{i}} \dd x^{\mu}  \otimes {{\textswab{b}}}_{i}$, where $({{\textswab{b}}}_{\mathfrak{1}} , \cdots , {{\textswab{b}}}_{\mathfrak{6}})$  is a basis of $\textswab{g} $. {Note that in the formulation of dreibein gravity,  the basis  of the $\mathfrak{so}(1,2)$-Lie algebra is denoted   $({{\textswab{b}}}_{\mathfrak{1}}^{(1,2)} , \cdots , {{\textswab{b}}}_{\mathfrak{\mathfrak{3}}}^{(1,2)})$.   We  induce a  connection on   associated bundles $ {{\PB}}   \times_{\rho } {\cal V}  $ via a representation $\rho$ of the    ${SO}(1,3)$ group,  see \cite{kobayaya01,Kol01}.  The image $\rho (\omega)$ of the gauge potential $\omega$ via the representation $\rho$ gives the matrix connection  $\rho(\omega) = \rho (  {\mathfrak{b}}_{i}   \omega_{\mu}^{{i}} \dd  x^{\mu}) =  \omega^{{i}}_{\mu} \rho ( {\mathfrak{b}}_{i} ) \dd  x^{\mu} =   \omega^{{i}}_{\mu}  (\rho(\mathfrak{b}_{i}))^{I}_{J} \dd x^{\mu} $. We denote $\rho ({\textswab{b}}_{i}) := \Delta_{i} = (\Delta_{i})^{I}_{J}$, where $0 \leq I, J \leq 3$ are Lorentz Lie algebra indices.   Working in a given representation, we     simply denote   the matrix elements by $\omega^{I}_{J} = \omega^{I}_{\mu J} \dd x^{\mu}$ with $ \omega^{I}_{\mu J}  = \omega^{i}_{\mu} ({{\Delta} }_{i})^{I}_{J}$. Alternatively,  in  section  \ref{sec:feozro9132}     the Lorentz spin connection is constructed on  the {vector}   bundle ${\cal V}$.  
         
       \subsubsection{{{Lorentz spin  connection: curvature and covariant exterior derivative}}}\label{sec:curvaexteriorcov}
  
The curvature $ \displaystyle  {{F}}^{{{\pmb{\omega}}}}     \in \Omega^{2}(\PB,\textswab{g}) $  of the connection   ${{{\pmb{\omega}}}} \in \Omega^1( {{{\cal{P}}}} ,  \textswab{g}) $  is written as $ \displaystyle  {{F}}^{{{\pmb{\omega}}}}   := {\dd}_{{{{\pmb{\omega}}}}} {{{\pmb{\omega}}}}  = \dd {{{\pmb{\omega}}}} + ({1}/{2}) {\pmb{[}}  {{{\pmb{\omega}}}}  ,  {{{\pmb{\omega}}}}   {\pmb{]}}  $,  where  for any ${\pmb{\lambda}} = (1/n!) {\pmb{\lambda}}^I_{\mu_1 \cdots \mu_n }   {{\bf e}}^{\mu_1}  \wedge \cdots \wedge {{\bf e}}^{\mu_n} \otimes  {{\textswab{b}}}_{I}   \in   \Omega^{n} ( \PB  , {\textswab{g}}) $,
 \bee\label{extcovariant2}
\left.
\begin{array}{rcl}
 \displaystyle   {\dd}_{{{{\pmb{\omega}}}}}   :      \Omega^{n} ( \PB  , {\textswab{g}})   & \longrightarrow &   \Omega^{n+1} ( \PB  , {\textswab{g}})   ,
  \\
\displaystyle  {\pmb{\lambda}}   & \longmapsto &  \displaystyle   \dd_{\pmb{\omega}} {\pmb{\lambda}},
\end{array}
\right.
\eee
is the  covariant exterior derivative relative to   ${{{\pmb{\omega}}}} $.  The   pullback by a section $ \sigma^{(\alpha)}$   gives the local expression of the connection  form        $ \omega =  (\sigma^{(\alpha)})^\star ( {{{\pmb{\omega}}}} ) \in T^\star{{\cal{X}}} \otimes {\textswab{g}} $  and  the curvature $2$-form  $  {F}^{{{{\pmb{\omega}}}}}   \in \Omega^{2} ({\PB} , \textswab{g})$, {\em i.e}   ${F}^{\omega} = (\sigma^{(\alpha)})^\star {F}^{{{\pmb{\omega}}}} \in  \Omega^{2} ({{\cal{X}}} , \textswab{g})$.  The  Lie algebra-valued $2$-form on space-time  ${F}^{\omega}_{\mu\nu} = {F}_{\mu\nu} $  is written  as $ {F} =  (1/2) {F}^{i}_{\mu\nu} {\textswab{b}}_{i} \otimes \dd x^{\mu} \wedge \dd x^{\nu}$.  The curvature $2$-form on the associated bundle is    $  \rho ({F}) \in \Omega^{2} ({{\cal{X}}} , {\hbox{\sffamily End}} ({\cal V}))$. In that case  $\displaystyle {F} = ({1}/{2}) {F}_{\mu\nu} \dd x^{\mu} \wedge \dd x^{\nu}$, where ${F}_{\mu\nu} = {F}_{\mu\nu}^{i} \Delta_{i} $. We denote ${F}^{I}_{\mu\nu J}  := {F}^{i}_{\mu\nu} (\Delta_{i})^{I}_{J}$.  The curvature  of the   spin connection  $\omega_\mu^{IJ}$   is written as   \cite{Baezbook,Rovelli001} 
 \bee\label{fowpq0923}
 \displaystyle {F}_{\mu\nu}^{IJ} [\omega]    =  2    \partial_{[\mu} \omega_{\nu]}^{IJ} + [\omega_\mu , \omega_\nu]^{IJ}
=
\partial_\mu \omega_{\nu}^{IJ} - \partial_\nu \omega_{\mu}^{IJ}  + \omega^{I} _{\mu K} \omega_{\nu}^{KJ}  -  \omega^{I} _{\nu K} \omega_{\mu}^{KJ} .
\eee
Note that the  covariant exterior derivative  $
 \dd_{\omega}   \omega =  {{\textswab{b}}}_{i}{\cal D} \omega^{i} $, or equivalently  $  \dd_{\omega}   \omega = {\Delta}_{IJ} {\cal D} \omega^{ IJ} 
 $, is given by   means of the object  
\begin{equation}\label{cneozcnzjcwqxs}
\left.
\begin{array}{ll}
  \displaystyle {\cal D} \omega^{IJ}    & =      \dd {\omega}^{IJ}  + {\omega}^{I}_K \wedge {\omega}^{KJ}  + {\omega}^{J}_K \wedge {\omega}^{IK}
=
 \dd {\omega}^{IJ}  + {\omega}^{I}_K \wedge {\omega}^{KJ}  - {\omega}^{J}_K \wedge {\omega}^{KI} ,
    \end{array}
\right.
\end{equation}
written in components $\left({\cal D}  \mathbf{\omega}\right)^{IJ}_{\mu \nu}        =     
2 \partial_{[\mu} \omega^{IJ}_{\nu]}  +  2 {\omega_{[\mu}}^I_K \omega^{KJ}_{\nu ]} - 2 {\omega_{[\mu}}^J_K \omega^{KI}_{\nu ]}   $. The variation $\delta {F}^{IJ}_{\mu\nu}$ of the curvature of the Lorentz spin connection is expressed via the covariant exterior derivative  $\delta {F}^{IJ}_{\mu\nu} = 2{\cal D}_{[\mu} \delta\omega_{\nu]}^{IJ}$, see {\em e.g.} \cite{Baezbook,Rovelli001}.

 \subsubsection{{{The pullbacks  $ {\textsf{g}} = e^{\star} {\textsf{h}}$ and $ \nabla = {e}^\star {\dttD}$}}}\label{sec:feozro9132}

    If we have a metric ${\textsf{h}}$ on ${\cal V}$, then we obtain a metric on ${{\cal{X}}}$ by pullback  ${\textsf{g}}  =  {e} ^\star {\textsf{h}}$, where  $\forall x \in {{\cal{X}}}, \forall \xi,  \zeta \in T_x{{\cal{X}}}$:  $( {e} ^\star {\textsf{h}})_x(\xi, \zeta) = {\textsf{g}}_x ( {e}_x(\xi), {e}_x(\zeta) )$. In this case, the vector space ${\cal V}$ is equipped  with a connection ${\dttD}$,  so that we obtain the   connection $\nabla = {e}^\star {\dttD}$ on $T{{\cal{X}}}$   described as follows:  $\forall \xi, \sigma \in {\Gamma}({{\cal{X}}}) = T{{\cal{X}}}, \ \nabla_\xi \sigma = {e}^\star ( {\dttD}_\xi  {e}(\sigma))$, where ${\dttD}: {\Gamma} ({{\cal{X}}}) \times \Gamma ({{\cal{X}}} , {\cal V}) \longrightarrow \Gamma ({{\cal{X}}}, {\cal V}): (X,\sigma) \mapsto {\dttD}_{X} \sigma $.    The set of 1-forms $\omega^J_I$ defined on an open subset ${\cal U}_{(\alpha)} \subset {{\cal{X}}}$  by $\omega^J_I =  \omega^J_{\mu I} \dd x^{\mu}$   gives, for any $ \xi \in {{{\mathfrak{X}}}} ({\cal U}_{(\alpha)} )$, $ \xi^\mu \omega^J_{\mu I}  = \omega^J_I (\xi)$. Then  
$
{\dttD}_\xi  s = {\dttD}_\xi (\sigma^I e_I) = \dd s^I (\xi) e_I +  \omega^J_{I} (\xi)  \sigma^I  {\textswab{e}}_{J}  
$. We have $({\dttD}_\mu \sigma)^I = \partial_\mu \sigma^I +  \omega^I_{\mu J} \sigma^J$. Now, using the solder form  we   obtain a connection on $T{{\cal{X}}}$. Pulling back the connection on ${\cal V}$ via $\nabla_\xi \sigma = {e}^\star ( {\dttD}_\xi  {e}(\sigma)) $, we get the covariant derivative's components:
\bee
(\nabla_\mu \xi)^\nu = \partial_\mu \xi^\nu + (e^\nu_I \partial_\mu e^I_\rho + e^\nu_I \omega^I_{\mu J} e^J_\rho ) \xi^\rho .
\eee
However, we have also
$
(\nabla_\mu \xi)^\nu  = \partial_\mu \xi^\nu + \Gamma^\nu_{\mu \rho} \xi^\rho
$. Therefore, 
 $
\Gamma^\nu_{\mu \rho} =  e^\nu_I \partial_\mu e^I_\rho + e^\nu_I \omega^I_{\mu J} e^J_\rho  
$ 
and  we reproduce the well known relation between the spin connection coefficients and the Christoffel symbol $\Gamma_{\mu\nu}^{\rho} $: $  \partial_\mu e^I_\nu + e^K_\nu \omega^I_{\mu K}  - \Gamma_{\mu\nu}^{\rho} e^I_\rho  = 0 $. We summarize the two pull-backs of interest:
\begin{center}
 $\xygraph{
!{<0cm,0cm>;<4.0cm,0cm>:<0cm,1.0cm>::}
!{(2,3.0) }*+{  {\textsf{h}} \  \hbox{on} \ {\cal V}  ,}="bbbb"
!{(0,3.0) }*+{  {\textsf{g}} = e^{\star} {\textsf{h}}  \ \hbox{on}  \ T{{\cal{X}}}     }="cccc"
!{(0,2) }*+{  \nabla = {e}^\star {\dttD} \ \hbox{on}  \ T{{\cal{X}}}    }="aqq1"
!{(2,2) }*+{   {\dttD} \  \hbox{on} \ {\cal V}, }="aqq"
"aqq":@{->}_{ \displaystyle  \quad \quad \nabla_\xi \sigma = {e}^\star ( {\dttD}_\xi  {e}(\sigma))   } "aqq1"
 "bbbb":@{->}_{  \displaystyle  \ \quad \quad   {\textsf{g}}_m ( {e}_m(\xi), {e}_m(\sigma) ) = ( {e} ^\star {\textsf{h}})_m(\xi, \sigma)   } "cccc"
}$
\end{center}
 which are related to the metric   and to the spin connection, respectively.  The bundle isomorphism  gives a correspondence between objects on the tangent bundle $T{{\cal{X}}}$ and the internal bundle ${\cal V}$. The curvature of the connection ${\dttD}$ is the    $2$-form   ${F}^{IJ} = \dd \omega^{IJ} + \omega^{I}_{K} \wedge \omega^{KJ}$, written in components as    ${F}^{IJ}_{\mu\nu} = \omega^{IJ}_{[\mu,\nu]} + \omega^{I}_{[\mu K} \omega^{KJ}_{\nu]}$. The bundle isomorphism $e$ maps the curvature of ${\dttD}$ to that of ${\nabla}$ with the relation ${R}_{\mu\nu}^{\rho\sigma} = {F}^{IJ}_{\mu\nu} e^{\rho}_{I} e^{\sigma}_{J}$. Finally, we recall   the   expressions of the Ricci tensor ${{R}_{\mu}}^{\nu} = {F}^{IJ}_{\mu\sigma} e^{\sigma}_{I} e^{\nu}_{J} $ and the  scalar  curvature ${R} = {R}_{\mu}^{\mu} = e^{\sigma}_{I} e^{\rho}_{J} {F}^{IJ}_{\sigma\rho}$.

     \subsection{Configuration space}\label{subsec:confplace}

In section \ref{subsec:001}, we first  briefly present two  {\em fully}  covariant  formulations {\em i.e.} that does not rely on any choice  of trivialization of some principal  bundle.  Then, in section \ref{subsec:002}, we present the  less sophisticated  configuration space that we will use   in      sections  \ref{sec:section2} - \ref{sec:section18}. The latter being dependent of a given trivialization  of the principal bundle $({{{\cal{P}}}} ,  {{\cal{X}}} , \pi , {{SO}}(1,3) )$. 

\subsubsection{Fully covariant configuration space}\label{subsec:001}
      
We mention two formalisms to  take into account the viewpoint of the geometry of the principal bundle $({{{\cal{P}}}} ,  {{\cal{X}}} , \pi , {SO}(1,3) )$.    The first is related to the  Gauge Natural Bundle approach, see Nijenhuis \cite{Nij01},   Eck \cite{Ek-01},     Kol\'{a}\v{r},  Michor and   Slov\'{a}k \cite{Kol01},   Fatibene and  Francaviglia \cite{Fatibene}. We construct  the gauge natural     bundle ${\PB}_{\rho} := {\big{(}} 
 {{{\cal{P}}}} \times_{{\cal{X}}} {L}  ({{\cal{X}}})
 {\big{)}} \times  {GL}(n)  $ associated     to the ${SO}(1,3)$-principal bundle  ${{{\cal{P}}}}$, see \cite{Fatibene,Matteucci}. We denote by ${{{\cal{{\cal{Y}}}}}}^{\hbox{\sffamily\tiny GNB}}_{{\hbox{\sffamily\tiny purely-frame}}} :=  {{{\cal{P}}}}_{\rho}  $  the {covariant configuration space}  of  the purely-frame gravitational theory. In the frame-affine framework, {\em i.e.} based on the  Palatini action of vielbein gravity, the {covariant configuration space}     is ${{{\cal{{\cal{Y}}}}}}^{\hbox{\sffamily\tiny GNB}}_{{\hbox{\sffamily\tiny frame-affine}}} :=  {{{\cal{P}}}}_{\rho}  \times  {{{\cal{{\cal{Y}}}}}}_{{\cal P}}$, where $ {{{\cal{{\cal{Y}}}}}}_{{\cal P}}$ is the space of connection of the principal ${{SO}}(1,3) $-bundle.   
  This fruitful approach  has been used in the context of gravity and Einstein-Cartan gravity   by Fatibene and Francaviglia \cite{Fatibene,Fatibene1},  and Matteucci \cite{Matteucci}. Afterward, the gauge natural approach blends with the multisymplectic viewpoint   in the papers by  Bruno,  Cianci and  Vignolo \cite{bruno,bruno1}. We refer also to   \cite{bruno2,bruno3} for the similar treatment of the Yang-Mills fields.  In this framework, the gauge symmetry  is obtained  via some reduction of the geometry of connections on the  principal bundle.  
 
Another fully covariant multisymplectic formulation   for the Yang-Mills fields is given by  H\'elein    \cite{pataym}.   
We give  a brief idea of the corresponding    multimomentum phase space for     vielbein gravity, following this  line of thought. 
Let   $\textswab{p} :=  \mathfrak{iso}(1,3) =   {\mathfrak{so}(1,3)}  \ltimes \Bbb{R}^{1,3} $ be the Poincar\'e Lie algebra.  We consider a $\textswab{p}$-valued connection $1$-form ${\pmb{\eta}} \in \Omega^{1} ({{{\cal{P}}}} , \textswab{p} )$ defined on the   principal fiber bundle  $({{{\cal{P}}}} ,  {{\cal{X}}} , \pi ,  {{SO}}(1,3)    )$   which  satisfies some normalization and {equivariance} conditions.   The covariant configuration space is ${{{\cal{{\cal{Y}}}}}}^{\hbox{\tiny{\sffamily cov}}} :=  \textswab{p} \otimes T^{\star } {{{\cal{P}}}}  \rightarrow {{{\cal{P}}}}   $. The multimomentum phase space is  ${{\cal M}}_{\hbox{\tiny{\sffamily DW}}}^{\hbox{\tiny{\sffamily cov}}}  = \Lambda^{m}_{\mathfrak{1}} T^{\star} (\textswab{p} \otimes T^{\star } {{{\cal{P}}}}) $,   the  {\sffamily DW} multisymplectic manifold  fibered  over $ \textswab{p} \otimes T^{\star } {{{\cal{P}}}} $. We refer to \cite{pataym} for more details on  the dimension $m =  n + r$}, where $n = \hbox{dim}({\cal{X}})$, and $r = \hbox{dim}({\textswab{p}})$.

     \subsubsection{Trivialization dependent covariant configuration space}\label{subsec:002}

Any   connection  ${\dttD} $ on   the internal bundle ${\cal V}$ can be  written as ${\dttD} = {\dttD}^{\circ} + {\omega}$, where $\omega \in \Omega^{1} ({{\cal{X}}} , \hbox{\sffamily End}({\cal V}))$ is the {matrix connection} and         ${\dttD}^{\circ} : {\Gamma} ({{\cal{X}}}) \otimes \Gamma({\cal V}) \rightarrow \Gamma({\cal V})$ is the {standard flat connection}. Note that ${\dttD}^{\circ}_{\zeta} \sigma^{(\alpha)} = \zeta({\sigma}^{(\alpha)} ) = \zeta((\sigma^{\alpha})^{I}) {\textswab{e}}_{I}$ is  trivialization dependent.     
We restrict ourselves, as suggested in \cite{HK-02}, to this  local approach which depends on a particular choice of   trivialization  of the principal bundle $({{{\cal{P}}}} ,  {{\cal{X}}} , \pi , {{SO}}(1,3) )$.      The   {\em    covariant  configuration space} is   the bundle ${\cal Y} :=   \mathfrak{iso}(1,3)   \otimes T^{\star} {\cal X}  $  over ${\cal X}$.  Albeit non fully covariant from the viewpoint of the geometry of gauge fields,  we  nevertheless    use  this approach   in      sections  \ref{sec:section2} - \ref{sec:section18}.  
     
\section{{{DW formulation of vielbein gravity}}}\label{sec:section2}

In this section  we describe the  {\sffamily DW} Hamiltonian formulation of the first order Palatini action   of vielbein gravity.  First, let us begin with the notations and the geometrical background  related to  the    covariant configuration space  used in the paper. 
\subsection{{{Geometrical setting and notations}}}\label{subsec:2.2}

Two independent  dynamical fields are $e \in {\cal V} \otimes T^\star {{\cal{X}}}$ and $ \omega \in \mathfrak{so}(1,3) \otimes  T^\star{{\cal{X}}} :=  \textswab{g} \otimes  T^\star{{\cal{X}}}$.  The former is  the solder form (or co-frame field), locally  seen as a $\Bbb{R}^{(1,3)}$-valued $1$-form, whereas  the latter is    the Lorentzian  spin connection,      a $\textswab{g} $-valued  $1$-form.  Let  ${{\cal{{\cal{Y}}}}} = \textswab{p} \otimes   T^\star {{\cal{X}}}$ be the bundle of $\textswab{p}:=\mathfrak{iso}(1,3)$-valued $1$-forms over the space-time manifold ${\cal X}$, {\em i.e.}    the     covariant configuration space. A point in ${{\cal{{\cal{Y}}}}} = \textswab{p} \otimes   T^\star {{\cal{X}}}$  is denoted  as $(x,e_x,\omega_x)$, where $x \in {\cal X}$,  $e_x \in  {{\cal{{\cal{Y}}}}}^{e}_x := \Bbb{R}^{1,3} \otimes T_{x}^{\star} {\cal X}$ and $\omega_x  \in {{\cal{{\cal{Y}}}}}^{\omega}_{x} := \textswab{g}   \otimes T_{x}^{\star} {\cal X} $.    
     \begin{figure}   [h!] 
\centering
$$    
   \xygraph{
!{<0cm,0cm>;<2.0cm,0cm>:<0cm,0.7cm>::}
!{(2,2) }*+{  {{\cal{X}}} }="bioxx2"
!{(2.7,2.5) }*+{ [\mathfrak{1}] }="oo"
 !{(2,4) }*+{   {\cal Y}^e   =    {\Bbb{R}}^{1,3} \otimes  T^\star  {{\cal{X}}}    }="bioxx3"
 "bioxx3" :@{->}^{    \pi_{{\cal{X}}}^{e}  } "bioxx2"
  "bioxx2" :@/^{0.50cm}/^{ {{e}}   } "bioxx3"
 }
\quad \quad\quad 
 \xygraph{
!{<0cm,0cm>;<2.0cm,0cm>:<0cm,0.7cm>::}
!{(2,2) }*+{  {{\cal{X}}} }="bioxx2"
!{(2.7,2.5) }*+{ [\mathfrak{2}] }="oo"
 !{(2,4) }*+{   {\cal Y}^{\omega}  =    \mathfrak{so}(1,3) \otimes  T^\star  {{\cal{X}}}    }="bioxx3"
 "bioxx3" :@{->}^{    \pi_{{\cal{X}}}^{\omega}  } "bioxx2"
  "bioxx2" :@/^{0.50cm}/^{ {{\omega}}   } "bioxx3"
 }
  \quad \quad  \quad
  \xygraph{
!{<0cm,0cm>;<2.0cm,0cm>:<0cm,0.7cm>::}
!{(2,2) }*+{  {{\cal{X}}} }="bioxx2"
!{(2.7,2.5) }*+{ [\mathfrak{3}] }="oo"
!{(2,4) }*+{   {{\cal{{\cal{Y}}}}} = \mathfrak{iso}(1,3) \otimes   T^\star {{\cal{X}}}   }="bioxx3"
"bioxx3" :@{->}^{    \pi_{{\cal{X}}}  } "bioxx2"
"bioxx2" :@/^{0.50cm}/^{ {{(e,\omega)}}   } "bioxx3"
}
    $$
\caption{$[\mathfrak{1}]$ The fiber bundle ${\cal Y}^{e} :=  {\Bbb{R}}^{1,3} \otimes  T^\star  {{\cal{X}}}    $ over ${\cal X}$. $[\mathfrak{2}]$  The fiber bundle  ${\cal Y}^{\omega} :=       \mathfrak{so}(1,3) \otimes  T^\star  {{\cal{X}}}   $ over ${\cal X}$. $[\mathfrak{3}]$ The covariant configuration space is   the fiber bundle ${{{\cal{{\cal{Y}}}}}} := \textswab{p} \otimes T^\star {{\cal{X}}}  = \mathfrak{iso}(1,3) \otimes   T^\star {{\cal{X}}} $  over the space-time manifold  ${\cal X}$.}
\label{fig:conftrivial}
\end{figure}
Let us consider the   maps  $e: {{\cal{X}}} \rightarrow {{\cal{{\cal{Y}}}}}^e =  \Bbb{R}^{(1,3)} \otimes T^\star {{\cal{X}}}$ and
$\omega : {{\cal{X}}} \mapsto {{\cal{{\cal{Y}}}}}^{\omega} =  \textswab{g}  \otimes T^\star {{\cal{X}}}$ written as 
\begin{equation}\label{0.h}
    \left.
\begin{array}{ccl}
\displaystyle  {{\cal{X}}} & \rightarrow & \displaystyle {{\cal{{\cal{Y}}}}}^{e} =  {\Bbb{R}^{(1,3)}} \otimes T^\star {{\cal{X}}},
\\ 
\displaystyle x & \mapsto & \displaystyle ( x , e(x)  ) = ( x,   e_\mu^I(x) \dd x^{\mu} \otimes  {\textswab{e}}_{I} ),
\end{array}
\right.
\
\hbox{and} 
\
   \left.
\begin{array}{ccl}
\displaystyle  {{\cal{X}}} & \rightarrow & \displaystyle {{\cal{{\cal{Y}}}}}^{\omega} =  \textswab{g}  \otimes T^\star {{\cal{X}}},
\\ 
\displaystyle x & \mapsto & \displaystyle ( x ,  \omega(x) )  = ( x, \omega_\mu^{IJ}(x) \dd x^{\mu} \otimes   \Delta_{IJ} ).
\end{array}
\right.
\nonumber
\end{equation}
These maps are
equivalently thought of as sections of ${\cal Y}^{e}$ and ${\cal Y}^{\omega}$ (see figure \ref{fig:conftrivial}-$[\mathfrak{1}]$ and \ref{fig:conftrivial}-$[\mathfrak{2}]$, respectively). 
 We   introduce also the map $ (e,\omega) : {\cal X} \rightarrow {\cal Y} $, that is written as
\begin{equation}\label{0.h22}
    \left.
\begin{array}{ccl}
\displaystyle  {{\cal{X}}} & \rightarrow & \displaystyle {{\cal{{\cal{Y}}}}}  =  \textswab{p}   \otimes  T^\star {{\cal{X}}},
\\ 
\displaystyle x & \mapsto & \displaystyle (x,e(x),\omega(x) ) =  (x ,  e_\mu^I(x) \dd x^{\mu} \otimes  {\textswab{e}}_{I} , \omega_\mu^{IJ}(x) \dd x^{\mu} \otimes   \Delta_{IJ} ) .
\end{array}
\right.
  \end{equation}
    Any choice as $(e (x),\omega(x))$ is equivalent to the data of an $n$-dimensional submanifold of the fiber bundle ${{{\cal{{\cal{Y}}}}}}$  and  is equivalently thought   of as   a section ${{\sigma}}^{(\alpha)} : {\cal U}^{(\alpha)} \subset {\cal X}\rightarrow {\cal Y}$, where ${{\sigma}} : x \mapsto {{\sigma}} (x) = (x , e(x) , \omega(x))  $, see figure \ref{fig:conftrivial}-$[\mathfrak{3}]$. Finally, the set of  local coordinates  in the covariant configuration  bundle ${{\cal{{\cal{Y}}}}}$  is equivalently denoted as $(x^\mu,e^I_\mu, \omega^{IJ}_\mu)$.

\subsection{{{The bundle  ${{\cal P}} =  T {{\cal{{\cal{Y}}}}} \otimes_{{{\cal{{\cal{Y}}}}}}   T^\star {{\cal{X}}}$}}}\label{21}

For any point $x \in {\cal X}$  the differential  $(\dd e)_{x} : T_x{\cal X} \mapsto T_{(x,e_x)} {\cal Y}^{e}$ is seen as an element of $T^{\star}_x{\cal X} \otimes T_{(x,e_x)} (T^{\star} {\cal X} \otimes {\cal V})$  canonically identified with $T_x^{\star}{\cal X} \otimes T_x^{\star}{\cal X} \otimes {\cal V}$. Analogously,  $(\dd \omega)_{x} : T_x{\cal X} \mapsto T_{(x,\omega_x)} {\cal Y}^{\omega}$ is seen as an element of $T^{\star}_x{\cal X} \otimes T_{(x,\omega_x)} (T^{\star} {\cal X} \otimes \textswab{g} )$  canonically identified with $T_x^{\star}{\cal X} \otimes T_x^{\star}{\cal X} \otimes \textswab{g} $. Let us consider   the bundles ${\cal P}^{e} := e^\star T {{\cal{{\cal{Y}}}}}^{e} \otimes   T^\star {{\cal{X}}} $  and   ${{\cal P}}^{\omega} := \omega^\star T {{\cal{{\cal{Y}}}}}^{\omega} \otimes  T^\star {{\cal{X}}}$  over the space-time manifold ${{\cal{X}}}$. These bundles   enable us   to describe  the differentials $\hbox{d}e$ and $\hbox{d}\omega$  of the map $e$ and $\omega$ as sections of the bundles ${{\cal P}}^{e}$ and ${{\cal P}}^{\omega}$, respectively.  
In particular,  the points $(x,v^{e}) \in {\cal P}^{e}$ and  $(x,v^{\omega}) \in {\cal P}^{\omega}$   are described by 
\bee
 v^{e}  =   \sum_{ \mu,\nu }   \sum_{  I } v^{I}_{\mu\nu} \dd x^{\mu} \otimes   \dd x^{\nu}  \otimes {\textswab{e}}_I  ,
\quad \quad \quad \quad
  v^{\omega}  =  \sum_{ \mu,\nu }    \sum_{  I < J } v^{IJ}_{\mu\nu}  \dd x^{\mu}  \otimes   \dd x^{\nu} \otimes \Delta_{IJ} ,
\eee
where $ v^{I}_{\mu\nu} := \partial_{\mu} e^{I}_{\nu}$  and $ v^{IJ}_{\mu\nu} := \partial_{\mu} \omega^{IJ}_{\nu}$, respectively. Local coordinates  on ${{\cal P}}^{e}$ and ${{\cal P}}^{\omega}$ are  denoted  by $(x^{\mu},{v}^{I}_{\mu\nu} )$ and $(x^{\mu},{v}^{IJ}_{\mu\nu} )$, respectively. Using the map \eqref{0.h22}, we introduce the bundle $ {{\cal P}}^{(e,\omega)}  := ( e , \omega)^\star T {{\cal{{\cal{Y}}}}} \otimes  T^\star {{\cal{X}}} $ over ${\cal X}$. Note that  $  {{\cal P}}^{(e,\omega)}  \subset {{\cal P}} :=    T {{\cal{{\cal{Y}}}}} \otimes_{\cal Y}  T^\star {{\cal{X}}}   $. This bundle is the  bundle over $ {\cal Y} := \textswab{p} \otimes T^{\star}{\cal X}$, such that the fiber over $(x,e_x,\omega_x)$ is $T_{(x,e_x,\omega_x)}(\textswab{p} \otimes T^{\star}{\cal X}) \otimes T_x^{\star} {\cal X}  $. In terms of local coordinates:
\bee 
 {{\cal P}}   = \left\{ (x^{\mu},e^{I}_{\mu},\omega^{I}_{\mu} , v^{I}_{\mu\nu} , v^{IJ}_{\mu\nu})    \ / \ (x^{\mu},e^{I}_{\mu},\omega^{I}_{\mu}  ) \in {\cal Y}, \  (v^{I}_{\mu\nu} , v^{IJ}_{\mu\nu}  ) \in   T_{(x,e_x,\omega_x)}  {\cal Y} \otimes    T^\star_{x} {{{\cal{X}}}}  \right\}. 
 \eee 
  Subsequently,     the covariant exterior derivatives    $ \dd_{\omega}e$ and $ \dd_{\omega}\omega$   are described as sections of the bundle ${{\cal P}}$. 
Recall that
\begin{equation}\label{exteriorderiv}
\left.
\begin{array}{ccl}
\displaystyle  \dd_{\omega}e  & = &  (1/2) \left(   \dd_{\omega}  e \right)^{I}_{\mu\nu}  \dd x^{\mu} \wedge \dd x^{\nu} \otimes   {\textswab{e}}_{I} = (1/2)  \left( \partial_{\mu} e^I_{\nu} +  {\omega_{\mu}}^I_J e^{J}_{\nu}  \right)  \dd x^{\mu} \wedge \dd x^{\nu} \otimes   {\textswab{e}}_{I},
\\ 
\displaystyle  \dd_{\omega} \omega  & = &  (1/2)  \left(   \dd_{\omega}  \omega  \right)^{IJ}_{\mu\nu} \dd x^{\mu} \wedge \dd x^{\nu} \otimes  {\Delta}_{IJ} =   (1/2) \left(    \partial_{\mu} \omega^{IJ}_{\nu}  +  {\omega_{\mu}}^I_K \omega^{KJ}_{\nu} -{\omega_{\mu}}^J_K \omega^{KI}_{\nu}   \right)  \otimes  {\Delta}_{IJ} .
\end{array}
\right.
\nonumber
\end{equation}
   We now consider the fiber bundle of $n$-vector fields  $ \Lambda^n_{\textsc{n}} T  {{\cal{{\cal{Y}}}}}$ over  $ {{\cal{{{Y}}}}} $. For any $ (x,e_x,\omega_x)  \in    {{{{{\cal{Y}}}}}}$  the fiber $ \Lambda^n_{\textsc{n}} T_{(x,e_x,\omega_x)}( {\textswab{p}} \otimes T^\star{{\cal{X}}}) = \Lambda^n_{\textsc{n}} T_{(x,e_x,\omega_x)} {{\cal{Y}}} $ can be identified with $ {{\cal P}}_{(x,e_x,\omega_x)} $ via  the diffeomorphism: 
\begin{equation}\label{diff}
\left.
\begin{array}{ccl}
\displaystyle T_{(x,e_x,\omega_x)}(\textswab{p} \otimes T^{\star}{\cal X}) \otimes T_x^{\star} {\cal X}    & \rightarrow & \displaystyle \Lambda^n_{\textsc{n}} T_{(x,e_x,\omega_x)}  ( {\textswab{p}}\otimes T^\star {{\cal{X}}}) ,
\\ 
\displaystyle  (\sum_{\mu, \nu} \sum_{I} \left(   \dd_{\omega}  e  \right)^{I}_{\mu\nu}  \dd x^{\mu} \otimes \dd x^{\nu} \otimes {\textswab{e}}_I ,  \sum_{\mu, \nu} \sum_{I<J}  \left(   \dd_{\omega}  \omega  \right)^{IJ}_{\mu\nu}   \dd x^{\mu} \otimes \dd x^{\nu} \otimes  \Delta_{IJ} )& \mapsto & \displaystyle  z = z_1 \wedge ... \wedge z_n ,
\end{array}
\right.
\nonumber
\end{equation}
where for any $1 \leq \alpha \leq n $, $\displaystyle  z_\alpha =   \frac{\partial}{\partial x^\alpha} + \sum_{1 \leq \beta \leq n}   z_{\alpha\beta}^{I}     \frac{\partial}{\partial e_\beta^I}  + \sum_{1 \leq \beta \leq n} z_{\alpha\beta}^{IJ} \frac{\partial}{\partial \omega_{\beta}^{IJ}} $, 
\bee\label{nvectorzz}
\left.
\begin{array}{ccl}
\displaystyle z_{\alpha\beta}^{I} & := & \displaystyle  \partial_{\alpha} e^I_{\beta} +  {\omega_{\alpha}}^I_J e^{J}_{\beta} ,
\\ 
\displaystyle  z_{\alpha\beta}^{IJ}  & := & \displaystyle   \partial_{\alpha} \omega^{IJ}_{\beta} +  {\omega_{\alpha}}^I_K  {\omega_{\beta}}^{KJ} - {\omega_{\alpha}}^J_K  {\omega_{\beta}}^{KI} .
\end{array}
\right.
\eee

\begin{figure}   [h!]
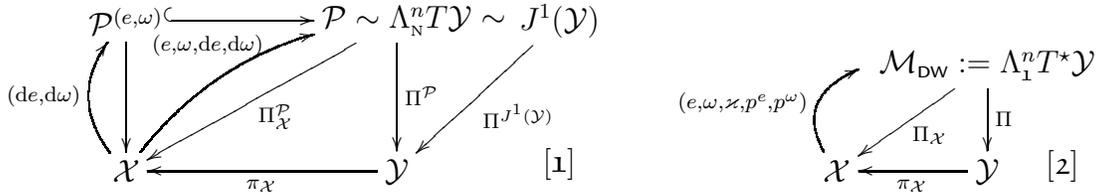
 
\centering
$$ 
 \xygraph{
!{<0cm,0cm>;<3.6cm,0cm>:<0cm,1.0cm>::}
!{(2,2) }*+{    {{\cal{{\cal{Y}}}}}   }="bioxx2"
 !{(1,2) }*+{  {\cal X} }="bioxx29"
 !{(1.6,3.6) }*+{     }="bioxx333"
 !{(2.6,2.1) }*+{ [\mathfrak{1}] }="oo"
!{(2,4) }*+{ {{\cal P}}  \sim \Lambda^n_{\textsc{n}} T  {\cal Y}   }="bioxx3"
!{(2.6,4) }*+{  \dttJ^{1} (  {{\cal{{\cal{Y}}}}}  ) }="bioxx2azer"
!{(2.3,4) }*+{ \quad \sim }="bioxx2azer22"
!{(1,4) }*+{   {{\cal P}}^{(e,\omega)} }="ooo"
"bioxx3" :@{->}^{     \Pi^{{{\cal P}}}   } "bioxx2"
"bioxx2azer" :@{->}^{  \Pi^{  \tiny{ \dttJ^{1} (  {{\cal{{\cal{Y}}}}}  ) } } } "bioxx2"
"ooo" :@{->}^{     } "bioxx29"
 "bioxx2" :@{->}^{    \pi_{{\cal{X}}}    } "bioxx29"
 "ooo" :@{^{(}->}^{       } "bioxx3"
  "bioxx3" :@{->}^{    \Pi_{{\cal{X}}}^{{{\cal P}}}    } "bioxx29"
 "bioxx29" :@/^{0.50cm}/^{ (  {\hbox{\scriptsize\dd}}  e , {\hbox{\scriptsize\dd}}  \omega )   } "ooo"
 "bioxx29" :@/^{0.50cm}/^{ (e , \omega ,  {\hbox{\scriptsize\dd}}  e , {\hbox{\scriptsize\dd}}  \omega )   } "bioxx3"
}
\quad \quad
   \xygraph{
!{<0cm,0cm>;<2.0cm,0cm>:<0cm,0.7cm>::}
!{(2,2) }*+{  {\cal Y} }="bioxx2"
!{(1.2,3.8) }*+{    }="bioxx28"
!{(1,2) }*+{  {\cal X} }="bioxx29"
 !{(2.5,2.1) }*+{ [\mathfrak{2}] }="oo"
!{(2,4) }*+{ {{\cal M}}_{\hbox{\tiny{\sffamily DW}}}  := \Lambda^{n}_{\mathfrak{1}}T^{\star} {\cal{{{Y}}}}     }="bioxx3"
!{(1.2,4) }*+{     }="bioxx333"
"bioxx3" :@{->}^{    \Pi   } "bioxx2"
"bioxx2" :@{->}^{    \pi_{{\cal{X}}}    } "bioxx29"
"bioxx3" :@{->}^{    \Pi_{{\cal{X}}}    } "bioxx29"
"bioxx29" :@/^{0.50cm}/^{ (e , \omega , \varkappa , p^{e} , p^{\omega}  )   } "bioxx333"
}
$$    
\caption{$[\mathfrak{1}]$ The  fiber bundles   ${{\cal P}}^{( e , \omega)} = ( e , \omega)^\star T {{\cal{{\cal{Y}}}}} \otimes_{{\cal X}}   T^\star {{\cal{X}}}   $ and $ {{\cal P}} :=  T {{\cal{{\cal{Y}}}}} \otimes_{\cal Y}   T^\star {{\cal{X}}}  $,   on which the Lagrangian density is defined. The latter is   identified with the bundle of decomposable $n$-vector fields  $  \Lambda^n_{\textsc{n}} T ({{\cal{{\cal{Y}}}}}) $  on the covariant    configuration space $   {\cal{Y}} =  \textswab{p}  \otimes   T^\star {{\cal{X}}}$.  $[\mathfrak{2}]$ The {\sffamily DW} multimomentum bundle $  {{\cal M}}_{\hbox{\tiny{\sffamily DW}}}  := \Lambda^{n}_{\mathfrak{1}}T^{\star} {{{{\cal{{\cal{Y}}}}}}}  $ as a fiber bundle over ${\cal{Y}}$.  }
\label{fig:image1za}
\end{figure}

Now  we  consider the   first order Palatini   density $L_{\tiny{\hbox{\sffamily Palatini}}} [e,\omega] =  \kappa {\hbox{\sffamily  vol}}  (e) e^\mu_I e^\nu_J {F}^{IJ}_{\mu\nu} [\omega]   $, equivalently   written as $    {L}_{\tiny{\hbox{\sffamily Palatini}}} [e,\omega]     \vol =  ({\kappa}/4)   \epsilon_{IJKL}{\epsilon}^{\mu\nu\rho\sigma} e^I_\mu  e^J_\nu {F}_{\rho\sigma}^{KL} [\omega]  \vol 
$ (see  appendix  \ref{app:lagrang} for details).  We  now set the constant  $\kappa := 1/2$, so that 
 \begin{equation}\label{lagrangiandensitypalatini}
\left.
\begin{array}{ccl}
{ {{L}}}_{\tiny{\hbox{\sffamily Palatini}}} [e,\omega]    & = &  \displaystyle
(1/8)   \epsilon_{IJKL}{\epsilon}^{\mu\nu\rho\sigma} e^I_\mu  e^J_\nu  {\big{(}} \partial_{\rho} \omega_{\sigma}^{KL} -     \partial_{\sigma} \omega_{\rho}^{KL} +  \omega^{K} _{\rho M} \omega_{\sigma}^{ML}  -  \omega^{K} _{\sigma M} \omega_{\rho}^{ML}  {\big{)}},
\\  
 \displaystyle    & = &  \displaystyle  (1/2)   e \left( e^{\mu}_{I} e^{\nu}_{J} -  e^{\nu}_{I} e^{\mu}_{J}  \right) ( {  \partial_{\mu}} \omega^{IJ}_{\nu}   + \omega^{I} _{\mu K} \omega_{\nu}^{KJ}   )  ,  
 \\ 
 \displaystyle    & = &  \displaystyle   \dt^{[\mu}_{I} e^{\nu]}_{J} ( {  \partial_{\mu}} \omega^{IJ}_{\nu}   + \omega^{I} _{\mu K} \omega_{\nu}^{KJ}   ) ,     
\end{array}
\right.\end{equation}
where we used the identity $   \dt^{[\mu}_{I} e^{\nu]}_{J} :=    e e^{[\mu}_I e^{\nu]}_J    =   ({1}/{4}) \epsilon_{IJ KL}    e^{K}_{\rho}      e^{L}_{\sigma}    \epsilon^{ \mu \nu \rho\sigma}$, see appendix  \ref{app:generalizedVD}. The Lagrangian density $L[e,\omega] : {{\cal P}} \rightarrow \Bbb{R}$  is thought of as  a function  defined on the bundle  ${{\cal P}}  $, {\em i.e.} the bundle over ${\cal Y}$ with the fiber over a point $(x,e_x,\omega_x) \in {\cal Y}$ given by $T_{(x, e_x,\omega_x)} {{\cal{Y}}} \otimes_{\cal{Y}}   T^\star_{x} {{\cal{X}}}$.   Then, the set of    local   coordinates $(x^\mu,e^I_\mu, \omega^{IJ}_\mu,  {v}^I_{\mu \nu} ,  {v}^{IJ}_{\mu \nu} )$  on ${\cal P} $ is equivalently described, using the definitions \eqref{nvectorzz}, by the set $(x^\mu,e^I_\mu, \omega^{IJ}_\mu, z^I_{\mu \nu} ,  z^{IJ}_{\mu \nu})$  on $ \Lambda^n_{\textsc{n}} T  {\cal Y}    $.  Alternatively, we can use    the set of coordinates on the first jet bundle ${\dttJ}^1({{\cal{{\cal{Y}}}}})$, see for example  \cite{bruno,bruno1,Espo}. We summarize   these constructions in   figure \ref{fig:image1za}-$[\mathfrak{1}]$.

 \subsection{{{DW multisymplectic manifold and   Legendre correspondence}}}\label{subsec:2.3}

 Now   we describe the {DW} multisymplectic manifold for the Palatini action of vielbein gravity. The multimomentum phase space  is constructed on the covariant configuration space ${{\cal{{\cal{Y}}}}} := \textswab{p} \otimes T^{\star}{{\cal{X}}}$, see the construction   in figure \ref{fig:image1za}-$[\mathfrak{2}]$.  We  present the notations used for the   {DW} submanifold     ${{\cal M}}_{\hbox{\tiny{\sffamily DW}}} \subset {{{\cal M}}}  :=  \Lambda^n T^\star ({\textswab{p}} \otimes T^\star{{\cal{X}}})  $, as introduced in section \ref{subsec:DWmultimomentum}.  The   {DW} manifold is
   \bee
{{\cal M}}_{\hbox{\tiny{\sffamily DW}}} =  \left\{ (x,e,\omega,p) / x \in {{\cal{X}}},  e \in \Bbb{R}(1,3) \otimes T^\star_{x} {{\cal{X}}} ; \omega \in \textswab{g} \otimes T^\star_{x} {{\cal{X}}}  ,p \in \Lambda^n_{\mathfrak{1}} T^\star ({\textswab{p}} \otimes T^\star{{\cal{X}}})  \right\}, 
\eee

 In the {DW} formulation   we consider   all the components  of the Poincar\'{e}-Cartan form, see \eqref{PCmulti002}, equal to zero except  $p_{1 ... n } := \varkappa$,  $ p_{1 ... (\nu-1) (e_\mu^{IJ}) (\nu+1) ... n } :=  {{p}}^{e_{\mu}\nu}_{IJ}$,  and $ p_{1 ... (\nu-1) ( \omega_\mu^{IJ}) (\nu+1), ... n } :=  {{p}}^{\omega_{\mu}\nu}_{IJ}$. Thus,  we restrict ourselves to $n$-forms $ p \in  \Lambda^nT^\star_{(x,e_x,\omega_x)} {{\cal{{\cal{Y}}}}} $ such that $  
\partial_{e_\mu^I} \wedge \partial_{e_\nu^I} \iN p$,  $ \partial_{\omega_\mu^{IJ}} \wedge \partial_{\omega_\nu^{IJ}} \iN p$, and  $\partial_{e_\mu^I} \wedge \partial_{\omega_\nu^{IJ}} \iN p  $  are identically vanishing.
 Equivalently,  the {DW} multisymplectic manifold is specified as
    \bee
{{\cal M}}_{\hbox{\tiny{\sffamily DW}}} =  \left\{   (x,e,\omega,  \varkappa  {\beta}  + {{p}}^{e_{\mu}\nu}_I
  \hbox{d} e_{\mu}^I \wedge   {\beta}_{\nu} + {{p}}^{\omega_{\mu}\nu}_{IJ}
  \hbox{d} \omega_{\mu}^{IJ} \wedge  {\beta}_{\nu} )  / (x,e,\omega) \in {{\cal{Y}}},  \varkappa , {{p}}^{e_{\mu}\nu}_I ,  {{p}}^{\omega_{\mu}\nu}_{IJ}  \in \Bbb{R}  \right\}. 
  \nonumber
\eee
   We consider the following Poincar\'e-Cartan ${{\theta}}^{{\tiny{\hbox{\sffamily DW}}}}_{(q,p)}$ $n$-form, for any $(q,p) \in {{\cal M}}_{\hbox{\tiny{\sffamily DW}}}  := \Lambda^{n}_{\mathfrak{1}}T^{\star} {\cal{{{Y}}}}    $ 
   \bee
{{\theta}}^{{\tiny{\hbox{\sffamily DW}}}}_{(q,p)}:= \varkappa  {\beta}  + {{p}}^{e_{\mu}\nu}_I
  \hbox{d} e_{\mu}^I \wedge   {\beta}_{\nu} + {{p}}^{
\omega_{\mu}\nu}_{IJ}
  \hbox{d} \omega_{\mu}^{IJ} \wedge  {\beta}_{\nu}.
\eee
Now, we describe   the Legendre correspondence for the {DW} formulation of    the first order Palatini   Lagrangian $L_{\tiny{\hbox{\sffamily Palatini}}} [e,\omega] =     \dtt^{[\rho\sigma]}_{IJ}  {\big{(}} \partial_{\rho} \omega_{\sigma}^{IJ}  +  \omega^{I} _{\rho M} \omega_{\sigma}^{MJ}    {\big{)}}    $, where  we denote  $ \dtt^{[\rho\sigma]}_{IJ} :=  \dtt^{[\rho}_{I}  e^{\sigma]}_{J}$, see appendix  \ref{app:generalizedVD}. 
The Legendre correspondence $(q,v)   \    {\pmb{\leftrightarrow}}    \ (q,p)   $ for the  formulation of vielbein gravity is  given by  
\begin{align}
   {{\cal P}}  \sim  \Lambda^{n}_{\textsc{n}}    T({\textswab{p}} \otimes T^\star{{\cal{X}}}) =   T {{\cal{{\cal{Y}}}}} \otimes_{{{\cal{{\cal{Y}}}}}}   T^\star {{\cal{X}}}   \ &  \ {\pmb{\leftrightarrow}}   \  {{\cal M}}_{\hbox{\tiny{\sffamily DW}}}  := \Lambda^{n}_{\mathfrak{1}}T^{\star} ({\textswab{p}} \otimes T^\star{{\cal{X}}})  = \Lambda^{n}_{\mathfrak{1}}T^{\star} {{{{\cal{{\cal{Y}}}}}}} ,  \nonumber \\ 
(q,v) \sim     (x^\mu,e^I_\mu, \omega^{IJ}_\mu, z^I_{\mu \nu} ,  z^{IJ}_{\mu \nu} ) \ & \  {\pmb{\leftrightarrow}}    \ (q,p) = (x^{\mu} , e^{I}_{\mu},   \omega^{IJ}_{\mu} , \varkappa ,  {{p}}^{e_{\mu}\nu}_{I} ,  {{p}}^{\omega_{\mu}\nu}_{IJ}  ).
\end{align}
In particular, the construction of the Legendre correspondence involves  the relation 
\begin{equation}\label{LCEQUATION}
(q,v)   {\pmb{\leftrightarrow}}  (q,p) \quad   \Longleftrightarrow \quad   \frac{\partial \langle p , v  \rangle  }{\partial v} =  \frac{\partial { { {{L}}}}(q,v)  }{\partial v} ,
\end{equation}
 between $(q,v)$ and $(q,p)$, where  we denote  $
\langle p , v  \rangle :={{\theta}}^{{\tiny{\hbox{\sffamily DW}}}}_{(q,p)} (\cal Z) $ and ${\cal Z}  \in\Lambda^n_{\textsc{n}} T  {\cal Y}   $. We consider a decomposable multivector field $ {\cal Z} = {\cal Z}_1 \wedge {\cal Z}_2 \wedge {\cal Z}_3 \wedge {\cal Z}_4   \in\Lambda^4_{\textsc{n}} T  {\cal Y}$, where for any $1 \leq \mu \leq 4$:
 \begin{equation}\label{youiu45678}
\left.
\begin{array}{ccl}
{\cal Z}_\mu   & = & \displaystyle  \frac{\partial}{\partial x^\mu} +{\cal Z}_{\mu\nu}^I \frac{\partial}{\partial e_\nu^I}  + {\cal Z}_{\mu\nu}^{IJ} \frac{\partial}{\partial \omega_\nu^{IJ}}   ,
\\
\displaystyle    & = & \displaystyle   \frac{\partial}{\partial x^\mu} +  {\big{(}} \partial_{\mu} e^I_{\nu} +  {\omega_{\mu}}^I_J e^{J}_{\nu} {\big{)}}    \frac{\partial}{\partial e_\nu^I}  + {\big{(}}   \partial_{\mu} \omega^{IJ}_{\nu} +  {\omega_{\mu}}^I_K  {\omega_{\nu}}^{KJ} - {\omega_{\mu}}^J_K  {\omega_{\nu}}^{KI}  {\big{)}}  \frac{\partial}{\partial \omega_\nu^{IJ}}  .
\end{array}
\right.
\end{equation}
Let us note that  the multivector field ${\cal Z}$ is written as 
\bee\left.
\begin{array}{ccl}
{\cal Z}    & = & \displaystyle   \sum_{ {{\pmb{\mu}}_1} < \cdots < {\pmb{\mu}}_4 } {\cal Z}^{{{\pmb{\mu}}_1} \cdots {{\pmb{\mu}}_4} }_{1 \cdots 4}
\frac{\partial}{\partial q^{{{\pmb{\mu}}_1}}} \wedge \cdots \wedge \frac{\partial}{\partial q^{{{\pmb{\mu}}_4}}} :=  \sum_{ {{\pmb{\mu}}_1} < \cdots < {\pmb{\mu}}_4 }
\left| \begin{array}{ccc}
{ {\cal Z}^{{\pmb{\mu}}_1}_{1} } & \cdots & { {\cal Z}^{{\pmb{\mu}}_1}_{4}   }\\
\vdots & & \vdots \\
{{\cal Z}^{{\pmb{\mu}}_4}_{1} } & \cdots & {{\cal Z}^{{\pmb{\mu}}_4}_{4} }
\end{array}\right|  
\frac{\partial}{\partial q^{{{\pmb{\mu}}_1}}} \wedge \cdots \wedge \frac{\partial}{\partial q^{{{\pmb{\mu}}_4}}} .
\end{array}
\right.
\nonumber
\eee
Now, for any $(q,p) \in   \Lambda^{n}_{\mathfrak{1}}T^{\star} {\cal{{{Y}}}}    $  and ${\cal Z} \in \Lambda^n_{\textsc{n}} T  {\cal Y}$, we make the straightforward calculation
\bee
\left.
\begin{array}{ccl}
\langle p , v  \rangle     & = &  \displaystyle  {{\theta}}^{\tiny{\hbox{\sffamily DW}}}_p ({\cal Z}) = \varkappa \beta ({\cal Z})   + {{p}}^{e_{\mu}\nu}_I
\dd e_{\mu}^I \wedge \beta_{\nu}  ({\cal Z}) + {{p}}^{
\omega_{\mu}\nu}_{IJ}
\dd \omega_{\mu}^{IJ} \wedge  \beta_{\nu}  ({\cal Z})  ,
\\  
 \displaystyle    & = &  \displaystyle     \varkappa   + {{p}}^{e_{\mu}\nu}_I
{\cal Z}_{\nu\mu}^{I} + {{p}}^{
\omega_{\mu}\nu}_{IJ}
{\cal Z}_{\nu\mu}^{IJ}.  
\end{array}
\right.
\end{equation}
 Let us compute the two parts involved in the Legendre correspondence. We calculate the partial derivatives with respect to the field derivatives $\partial_{\nu} e_{\mu}^{I}$ and $\partial_{\nu} \omega_{\mu}^{IJ}$  
\begin{equation}
\left.
\begin{array}{ccc}
\displaystyle  \frac{\partial \langle p , v  \rangle }{\partial (\partial_\nu \omega_{\mu}^{IJ})}    & = & \displaystyle    {{p}}^{\omega_{\mu}\nu}_{IJ},  
\\ 
\displaystyle    \frac{\partial \langle p , v  \rangle }{\partial (\partial_\nu e_{\mu}^{I})}    & = & \displaystyle      {{p}}^{e_{\mu}\nu}_I,  
\end{array}
\right.
\ \
\left.
\begin{array}{lcl}
\displaystyle  \frac{\partial  {{{L}}}_{\tiny{\hbox{\sffamily Palatini}}} [e,\omega]  }{\partial (\partial_\mu \omega_{\nu}^{IJ})}   & = & \displaystyle  \frac{\partial}{\partial (\partial_\mu \omega_{\nu}^{IJ})}  {\Big{(}} e e^{\mu}_{I} e^{\nu}_{J}( {  \partial_{[\mu}} \omega^{IJ}_{\nu]} + \omega^{I} _{[\mu K} \omega_{\nu]}^{KJ}) {\Big{)}}  =     \dtt^{[\mu}_{I} e^{\nu]}_{J}  ,
\\ 
\displaystyle  \frac{\partial { {{L}}}_{\tiny{\hbox{\sffamily Palatini}}} [e,\omega]  }{\partial (\partial_\mu e_{\nu}^{I})}  & = & \displaystyle \frac{\partial}{\partial (\partial_\mu e_{\nu}^{I})}  {\Big{(}} e e^{\mu}_{I} e^{\nu}_{J}( {  \partial_{[\mu}} \omega^{IJ}_{\nu]} + \omega^{I} _{[\mu K} \omega_{\nu]}^{KJ}) {\Big{)}} =  0    .
\end{array}
\right.
\nonumber
\end{equation}
Therefore, the Legendre correspondence yields
 \begin{equation}\label{gliodxd55}
  {{p}}^{\omega_{\mu}\nu}_{IJ}   = -   \dtt^{[\mu}_{I} e^{\nu]}_{J}  =  - (1/4) \epsilon_{IJKL} \epsilon^{\mu\nu\rho\sigma} e^{K}_{\rho} e^{L}_{\sigma},
\quad \quad \quad    
 {{p}}^{e_{\mu}\nu}_I  =   0  ,   
\end{equation}
for  the multimomenta   related to    $\omega_{\mu}^{IJ}$ and $e^{I}_{\mu}$,    respectively.      Then, the Legendre correspondence  yields  ${p}^{\omega_\nu \mu}_{IJ}  +  {p}^{\omega_\mu \nu}_{IJ} = 0$ and ${{p}}^{e_\mu \nu}_{I} = 0$. It is an example of the set of Dirac primary constraints in the {DW} multisymplectic formalism. Therefore, we
shall be restricted to the submanifold $  {{\cal C}} \subset  {{\cal M}}_{\hbox{\tiny{\sffamily DW}}}  $ for taking into account the primary constraints:  
  \begin{equation}\label{kiqlapw}
\left.
\begin{array}{ll}
& 
\displaystyle   {{\cal C}} \subset  {{\cal M}}_{\hbox{\tiny{\sffamily DW}}}  =   \{ (x,e,\omega,p) \in {{\cal M}}_{\hbox{\tiny{\sffamily DW}}} \   /   \    {{p}}^{\omega_{\mu}\nu}_{IJ}   = -  \dtt^{[\mu}_{I} e^{\nu]}_{J},   \quad {{p}}^{e_\mu \nu}_{I} = 0   \}   .
\end{array}
\right.
\end{equation}
  The Legendre transformation is   degenerate since we cannot express arbitrary field derivative via multimomenta. Let us note that the multimomenta $ {{p}}^{\omega_\mu \nu}_{IJ} := {{p}}^{\omega_\mu \nu}_{IJ} (x,e)$ are functions of the vierbein.

\subsection{{{DW Hamiltonian of the Palatini action}}}\label{subsec:Ham093U984}

Now we present  the    {DW} Hamiltonian function   of  the Palatini action of vielbein gravity.   The Legendre correspondence is generated by the function  $  {W}^{{\tiny\hbox{\sffamily DW}}} (q,v,p)  :=  \langle p , v  \rangle - { { {{L}}}}(q,v) $, i.e.
\[
\left.
\begin{array}{rcl}
 \displaystyle   {W}^{{\tiny\hbox{\sffamily DW}}} (q,v,p)   & = &  \displaystyle  =   \displaystyle  \varkappa   + {{p}}^{e_{\mu}\nu}_I
{\cal Z}_{\nu\mu}^{I} + {{p}}^{
\omega_{\mu}\nu}_{IJ}
{\cal Z}_{\nu\mu}^{IJ} -  \dt^{[\mu}_{I} e^{\nu]}_{J}{\big{(}} \partial_{\mu}{\omega}^{IJ}_{\nu} + \omega^{I} _{\mu K} \omega_{\nu}^{KJ} {\big{)}},
  \\
\displaystyle     & = &  \displaystyle  \varkappa   + {{p}}^{e_{\mu}\nu}_I
{\cal Z}_{\nu\mu}^{I}+  {{p}}^{
\omega_{\mu}\nu}_{IJ} 
{\big{(}} 
 \partial_{\nu} \omega^{IJ}_{\mu} +  {\omega_{\nu}}^I_K  {\omega_{\mu}}^{KJ} - {\omega_{\nu}}^J_K  {\omega_{\mu}}^{KI} {\big{)}}  ,
  \\
\displaystyle     &   &  \displaystyle 
  -   \dt^{[\nu}_{I} e^{\mu]}_{J}  {\big{(}} \partial_{\nu}{\omega}^{IJ}_{\mu} + \omega^{I} _{\nu K} \omega_{\mu}^{KJ}  
 {\big{)}}.
\end{array}
\right.
\]
 Let us work on ${{\cal C}} \subset {{\cal M}}_{\hbox{\tiny{\sffamily DW}}}$. We introduce the   Hamiltonian function ${\cal{H}} :  {{\cal M}}_{\hbox{\tiny{\sffamily DW}}} \rightarrow \Bbb{R}$  defined by ${\cal{H}} = \langle p , \nu (q,p)  \rangle - { { {{L}}}}(q,p, \nu(q,p))$, where $\nu(q,p)$ is  such that $(q,\nu(q,p))  
\ {\pmb{\leftrightarrow}}   \      (q,p)$. For any  $v \in T_{(x,e_x,\omega_x)} {{\cal{{\cal{Y}}}}} \otimes T^{\star}_x {{\cal{X}}}$  the equation \eqref{LCEQUATION} has a solution $p \in  {{\cal M}}_{\hbox{\tiny{\sffamily DW}}} $ if  and only if  $p \in {{\cal C}} $ with 
  \begin{equation}\label{kiqfdfdlapw}
\left.
\begin{array}{ll}
& 
\displaystyle {{\cal C}}  =  \{ (x,e,\omega,  \varkappa {\beta} -       e e^{[\mu}_{I} e^{\nu]}_{J}
\hbox{d}\omega_{\mu}^{IJ} \wedge  {\beta}_{\nu}  )   \   /   \    (x,e,\omega) \in  {{\cal{{\cal{Y}}}}} = \textswab{p} \otimes   T^\star {{\cal{X}}}, \  \varkappa \in \Bbb{R}  \}   .
\end{array}
\right.
\nonumber
\end{equation}
  The use of the constraint \eqref{gliodxd55},  {\em i.e.} ${{p}}^{e_{\mu}\nu}_I = 0 $ and ${{p}}^{
\omega_{\mu}\nu}_{IJ}   =    -   e e^{[\mu}_I e^{\nu]}_J $, leads to the expression of  the  Hamiltonian function  restricted   to the hypersurface of constraints $ {{\cal C}}$. Thus, $ {\cal{H}}^{{\tiny\hbox{\sffamily Palatini}}}  (q,p) :=  \iota^{\star}  {\cal{H}}^{{\tiny\hbox{\sffamily DW}}} (q,p)$ is written as
     \[
\left.
\begin{array}{rcl}
   \displaystyle     {\cal{H}}^{{\tiny\hbox{\sffamily Palatini}}}  (q,p)       & = &  \displaystyle 
\varkappa  +  {{p}}^{
\omega_{\mu}\nu}_{IJ} 
{\big{(}} 
 \partial_{\nu} \omega^{IJ}_{\mu} +  {\omega_{\nu}}^I_K  {\omega_{\mu}}^{KJ} - {\omega_{\nu}}^J_K  {\omega_{\mu}}^{KI} {\big{)}}   -   {{p}}^{
\omega_{\mu}\nu}_{IJ}     {\big{(}} \partial_{\nu}{\omega}^{IJ}_{\mu} + \omega^{I} _{\nu K} \omega_{\mu}^{KJ}  
 {\big{)}} ,
\\
\displaystyle     & = &  \displaystyle      \varkappa  -  {{p}}^{
\omega_{\mu}\nu}_{IJ} 
 {\omega_{\nu}}^J_K  {\omega_{\mu}}^{KI}  = \varkappa -  e e^{[\mu}_I e^{\nu]}_J
 {\omega_{\mu}}^J_K  {\omega_{\nu}}^{KI} .
\end{array}
\right.
\]  
The  Hamiltonian function $  {\cal{H}}^{{\tiny\hbox{\sffamily Palatini}}}  (q,p)  : {{\cal C}}  \rightarrow \Bbb{R} $  is equivalently written as
  $ {\cal{H}}^{{\tiny\hbox{\sffamily Palatini}}}  (q,p) =     \varkappa + H^{{\tiny\hbox{\sffamily Palatini}}} (q,p)  $,  where   $H^{{\tiny\hbox{\sffamily Palatini}}} (e,\omega)  := \iota^{\star}  H^{{\tiny\hbox{\sffamily DW}}} (q,p)$ is the {DW} Hamiltonian   \cite{kannnnnnnaa,kannnnnnnaa1} evaluated on the constraint hypersurface ${{\cal C}}$.  In    section  \ref{sec:section17}, we   explore  the $n$-phase space approach, we fix $\varkappa  = e e^{\mu}_{I} e^{\nu}_{J}    \omega^{J} _{[\mu K} \omega_{\nu]}^{KI} $. Note that we can always choose $\varkappa(x)$ such that ${\cal{H}} (x,e(x),\omega(x), \varkappa(x), {{p}}(x))$ is constant, see \cite{HK-02,HK-03}.

 \subsection{{{Exterior derivative of the DW Hamiltonian}}  
 }\label{subsec:ext73645}

In this section we derive the exterior derivative of the {DW} Hamiltonian  function  for the formulation of dreibein and vierbein gravity.
First, let us consider the case of dreibein gravity.  We denote the exterior derivative by  $\hbox{d} {\cal{H}}^{{\tiny\hbox{\sffamily Palatini}}}_{\tiny{\hbox{\sffamily\tiny 3D}}}$. We have
\bee\label{extextYY}
\hbox{d} {\cal{H}}^{{\tiny\hbox{\sffamily Palatini}}}_{\tiny{\hbox{\sffamily 3D}}}  (q,p) =   \hbox{d} \varkappa    -   \dtt^{[\mu}_{I} e^{\nu]}_{J} \hbox{d}  \left( \omega^{J} _{\mu K} \omega_{\nu}^{KI}   \right)    - \hbox{d}  \left(     \dtt^{[\mu}_{I} e^{\nu]}_{J} \right) \omega^{J} _{\mu K} \omega_{\nu}^{KI} .
\eee
When the dimension of the space-time manifold is $n = 3$, we have the  algebraic relation   $\displaystyle  -   \dtt^{[\mu}_{I} e^{\nu]}_{J} =  -    e e^{[\mu}_{I} e^{\nu]}_{J} = - (1/2) \epsilon^{\mu\nu\rho} \epsilon_{IJK} e^{K}_{\rho} $.  Then the second term in  \eqref{extextYY} takes the form
\[
\left.
\begin{array}{rcl}
    \displaystyle     -  \dtt^{[\mu}_{I} e^{\nu]}_{J}   \hbox{d} \left( \omega^{J} _{\mu K} \omega_{\nu}^{KI}  \right)    & = &  \displaystyle -   (1/2) \epsilon^{\mu\nu\rho} \epsilon_{IJK} e^{K}_{\rho}  \left( \omega^{J} _{\mu K} \hbox{d} \omega_{\nu}^{KI} +  \hbox{d} ( \omega^{JM}_{\mu} {\textsf{h}}_{MK})  \omega_{\nu}^{KI}  \right) ,
    \\
    \displaystyle    & = &  \displaystyle  -    \epsilon_{IJM}  \epsilon^{\mu\nu\lambda} e_\lambda^M     \omega^{J} _{\nu K} \hbox{d} \omega_{\mu}^{KI}   =  -   \epsilon_{IJK}  \epsilon^{\mu\nu\rho}  e_{\rho}^{K} \omega^{J} _{\nu M} \hbox{d} \omega_{\mu}^{MI} ,
     \end{array} 
\right.
\]
 where we have used  the algebraic relation    
  \bee\label{blvo00dsd93322}
  \epsilon^{\mu\rho\sigma}   \epsilon_{IJK} e^{I}_\mu {{\omega_\sigma}^J}_M \dd \omega_\rho^{MK} = -  {1}/{2} \epsilon^{\mu\rho\sigma}   \epsilon_{LJK} e^{I}_\mu {{\omega_\sigma}^L}_I \dd \omega_\rho^{JK}  =   {1}/{2}    \epsilon^{\mu\nu\rho}   \epsilon_{LJI} e^{K}_\rho {{\omega_\nu}^L}_K \dd \omega_\mu^{IJ}.
\eee
  This relation is used also     to  decompose the interior product $X^{\footnotesize{{{\cal C}}}} \iN {\pmb{\omega}}^{{\tiny\hbox{\sffamily Palatini}}}   $ in  the basis $1$-forms $\dd\omega^{IJ}_{\mu}$. 
   Also,  since $ \displaystyle
 \hbox{d} (e e^{\mu}_I e^{\nu}_J ) =   \hbox{d}  (  \epsilon_{IJM}  \epsilon^{\mu\nu\lambda} e_\lambda^M  ) =   \epsilon_{IJM}  \epsilon^{\mu\nu\lambda} \dd e_{\lambda}^M
$,  the third  term in \eqref{extextYY}   is written as
$(1/2)  \epsilon_{IJM}  \epsilon^{\mu\nu\lambda} \omega^{J} _{\mu K} \omega_{\nu}^{KI} \dd e_{\lambda}^M $. Now, we obtain the expression for the $1$-form $\hbox{d} {\cal{H}}^{{\tiny\hbox{\sffamily Palatini}}}_{{\tiny\hbox{\sffamily 3D}}} (q,p)$, namely 
\begin{equation} \label{dH3D}
\hbox{d} {\cal{H}}^{{\tiny\hbox{\sffamily Palatini}}}_{{\tiny\hbox{\sffamily 3D}}} (q,p) = \hbox{d} \varkappa + (1/2)  \epsilon_{IJM}  \epsilon^{\mu\nu\lambda} \omega^{J} _{\mu K} \omega_{\nu}^{KI} \hbox{d}  e_{\lambda}^M + (1/2)  \epsilon^{\mu\nu\rho}   \epsilon_{LJI} e^{K}_\rho {{\omega_\nu}^L}_K \dd \omega_\mu^{IJ}.     
\end{equation}
When $n=4$,     $ \displaystyle
\dtt^{[\mu\nu]}_{IJ}  =   
(1/4) \epsilon_{IJ KL}    e^{K}_{\rho}      e^{L}_{\sigma}    \epsilon^{ \mu \nu \rho\sigma}   
 $, therefore  $\dd  {\big{(}}   \dtt^{[\mu\nu]}_{IJ}  {\big{)}}   = ({1}/{2})  \epsilon_{IJ KL}      \epsilon^{ \mu \nu \rho\sigma}   e^{K}_{\rho}     \dd e^{L}_{\sigma} $.
 Thus, we have  
  \bee\label{ext-der-vier}
 \left.
\begin{array}{rcl}
 \displaystyle   \hbox{d} {\cal{H}}^{{\tiny\hbox{\sffamily Palatini}}}_{{\tiny\hbox{\sffamily 4D}}}    & = &  \displaystyle 
   \hbox{d}  \varkappa  - \hbox{d} {\big{(}}    \dtt^{[\mu}_{I} e^{\nu]}_{J} {\big{)}} (    \omega^{J} _{\mu K} \omega_{\nu}^{KI}  ) -    \dtt^{[\mu}_{I} e^{\nu]}_{J}  \hbox{d} {\big{(}}   \omega^{J} _{\mu K} \omega_{\nu}^{KI} {\big{)}} ,
  \\
 \displaystyle      & = &  \displaystyle  \hbox{d}  \varkappa  - (1/2)   \epsilon_{IJKL} \epsilon^{\mu\nu\rho\sigma} e^{K}_{\rho}   \omega^{J} _{\mu M} \omega_{\nu}^{MI}  \hbox{d}    e^{L}_{\sigma} ,
   \\
 \displaystyle      &  &  \displaystyle
 +  (1/2)  \epsilon_{IJKL} \epsilon^{\mu\nu\rho\sigma} e^{K}_{\rho}     e^{L}_{\sigma}  \omega^{J} _{\nu M} \hbox{d} \omega_{\mu}^{MI}    .
 \end{array}
\right.
\eee
Using  the algebraic relation
\bee\label{blvo00933224}
    \epsilon^{\mu\nu\rho\sigma}   \epsilon_{IJKL} e^{I}_\mu e^{J}_\nu {{\omega_\sigma}^K}_M \dd \omega_\rho^{ML} = -  \epsilon^{\mu\nu\rho\sigma}   \epsilon_{INKL} e^{I}_\mu  e^{J}_\nu {{\omega_\sigma}^N}_J \dd \omega_\rho^{KL},
  \eee
 see   \cite{bruno,bruno1},  the last term in  \eqref{ext-der-vier}   is equivalently written as 
\[
\left.
\begin{array}{rcl}
\displaystyle \epsilon_{IJKL} \epsilon^{\mu\nu\rho\sigma} e^{K}_{\rho}     e^{L}_{\sigma}  \omega^{J} _{\nu M} \hbox{d} \omega_{\mu}^{MI}    & = &   \displaystyle          -  \epsilon_{IJKL} \epsilon^{ \mu\nu\rho \sigma} e^{I}_{\mu}     e^{J}_{\nu}  \omega^{K} _{\sigma M} \hbox{d} \omega_{\rho}^{ML}  =   \epsilon^{\mu\nu\rho\sigma}   \epsilon_{INKL} e^{I}_\mu  e^{J}_\nu {{\omega_\sigma}^N}_J \dd \omega_\rho^{KL}  ,
 \\
 \displaystyle      & = &   \displaystyle   - \epsilon^{\mu \nu \rho\sigma}   \epsilon_{IJKN} e^{K}_\rho  e^{L}_\nu {{\omega_\sigma}^N}_L \dd \omega_\mu^{IJ}   =  - \epsilon^{\mu \nu \rho\sigma}   \epsilon_{IJKL} e^{K}_\rho  e^{N}_\nu {{\omega_\sigma}^L}_N \dd \omega_\mu^{IJ}      .
 \end{array}
\right.
\]
Therefore, the exterior derivative of the {DW} Hamiltonian function related to the Palatini action of vierbein gravity is given by  
\begin{equation}\label{ham4D}
\hbox{d} {\cal{H}}^{{\tiny\hbox{\sffamily Palatini}}}_{{\tiny\hbox{\sffamily 4D}}} (q,p) =  \hbox{d}  \varkappa   -   (1/2) \epsilon_{IJKL} \epsilon^{\mu\nu\rho\sigma} e^{K}_{\rho}   \omega^{J} _{\mu M} \omega_{\nu}^{MI}  \hbox{d}    e^{L}_{\sigma}  +  (1/2)  \epsilon_{IJKL}  \epsilon^{\mu \nu \rho\sigma}  e^{K}_\rho  e^{N}_\nu {{\omega_\sigma}^L}_N \dd \omega_\mu^{IJ}.
  \end{equation}
 
\subsection{{Primary constraints and the extended Hamiltonian}}\label{subsec:saapsjpj89} 

  The set of primary constraints that weakly vanish on the constraint hypersurface, following the terminology of Dirac,  are 
$
{{p}}^{e_{\mu}\nu}_{I}  
\approx 0$ and  $ {{p}}^{\omega_{\mu}\nu}_{IJ}    \approx  -  e e^{[\mu}_{I} e^{\nu]}_{J}$.  An extension of the traditional method developed by   Dirac in the {DW} formulation involves the  construction of  an {extended} Hamiltonian,  
\[
{\cal{H}}^{\hbox{\tiny{\sffamily Ext}}} = \varkappa  -    e e^{[\mu}_I e^{\nu]}_J {\omega_{\mu}}^J_K  {\omega_{\nu}}^{KI} + \lambda^{I}_{\nu\mu} {{p}}^{e_{\mu}\nu}_{I} 
 + 
\lambda^{IJ}_{\nu\mu}   {\big{(}} {{p}}^{\omega_{\mu}\nu}_{IJ}  +    e e^{[\mu}_I e^{\nu]}_J {\big{)}} .
\]
 The extended {DW} Hamiltonian   is   
$
{\cal{H}}^{\hbox{\tiny{\sffamily Ext}}} =  {\cal{H}}^{\hbox{\tiny{\sffamily Palatini}}} +  \lambda^{I}_{\nu\mu} {{p}}^{e_{\mu}\nu}_{I} 
 + 
\lambda^{IJ}_{\nu\mu}   {\big{(}} {{p}}^{\omega_{\mu}\nu}_{IJ}  +    e e^{[\mu}_I e^{\nu]}_J {\big{)}}    
$. Here, $   \lambda^{I}_{\nu\mu} $ and $\lambda^{IJ}_{\nu\mu}$ are   Lagrange multipliers. We postulate, since there is no reason to assume they are   valid {\em a priori}, the {DW}  Hamilton equations
\bee
  \left.
\begin{array}{ccl}
\displaystyle  \frac{\partial  \omega^{IJ}_\mu }{ \partial x^{\nu} } (x) &   =  
& \displaystyle   \frac{\partial {\cal{H}}^{\hbox{\tiny{\sffamily Ext}}} }{\partial   {{p}}^{\omega_{\mu}\nu}_{IJ}  } (x,e,\omega, \varkappa , p),
\\
\displaystyle   \frac{\partial  e^{I}_\mu }{ \partial x^{\nu} }    (x)   &
=     & \displaystyle    \frac{\partial {\cal{H}}^{\hbox{\tiny{\sffamily Ext}}} }{\partial   {{p}}^{\omega_{\mu}\nu}_{IJ}  } (x,e,\omega, \varkappa , p),
\end{array}
\right.
\quad      
  \left.
\begin{array}{ccl}
 \displaystyle   \sum_\nu\frac{\partial      {{p}}^{\omega_{\mu}\nu}_{IJ}  }{\partial x^\nu}(x)    
   &
=    & \displaystyle   - \frac{\partial  {\cal{H}}^{\hbox{\tiny{\sffamily Ext}}}  }{\partial \omega^{IJ}_\mu  }  (x,e,\omega, \varkappa , p),
\\
 \displaystyle   \sum_\nu\frac{\partial     {{p}}^{e_{\mu}\nu}_{I}   }{\partial x^\nu}(x)    
   &
=    & \displaystyle   - \frac{\partial  {\cal{H}}^{\hbox{\tiny{\sffamily Ext}}}  }{\partial e^{I}_\mu  }  (x,e,\omega, \varkappa , p) ,
\end{array}
\right.
\eee
 In the context of the polysymplectic formalism   \cite{Kana-01}, the extended {DW} Hamiltonian function  is   written as 
$ \displaystyle {{H}}^{\hbox{\tiny{\sffamily Ext}}} =  {H}^{\hbox{\tiny{\sffamily Palatini}}}
 + \lambda^{I}_{\nu\mu} {{p}}^{e_{\mu}\nu}_{I} 
 + 
\lambda^{IJ}_{\nu\mu}   {\big{(}} {{p}}^{\omega_{\mu}\nu}_{IJ}  +    e e^{[\mu}_I e^{\nu]}_J {\big{)}} 
$. Then,  the system of {DW} Hamilton equation is given as
\bee
  \left.
\begin{array}{ccl}
\displaystyle   {\partial_{\nu}  \omega^{IJ}_\mu }  (x) &   =  
& \displaystyle    \lambda^{IJ}_{\nu\mu},
\\
\displaystyle    {\partial_{\nu}  e^{I}_\mu }    (x)   &
=     & \displaystyle   \lambda^{I}_{\nu\mu},
 \end{array}
\right.
\quad \quad    
  \left.
\begin{array}{ccl}
 \displaystyle     {\partial_\nu      {{p}}^{\omega_{\mu}\nu}_{IJ}  } (x)    
   &
=    & \displaystyle   -   ({\partial  {{H}}^{\hbox{\tiny{\sffamily Ext}}}  }/{\partial \omega^{IJ}_\mu  } )  (x,e,\omega,  p),
\\
 \displaystyle   {\partial_\nu     {{p}}^{e_{\mu}\nu}_{I}   } (x)    
   &
=    & \displaystyle   - ({\partial  {{H}}^{\hbox{\tiny{\sffamily Ext}}}  }/{\partial e^{I}_\mu  } )  (x,e,\omega,   p).
\end{array}
\right.
\eee
For a detailed analysis of constraints within the  polysymplectic approach to the {DW} Hamiltonian formalism, we refer to  \cite{kannnnnnnaa,kannnnnnnaa1}. Note that our conventions here differ from those of Kanatchikov:   the    polymomenta have opposite sign.

\subsection{{{DW Hamilton equations on $(  {{\cal C}}  , {\pmb{\omega}}^{{\tiny\hbox{\sffamily Palatini}}}) $}}}\label{subsec:2.4}
 
The canonical {DW} multisymplectic $(n+1)$-form ${\pmb{\omega}}^{\tiny{\hbox{\sffamily DW}}}  = \hbox{d}   {{\theta}}^{{\tiny{\hbox{\sffamily DW}}}}$  previously introduced in \eqref{ZZZ01}  is written as ${\pmb{\omega}}^{{\tiny\hbox{\sffamily DW}}} =  \hbox{d} \varkappa \wedge {\beta} + \hbox{d} {{p}}^{
\omega_{\mu}\nu}_{IJ}  \wedge
\hbox{d} \omega_{\mu}^{IJ} \wedge  {\beta}_{\nu}  
$. Let us introduce the $(n+1)$-form   ${\pmb{\omega}}^{{\tiny\hbox{\sffamily  Palatini}}} :=       \iota^{\star}   {\pmb{\omega}}^{{\tiny\hbox{\sffamily  DW}}}$,  where $\iota : {{\cal C}} \hookrightarrow  {\cal M}_{{\tiny\hbox{\sffamily  DW}}}$ is the canonical inclusion.  In local coordinates,
\begin{equation}\label{PPAA}
{\pmb{\omega}}^{{\tiny\hbox{\sffamily Palatini}}} =  \hbox{d} \varkappa \wedge {\beta} -  \hbox{d}  \left( e e^{[\mu}_I e^{\nu]}_J  \right) \wedge
\hbox{d} \omega_{\mu}^{IJ} \wedge  {\beta}_{\nu}   .
\end{equation}
Using \eqref{PPAA} we can now describe  the  Einstein equations in the  {DW} Hamilton formulation, where   the {DW} Hamilton equations in geometric form are written as
 \bee\label{sqdnzoaj}
X \iN {\pmb{\omega}}^{{\tiny\hbox{\sffamily Palatini}}}   =  (-1)^n \hbox{d} {\cal{H}}^{{\tiny\hbox{\sffamily Palatini}}}  .
\eee
Let      $    \Xi^{\tiny{\hbox{\sffamily DW}}}  \in  \Gamma ( {{\cal M}}_{\hbox{\tiny{\sffamily DW}}}  , T  {{\cal M}}_{\hbox{\tiny{\sffamily DW}}} ) $   be a vector field  on  $ {{\cal M}}_{\hbox{\tiny{\sffamily DW}}}$ and   $    X^{\tiny{\hbox{\sffamily DW}}}  \in  \Gamma ( {{\cal M}}_{\hbox{\tiny{\sffamily DW}}}  , \Lambda^{n} T  {{\cal M}}_{\hbox{\tiny{\sffamily DW}}} ) $   be a $n$-vector field on  $ {{\cal M}}_{\hbox{\tiny{\sffamily DW}}}$. Then, we  construct on the constraint hypersurface ${{\cal C}}$   the vector  field $  \Xi^{\footnotesize{{{\cal C}}}} := {\pmb{\pi}}_{\star}   \Xi^{\tiny{\hbox{\sffamily DW}}}  \in  \Gamma ( {{\cal C}}  , T  {{\cal C}}  )   $ and the  $n$-vector field $  X^{\footnotesize{{{\cal C}}}}  := {\pmb{\pi}}_{\star}   \Xi^{\tiny{\hbox{\sffamily DW}}}\in  \Gamma ( {{\cal C}}  , \Lambda^{n} T  {{\cal C}}  ) $, respectively. We have denoted  by ${\pmb{\pi}}$   the canonical projection  ${\pmb{\pi}} : {{\cal M}}_{\hbox{\tiny{\sffamily DW}}}   \rightarrow {{\cal C}}   $ such that ${\pmb{\pi}} \circ \iota = \hbox{{\sffamily Id}}_{{{\cal C}} }$.     

Note that, because of the primary constraints, there is no reason {\em a priori}  that the set of  {DW} Hamilton equations is in a one-to-one correspondence  with the Euler-Lagrange system of equations. Nevertheless,  working on ${{\cal C}} \hookrightarrow {{\cal M}}_{\hbox{\tiny{\sffamily DW}}}  $,  the {DW} Hamilton equations  in geometric form    \eqref{sqdnzoaj} reproduces the Einstein system.  The   {DW} Hamilton equations   $X^{\footnotesize{{{\cal C}}}}   \iN  ( \iota^{\star}   {\pmb{\omega}}^{{\tiny\hbox{\sffamily  DW}}}) =  (-1)^n \hbox{d}  (  \iota^{\star}   {{{{\cal{H}}}}}^{{\tiny\hbox{\sffamily  DW}}} ) $ are presented for dreibein and vierbein gravity in section \ref{subsec:2.4.1} and \ref{subsec:2.4.2}, respectively.

\subsubsection{{{DW Hamilton equations of dreibein gravity}}}\label{subsec:2.4.1}

 First, we consider  the {DW} Hamilton equations for the Palatini action of dreibein gravity.  Let   $ X^{\footnotesize{{{\cal C}}}}   = X^{\footnotesize{{{\cal C}}}}_1 \wedge X^{\footnotesize{{{\cal C}}}}_2 \wedge X^{\footnotesize{{{\cal C}}}}_3 \in \Lambda^{3} T  {{\cal C}} $ be a decomposable $3$-vector field, where for any $1 \leq \nu \leq 3$, 
\bee
  X^{\footnotesize{{{\cal C}}}}_\nu = \frac{\partial}{\partial x^\nu} + \Theta_{\nu {\mu}}^{I}   \frac{\partial}{\partial e^I_{\mu}} + \Theta_{\nu {\mu}}^{IJ}   \frac{\partial}{\partial \omega^I_{\mu}}  + {\Upsilon}_{\nu} \frac{\partial }{ \partial \varkappa} .
\eee
 First, we re-express ${\pmb{\omega}}^{{\tiny\hbox{\sffamily Palatini}}}$ as follows:
\[
\left.
\begin{array}{rcl}
 \displaystyle  {\pmb{\omega}}^{{\tiny\hbox{\sffamily Palatini}}}   & = &  \displaystyle \hbox{d} \varkappa \wedge \beta -    \hbox{d}    {\big{(}}    \dt^{[\mu}_{I} e^{\nu]}_{J}   {\big{)}} \wedge
\dd \omega_{\mu}^{IJ} \wedge  \beta_{\nu}    =  \hbox{d} \varkappa \wedge \beta - (1/2)      \epsilon_{IJM}     \epsilon^{\mu\nu\lambda}   \hbox{d}    e_\lambda^M  \wedge
\dd \omega_{\mu}^{IJ} \wedge  \beta_{\nu} .
\end{array}
\right.
\]  
The left  hand side of \eqref{sqdnzoaj} is given by the interior product  $X^{\footnotesize{{{\cal C}}}} \iN {\pmb{\omega}}^{{\tiny\hbox{\sffamily Palatini}}}$.    Then,  
\[
\left.
\begin{array}{rcl}
 \displaystyle  X^{\footnotesize{{{\cal C}}}} \iN {\pmb{\omega}}^{{\tiny\hbox{\sffamily Palatini}}}      & = &  \displaystyle 
 - (1/2)    {\epsilon}_{IJL} \epsilon^{\mu\nu\alpha}   {\big{(}} (\dd \omega_{\mu}^{IJ} \wedge  {\beta}_{\nu} ) (X) \dd e_\alpha^L -  (\dd e_\alpha^L \wedge
 \beta_{\nu}   ) (X) \dd \omega_{\mu}^{IJ}   {\big{)}} ,
  \\
 \displaystyle     & &  \displaystyle - (1/2)    {\epsilon}_{IJL} \epsilon^{\mu\nu\alpha}   {\big{(}}  ( \dd e_\alpha^L \wedge
\dd \omega_{\mu}^{IJ}\wedge  \beta_{\rho\nu}  ) (X) \dd x^{\rho}   {\big{)}},
\\
 \displaystyle     & = &   
  \displaystyle  \dd \varkappa  - (\dd \varkappa \wedge  \beta_\rho) (X) \dd x^{\rho}   - (1/2)     {\epsilon}_{IJL} \epsilon^{\mu\nu\alpha}   ( \dd e_\alpha^L \wedge
\dd \omega_{\mu}^{IJ}\wedge  \beta_{\rho\nu}  ) (X) \dd x^{\rho}  , \\
 \displaystyle     &  &  \displaystyle
-  (1/2)       {\epsilon}_{IJL} \epsilon^{\mu\nu\alpha}  (\dd \omega_{\mu}^{IJ} \wedge  \beta_{\nu} ) (X) \dd e_\alpha^L  +   (1/2)      {\epsilon}_{IJL} \epsilon^{\mu\nu\alpha}   (\dd e_\alpha^L \wedge
 \beta_{\nu}   ) (X) \dd \omega_{\mu}^{IJ}  .
\end{array}
\right.
\]
Finally, the expression   becomes 
\bee\label{qcnodvnoq00}
\left.
\begin{array}{rcl}
 \displaystyle X^{\footnotesize{{{\cal C}}}}  \iN {\pmb{\omega}}^{{\tiny\hbox{\sffamily Palatini}}}      & = &  \displaystyle   \hbox{d}\varkappa - \Upsilon_\rho \dd x^{\rho}  -  (1/2)   {\epsilon}_{IJL} \epsilon^{\mu\nu\alpha}  {\big{(}}\Theta^{IJ}_{\nu\mu}      \Theta^{L}_{\rho\alpha}  - \Theta^{IJ}_{\rho\mu} \Theta^{L}_{\nu\alpha} {\big{)}} \dd x^{\rho}, 
  \\
  \displaystyle    & &  \displaystyle   - (1/2)  {\epsilon}_{IJL} \epsilon^{\mu\nu\alpha}     \Theta^{IJ}_{\nu\mu} \dd e_\alpha^L + (1/2)   {\epsilon}_{IJL} \epsilon^{\mu\nu\alpha}      \Theta^L_{\nu\alpha} \dd \omega_{\mu}^{IJ}.
\end{array}
\right.
\eee
 which is equal to   the right hand side of \eqref{sqdnzoaj} 
\begin{equation}\label{qcnodvnoq01}
\hbox{d} {\cal{H}}^{{\tiny\hbox{\sffamily Palatini}}}_{{\tiny\hbox{\sffamily 3D}}} (q,p) = \hbox{d} \varkappa +   (1/2)  \epsilon_{IJM}  \epsilon^{\mu\nu\lambda} \omega^{J} _{\mu K} \omega_{\nu}^{KI} \hbox{d}  e_{\lambda}^M + (1/2)  \epsilon^{\mu\nu\rho}   \epsilon_{LJI} e^{K}_\rho {{\omega_\nu}^L}_K \dd \omega_\mu^{IJ}  .
\end{equation}
 The equality between     \eqref{qcnodvnoq00} and \eqref{qcnodvnoq01} leads to the {DW} Hamilton system of equations  
 \begin{equation}\label{fdoxx65} 
\left.
\begin{array}{rcl}
\displaystyle
 \epsilon_{IJL}    \epsilon^{\mu\nu\alpha}   {\big{(}}  \Theta^{IJ}_{\nu \mu}  +    \omega^{J} _{\mu K} \omega_{\nu}^{KI}  {\big{)}}    &  =  &    \displaystyle 0,
\\ 
\displaystyle     \epsilon_{IJL}  \epsilon^{\mu\nu\alpha}{\big{(}}   \Theta^{L}_{\nu \alpha}    +      e^{K}_\alpha {{\omega_\nu}^L}_K  {\big{)}}   
 &  =  &   \displaystyle 0  ,
 \\
 \displaystyle  
  -  \epsilon_{IJL} \epsilon^{\mu\nu\alpha}  {\big{(}}\Theta^{IJ}_{\nu\mu}      \Theta^{L}_{\rho\alpha}  - \Theta^{IJ}_{\rho\mu} \Theta^{L}_{\nu\alpha} {\big{)}}   &  =  &   \displaystyle  \Upsilon_\rho  .
 \end{array}
\right.
\end{equation}
  The system \eqref{fdoxx65}  is the  {DW}  Hamilton equations  associated to the first order   Palatini action of dreibein gravity and is written as
\begin{equation}\label{HEPA3D}
 \epsilon_{IJK}    {F}^{JK}    = 0  ,
\quad \quad   \quad \quad   \epsilon_{IJK} \dd_{\omega} e^{K} =  0 ,
 \end{equation}
with the additional equation $\Upsilon_\rho = \partial_\rho \varkappa =   -  \epsilon_{IJL} \epsilon^{\mu\nu\alpha}  {\big{(}}\Theta^{IJ}_{\nu\mu}      \Theta^{L}_{\rho\alpha}  - \Theta^{IJ}_{\rho\mu} \Theta^{L}_{\nu\alpha} {\big{)}}  $.

\subsubsection{{{DW Hamilton equations of vierbein gravity}}}\label{subsec:2.4.2}

Now, we are interested in the {DW} Hamilton equations for the Palatini action of vierbein gravity.  We consider the   $5$-form 
\bee\label{ddddddslqlq}
 {\pmb{\omega}}^{{\tiny\hbox{\sffamily Palatini}}}   =  \hbox{d} \varkappa \wedge {\beta} -  \hbox{d} (   e e^{[\mu}_I e^{\nu]}_J ) \wedge
\hbox{d} \omega_{\mu}^{IJ} \wedge  {\beta}_{\nu}   = \hbox{d} \varkappa \wedge {\beta} -  (1/2)  \epsilon_{IJKL} \epsilon^{\mu\nu\rho\sigma} e^{K}_{\rho} \hbox{d} e^{L}_{\sigma} \wedge
\hbox{d} \omega_{\mu}^{IJ} \wedge  {\beta}_{\nu} .
\eee
Let us consider a multivector field  $X^{\footnotesize{{{\cal C}}}} = X^{\footnotesize{{{\cal C}}}}_1 \wedge X^{\footnotesize{{{\cal C}}}}_2 \wedge X^{\footnotesize{{{\cal C}}}}_3  \wedge X^{\footnotesize{{{\cal C}}}}_4 \in \Lambda^{4} T{{\cal C}}$, where for any  $1 \leq  \nu \leq 4  $,  
\bee\label{iodpsosss}
X^{\footnotesize{{{\cal C}}}}_\nu= \frac{\partial}{\partial x^\nu} + \Theta_{\nu {\mu}}^{I}   \frac{\partial}{\partial e^I_{\mu}} + \Theta_{\nu {\mu}}^{IJ}   \frac{\partial}{\partial \omega^I_{\mu}}  + {\Upsilon}_{\nu} \frac{\partial }{ \partial \varkappa}.
\eee
Then  
\[
\left.
\begin{array}{rcl}
 \displaystyle     X^{\footnotesize{{{\cal C}}}}  \iN {\pmb{\omega}}^{{\tiny\hbox{\sffamily Palatini}}}  & = &  \displaystyle  X \iN ( \dd \varkappa \wedge \beta ) -  (1/2)   \epsilon_{IJKL} \epsilon^{\mu\nu\rho\sigma} e^{K}_{\rho}    X \iN (  \hbox{d}    e^{L}_{\sigma}    \wedge
 \dd \omega_{\mu}^{IJ} \wedge  \beta_{\nu} ) ,
 \\
 \displaystyle     & = &  \displaystyle \beta (X) \dd \varkappa - (\dd \varkappa \wedge  \beta_\rho) (X) \dd x^{\rho},
 \\
 \displaystyle     & &  \displaystyle  - (1/2) \epsilon_{IJKL} \epsilon^{\mu\nu\rho\sigma} e^{K}_{\rho}   {\big{(}} (\dd \omega_{\mu}^{IJ} \wedge  \beta_{\nu} ) (X) \dd e_\sigma^L -  (\dd e_\sigma^L \wedge
 \beta_{\nu}   ) (X)\dd \omega_{\mu}^{IJ}  
  {\big{)}} ,
 \\
 \displaystyle     & &  \displaystyle   - (1/2)  \epsilon_{IJKL} \epsilon^{\mu\nu\rho\sigma} e^{K}_{\rho}   {\big{(}} ( \dd e_\sigma^L \wedge
\dd \omega_{\mu}^{IJ}\wedge  \beta_{\lambda\nu}  ) (X) \dd x^{\lambda}  {\big{)}} .
\end{array}
\right.
\]
Since,  $ (\dd \omega_{\mu}^{IJ} \wedge  \beta_{\nu} ) (X) = \Theta^{IJ}_{\nu\mu} $, $  (\dd e_\sigma^L \wedge
 \beta_{\nu}   ) (X) =  \Theta^{I}_{\nu\mu} $ and   $( \dd e_\sigma^L \wedge
\dd \omega_{\mu}^{IJ}\wedge \beta_{\lambda\nu}  ) (X)  = {{(}}\Theta^{IJ}_{\nu\mu}      \Theta^{L}_{\lambda\sigma}  - \Theta^{IJ}_{\lambda\mu} \Theta^{L}_{\nu\sigma} {{)}}$, the left hand side of \eqref{sqdnzoaj} is written as     
\[
X^{\footnotesize{{{\cal C}}}} \iN {\pmb{\omega}}^{{\tiny\hbox{\sffamily Palatini}}} =  \dd \varkappa  - \Upsilon_\rho  \dd x^{\rho}  -  (1/2)   \epsilon_{IJKL} \epsilon^{\mu\nu\rho\sigma} e^{K}_{\rho} {\big{(}}   \Theta^{IJ}_{\nu\mu} \dd e_\sigma^L    - \Theta^L_{\nu\sigma} \dd \omega_{\mu}^{IJ} +{{(}}\Theta^{IJ}_{\nu\mu}      \Theta^{L}_{\lambda\sigma}  - \Theta^{IJ}_{\lambda\mu} \Theta^{L}_{\nu\sigma} {{)}} \dd x^{\lambda} {\big{)}}.
\]
The {DW} Hamilton equations   \eqref{sqdnzoaj}  are obtained by equalizing the interior product $X^{\footnotesize{{{\cal C}}}}  \iN {\pmb{\omega}}^{{\tiny\hbox{\sffamily Palatini}}} $ with the expression of  $\hbox{d} {\cal{H}}^{{\tiny\hbox{\sffamily Palatini}}}_{{\tiny\hbox{4D}}} (q,p) $ found in \eqref{ham4D}. We obtain 
\begin{equation} 
\left.
\begin{array}{rcl}
\displaystyle    -    \epsilon_{IJKL} \epsilon^{\mu\nu\rho\sigma} e^{K}_{\rho}   \Theta^{IJ}_{\nu\mu}     & =  &  \displaystyle      -    \epsilon_{IJKL} \epsilon^{\mu\nu\rho\sigma} e^{K}_{\rho}    \omega^{J} _{\mu M} \omega_{\nu}^{MI}  ,
\\
 \displaystyle     \epsilon_{IJKL} \epsilon^{\mu\nu\rho\sigma} e^{K}_{\rho}   \Theta^L_{\nu\sigma}    & =  &  \displaystyle    \epsilon_{IJKL}  \epsilon^{\mu \nu \rho\sigma}  e^{K}_\rho  e^{N}_\nu {{\omega_\sigma}^L}_N       , 
\\
\displaystyle
  - \Upsilon_\lambda     -      (1/2)   \epsilon_{IJKL} \epsilon^{\mu\nu\rho\sigma} e^{K}_{\rho} {\big{(}}   \Theta^{IJ}_{\nu\mu}      \Theta^{L}_{\lambda\sigma}  - \Theta^{IJ}_{\lambda\mu} \Theta^{L}_{\nu\sigma}  {\big{)}}  
  & =  &  \displaystyle   0 .
\end{array}
\right.
\end{equation}
Therefore, we obtain the  {DW}  Hamilton system of equations
\begin{equation}\label{qlskdkq000}
\left.
\begin{array}{rcl}
\displaystyle   \epsilon_{IJKL} \epsilon^{\mu\nu\rho\sigma}   e^{K}_{\rho}   {\big{(}}     \Theta^{IJ}_{\mu \nu}  +  \omega^{I}_{\mu M} \omega_{\nu}^{MJ}  {\big{)}}    & =  &  \displaystyle   0   ,
\\ 
\displaystyle       \epsilon_{IJKL} \epsilon^{\mu\nu\rho\sigma} e^{K}_{\rho}   {\big{(}}  \Theta^L_{\nu\sigma}  +  {{\omega}}^{L}_{\nu N}   e^{N}_\sigma  {\big{)}}   & =  &  \displaystyle   0  ,
\\
\displaystyle
  - \Upsilon_ \lambda     -      (1/2)   \epsilon_{IJKL} \epsilon^{\mu\nu\rho\sigma} e^{K}_{\rho} {\big{(}}    \Theta^{IJ}_{\nu\mu}      \Theta^{L}_{\lambda\sigma}  - \Theta^{IJ}_{\lambda\mu} \Theta^{L}_{\nu\sigma}  {\big{)}}  
  & =  &  \displaystyle   0.   
\end{array}
\right.
\end{equation}
 We reproduce the results obtained by   Bruno,   Cianci and  Vignolo \cite{bruno,bruno1}. The equations of motion \eqref{qlskdkq000} are equivalent to the Einstein's equations \eqref{qmamqma01}  written as
\begin{equation}\label{HEPA3D}
 \epsilon_{IJKL} e^J \wedge {F}^{KL}   = 0  ,
\quad \quad   \quad \quad    \epsilon_{IJKL} e^I \wedge \dd_{\omega} e^J  =  0 ,
 \end{equation}
with the additional equation $\partial_\rho \varkappa :=   -      (1/2)   \epsilon_{IJKL} \epsilon^{\mu\nu\rho\sigma} e^{K}_{\rho} {\big{(}}    \Theta^{IJ}_{\nu\mu}      \Theta^{L}_{\lambda\sigma}  - \Theta^{IJ}_{\lambda\mu} \Theta^{L}_{\nu\sigma}   {\big{)}}    $.

 \section{{{$n$-phase space formulation of vielbein gravity }}}\label{sec:section17}

 In this section  we concentrate on the study of the  pre-multisymplectic space defined by the   constraint  $
 {\cal{H}}^{{\tiny\hbox{\sffamily Palatini}}}  (q,p) =   \varkappa - e e^{[\mu}_{I} e^{\nu]}_{J} (    \omega^{J} _{\mu K} \omega_{\nu}^{KI}  ) = 0$. This formulation is related to the   $n$-phase space framework introduced by  Kijowski and Szczyrba    \cite{JK-01,KS0,KS1,KS2}  and further developed by  H\'elein \cite{H-02}. Let us begin with some definitions, see \cite{H-02}:
    \begin{defin}\label{def02}
               A  $n$-multimomentum phase space  (or simply an $n$-phase space)   is a triple  $({\cal M}, {\pmb{\omega}}, \beta)$,  where ${\cal M}$ is a smooth manifold, ${\pmb{\omega}}$  is a closed $(n + 1)$-form and $\beta$ is an everywhere non-vanishing $n$-form.   
\end{defin}
 \begin{defin}\label{def04}
  A pre-multisymplectic manifold   is a pair  ${\big{(}}  {\cal M} , {\pmb{\omega}}{\big{)}}$, where ${\cal M}$ is    a smooth manifold $ {\cal M} $ and     $ {\pmb{\omega}} $ is  a closed  $(n+1)$-form on $ {\cal M}$.   
\end{defin}
In the  {DW} $n$-phase space formulation we express the {dynamical structure} on the {level set} of ${\cal{H}}$ {\em i.e.}   by   means of  the   constraint ${\cal{H}} = 0$.   We can  canonically  construct a  $n$-phase space  $(  {{\cal C}}_\circ   ,{\pmb{\omega}}|_{ {{\cal C}}_\circ   },{  {\beta}} = {  {\Xi}} \iN {\pmb{\omega}} |_{ {{\cal C}}_\circ   } )$, where  $
  {{\cal C}}_\circ  := {\cal{H}}^{-1}(0):= \{ (q,p) \in {{\cal M}}  / \ {\cal{H}}(q,p) =0\}
$ and ${ {\Xi}}$ is a vector field such that $\hbox{d}{\cal{H}}({ {\Xi}}) = 1$.
  The dynamical equations  in the pre-multisymplectic formulation, already presented in   geometrical form \eqref{geo-pre-ham-equ}, are equivalently written as
\begin{equation}\label{ZZZZoo} 
{\forall \Xi \in C^\infty (\pmb{\cal M}, T_m{{{\cal M}}})  ,\quad  \quad \left(\Xi \iN {\pmb{\omega}}^{} \right) {\big{|}}_{{\Gamma}} = 0 \quad \quad     \hbox{and}  \quad   \quad   \beta|_{{\Gamma}} \neq 0},
\end{equation}
see      \cite{FH-01,H-02}. We   denote by  $ {{\cal C}}_{\circ}   $  the hypersuface of constraints contained in the level set ${{\cal C}}_\circ $, {\em i.e.} we have the inclusion of spaces $  {{\cal C}}_{\circ} \subset {{\cal C}}_\circ \hookrightarrow {{\cal M}}_{\hbox{\tiny{\sffamily DW}}} $. Using the primary constraints, the hypersurface of constraints is now
   \begin{equation}\label{kiqlapw-biss}
\left.
\begin{array}{ll}
& 
\displaystyle    {{\cal C}}_{\circ}      := {{\{}}  (x,e,\omega,p) \in {{\cal M}}_{\hbox{\tiny{\sffamily DW}}} \   /   \      \varkappa = e e^{[\mu}_{I} e^{\nu]}_{J} (    \omega^{J} _{\mu K} \omega_{\nu}^{KI}  ),   \quad    {{p}}^{\omega_{\mu}\nu}_{IJ}   = -  $\hbox{\sffamily\bfseries E}$^{[\mu}_{I} e^{\nu]}_{J},   \quad {{p}}^{e_\mu \nu}_{I} = 0   {{\}}}    .
\end{array}
\right.
\end{equation}
Now we give the pre-multisymplectic formulation of dreibein and vierbein gravity.  Note that we introduce  the canonical inclusion $ {\iota_\circ }: {{\cal C}}_{\circ}  {\hookrightarrow}   {{\cal M}}_{\hbox{\tiny{\sffamily DW}}}  $  and    the projection ${\pmb{\pi}}_{\circ} :  {{\cal M}}_{\hbox{\tiny{\sffamily DW}}}  \rightarrow   {{\cal C}}_{\circ}  $. Then, we consider $n$-vector fields $ X^{\footnotesize{{{\cal C}}_{\circ}}} \in \Gamma    ({{\cal C}}_{\circ}, \Lambda^{n} {{\cal C}}_{\circ})$  obtained by the push-forward  $ X^{\footnotesize{{{\cal C}}_{\circ}}} =({\pmb{\pi}}_{\circ} )_{\star}   X^{\tiny{\hbox{\sffamily DW}}} $.

 \subsection{{{Pre-Multisymplectic  formulation  of dreibein gravity}}}\label{subsec:3Dpremulti}

In this section, we consider  the    first order Palatini   functional   of dreibein gravity $ 
{{{\cal S}}}_{\tiny{\hbox{\sffamily Palatini}}} [e,\omega] =  \int  \epsilon_{IJK} e^I \wedge  {F}^{JK} $, where  $ {F}^{JK} = \dd\omega^{JK} + {\omega^J}_L \wedge \omega^{LK} $  is the curvature $2$-form.

 \subsubsection{{{Canonical forms}}}\label{subsec:3Dpremulti1}

Since $e^I = e^I_\mu \dd x^\mu$ and $\omega^{JK} = \omega_{\mu}^{JK} \dd x^\mu$, we obtain the following expression for the Poincar\'e-Cartan $3$-form, identified with the Palatini action $3$-form itself {\em i.e.} $ \epsilon_{IJK} e^I \wedge  {F}^{JK}$:
\begin{equation}\label{fifi000}
{{\theta}}^\circ =       \ \epsilon_{IJK} \epsilon^{\mu\rho\sigma}  {\big{(}} e^I_\mu   \dd  \omega^{JK}_\sigma \wedge {\beta}_\rho    +  e^I_\mu {{\omega_\rho}^J}_L   \omega^{LK}_\sigma   {\beta}   {\big{)}}.
\end{equation}
   We demonstrate \eqref{fifi000} by direct calculation  $    {{\theta}}^\circ   =               \epsilon_{IJK} e^I_\mu  \dd x^\mu \wedge \dd  \omega^{JK}_\sigma \wedge \dd x^\sigma  +        \epsilon_{IJK} e^I_\mu   {{\omega_\rho}^J}_L  \omega^{LK}_\sigma \dd x^\mu \wedge    \dd x^\rho \wedge  \dd x^ \sigma .
$
The Poincar\'e-Cartan $3$-form  is written as $ {{\theta}}^\circ = {{\theta}}^\circ_{\mathfrak{1}} +  {{\theta}}^\circ_{\mathfrak{2}}   $, where 
\begin{equation}\label{aa1}
 {{\theta}}^\circ_{\mathfrak{1}}   =           \epsilon_{IJK} e^I_\mu   {{\omega_\rho}^J}_L  \omega^{LK}_\sigma \dd x^{\mu} \wedge    \dd x^{\rho} \wedge  \dd x^{\sigma}  ,
\quad     
 {{\theta}}^\circ_{\mathfrak{2}} =      \epsilon_{IJK} e^I_\mu  \dd x^{\mu} \wedge \dd \omega^{JK}_\sigma \wedge \dd x^{\sigma}  .  \end{equation}
We re-express the terms $ {{\theta}}^\circ_{\mathfrak{1}}$ and $ {{\theta}}^\circ_{\mathfrak{2}}$ using  the following lemma.
\begin{lemm}\label{lem:hhhhhhhdd01}
  The terms $ {{\theta}}^\circ_{\mathfrak{1}} $ and $ {{\theta}}^\circ_{\mathfrak{2}} $ are given by  {\em
\begin{equation}\label{fifi020}
\left.
\begin{array}{rcl}
 \displaystyle   {{\theta}}^\circ_{\mathfrak{1}}     & = &  \displaystyle       \epsilon_{IJK}  \epsilon^{\mu\rho\sigma}  e^I_\mu   {{\omega_\rho}^J}_L  \omega^{LK}_\sigma \beta , 
  \\
  \displaystyle    {{\theta}}^\circ_{\mathfrak{2}}  & =  &  \displaystyle    -      \epsilon_{IJK} \epsilon^{\mu\rho\sigma} e^I_\mu   \dd  \omega^{JK}_\rho \wedge  \beta_\sigma   .
\end{array}
\right.
\end{equation}}
  \end{lemm}
 {\sffamily Proof}.  The formula for $ {{\theta}}^\circ_{\mathfrak{2}} $  is straightforward. Since   $ {\beta}_1 = \dd x^{2} \wedge \dd x^{3}, \ \beta_2 =  - \dd x^{1} \wedge \dd x^{3}$, and $ \beta_3 = \dd x^{1} \wedge \dd x^{2} $  we find 
$
   \epsilon_{IJK} \epsilon^{\mu\rho\sigma} e^I_\mu   \dd  \omega^{JK}_\rho \wedge \beta_\sigma    = -     \epsilon_{IJK} e^I_\mu  \dd x^{\mu} \wedge \dd  \omega^{JK}_\sigma \wedge \dd x^{\sigma} = -   {{\theta}}^\circ_{\mathfrak{2}}  
$. Now we focus on the first term  $ {{\theta}}^\circ_{\mathfrak{1}} $.   Using $ \displaystyle {\beta} = \dd x^{1} \wedge \dd x^{2} \wedge \dd x^{3}  =  ({1}/{3!}) \epsilon_{\alpha\beta\gamma} \dd x^{\alpha} \wedge \dd x^{\beta} \wedge \dd x^{\gamma}$, we have 
\[
\left.
\begin{array}{rcl}
\displaystyle     \epsilon_{IJK}  \epsilon^{\mu\rho\sigma}  e^I_\mu   {{\omega_\rho}^J}_L  \omega^{LK}_\sigma \beta   & = &  \displaystyle        \epsilon_{IJK}  \epsilon^{\mu\rho\sigma}  e^I_\mu   {{\omega_\rho}^J}_L  \omega^{LK}_\sigma \beta ,
  \\
 \displaystyle     & = &  \displaystyle        ({1}/{3!})  \epsilon_{IJK}  \epsilon^{\mu\rho\sigma}  \epsilon_{\alpha\beta\gamma}  e^I_\mu   {{\omega_\rho}^J}_L  \omega^{LK}_\sigma   \dd x^{\alpha} \wedge \dd x^{\beta} \wedge \dd x^{\gamma}  .
\end{array}
\right.
\]
Using     the formula 
$    \epsilon^{\mu\rho\sigma}  \epsilon_{\alpha\beta\gamma}  =    3! \delta^{[\mu}_\alpha \delta^\rho_\beta \delta^{\sigma]}_\gamma =   ({3!}/{3}) \left( \delta^{\mu}_\alpha \delta^{[\rho}_\beta \delta^{\sigma]}_\gamma     -  \delta^{\rho}_\alpha  \delta^{[\mu}_\beta   \delta^{\sigma]}_\gamma +  \delta^{\sigma}_\alpha   
\delta^{[\mu}_\beta  \delta^{\rho]}_\gamma\right)  $,  
 see  appendix   \ref{app:algebraicvielbein},
 we obtain
 \[
 \left.
\begin{array}{rcl}
\displaystyle    \epsilon_{IJK}  \epsilon^{\mu\rho\sigma}  e^I_\mu   {{\omega_\rho}^J}_L  \omega^{LK}_\sigma \beta   & = &   \displaystyle
      \    \epsilon_{IJK}  e^I_\mu   {{\omega_\rho}^J}_L  \omega^{LK}_\sigma  \dd x^{\mu} \wedge \dd  x^{\rho} \wedge \dd  x^{\sigma}      .
\end{array}
\right.
\]
  Note  that 
$ \displaystyle \epsilon^{\mu\rho\sigma}  \beta =  ({1}/{3!}) \epsilon^{\mu\rho\sigma}  \epsilon_{\alpha\beta\gamma} \dd x^{\alpha} \wedge \dd x^{\beta} \wedge \dd x^{\gamma} = \dd x^{\mu} \wedge \dd x^{\rho} \wedge \dd x^{\sigma} $. Using lemma \ref{lem:hhhhhhhdd01}, the Poincar\'e-Cartan 3-form  \eqref{fifi000} is written as  
$
\displaystyle
{{\theta}}^\circ   =      \epsilon_{IJK} \epsilon^{\mu\rho\sigma}  {\big{(}} e^I_\mu   \dd \omega^{JK}_\sigma \wedge \beta_\rho    +  e^I_\mu {{\omega_\rho}^J}_L   \omega^{LK}_\sigma   \beta   {\big{)}} $.
  We are now interested in the exterior derivative  $\dd {{\theta}}^\circ $.   The exterior derivative   is decomposed in two terms $  \dd {{\theta}}^\circ   =  \dd {{\theta}}^\circ_{\mathfrak{1}}   +  \dd {{\theta}}^\circ_{\mathfrak{2}} $, where 
  \begin{equation}\label{fifi001}
  \left.
\begin{array}{rcl}
 \displaystyle \dd {{\theta}}^\circ_{\mathfrak{1}}     & = &  \displaystyle        \epsilon_{IJK} \epsilon^{\mu\rho\sigma} {{\omega_\rho}^J}_L   \omega^{LK}_\sigma \dd e^I_\mu    \wedge {\beta} +        \epsilon_{IJK} \epsilon^{\mu\rho\sigma} e^I_\mu {\big{(}}  \dd {{\omega_\rho}^J}_L      \omega^{LK}_\sigma +  {{\omega_\rho}^J}_L \dd  \omega^{LK}_\sigma  {\big{)}} \wedge \beta,
 \\
\displaystyle   \dd  {{\theta}}^\circ_{\mathfrak{2}}   & = &  \displaystyle       \epsilon_{IJK} \epsilon^{\mu\rho\sigma} \hbox{d}   e^I_\mu  \wedge  \dd  \omega^{JK}_\sigma \wedge \beta_\rho .  
\end{array}
\right.
\eee
 Note that   the exterior derivative  $\dd {{\theta}}^\circ_{\mathfrak{1}} $ is given as
    \begin{equation}\label{fifi002}
  \left.
\begin{array}{rcl}
 \displaystyle \dd {{\theta}}^\circ_{\mathfrak{1}}     & = &  \displaystyle     \hbox{d}  {{(}}       \epsilon_{IJK} \epsilon^{\mu\rho\sigma}  e^I_\mu {{\omega_\rho}^J}_L   \omega^{LK}_\sigma   \beta    {{)}}  =     \epsilon_{IJK} \epsilon^{\mu\rho\sigma} \hbox{d}  {{(}} e^I_\mu {{\omega_\rho}^J}_L   \omega^{LK}_\sigma  {{)}} \wedge \beta,
 \\
\displaystyle    & = &  \displaystyle       \epsilon_{IJK} \epsilon^{\mu\rho\sigma} {{\omega_\rho}^J}_L   \omega^{LK}_\sigma \dd e^I_\mu    \wedge {\beta} +        \epsilon_{IJK} \epsilon^{\mu\rho\sigma} e^I_\mu {{(}}  \dd {{\omega_\rho}^J}_L      \omega^{LK}_\sigma +  {{\omega_\rho}^J}_L \dd  \omega^{LK}_\sigma  {{)}} \wedge \beta  .
\end{array}
\right.
\eee
where we have used $ \hbox{d}     {{(}} e^I_\mu {{\omega_\rho}^J}_L   \omega^{LK}_\sigma  {{)}} = \hbox{d} {{(}} e^I_\mu {{)}} {{\omega_\rho}^J}_L   \omega^{LK}_\sigma + e^I_\mu \hbox{d} {{(}} {{\omega_\rho}^J}_L {{)}}    \omega^{LK}_\sigma + e^I_\mu{{\omega_\rho}^J}_L \hbox{d} {{(}}  \omega^{LK}_\sigma {{)}}  $.

 Using \eqref{fifi001},   the multisymplectic $4$-form $  {\pmb{\omega}}^\circ   =    \hbox{d} {{\theta}}^\circ   =  \dd {{\theta}}^\circ_{\mathfrak{1}}      + \dd {{\theta}}^\circ_{\mathfrak{2}}       
$ is now written as
   \begin{equation}\label{fifi003}
 \left.
\begin{array}{rcl}
\displaystyle  {\pmb{\omega}}^\circ      & = &  \displaystyle
      \epsilon_{IJK} \epsilon^{\mu\rho\sigma}    \dd e^I_\mu \wedge  \dd  \omega^{JK}_\sigma \wedge {\beta}_\rho    +         \epsilon_{IJK} \epsilon^{\mu\rho\sigma}   ({{\omega_\rho}^J}_L   \omega^{LK}_\sigma  ) \dd e^I_\mu \wedge {\beta}    ,
\\
\displaystyle     &  &  \displaystyle  -     
    \epsilon_{LJK} \epsilon^{\mu\rho\sigma}   (  e^{I}_\mu {{\omega_\sigma}^L}_I  ) \dd \omega_\rho^{JK}  \wedge {\beta}     .
\end{array}
\right.
\eee

 \subsubsection{{{DW Hamilton equations}}}\label{subsec:3Dpremulti2}

In the  pre-multisymplectic formulation, we work on the level set $  {{\cal C}}_\circ := {\cal{H}}^{-1} (0)   $. The submanifold of interest is  the constraint hypersurface ${{\cal C}}_{\circ} \subset  {{\cal C}}_\circ$. The {DW} Hamilton equations  are written  in geometric form   as $ X^{\footnotesize{{{\cal C}}_{\circ}}}  \iN {\pmb{\omega}}^\circ|_{{\Gamma}} = 0$. We  evaluate the interior product of the vector field $X^{\footnotesize{{{\cal C}}_{\circ}}}  $ with  the terms $ \dd {{\theta}}^\circ_{\mathfrak{1}} $ and $ \dd {{\theta}}^\circ_{\mathfrak{2}} $, respectively. First, we find the term
 \begin{equation}\label{fifi012}
\left.
\begin{array}{rcl}
 \displaystyle   X^{\footnotesize{{{\cal C}}_{\circ}}}   \iN   \dd {{\theta}}^\circ_{\mathfrak{1}}       & = &  \displaystyle 
  X^{\footnotesize{{{\cal C}}_{\circ}}}   \iN   {\big{(}}       \epsilon_{IJK} \epsilon^{\mu\rho\sigma}    {{\omega_\rho}^J}_L   \omega^{LK}_\sigma   \dd e^I_\mu \wedge \beta    
-     \epsilon_{LJK} \epsilon^{\mu\rho\sigma} e^{I}_\mu {{\omega_\rho}^L}_I \dd \omega_\sigma^{JK}  \wedge {\beta}  {\big{)}} ,
  \\
\displaystyle     & = &  \displaystyle       \epsilon_{IJK} \epsilon^{\mu\rho\sigma}    {{\omega_\rho}^J}_L   \omega^{LK}_\sigma    {\big{(}}    \beta (X )  \dd e^I_\mu  - (\dd e^I_\mu \wedge  \beta_\lambda) (X) \dd x^\lambda  {\big{)}} ,
\\
\displaystyle     &   &  \displaystyle    -      \epsilon_{IJK} \epsilon^{\mu\rho\sigma}  {\big{(}}   e^L_\mu {{\omega_\rho}^I}_L      \beta (X )   \dd \omega^{JK} _{\sigma}  - ( \dd \omega^{JK} _{\sigma} \wedge  \beta_\lambda) (X) \dd x^\lambda  {\big{)}} ,
\\
\end{array}
\right.
\end{equation}
where we have used ${\beta}(X)    =  1$. Then, we find  the other term
\begin{equation}\label{fifi011}
\left.
\begin{array}{rcl}
 \displaystyle  X^{\footnotesize{{{\cal C}}_{\circ}}}   \iN   \dd {{\theta}}^\circ_{\mathfrak{2}}       & = &  \displaystyle  X^{\footnotesize{{{\cal C}}_{\circ}}}   \iN {\big{(}}     \epsilon_{IJK} \epsilon^{\mu\rho\sigma}  \dd e^I_\mu \wedge  \dd  \omega^{JK}_\sigma \wedge \beta_\rho  {\big{)}} ,
 \\
\displaystyle     & = &  \displaystyle        \epsilon_{IJK} \epsilon^{\mu\rho\sigma}     {\big{(}}   {\big{(}}    \dd  \omega^{JK}_\sigma \wedge {\beta}_\rho   {\big{)}} (X)  \dd e^I_\mu  -  {\big{(}}  \dd e^I_\mu   \wedge {\beta}_\rho  {\big{)}} (X)    \dd  \omega^{JK}_\sigma  {\big{)}}  ,
 \\
\displaystyle     &   &  \displaystyle + 
      \epsilon_{IJK} \epsilon^{\mu\rho\sigma}   {\big{(}}  \dd e^I_\mu \wedge  \dd  \omega^{JK}_\sigma \wedge {\beta}_{\lambda\rho}   {\big{)}}   (X) \dd x^\lambda  {\big{)}}  .
  \end{array}
\right.
\eee
    Now, using the equations \eqref{fifi011} and \eqref{fifi012},    
\[
X^{\footnotesize{{{\cal C}}_{\circ}}}   \iN {\pmb{\omega}}^\circ =      \epsilon_{IJK} \epsilon^{\mu\rho\sigma}   {\Big{(}}   {\big{(}}   \Theta^{JK}_{\rho \sigma} +   {{\omega_\rho}^J}_L   \omega^{LK}_\sigma   {\big{)}}  \dd e^I_\mu  -  {\big{(}}  \Theta^{I}_{\rho\mu} + e^L_\mu {{\omega_\rho}^I}_L {\big{)}} \dd \omega^{JK} _{\sigma}  {\big{)}}  +   {\Upsilon}_{\lambda}  \dd x^\lambda {\Big{)}} ,
\]
with
$  {\Upsilon}_{\lambda}  
=  e^L_\mu {{\omega_\rho}^I}_L  \Theta^{JK}_{\lambda\sigma}   -   {{\omega_\rho}^J}_L   \omega^{LK}_\sigma \Theta^{I}_{\lambda\mu} +  {\big{(}}\Theta^{JK}_{\lambda\sigma}      \Theta^{I}_{\rho\mu}  - \Theta^{JK}_{\rho\sigma} \Theta^{I}_{\lambda\mu} {\big{)}}  
$.  Then, the {DW} Hamilton equations in the pre-multisymplectic formulation ({\em i.e.}  $ X \iN {\pmb{\omega}}^\circ|_{{\Gamma}} = 0$) are given by  
\begin{equation}\label{HE3Dpre}
\left.
\begin{array}{rcc}
\displaystyle   \epsilon_{IJK} \epsilon^{\mu\rho\sigma}   {\big{(}} \Theta^{JK}_{\rho \sigma}  +    {{\omega_\rho}^J}_L   \omega^{LK}_\sigma  {\big{)}} & = & 0  ,
\\
 \displaystyle     \epsilon_{IJK}   \epsilon^{\mu\rho\sigma}  {\big{(}}    \Theta^I_{\rho\mu}  +  e^L_\mu {{\omega_\rho}^I}_L   {\big{)}} & = & 0,
 \\
 \displaystyle    \epsilon_{IJK}   \epsilon^{\mu\rho\sigma}     {\Upsilon}_{\lambda}  & = & 0
,\end{array}
\right.
\end{equation}
 ${\hbox{\sffamily{Remarks}}}$: $(\mathfrak{1})$ Note  that if the first two conditions in $\eqref{HE3Dpre}$ are satisfied, then the last one    is automatically verified.   
\[ 
\left.
\begin{array}{rcl}
\displaystyle      \epsilon_{IJK}   \epsilon^{\mu\rho\sigma}    { \Upsilon}_{\lambda}     & = &  \displaystyle 
 \epsilon_{IJK}   \epsilon^{\mu\rho\sigma}  e^L_\mu {{\omega_\rho}^I}_L  \Theta^{JK}_{\lambda\sigma}   -   \epsilon_{IJK}   \epsilon^{\mu\rho\sigma}  {{\omega_\rho}^J}_L   \omega^{LK}_\sigma \Theta^{I}_{\lambda\mu} +  {\big{(}}\Theta^{JK}_{\lambda\sigma}      \Theta^{I}_{\rho\mu}  - \Theta^{JK}_{\rho\sigma} \Theta^{I}_{\lambda\mu} {\big{)}}  ,
  \\
\displaystyle     & = &  \displaystyle    - \Theta^{I}_{\rho\mu}    \Theta^{JK}_{\lambda\sigma}   + \Theta^{JK}_{\rho\sigma}  \Theta^{I}_{\lambda\mu} +  {\big{(}}\Theta^{JK}_{\lambda\sigma}      \Theta^{I}_{\rho\mu}  - \Theta^{JK}_{\rho\sigma} \Theta^{I}_{\lambda\mu} {\big{)}}  = 0.
\end{array}
\right.
\]
 $(\mathfrak{2})$ The system \eqref{HE3Dpre} reproduces the Einstein's equations and is equivalently written as the following two equations: $ \epsilon_{IJK} {F}^{JK} $ and $  \epsilon_{IJK}  \dd_{\omega} e^I  = 0$.

  {\sffamily{Proof}} Note that  $\displaystyle \epsilon^{\rho\sigma\mu} \beta_ {\mu} = \dd x^{\rho} \wedge \dd x^{\sigma}$, where $ \beta_ {\mu}  \neq 0$. We straightforwardly obtain 
 \[
 \left.
\begin{array}{rcl}
\displaystyle    \epsilon_{IJK} {F}^{JK}    & = &  \displaystyle
\epsilon_{IJK} {\big{(}}  \partial_{[\rho} \omega_{\sigma]}^{JK} + {\omega_{[\rho}^J}_L \omega_{\sigma]}^{LK}   {\big{)}}  \dd x^{\rho} \wedge \dd x^{\sigma}    
 =  \epsilon_{IJK}  \epsilon^{\mu\rho\sigma}  {\big{(}} \Theta^{JK}_{\sigma\rho}  + {\omega_{\rho}^J}_L \omega_{\sigma}^{LK}   {\big{)}}  \beta_ {\mu},
 \\
 \displaystyle     \epsilon_{IJK}   \dd_{\omega} e^K   & = &  \displaystyle   \epsilon_{IJK} {\big{(}}  \partial_{[\rho} e_{\sigma]}^{K} + {\omega_{[\rho}^K}_L e_{\sigma]}^{L}   {\big{)}}  \dd x^{\rho} \wedge \dd x^{\sigma}      =  \epsilon_{IJK}  \epsilon^{\rho\mu \sigma}  {\big{(}} \Theta^{K}_{\mu\rho}  + {\omega_{\rho}^K}_L e_{\mu}^{L}   {\big{)}}  \beta_ {\sigma}   .
\end{array}
\right.
\]

 \subsection{{{Pre-multisymplectic formulation of  vierbein gravity}}}\label{subsec:4Dpremulti}

 In this section we are interested in the pre-multisymplectic formulation of  vierbein gravity.  Here we will reproduce some results found in    Bruno {\em et al.}    \cite{bruno,bruno1} and Rovelli     \cite{Rovelli002,Carlo1}. 
 
 Let us     consider the action  functional 
$   {{\cal S}}_{\tiny{\hbox{\sffamily  Palatini}}} [e,\omega] = (1/2)   \int  \epsilon_{IJKL} e^I \wedge e^J \wedge {F}^{KL}   ,
$ where ${F}^{KL}  = \dd \omega^{KL} + {\omega^K}_M \wedge \omega^{ML} $. 

 \subsubsection{{{Canonical forms}}}\label{subsec:34premulti1}

Since $e^I := e^I_\mu \dd x^{\mu}$ and  $\omega^{KL} := \omega_{\mu}^{KL} \dd x^{\mu}$,   we obtain the following expression for the Poincar\'e-Cartan   $4$-form  ${{\theta}}^\circ   =  (1/2) \left(  \epsilon_{IJKL} \epsilon^{\mu\nu\rho\sigma} e^I_\mu e^J_\nu  \dd  \omega^{KL}_\rho \wedge {\beta}_\sigma  +    \epsilon_{IJKL} \epsilon^{\mu\nu\rho\sigma} e^I_\mu e^J_\nu   {{\omega_\sigma}^K}_M  \omega^{ML}_\rho \beta \right)$. By direct calculation
\begin{equation}\label{fifi017}
\left.
\begin{array}{ccl}
\displaystyle {{\theta}}^\circ  & = & \displaystyle  (1/2) \epsilon_{IJKL} e^I_\mu \dd x^{\mu} \wedge e^J_\nu \dd x^{\nu} \wedge ( \dd ( \omega^{KL}_\rho \dd x^{\rho} ) + {{\omega_\rho}^K}_M  \dd x^{\rho} \wedge \omega^{ML}_\sigma \dd x^{\sigma}  ), 
\\
\displaystyle  & = &  \displaystyle  (1/2)   \epsilon_{IJKL} e^I_\mu e^J_\nu \dd x^{\mu} \wedge  \dd x^{\nu} \wedge \dd  \omega^{KL}_\sigma \wedge \dd x^{\sigma} ,
\\
\displaystyle  &   &  \displaystyle   +(1/2) 
  \epsilon_{IJKL} e^I_\mu e^J_\nu   {{\omega_\rho}^K}_M  \omega^{ML}_\sigma \dd x^{\mu} \wedge  \dd x^{\nu} \wedge  \dd x^{\rho} \wedge  \dd x^{\sigma}  .
\end{array}
\right.
\eee
The Poincar\'e-Cartan form is written  as  $  {{\theta}}^\circ  =  {{\theta}}^\circ_{\mathfrak{1}}    +   {{\theta}}^\circ_{\mathfrak{2}}$,  where 
\begin{equation}\label{aa1}
\left.
\begin{array}{rcl}
\displaystyle    {{\theta}}^\circ_{\mathfrak{1}}    & = & (1/2)   \epsilon_{IJKL} e^I_\mu  e^J_\nu  \dd x^{\mu} \wedge \dd x^{\nu} \wedge \dd  \omega^{JK}_\sigma \wedge \dd x^{\sigma}   ,
\\
 \displaystyle  {{\theta}}^\circ_{\mathfrak{2}}    & = &  (1/2)    \epsilon_{IJKL} e^I_\mu e^J_\nu  {{\omega_\rho}^K}_M  \omega^{ML}_\sigma \dd x^{\mu} \wedge    \dd x^{\nu} \wedge  \dd x^{\rho} \wedge  \dd x^{\sigma}  .
\end{array}
\right.
\end{equation}
 Since
 $ \displaystyle  \epsilon^{\mu\nu\rho\sigma}   {\beta}_\sigma  = \left( { 1! (4-1)! } / {3!} \right)  \delta_{\alpha}^{[\mu} \delta_\beta^\nu \delta_{\gamma}^{\rho]}  \dd x^{\alpha} \wedge \dd x^{\beta} \wedge \dd x^{\gamma} $, 
we   obtain $ \epsilon^{\mu\nu\rho\sigma}   {\beta}_\sigma =  \dd x^{\mu} \wedge \dd x^{\nu} \wedge \dd x^{\rho}$. 
Then, 
 $\dd x^{\mu} \wedge  \dd x^{\nu} \wedge \dd  \omega^{KL}_\sigma \wedge \dd x^{\sigma}  =  \epsilon^{\mu\nu\rho\sigma}  \dd  \omega^{KL}_\rho \wedge \vol_\sigma$. Hence,
 \begin{equation}\label{fifi030}
 \epsilon_{IJKL} e^I_\mu  e^J_\nu  \dd x^{\mu} \wedge \dd x^{\nu} \wedge \dd  \omega^{JK}_\sigma \wedge \dd x^{\sigma}    = (1/2)   \epsilon_{IJKL} \epsilon^{\mu\nu\rho\sigma} e^I_\mu e^J_\nu  \dd  \omega^{KL}_\rho \wedge {\beta}_\sigma  
 = {{\theta}}^\circ_{\mathfrak{1}}   .
\end{equation}
 Note that  the volume form  ${\beta} = \dd x^{1} \wedge \dd x^{2} \wedge \dd x^{3} \wedge \dd x^{4} $ is equivalently written $\beta =  ({1}/{4!}) \epsilon_{\alpha\beta\gamma\delta} \dd x^{\alpha} \wedge \dd x^{\beta} \wedge \dd x^{\gamma} \wedge \dd x^{\delta}$,  then    the second term in   \eqref{fifi017} is written as  
\begin{equation}\label{azdzzdzzddzda1}
\left.
\begin{array}{rcl}
\displaystyle    \epsilon_{IJKL} \epsilon^{\mu\nu\rho\sigma} e^I_\mu e^J_\nu   {{\omega_\sigma}^K}_M  \omega^{ML}_\rho \beta  & = &   
(1/2)    \epsilon_{IJKL} e^I_\mu e^J_\nu  {{\omega_\rho}^K}_M  \omega^{ML}_\sigma \dd x^{\mu} \wedge    \dd x^{\nu} \wedge  \dd x^{\rho} \wedge  \dd x^{\sigma} =  {{\theta}}^\circ_{\mathfrak{2}}  
\end{array}
\right.
\nonumber
\end{equation}
where we have   used the formula \eqref{algebraicvvveilbe} for the expression $ \epsilon^{\mu\nu\rho\sigma}  \epsilon_{\alpha\beta\gamma\delta}   $.
   
 Let us  compute the pre-multisymplectic $5$-form 
$
{\pmb{\omega}}^\circ = \hbox{d} {{\theta}}^\circ
$:
\[
\left.
\begin{array}{rcl}
   \displaystyle     {\pmb{\omega}}^\circ  & = &  \displaystyle    \epsilon_{IJKL} \epsilon^{\mu\nu\rho\sigma} e^I_\mu  {\big{(}}  \dd e^J_\nu  \wedge \dd  \omega^{KL}_\rho \wedge \beta_\sigma    +   {{\omega}}_{\sigma M}^K  \omega^{ML}_\rho \dd  e^J_\nu    \wedge \beta   {\big{)}} ,
\\
\displaystyle     &  &  \displaystyle 
+ (1/2)  \epsilon_{IJKL} \epsilon^{\mu\nu\rho\sigma} e^I_\mu e^J_\nu \omega^{ML}_\rho  \hbox{d}  {\big{(}} {{\omega}}_{\sigma M}^K  {\big{)}}     \wedge {\beta}  + (1/2) \epsilon_{IJKL} \epsilon^{\mu\nu\rho\sigma} e^I_\mu e^J_\nu   {{\omega_\sigma}^K}_M \hbox{d}  {\big{(}}   \omega^{ML}_\rho  {\big{)}}   \wedge {\beta} ,
\\
\displaystyle     &  = &  \displaystyle   \epsilon_{IJKL} \epsilon^{\mu\nu\rho\sigma} e^I_\mu  {\big{(}}  \dd e^J_\nu  \wedge \dd  \omega^{KL}_\rho \wedge \beta_\sigma    +   {{\omega}}_{\sigma M}^K   \omega^{ML}_\rho \dd e^J_\nu    \wedge \beta   {\big{)}} ,
\\
\displaystyle     &  &  \displaystyle  +  \epsilon_{IJKL} \epsilon^{\mu\nu\rho\sigma} e^I_\mu e^J_\nu    {{\omega_\sigma}^K}_M \hbox{d}   \omega^{ML}_\rho   \wedge \beta .
\end{array}
\right.
\]
Using  the algebraic relation 
   $   \epsilon^{\mu\nu\rho\sigma}   \epsilon_{IJKL} e^{I}_\mu e^{J}_\nu {{\omega_\sigma}^K}_M \dd \omega_\rho^{ML} = -  \epsilon^{\mu\nu\rho\sigma}   \epsilon_{INKL} e^{I}_\mu  e^{J}_\nu {{\omega_\sigma}^N}_J \dd \omega_\rho^{KL}$, the  pre-multisymplectic $5$-form is written as
    \bee\label{rayman01}
  \left.
\begin{array}{rcl}
\displaystyle    {\pmb{\omega}}^\circ     & = &  \displaystyle 
   \epsilon_{IJKL} \epsilon^{\mu\nu\rho\sigma} e^I_\mu   \dd e^J_\nu  \wedge \dd  \omega^{KL}_\rho \wedge \beta_\sigma       +    \epsilon_{IJKL} \epsilon^{\mu\nu\rho\sigma} e^I_\mu     {{\omega_\sigma}^K}_M  \omega^{ML}_\rho \dd e^J_\nu    \wedge \beta ,
   \\
\displaystyle     &   &  \displaystyle    - \epsilon^{\mu\nu\rho\sigma}   \epsilon_{INKL} e^{I}_\mu  e^{J}_\nu {{\omega_\sigma}^N}_J \dd \omega_\rho^{KL}     \wedge \beta     .
\end{array}
\right.
\eee

 \subsubsection{{{DW Hamilton equations}}}\label{subsec:4Dpremulti2}

In the  pre-multisymplectic setting  we work with the constraint ${\cal{H}} = 0$.  The dynamics is expressed on the level set $  {{\cal C}}_\circ := {\cal{H}}^{-1} (0)   $    and the {DW} Hamilton equations are written as
\begin{equation}\label{fifi035}
  X^{\footnotesize{{{\cal C}}_{\circ}}}   \iN {\pmb{\omega}}^\circ  {\big{|}}_{{\Gamma}}  = 0 .
\end{equation}
We now evaluate, for  vierbein gravity, the interior product of the multivector field $X^{\footnotesize{{{\cal C}}_{\circ}}}   \in \Lambda^{4} T{{\cal C}}_\circ $, with  the   three terms in \eqref{rayman01}. We choose  a $4$-vector $X^{\footnotesize{{{\cal C}}_{\circ}}}   = X^{\footnotesize{{{\cal C}}_{\circ}}}_1 \wedge X^{\footnotesize{{{\cal C}}_{\circ}}}_2 \wedge X^{\footnotesize{{{\cal C}}_{\circ}}}_3 \wedge X^{\footnotesize{{{\cal C}}_{\circ}}}_4$, where for any $1 \leq \alpha \leq 4$,  the vector field  $X_{\alpha} \in \mathfrak{X}^{1} ({{\cal C}}_\circ) $ is    
\[      
X^{\footnotesize{{{\cal C}}_{\circ}}}_\alpha   = \frac{\partial}{\partial x^ \alpha} + \Theta_{\alpha {\mu }}^{I}   \frac{\partial}{\partial e^I_{\mu }} + \Theta_{\alpha {\mu }}^{IJ}   \frac{\partial}{\partial \omega^I_{\mu }}   .
\]
The left side of   \eqref{fifi035} is written  as
\[
\left.
\begin{array}{rcl}
 \displaystyle   X^{\footnotesize{{{\cal C}}_{\circ}}}   \iN  {\pmb{\omega}}^\circ     & = &  \displaystyle     - \epsilon^{\mu\nu\rho\sigma}   \epsilon_{INKL} e^{I}_\mu  e^{J}_\nu {{\omega_\sigma}^N}_J \dd \omega_\rho^{KL}     \wedge \beta      -    \epsilon_{IJKL} \epsilon^{\mu\nu\rho\sigma} e^I_\mu     ( \dd e^J_\nu    \wedge {\beta}_\sigma)  (X)    ,
 \\
 \displaystyle     &  &  \displaystyle    -     \epsilon_{INKL} \epsilon^{\mu\nu\rho\sigma}  e^{I}_\mu  e^{J}_\nu {{\omega_\sigma}^N}_J  {\big{(}}   (\beta )(X)  \dd  \omega^{KL}_\rho    + (  \dd  \omega^{KL}_\rho \wedge \beta_\lambda )(X)   \dd  x^{\lambda} {\big{)}} ,
 \\
 \displaystyle     &  &  \displaystyle 
 +  \epsilon_{IJKL} \epsilon^{\mu\nu\rho\sigma}   e^I_\mu \left(  ( \dd e^J_\nu  \wedge  \dd  \omega^{KL}_\rho \wedge  \beta_{\lambda\sigma} )   (X)  \dd  x^{\lambda}  \right) ,
\\
\displaystyle     &  = &  \displaystyle     \epsilon_{IJKL} \epsilon^{\mu\nu\rho\sigma}  e^I_\mu  {\big{(}}  (\Theta^{KL}_{\sigma\rho} +  {{\omega_\sigma}^K}_M  \omega^{ML}_\rho ) \dd e^J_\nu - (e^{I}_\mu  e^{N}_\nu {{\omega_\sigma}^J}_N  + \Theta^{I}_{\sigma\nu}) \dd  \omega^{KL}_\rho  +  {\Upsilon}_{\lambda}    \dd x^{\lambda}{\big{)}} ,
\end{array}
\right.
\]
  where
$  {\Upsilon}_{\lambda}  
=       e^{N}_\nu {{\omega_\rho}^J}_N   \Theta^{KL}_{\lambda\sigma} -    {{\omega_\sigma}^K}_M  \omega^{ML}_\rho   \Theta^{J}_{\lambda\nu} +  {\big{(}}\Theta^{KL}_{\lambda\sigma}      \Theta^{J}_{\rho\nu}  - \Theta^{KL}_{\rho\sigma} \Theta^{J}_{\lambda \nu} {\big{)}}  
$. 
 In the pre-multisymplectic setting we find the {DW} Hamilton equations for the Palatini action 
\begin{equation}\label{HE3D} 
\left.
\begin{array}{ccc}
\displaystyle   \epsilon_{IJKL} \epsilon^{\mu\nu\rho\sigma} e^I_\mu  {\big{(}}\Theta^{KL}_{\sigma\rho} +  {{\omega_\sigma}^K}_M  \omega^{ML}_\rho {\big{)}} & = & 0  ,
\\ 
 \displaystyle      \epsilon_{IJKL} \epsilon^{\mu\nu\rho\sigma} e^I_\mu    {\big{(}}  \Theta^{J}_{\sigma\nu}     +       e^{N}_\nu {{\omega_\sigma}^J}_N    {\big{)}} & = & 0 ,
 \\
  \displaystyle    \epsilon_{IJKL}   \epsilon^{\mu\nu\rho\sigma}     {\Upsilon}_{\lambda}  & = & 0.
\end{array}
\right.
\end{equation}
Analogously to the dreibein case, see the end of the section \ref{subsec:3Dpremulti2}, we obtain the Einstein's system of equations in term of  differential forms. We have,  see also \eqref{HEPA3D},    $ \epsilon_{IJKL} e^{I}  \wedge \dd_{\omega} e^J  = 0 $ and $  \epsilon_{IJKL} e^I \wedge e^J \wedge {F}^{KL} = 0$, together with the equation $  \epsilon_{IJKL}   \epsilon^{\mu\nu\rho\sigma}     {\Upsilon}_{\lambda} = 0$.
      
\section{{{Hamiltonian $(n-1)$-forms and brackets}}}\label{sec:section18}

   \subsection{{{Hamiltonian $(n-1)$-forms, homotopy Lie algebra}}}\label{subsec:saison0010}
 
 We begin  this section with the definition of Hamiltonian $(n-1)$-forms and their  related   Hamiltonian vector fields, {\em c.f.}     
Cari\~{n}ena, Crampin and Ibort \cite{cari},      Kanatchikov  \cite{Kana-01,Kana-02,Kana-014}, Forger {\em et al.}  \cite{Forger0010,Forger015,Forger011}, H\'elein and Kouneiher \cite{HK-01,HK-02,HK-03}.  
 \begin{defin}\label{def:bracket} 
 Let $({\cal M} , {\pmb{\omega}} ) $ be a   multisymplectic manifold. An $(n-1)$-form ${\pmb{\varphi}}$ is called a  Hamiltonian   $(n-1)$-form if and only if there exists $\Xi_{\pmb{\varphi}} \in \mathfrak{X} ({\cal M})$ such that  {\em$\Xi_{\pmb{\varphi}} \iN {\pmb{\omega}} + \dd {\pmb{\varphi}} = 0 $.}
\end{defin}
We denote by ${\Omega}_{\tiny{\hbox{\sffamily Ham}}}^{{{n-1}}} ({\cal M})$ the set of all Hamiltonian   $(n-1)$-forms. 
 For any $  {\pmb{\varphi}}, {\pmb{\rho}} \in {\Omega}_{\tiny{\hbox{\sffamily Ham}}}^{{{n-1}}} ({\cal M}) $, let us   define the   bracket  
\begin{equation}\label{gigi0}
{\big{\{}} {\pmb{\varphi}} , {\pmb{\rho}} {\big{\}}} := \Xi_{\pmb{\varphi}} \wedge \Xi_{\pmb{\rho}} \iN {\pmb{\omega}} =    \Xi_{\pmb{\varphi}}   \iN \hbox{d} {\pmb{\rho}} = - \Xi_{\pmb{\rho}}   \iN \hbox{d} {\pmb{\varphi}},
\end{equation}
where  ${{\{}} {\pmb{\varphi}} , {\pmb{\rho}} {{\}}} \in  {\Omega}_{\tiny{\hbox{\sffamily Ham}}}^{{{n-1}}} ({\cal M})$. For any form ${\pmb{\eta}} \in \Omega^{\ast} ({\cal M})$ and any decomposable multivector field $\Xi := \Xi_1 \wedge \cdots \wedge \Xi_n \in \mathfrak{X}^{n} ({\cal M})$, we have $\Xi  \iN {\pmb{\eta}}  = (\Xi_1 \wedge \cdots \wedge \Xi_n) \iN {\pmb{\eta}} := \Xi_{n} \iN \cdots \iN \Xi_1 \iN {\pmb{\eta}}  $.   This definition is the natural analogue of  the  Poisson bracket in  classical mechanics.    The bracket  defined in \ref{def:bracket} satisfies the antisymmetry property: ${{\{}}  {\pmb{\varphi}}  ,  {\pmb{\rho}} {\big{\}}} + {\big{\{}}  {\pmb{\rho}}  ,  {\pmb{\varphi}} {{\}}} = 0 $, but  the  Jacobi condition is only satisfied modulo an exact term, see \cite{HK-01,Rogers}. For any $  {\pmb{\varphi}}, {\pmb{\rho}}, {\pmb{\eta}} \in \textswab{P}_{\circ}^{n-1} ({\cal M})$ 
\bee\label{antyJacobi01}
{\{}  \{ {\pmb{\rho}} , {\pmb{\eta}}  \} , {\pmb{\varphi}} {{\}}} +
  {{\{}}  \{ {\pmb{\eta}}   , {\pmb{\varphi}} \} , {\pmb{\rho}} {{\}}}+
    {{\{}}  \{  {\pmb{\varphi}} , {\pmb{\rho}} \} , {\pmb{\eta}}   {{\}}}    
    = \hbox{d} (\Xi_{{\pmb{\varphi}}} \wedge \Xi_{{\pmb{\rho}}} \wedge \Xi_{{\pmb{\eta}} } \iN {\pmb{\omega}}).
\eee
Using  the Cartan formula, {\em i.e.}   ${\cal{L}}_{\Xi} {\pmb{\omega}}  = \hbox{d} (\Xi \iN {\pmb{\omega}}) + \Xi \iN \hbox{d} {\pmb{\omega}} = 0$, we    define  a locally Hamiltonian vector field of $({\cal M}, {\pmb{\omega}})$ to be  a vector field $\Xi \in \Gamma({\cal M} , T{\cal M})$, such that 
$
{\cal{L}}_{\Xi} {\pmb{\omega}} = 0  
$ (since $\hbox{d} {\pmb{\omega}} = 0$).   We are looking for vector fields $\Xi \in \Gamma({\cal M} , T{\cal M}) $, such that $\hbox{d} (\Xi \iN {\pmb{\omega}})= 0$.  We denote by $ \mathfrak{X}_{\tiny{\hbox{\sffamily Ham}}}^{{{1}}} ({\cal M})   $ the set of locally Hamiltonian vector fields of the multisymplectic manifold $( {\cal M} , {\pmb{\omega}} ) $, {\em i.e.}
 \bee
\mathfrak{X}_{\tiny{\hbox{\sffamily Ham}}}^{{{1}}} ({\cal M})    = 
\left\{  \Xi \in \Gamma({\cal M} , T{\cal M}) \  /   \  \hbox{d} (\Xi \iN {\pmb{\omega}})  = 0   \right\}  =  \left\{ \Xi \in \Gamma({\cal M} , T{\cal M}) \  /   \ {\cal{L}}_{\Xi} {\pmb{\omega}} = 0  \right\}.
\eee 
  Although antisymmetric, the bracket   \eqref{gigi0}  nevertheless fails   to respect the  Jacobi property which is necessary  to obtain a strict     Lie    algebraic structure. Thus, $ {\big{(}} {\Omega}_{\tiny{\hbox{\sffamily Ham}}}^{{{n-1}}} ({\cal M})   ,  {\big{\{}}  \cdot , \cdot {\big{\}}}    {\big{)}}  $  is not a Lie algebra. The fact that this bracket satisfies the Jacobi identity only up to an exact form  was already noted by         Goldschmidt and Sternberg in \cite{HGSS}. This co-cycle obstruction reveals  the  connection with homotopy Lie algebra, see \cite{Lada,Lada1}.  We refer to the paper by   Baez and {\em al.} \cite{Baez1,Baez2}, where the Lie $2$-algebra is  used to
describe the dynamics of the classical bosonic string.  More generally,  the relation between {MG}  and  $L_{\infty}$-algebra is found in  Rogers \cite{Rogers,Rogers1}, Richter \cite{Richter1,Richter2}, and   Vitagliano \cite{Vit0}, where a $L_{\infty}$-algebra is a chain complex equipped with an  antisymmetric bracket operation that satisfies the Jacobi identity up to coherent homotopy  \cite{Baez1,Rogers1}.  
   
      \subsection{{{Hamiltonian forms, graded Poisson bracket}}}\label{subsec:saison001}
        
    In Kanatchikov's approach \cite{Kana-01,Kana-02,Kana-014,Kana-015}   the polysymplectic form $  {\pmb{\omega}}^{\hbox{\tiny\sffamily V}} =  \dd {{p}}^{\mu}_{i} \wedge \dd y^{i} \wedge \vol_\mu  $ is used to construct   the graded Poisson bracket    on    forms of arbitrary degrees. Let $ {\pmb{\varphi}} \in {\Omega}_{\tiny{\hbox{\sffamily Ham}}}^{{{p}}} ({\cal M})   $, ${\pmb{\rho}} \in {\Omega}_{\tiny{\hbox{\sffamily Ham}}}^{{{q}}} ({\cal M})  $ and  ${\pmb{\eta}} \in {\Omega}_{\tiny{\hbox{\sffamily Ham}}}^{{{r}}} ({\cal M}) $ (where $0 \leq p,q,r \leq n-1$) be Hamiltonian  forms, as defined in  \cite{Kana-01},  of degrees $\hbox{\sffamily deg} ({\pmb{\varphi}}) := p$, $\hbox{\sffamily deg} ({\pmb{\rho}}) := q$, and $\hbox{\sffamily deg} ({\pmb{\eta}}) := r$, respectively.    The  graded Poisson bracket on Hamiltonian $(p-1)$-forms  of arbitrary degrees   is 
 \bee\label{graded001} 
{{\{}}  \overset{p}{\pmb{\varphi}}   ,  \overset{q}{\pmb{\rho}}   {{\}}}  = 
(-1)^{n-p}  \Xi_{\pmb{\varphi}}   \iN   \Xi_{\pmb{\rho}}   \iN  \pmb{\omega}^{\hbox{\tiny\sffamily V}}  = (-1)^{n-p}  \Xi_{\pmb{\varphi}}   \iN  \dd^{\hbox{\tiny\sffamily V}}    \overset{q}{\pmb{\rho}} ,
\eee
where $  \dd^{\hbox{\tiny\sffamily V}}  $ is the vertical exterior derivative and    the respective  Hamiltonian multivector fields related to ${\pmb{\varphi}}  $  and  $ {\pmb{\rho}}  $  are  $ \Xi_{\pmb{\varphi}}  \in {\mathfrak{X}}^{ n-p  }_{\tiny{\hbox{\sffamily Ham}}} ( {\cal M} )    ,    \Xi_{\pmb{\rho}}    \in {\mathfrak{X}}^{ n-q  }_{\tiny{\hbox{\sffamily Ham}}} ( {\cal M} )     $. The   graded  Poisson bracket \eqref{graded001}  is graded antisymmetric, {\em i.e.}   
 \bee 
\left.
\begin{array}{ccl}
  \displaystyle    {{\{}}  \overset{p}{\pmb{\varphi}}   ,  \overset{q}{\pmb{\rho}}   {{\}}}    
   & = &   \displaystyle  - (-1)^{ (n-p  - 1) (n-q-1) }
 {\big{\{}}   \overset{q}{\pmb{\rho}}  , \overset{p}{\pmb{\varphi}}    {\big{\}}}  ,
  \end{array}
\right.
\eee
and satisfies the graded Jacobi identity 
\bee
(-1)^{  d_{\pmb{\varphi}} d_{\pmb{\eta}}   } {{\{}} \overset{p}{\pmb{\varphi}}  {{\{}}  \overset{q}{ {\pmb{\rho}}  }  ,  \overset{r}{ {\pmb{\eta}}  }  {{\}}}    {{\}}}    + 
(-1)^{ d_{{\pmb{\rho}}} d_{{\pmb{\varphi}}}    }
{{\{}}  \overset{p}{ {\pmb{\rho}}  } {{\{}}  \overset{q}{ {\pmb{\eta}}  }  ,    \overset{q}{\pmb{\varphi}}   {{\}}}    {{\}}}    + 
(-1)^{ d_{{\pmb{\eta}}}   d_{{\pmb{\rho}}}    }
{{\{}}  \overset{p}{ {\pmb{\eta}}  } {{\{}}  \overset{q}{ {\pmb{\varphi}}  }  ,  \overset{r}{ {\pmb{\rho}}  }  {{\}}}    {{\}}}      
= 0,
\eee
where we have denoted by  $d_{\pmb{\varphi} } := n - \hbox{\sffamily deg} ({\pmb{\varphi}}) -1$, $d_{\pmb{\eta} } := n - \hbox{\sffamily deg} ({\pmb{\eta}}) -1$,  and $d_{\pmb{\rho} } := n - \hbox{\sffamily deg} ({\pmb{\rho}}) -1$. Note that $\hbox{\sffamily deg} ({\pmb{\eta}})$ denote   the degree  of the  Hamiltonian form ${\pmb{\eta}}$.    The Poisson bracket of Hamiltonian forms  is obtained using  the Schouten-Nijenhuis bracket ${\pmb{{[}}}  ,  {\pmb{{]}}}  $ of the  related 
Hamiltonian multivector fields 
$
 - \dd  {{\{}}  {\pmb{\varphi}}   ,  {\pmb{\rho}}   {{\}}}  = {\pmb{{[}}}   \Xi_{\pmb{\varphi}}   ,   \Xi_{\pmb{\rho}}   {\pmb{{]}}}    \iN {\pmb{\omega}}^{\hbox{\tiny\sffamily V}} 
 $.
 The  Schouten-Nijenhuis bracket, see \cite{Nij01,Nij02,Schouten01}, {\em i.e.}  a  bilinear map  $ {\pmb{{[}}}  ,  {\pmb{{]}}}    : {\mathfrak{X}}^{\ast}_{\tiny{\hbox{\sffamily Ham}}} (  {\cal M}  ) \times {\mathfrak{X}}^{\ast}_{\tiny{\hbox{\sffamily Ham}}} ( {\cal M} )  \rightarrow {\mathfrak{X}}^{\ast}_{\tiny{\hbox{\sffamily Ham}}}  (  {\cal M}  )$, that obeys  the graded antisymmetric property  and the graded Leibniz rule
  \bee\label{qdksqflgj001}
\left.
\begin{array}{ccl}
   {\pmb{{[}}}   {\Xi}_{\mathfrak{1}}  ,   {\Xi}_{\mathfrak{2}}  {\pmb{{]}}}  & = & \displaystyle   - (-1)^{ (  \hbox{\sffamily\footnotesize deg} ({\Xi}_{1})  -1) (   \hbox{\sffamily\footnotesize deg} ({\Xi}_{2})  -1)   }   {\pmb{{[}}}  {\Xi}_{\mathfrak{2}}   ,  {\Xi}_{\mathfrak{2}}  {\pmb{{]}}},
    \\ 
  \displaystyle  {\pmb{{[}}}  {\Xi}_{\mathfrak{1}}   ,  {\Xi}_{\mathfrak{2}} \wedge {\Xi}_{\mathfrak{3}}   {\pmb{{]}}}    & = &  \displaystyle 
   {\pmb{{[}}}  {\Xi}_{\mathfrak{1}}   ,  {\Xi}_{\mathfrak{2}}   {\pmb{{]}}}  \wedge {\Xi}_{\mathfrak{3}}  
 + (-1)^{ (  \hbox{\sffamily\footnotesize deg} ({\Xi}_{1})  -1)  \hbox{\sffamily\footnotesize deg} ({\Xi}_{2})  } {\Xi}_{\mathfrak{2}} \wedge    {\pmb{{[}}}  {\Xi}_{\mathfrak{1}}   ,  {\Xi}_{\mathfrak{3}}   {\pmb{{]}}},
  \end{array}
\right.
\eee
 as well as the  graded Jacobi identity  
  \bee\label{qdksqflgj002}
\left.
\begin{array}{ccl}  
   \displaystyle  0    &  = &  \displaystyle
 (-1)^{d_{\mathfrak{1}}d_{\mathfrak{3}}}    {\pmb{{[}}}       {\Xi}_{\mathfrak{1}}   ,  {\pmb{{[}}}  {\Xi}_{\mathfrak{2}} ,   {\Xi}_{\mathfrak{3}}   {\pmb{{]}}} {\pmb{{]}}}    +
  (-1)^{d_{\mathfrak{2}}d_{\mathfrak{3}}}     {\pmb{{[}}}    {\Xi}_{\mathfrak{3}}  ,  {\pmb{{[}}}       {\Xi}_{\mathfrak{1}}  ,   {\Xi}_{\mathfrak{2}}   {\pmb{{]}}}  {\pmb{{]}}} 
      + 
 (-1)^{d_{\mathfrak{1}}d_{\mathfrak{2}}}     {\pmb{{[}}}   {\Xi}_{\mathfrak{2}}  ,    {\pmb{{[}}}   {\Xi}_{\mathfrak{3}} ,     {\Xi}_{\mathfrak{1}}   {\pmb{{]}}}  {\pmb{{]}}}  ,  \end{array}
\right.
 \eee
where  $d_{i} := \hbox{\sffamily deg} ({\Xi}_{i}) - 1 $ and $ \hbox{\sffamily deg} ({\Xi}_{i}) $ denote   the degrees of the respective multivector fields.
On vector fields, the Schouten-Nijenhuis bracket reduces to    the standard Lie bracket.         However,  the exterior product of two Hamiltonian forms  $\overset{p}{\pmb{\varphi}}   \wedge \overset{q}{ {\pmb{\rho}}}$ is not Hamiltonian in general.  Kanatchikov introduces   the co-exterior product $\bullet$   of horizontal forms  $ {\pmb{\varphi}} \bullet  {\pmb{\rho}} = \star^{-1} (\star {\pmb{\varphi}}     \wedge \star {\pmb{\rho}}) $,  see \cite{Kana-014}.      
      The space of Hamiltonian forms is closed with respect to the co-exterior product. Thus,    $ \textswab{A}^{{\hbox{\tiny\sffamily Poly}}}_{{\hbox{\tiny\sffamily DW}}}  = {{\{}}   {\Omega}_{\tiny{\hbox{\sffamily Ham}}}^{{{\ast}}}  (  {\cal M}^{{\hbox{\tiny\sffamily Poly}}}_{{\hbox{\tiny\sffamily DW}}}  ) ,   {{\{}}  , {{\}}}  , \bullet {{\}}}   $ is   a Gerstenhaber algebra \cite{Gerstenhaber01}.  
As an illustration   of the  use of the higher dimensional  algebraic  structures   in   field theory we refer to the  example of  the classical string.  The {DW} Hamiltonian formulation  of  Nambu-Goto string, using the polysymplectic formalism and   the Poisson-Gerstenhaber algebra \cite{Kana-015}, is given by   Kanatchikov in \cite{Kana-01,Kana-02}.
  
In  section \ref{subsec:SAOFex} and \ref{subsec:saison002} we will  consider   Hamiltonian  $(n-1)$-forms  ${\pmb{\varphi}} = {\pmb{\varphi}}^{\mu} \vol_{\mu} \in \Omega^{n-1}_{\hbox{\sffamily\tiny Ham}} ({\cal M} )$. In that case, the graded Poisson  structure reduces to a Poisson structure.  For any ${\pmb{\varphi}}, {\pmb{\rho}} \in \Omega^{n-1}  ({\cal M}_{{\hbox{\tiny\sffamily DW}}}^{{\hbox{\tiny\sffamily Poly}}})$, the  bracket    is  defined as   ${{\{}}  {\pmb{\varphi}}   ,  {\pmb{\rho}}   {{\}}}  := 
- \Xi_{\pmb{\varphi}}   \iN   \Xi_{\pmb{\rho}}   \iN  {\pmb{\omega}}^{\hbox{\tiny\sffamily V}}  = (-1)^{n-r}  \Xi_{\pmb{\varphi}}   \iN     \dd^{\hbox{\tiny\sffamily V}}    {\pmb{\rho}}   
$,  where $ \Xi_{\pmb{\varphi}}   ,    \Xi_{\pmb{\rho}}  \in   \mathfrak{X}_{\tiny{\hbox{\sffamily Ham}}}^{{{1}}} ({\cal M})  $. The Poisson bracket has the   antisymmetry  property ${{\{}}  {\pmb{\varphi}}  ,  {\pmb{\rho}} {{\}}} + {{\{}}  {\pmb{\rho}}  ,  {\pmb{\varphi}} {{\}}} = 0 $ and it  satisfies the Jacobi identity  $ {{\{}}  {\pmb{\varphi}}  {{\{}}   { {\pmb{\rho}}  }  ,   { {\pmb{\eta}}  }  {{\}}}    {{\}}}    +  {{\{}}  { {\pmb{\rho}}  } {{\{}}   { {\pmb{\eta}}  }  ,     {\pmb{\varphi}}   {{\}}}    {{\}}}    +  
{{\{}}   { {\pmb{\eta}}  } {{\{}}   { {\pmb{\varphi}}  }  ,  { {\pmb{\rho}}  }  {{\}}}    {{\}}}      
= 0$. 
  
  \subsection{{{Hamiltonian  $(n-1)$-forms}}}\label{subsec:SAOFex}
   In this section we consider     Hamiltonian  $(n-1)$-forms and their related Hamiltonian vector fields  on the {DW} manifold $   {{\cal M}}_{\tiny{\hbox{\sffamily DW}}} $.  We will work  with the multisymplectic manifold  $( {{\cal M}}_{\tiny{\hbox{\sffamily DW}}}  ,  {\pmb{\omega}}^{\tiny{\hbox{\sffamily DW}}} ) =  ( {{\cal M}}_{\tiny{\hbox{\sffamily DW}}}  ,  \iota_{\mathfrak{1}}^{\star} {\pmb{\omega}}  )  $ and with the pair $({{\cal C}} , \iota^{\star} \Omega^{\hbox{\sffamily\tiny DW}})$, respectively.
   
First, we  use  the results of H\'{e}lein and Kouneiher \cite{HK-03}, see, in particular,  section $5.2$, page $771$. We consider    the   general formula which describes  the Hamiltonian vector fields  and their related Hamiltonian  $(n-1)$-forms. In the terminology  by   H\'{e}lein and Kouneiher   those objects are termed  {\guillemotleft {\em algebraic observable $(n-1)$-forms}\guillemotright}  and  {\guillemotleft {\em infinitesimal symplectomorphisms}\guillemotright},   respectively (see \cite{HK-03}). This formulation corresponds to the algebraic structure  described in      section \ref{subsec:saison0010}. 
   
   Let   $ \Xi  \in \Gamma(  {{\cal M}}_{\tiny{\hbox{\sffamily DW}}}  , T   {{\cal M}}_{\tiny{\hbox{\sffamily DW}}}  )
$ be an arbitrary vector field  on $ {{\cal M}}_{\tiny{\hbox{\sffamily DW}}}$ written as
   \bee
\Xi :=    {\pmb{X}}^{\nu}   \frac{\partial}{\partial x^{\nu}}  +  {\pmb{\Theta}}^{M}_{\lambda}   \frac{\partial}{\partial e^M_{\lambda}}   + {\pmb{\Theta}}_{{\mu}}^{IJ}     \frac{\partial}{\partial \omega^{IJ}_{\mu}}  +   {\pmb{\Upsilon}}      \frac{\partial}{\partial \varkappa}  + { { {\pmb{\Upsilon}}^{e_{\mu}\nu}_{I}       \frac{\partial}{\partial {p}^{e_{\mu}\nu}_{I}} }}+  {\pmb{\Upsilon}}^{\omega_{\mu}\nu}_{IJ}       \frac{\partial}{\partial {p}^{\omega_{\mu}\nu}_{IJ}},
\eee
such that $\dd ( \Xi \iN  {{\Omega}}^{\tiny{\hbox{\sffamily DW}}}   ) = 0$. Note that  $   {\pmb{X}}^{\nu}  ,  {\pmb{\Theta}}^{M}_{\lambda}  , {\pmb{\Theta}}_{{\mu}}^{IJ}       ,  {\pmb{\Upsilon}}    ,     {\pmb{\Upsilon}}^{e_{\mu}\nu}_{I}  $ and $ {\pmb{\Upsilon}}^{\omega_{\mu}\nu}_{IJ}$ are smooth functions on $ {{\cal M}}_{\tiny{\hbox{\sffamily DW}}} $. 
 The set  of all  infinitesimal symplectomorphisms, {\em i.e.}   locally Hamiltonian vector fields, of $( {{\cal M}}_{\hbox{\tiny{\sffamily DW}}} ,   {{\Omega}}^{\tiny{\hbox{\sffamily DW}}}  )$ is  described by vector fields $ \Xi  =   \Xi (  \dttQ  )  +  \Xi (   \dttP     )  $, where 
 \bee
\left.
\begin{array}{ccl}
   \displaystyle   \Xi (   \dttQ     )     & =  & \displaystyle      {\Upsilon}  \frac{\partial}{\partial \varkappa}
      +{\Upsilon}^{e_{\mu}\alpha}_{I}  \frac{\partial}{\partial { {p}}^{e_\mu \alpha}_{I} } 
 +{\Upsilon}^{\omega_{\mu}\alpha}_{IJ}  \frac{\partial}{\partial { {p}}^{\omega_\mu \alpha}_{IJ} } ,
 \quad
  \hbox{with} \quad \frac{\partial {{\Upsilon}}  }{\partial  \omega^{IJ}_{\mu}}  - \frac{\partial  {{\Upsilon}}^{e_{\mu}\nu}_{I}}{\partial x^{\nu}}    - \frac{\partial  {{\Upsilon}}^{\omega_{\mu}\nu}_{IJ}}{\partial x^{\nu}}  =     0  ,
\\
  \displaystyle    \Xi (   \dttP     )     & =  & \displaystyle        X^{\nu}  \frac{\partial}{\partial x^{\nu}}  + \Theta^{M}_{\lambda} \frac{\partial}{\partial e^M_{\lambda}}   + \Theta_{{\mu}}^{IJ}   \frac{\partial}{\partial \omega^{IJ}_{\mu}}     -   
 \left(           \varkappa  ( \frac{\partial  X^{\nu} }{\partial x^{\nu}} )  + \frac{\partial  \Theta_{{\mu}}^{IJ} }{\partial x^{\nu}}  {p}^{\omega_\mu \nu}_{IJ}   + \frac{\partial  \Theta_{{\mu}}^{I} }{\partial x^{\nu}}  {p}^{e_\mu \nu}_{I}        \right)     
    \frac{\partial}{\partial \varkappa}  ,    
  \\
 \displaystyle    &   & \displaystyle    
 \quad +  
  \left(   {p}^{{e}_\rho \sigma}_{K} \delta^{\mu}_{\rho}    \left(  \delta_{  I}^{  K} \big{[}  (\frac{\partial  X^{\nu} }{\partial x^{\sigma}} )   -  \delta^{\nu}_{\sigma}     (\frac{\partial  X^{\lambda} }{\partial x^{\lambda}} )   \big{]}   -     ( \frac{\partial  \Theta_{{\sigma}}^{K} }{\partial {\omega}_{\nu}^{I}} )  \right)   -  \varkappa   ( \frac{\partial  X^{\nu} }{\partial {{e}^{I}_{\mu}}  }  )  \right)  
   \frac{\partial}{ \partial {p}^{e_\mu \nu}_{I} },
    \\
 \displaystyle    &   & \displaystyle   
\quad  +  
  \left(   {p}^{{\omega}_\rho \sigma}_{KL} \delta^{\mu}_{\rho}     \left( \delta_{  I}^{  K}\delta_{ J}^{ L} \big{[}  (\frac{\partial  X^{\nu} }{\partial x^{\sigma}} )   -  \delta^{\nu}_{\sigma}     (\frac{\partial  X^{\lambda} }{\partial x^{\lambda}} )   \big{]}   -     ( \frac{\partial  \Theta_{{\sigma}}^{KL} }{\partial {\omega}_{\nu}^{IJ}} )  \right)   -  \varkappa   ( \frac{\partial  X^{\nu} }{\partial {{\omega}^{IJ}_{\mu}}  }  )  \right)  
   \frac{\partial}{ \partial {p}^{\omega_\mu \nu}_{IJ} },
 \end{array}   
\right.
\nonumber
\eee
 and $   { {X}}^{\nu}  ,  { {\Theta}}^{M}_{\lambda}  , { {\Theta}}_{{\mu}}^{IJ}       ,  { {\Upsilon}}    ,     { {\Upsilon}}^{e_{\mu}\nu}_{I}  $ and $  { {\Upsilon}}^{\omega_{\mu}\nu}_{IJ}$ are smooth functions on ${\cal Y} $.
  We hope to present elsewhere \cite{Veygrav2}   a  detailed analysis  of all algebraic observable $(n-1)$-forms, {\em i.e.} of all Hamiltonian $(n-1)$-forms as defined in section \ref{subsec:saison0010},  in the {DW} formulation of vielbein gravity. 
  
  We now restrict ourselves to  simple examples  of Hamiltonian $(n-1)$-forms in ${\Omega}_{\tiny{\hbox{\sffamily Ham}}}^{{{n-1}}} ( {{\cal M}}_{\tiny{\hbox{\sffamily DW}}} )$.      
   Let us consider   the     $(n-1)$-forms  $ \dttQ_{e, \chi}     =  \dttQ^{I}_{e,\chi}   \otimes {\textswab{e}}_{I}$,    $  \dttQ_{\omega , \psi}  = \dttQ^{IJ}_{\omega , \psi}   \otimes  \Delta_{IJ}$,   $ \dttP_{IJ}^{\omega, \varphi} =  \dttP_{IJ}^{\omega, \varphi}  \otimes  \Delta^{IJ} $,  and ${\dttP}_{I}^{e,\zeta}     =  {\dttP}_{I}^{e,\zeta}     \otimes {\textswab{e}}^{I} $, where
   \bee\label{obs001}
\left.
\begin{array}{rcl}
 \displaystyle   \dttQ_{e , \chi}    & =  & \displaystyle     \chi^{\mu\nu}_{I} (x) e^{I}_{\mu}     \vol_{\nu},
\\
 \displaystyle     \dttQ_{\omega , \psi}    & =  & \displaystyle     \psi^{\mu\nu}_{IJ} (x) \omega^{IJ}_{\mu}     \vol_{\nu},
    \end{array}
\right.
\quad  
 \left.
\begin{array}{rcl} 
  \displaystyle   \dttP_{e, \zeta}  & =  & \displaystyle   \zeta^{I}_{\mu} (x) p^{e_\mu \nu}_{I} \vol_{\mu}  ,
\\
\displaystyle    \dttP_{\omega, \varphi} & =  & \displaystyle     \varphi_{\mu}^{IJ} (x) {{p}}^{\omega_{\mu}\nu}_{IJ}  \vol_{\nu} .
 \end{array}
\right.
  \eee  
  If we evaluate those different $(n-1)$-forms on the hypersurface of constraints ${{{\cal C}}} $ defined in section \ref{subsec:2.3}, we obtain    
\bee
 \left.
\begin{array}{rcl}
\displaystyle  \dttQ_{e , \chi}  \big |_{{{\cal C}}}      & =  & \displaystyle  \iota^{\star} \dttQ_{\omega , \psi}        =   \chi^{\mu\nu}_{I} (x) e^{I}_{\mu}     \vol_{\nu} =  \dttQ_{e , \chi}
\\
\displaystyle      \dttQ_{\omega , \psi}    \big |_{{{\cal C}}}    & =  & \displaystyle     \iota^{\star}     \dttQ_{\omega , \psi}    = \psi^{\mu\nu}_{IJ} (x) \omega^{IJ}_{\mu}     \vol_{\nu} =   \dttQ_{\omega , \psi}  ,
\\
  \displaystyle   \dttP_{e, \zeta}  \big |_{{{\cal C}}} & =  & \displaystyle  \iota^{\star}     \dttP_{e, \zeta}  = 0 ,
\\
\displaystyle    \dttP_{\omega, \varphi}  \big |_{{{\cal C}}} & =  & \displaystyle  \iota^{\star}   \dttP_{\omega, \varphi}  =      - (1/4) \varphi_{\mu} (x)   \epsilon_{IJKL} \epsilon^{\mu\nu\rho\sigma} e^{K}_{\rho} e^{L}_{\sigma}  \vol_{\nu} .
 \end{array}
\right.
\eee
  The exterior derivative of    $(n-1)$-forms $ \dttQ_{e , \chi}  $, $   \dttQ_{\omega , \psi}  $, $ \dttP_{e, \zeta}  $, and $ \dttP_{\omega, \varphi} $, are given by  
\begin{equation}\label{extderivobservables}
\left.
\begin{array}{ccl}
\displaystyle  \dd  \dttQ_{e , \chi}    & =  & \displaystyle      e^{I}_{\mu}    \partial_{\nu}     \chi^{\mu\nu}_{I} (x)     \vol +   \chi^{\mu\nu}_{I} (x)   \dd   e^{I}_{\mu}   \wedge     \vol_{\nu}  ,
\\  
 \displaystyle \dd    \dttQ_{\omega , \psi}    & =  & \displaystyle   
 \omega^{IJ}_{\mu}  \partial_{\nu} \psi^{\mu\nu}_{IJ}     (x)   \vol + \psi^{\mu\nu}_{IJ} (x) \dd  \omega^{IJ}_{\mu} \wedge     \vol_{\nu}     ,
 \\
 \displaystyle  \dd \dttP_{e, \zeta}  & =  & \displaystyle   \zeta^{I}_{\mu} (x)     \dd   p^{e_\mu \nu}_{I}   \wedge \vol_{\nu}    +  p^{e_\mu \nu}_{I}   \partial_{\nu}  \zeta^{I}_{\mu} (x)       \vol   ,
\\
\displaystyle  \dd   \dttP_{\omega, \varphi} & =  & \displaystyle     \varphi_{\mu}^{IJ} (x)       \dd   {{p}}^{\omega_{\mu}\nu}_{IJ}   \wedge \vol_{\nu}    + {{p}}^{\omega_{\mu}\nu}_{IJ}  \partial_{\nu}      \varphi_{\mu}^{IJ}    (x)    \vol   .
\\
 \end{array}
\right.
 \end{equation}
 The  Hamiltonian   $(n-1)$-form $  \dttQ_{\omega , \psi}   $ is equivalently written as  
$   \dttQ_{\omega , \psi}   =  ({1}/{2}) \psi^{\mu\nu}  (x) \omega^{IJ} \wedge \vol_{\mu\nu} ,
$
 where $\psi^{\mu\nu} (x)$ is a real function such that $\psi^{\mu\nu} = - \psi^{\nu\mu}$. Then,   
$
 \dttQ_{\omega , \psi}    =   ({1}/{2})  \psi^{\mu\nu} (x) \omega^{IJ}_{\rho} \dd x^{\rho} \wedge    \vol_{\mu\nu} =     ({1}/{2})  \psi^{\mu\nu} (x) {\big{(}} \omega^{IJ}_{\mu}     \vol_{\nu} - \omega^{IJ}_{\nu}    \vol_{\mu} {\big{)}} =  \psi^{\mu\nu} (x) \omega^{IJ}_{\mu}     \vol_{\nu} 
$. The exterior derivative  of the   $(n-1)$-forms is 
\bee
\left.
\begin{array}{ccl}
\displaystyle   
\dd   \dttQ_{\omega , \psi}
    & = &  \displaystyle    \dd {{(}} \psi^{\mu\nu}(x) \omega^{IJ}_{\mu}     \vol_{\nu}  {{)}} 
    =\omega^{IJ}_{\mu}  {\partial_\sigma  \psi^{\mu\nu}_{IJ}}  (x) \dd x^{\sigma}  \wedge     \vol_{\nu}
 + \psi^{\mu\nu}_{IJ}(x) \dd  \omega^{IJ}_{\mu} \wedge     \vol_{\nu}  ,
 \\
 \displaystyle   
    & = &  \displaystyle
 \omega^{IJ}_{\mu} {\partial_\nu \psi^{\mu\nu}_{IJ}}    (x)   \vol + \psi^{\mu\nu}_{IJ} (x) \dd  \omega^{IJ}_{\mu} \wedge     \vol_{\nu} ,
 \end{array}
\right.
\eee
whereas the exterior derivative of the   $(n-1)$-form $ \dttP_{\omega, \varphi} $  is written as
 \bee
\left.
\begin{array}{lll}
\displaystyle \dd  \dttP_{\omega, \varphi}  & = & \displaystyle   \dd {\big{(}}     \varphi_{\mu}^{IJ} (x) {{p}}^{\omega_{\mu}\nu}_{IJ}  \vol_{\nu}    {\big{)}} 
=    \varphi_{\mu}^{IJ} (x)       \dd   {{p}}^{\omega_{\mu}\nu}_{IJ}   \wedge \vol_{\nu}    + {{p}}^{\omega_{\mu}\nu}_{IJ} \dd      \varphi_{\mu}^{IJ} (x) \wedge  \vol_{\nu}  ,
\\
\displaystyle   & = & \displaystyle    \varphi_{\mu}^{IJ} (x)       \dd   {{p}}^{\omega_{\mu}\nu}_{IJ}   \wedge \vol_{\nu}    + {{p}}^{\omega_{\mu}\nu}_{IJ}  \partial_{\nu}      \varphi_{\mu}^{IJ}    (x)    \vol       .
\end{array}
\right.
\eee
 Using the constraints \eqref{kiqlapw},  the exterior derivatives of the Hamiltonian $(n-1)$-forms of type $\dttQ_{\omega , \psi}   \big |_{{{\cal C}}}  $ and   $ \dttP_{\omega, \varphi}  \big |_{{{\cal C}}}    $   are now written as   
\bee
\left.
\begin{array}{lll}
\displaystyle   \dd  \dttQ_{\omega , \psi}   \big |_{{{\cal C}}}   & = & \displaystyle  \omega^{IJ}_{\mu} {\partial_\nu \psi^{\mu\nu}_{IJ} }    (x)   \vol + \psi^{\mu\nu}_{IJ}  (x) \dd  \omega^{IJ}_{\mu} \wedge     \vol_{\nu}  =   \dd  \dttQ_{\omega , \psi},
\\
\displaystyle  \dd    \dttP_{\omega, \varphi}  \big |_{{{\cal C}}}    & = & \displaystyle     -  (1/4)
 \epsilon_{IJKL} \epsilon^{\mu\nu\rho\sigma}  e^{K}_{\rho} e^{L}_{\sigma}   \dd    \varphi_{\mu}^{IJ} (x)   \wedge  \vol_{\nu}  
  -  (1/2) \varphi_{\mu}^{IJ} (x)   \epsilon_{IJKL} \epsilon^{\mu\nu\rho\sigma} e^{K}_{\rho} \dd    e^{L}_{\sigma}  \wedge \vol_{\nu}  .
  \end{array}
\right.
\nonumber
\eee
 
\begin{lemm}\label{lem:observable12}
  The  Hamiltonian vector fields   related to the Hamiltonian $(n-1)$-forms {\em $ \dttQ_{e , \chi}  $, $   \dttQ_{\omega , \psi}  $, $ \dttP_{e, \zeta}  $, and  $ \dttP_{\omega, \varphi} $}, which  are denoted as       {\em $  \Xi (  \dttQ_{e , \chi}  )  $, $\Xi(   \dttQ_{\omega , \psi})$, $\Xi(\dttP_{e, \zeta})$,} and  {\em $\Xi(\dttP_{\omega, \varphi})$},     are given by    {\em 
\begin{equation}\label{extderivobservables}
\left.
\begin{array}{rcl}
\displaystyle  \Xi (  \dttQ_{e , \chi}  )  & =  & \displaystyle       -   e^{I}_\mu \left(   {\partial_{\nu} \chi^{\mu\nu}_{I}}   \right)    \frac{\partial}{\partial \varkappa} 
-   \chi^{\mu\nu}_{I}       \frac{\partial}{\partial {{p}}^{e_{\mu}\nu}_{I}}    ,
\\  
 \displaystyle \Xi  (    \dttQ_{\omega , \psi}  )   & =  & \displaystyle    -   \omega^{IJ}_\mu   \left(  {\partial_{\nu} \psi^{\mu\nu}_{IJ}}   \right)   \frac{\partial}{\partial \varkappa} 
-   \psi^{\mu\nu}_{IJ}   \frac{\partial}{\partial {{p}}^{\omega_{\mu}\nu}_{IJ}}     ,
\\
 \end{array}
\right.
\quad
\left.
\begin{array}{rcl}
 \displaystyle  \Xi ( \dttP_{e, \zeta} ) & =  & \displaystyle     \zeta_{\mu}^{I}   \frac{\partial}{\partial {e}_{\mu}^{I } }      -    {{p}}^{e_{\mu}\nu}_{I }   \left(   {\partial_{\nu}  \zeta_{\mu}^{I}   }  \right)     \frac{\partial}{\partial \varkappa} ,
 \\
\displaystyle  \Xi (   \dttP_{\omega, \varphi} ) & =  & \displaystyle      \varphi_{\mu}^{IJ}  \frac{\partial}{\partial {\omega}_{\mu}^{IJ} }      -  {{p}}^{\omega_{\mu}\nu}_{IJ}   \left(  {\partial_{\nu}  \varphi_{\mu}^{IJ}   }   \right)       \frac{\partial}{\partial \varkappa}  .
\\
 \end{array}
\right.
\nonumber
 \eee
 }
\end{lemm}
{\sffamily Proof}.  Let us compute the contractions   on the multisymplectic manifold $({{\cal M}}_{\hbox{\tiny{\sffamily DW}}} , {\pmb{\omega}}^{\tiny{\hbox{\sffamily DW}}})$, where the  vector field $\Xi (  \dttP_{\omega, \varphi} )  $ on ${{\cal M}}_{\hbox{\tiny{\sffamily DW}}}$ is given as in lemma \ref{lem:observable12}. By the straightforward calculation,
  \[
\left.
\begin{array}{lll}
\displaystyle  \Xi (  \dttP_{\omega, \varphi}  ) \iN {\pmb{\omega}}^{\tiny{\hbox{\sffamily DW}}} & = & 
\displaystyle      {\big{(}}     \varphi_{\mu} (x)  {\partial}/{ \partial \omega^{IJ}_{\mu} }
-     {\big{(}} 
 {\partial_\nu \varphi_{\mu} }  (x) {{p}}^{\omega_{\mu} \nu}_{IJ} {\big{)}}  {\partial}/{\partial \varkappa}
{\big{)}}
  \iN {\big{(}}  \hbox{d} \varkappa \wedge {\beta}  +
  \dd
   {{p}}^{
\omega_{\mu}\nu}_{IJ} \wedge 
  \hbox{d} \omega_{\mu}^{IJ} \wedge  {\beta}_{\nu}  {\big{)}} ,
\\ 
 \displaystyle   & = & \displaystyle -  {\big{(}} 
 {\partial_{\nu} \varphi_{\mu} } (x) {{p}}^{\omega_{\mu} \nu}_{IJ} {\big{)}} 
\vol  -     \varphi_{\mu} (x)     \dd
   {{p}}^{\omega_{\mu}\nu}_{IJ}  \wedge  {\beta}_{\nu}  = - \dd \dttP_{\omega, \varphi} ,
\end{array}
\right.
\]
 \[
\left.
\begin{array}{ccl}
\displaystyle   
\Xi (     \dttQ_{\omega , \psi}     )  \iN {\pmb{\omega}}^{\tiny{\hbox{\sffamily DW}}} & = &  \displaystyle  -  \left(  (\omega^{IJ}_\mu  {\partial_\nu \psi^{\mu\nu}}   )   {\partial}/{\partial \varkappa} 
+ \psi^{\mu\nu} (x)  {\partial}/{\partial   {{p}}^{\omega_{\mu}\nu}_{IJ}   }  \right)  \iN   \left(  \hbox{d}  \varkappa \wedge {\beta} +  \hbox{d}    {{p}}^{\omega_{\mu}\nu}_{IJ}   \wedge 
\dd \omega^{IJ}_\mu \wedge  {\beta}_{\nu}\right),
\\
\displaystyle & = & \displaystyle  -  \omega^{IJ}_{\mu} \left(   {\partial_\nu \psi^{\mu\nu}}    (x)  \right) \vol 
 -
\psi^{\mu\nu} (x) \dd  \omega^{IJ}_{\mu} \wedge     \vol_{\nu}  = - \dd   \dttQ_{\omega , \psi}     .
\end{array}
\right.
\]
Analogously,   a straightforward  calculation  yields  the Hamiltonian vector fields on the constraints  hypersurface    ${{{\cal C}}} $ defined in section \ref{subsec:2.3}. More precisely,  working on $({{\cal C}} , \iota^{\star} \Omega^{\hbox{\sffamily\tiny DW}})$  we obtain: 
\begin{lemm}\label{lem:observable1}
The  Hamiltonian vector fields   related to the Hamiltonian $(n-1)$-forms {\em $ \dttQ_{e , \chi}  \big |_{{{\cal C}}}     $, $ \dttP_{e, \zeta}  \big |_{{{\cal C}}}     $, $   \dttQ_{\omega , \psi}  \big |_{{{\cal C}}}    $}, and  {\em $ \dttP_{\omega, \varphi}   \big |_{{{\cal C}}}    $}  are given by    {\em 
\begin{equation}\label{extderivobservables}
\left.
\begin{array}{rcl}
 \displaystyle  \Xi (  \dttQ_{e , \chi}  ) \big |_{{{\cal C}}}    & =  & \displaystyle       -   e^{I}_\mu \left(   {\partial_{\nu} \chi^{\mu\nu}_{I}}   \right)    \frac{\partial}{\partial \varkappa} 
+   \chi^{\mu\nu}_{I}       \frac{\partial}{\partial {{p}}^{e_{\mu}\nu}_{I}}      ,
 \\  
 \displaystyle       \Xi ( \dttP_{e, \zeta} ) \big |_{{{\cal C}}}     & =  & \displaystyle          \zeta_{\mu}^{I} (x)   \frac{\partial}{\partial {e}_{\mu}^{I } }       ,
 \\  
 \displaystyle \Xi  (    \dttQ_{\omega , \psi}   )   \big |_{{{\cal C}}}   & =  & \displaystyle   -   {\big{(}} \omega^{IJ}_\mu \frac{\partial \psi^{\mu\nu}}{\partial x^{\nu}}     {\big{)}}    \frac{\partial}{\partial \varkappa}         -    \left( \frac{1}{6}   \psi^{\mu\nu} (x)   \epsilon^{IJKL} \epsilon_{\mu\nu\rho\sigma}    e_{K}^{\rho} \right)   \frac{\partial  }{\partial e_{\sigma}^{L} }  ,
 \\
 \displaystyle  \Xi (   \dttP_{\omega, \varphi} )   \big |_{{{\cal C}}} & =  & \displaystyle      \varphi_{\mu}^{IJ}  \frac{\partial}{\partial {\omega}_{\mu}^{IJ} }    +  \left(
\frac{1}{4} \epsilon_{IJOP} \epsilon^{\mu\nu\alpha\beta} \frac{\partial \varphi_{\mu} }{\partial x^{\nu} }  e^{O}_{\alpha} e^{P}_{\beta}  \right) \frac{\partial}{\partial \varkappa}  .
\\
 \end{array}
\right.
   \eee
 }
  \end{lemm}
We  present  the explicit calculation  for the $(n-1)$-forms $\dttQ_{\omega , \psi} $ and $ \dttP_{\omega, \varphi}$.  The interior product $  \Xi (    \dttP_{\omega, \varphi}  \big |_{{{\cal C}}}     ) \iN {\pmb{\omega}}^{\hbox{\tiny{\sffamily Palatini}}} $ yields
\[
\left.
\begin{array}{lll}
\displaystyle  \Xi (    \dttP_{\omega, \varphi}  \big |_{{{\cal C}}}     ) \iN {\pmb{\omega}}^{\hbox{\tiny{\sffamily Palatini}}} & = & \displaystyle   \left(    \varphi_{\mu} (x)  {\partial}/{ \partial \omega^{IJ}_{\mu} }
+ {\big{(}} 
(1/4) \epsilon_{IJOP} \epsilon^{\mu\nu\alpha\beta}  {\partial_\nu \varphi_{\mu} }   e^{O}_{\alpha} e^{P}_{\beta}  {\big{)}}  {\partial}/{\partial \varkappa}
 \right),
\\
 \displaystyle   &  & \displaystyle
  \iN {\big{(}}  \hbox{d} \varkappa \wedge {\beta} -   (1/2) \epsilon_{IJKL} \epsilon^{\mu\nu\rho\sigma} e^{K}_{\rho} \hbox{d} e^{L}_{\sigma} \wedge
\hbox{d} \omega_{\mu}^{IJ} \wedge  {\beta}_{\nu}  {\big{)}},
\\
 \displaystyle   & = & \displaystyle      (1/2) \epsilon_{IJKL} \epsilon^{\mu\nu\rho\sigma} \varphi_{\mu} (x)  e^{K}_{\rho} \hbox{d} e^{L}_{\sigma}   \wedge  {\beta}_{\nu}
 + (1/4)
 \epsilon_{IJKL} \epsilon^{\mu\nu\rho\sigma}  e^{K}_{\rho} e^{L}_{\sigma}   \partial_{\nu}    \varphi_{\mu} (x)     \vol  .
\end{array}
\right.
\]
Therefore, 
$ \displaystyle
  \Xi (    \dttP_{\omega, \varphi}  )  \big |_{{{\cal C}}}      \iN {\pmb{\omega}}^{\hbox{\tiny{\sffamily Palatini}}}   = -    \dd   \dttP_{\omega, \varphi}  \big |_{{{\cal C}}} 
$.  Now, we calculate $ \Xi (      \dttQ_{\omega , \psi}   )    \big |_{{{\cal C}}}      \iN {\pmb{\omega}}^{\hbox{\tiny{\sffamily Palatini}}} $.      Let us contract  both   sides of    $ {{p}}^{\omega_{\mu}\nu}_{IJ}    = - (1/4) \epsilon_{IJMN} \epsilon^{\mu\nu\alpha\beta} e^{M}_{\alpha} e^{N}_{\beta}   $ with $\epsilon^{IJKL} \epsilon_{\mu\nu\rho\sigma} e^{\rho}_{K} $. We obtain 
   \[
 \underbrace{
\epsilon^{IJKL} \epsilon_{\mu\nu\rho\sigma}
 {{p}}^{\omega_{\mu}\nu}_{IJ}   e^{\rho}_{K} 
 }_{ ({\mathfrak{1}})  }
  = 
   \underbrace{
   -  (1/4) \epsilon^{IJKL} \epsilon_{\mu\nu\rho\sigma} \epsilon_{IJMN}  \epsilon^{\mu\nu\alpha\beta} e^{M}_{\alpha} e^{N}_{\beta} e^{\rho}_{K}  
    }_{ ({\mathfrak{2}})  } ,
 \]  
where
 \[
\left.
\begin{array}{rcl}
 \displaystyle   ({\mathfrak{2}})  & = &  \displaystyle  -  (1/4)  \epsilon_{IJMN}  \epsilon^{IJKL} \epsilon_{\mu\nu\rho\sigma}  \epsilon^{\mu\nu\alpha\beta} e^{M}_{\alpha} e^{N}_{\beta} e^{\rho}_{K}    
=
  - (1/4) (2!) (2!) \delta^{[K}_{M}  \delta^{L]}_{N}     (2!) (2!) \delta^{[\alpha}_{\rho}  \delta^{\beta]}_{\sigma}   e^{M}_{\alpha} e^{N}_{\beta} e^{\rho}_{K}   ,
 \\
 \displaystyle     & = &  \displaystyle   - 2  \delta^{[K}_{M}  \delta^{L]}_{N}     (  e^{M}_{\rho} e^{N}_{\sigma} e^{\rho}_{K}   - e^{M}_{\sigma} e^{N}_{\rho} e^{\rho}_{K} ) 
 = - 2  \delta^{[K}_{M}  \delta^{L]}_{N}     (  \delta^{M}_{K} e^{N}_{\sigma}     - e^{M}_{\sigma} \delta^{N}_{K}   ),
  \\
\displaystyle     & = &  \displaystyle -    {\big{(}} (  \delta^{K}_{K} e^{L}_{\sigma}     - e^{K}_{\sigma} \delta^{L}_{K}   )   -    (  \delta^{L}_{K} e^{K}_{\sigma}     - e^{L}_{\sigma} \delta^{K}_{K}   ) {\big{)}}
= 
 -      {\big{(}}  (  4 e^{L}_{\sigma}     - e^{L}_{\sigma}    )   -    (   e^{L}_{\sigma}     - 4 e^{L}_{\sigma}    )  {\big{)}}
 = - 6 e^{L}_{\sigma}   .
\end{array}
\right.
\]
Then, we obtain 
  \bee\label{gliopospsp99}
 e^{L}_{\sigma} = -   (1/3!) \cdot ({\mathfrak{1}})  =   -   (1/6)   \epsilon^{IJKL} \epsilon_{\mu\nu\rho\sigma}
 {{p}}^{\omega_{\mu}\nu}_{IJ}   e^{\rho}_{K}    .
 \eee 
We directly verify this result by the   straightforward calculation: 
 \[
\left.
\begin{array}{rcl}
 \displaystyle          -   (1/6)     \epsilon^{IJKL} \epsilon_{\mu\nu\rho\sigma}  {{p}}^{\omega_{\mu}\nu}_{IJ}  e_{K}^{\rho}   & = &  \displaystyle 
   (1/4)   (1/6)    \epsilon^{IJKL} \epsilon_{IJOP} \epsilon_{\mu\nu\rho\sigma}   \epsilon^{\mu\nu\alpha\beta} ( e^{O}_{\alpha} e^{P}_{\beta} )    e_{K}^{\rho} ,
 \\
 \displaystyle     & = &  \displaystyle      (1/24)    \bl (2!) (2!) \delta^{[K}_{O} \delta^{L]}_{P} \br \bl  (2!) (2!) \delta^{[\alpha}_{\rho} \delta^{\beta]}_{\sigma}  \br      e^{O}_{\alpha} e^{P}_{\beta}     e_{K}^{\rho}
 \\
  \displaystyle     & = &  \displaystyle    (1/6)     \bl  (   e^{K}_{\rho} e^{L}_{\sigma}   e_{K}^{\rho} -   e^{K}_{\sigma} e^{L}_{\rho}   e_{K}^{\rho} ) -  (   e^{L}_{\rho} e^{K}_{\sigma}   e_{K}^{\rho} -   e^{L}_{\sigma} e^{K}_{\rho}   e_{K}^{\rho} ) \br 
   \\
  \displaystyle     & = &  \displaystyle  
    2    (1/6)    (   e^{K}_{\rho} e^{L}_{\sigma}   e_{K}^{\rho} -     e^{K}_{\sigma} e^{L}_{\rho}   e_{K}^{\rho} )  =    2  (1/6)       (   \delta^{\rho}_{\rho} e^{L}_{\sigma}     -     \delta^{\rho}_{\sigma} e^{L}_{\rho}   )    ,
   \\
 \displaystyle     & = &  \displaystyle 
  2  (1/6)      (  4 e^{L}_{\sigma}     -      e^{L}_{\sigma}   )
   = 6 \cdot (1/6)     e^{L}_{\sigma}  = e^{L}_{\sigma}  .
\end{array}
\right.
\]
  Using   \eqref{gliopospsp99}, we obtain
$ \displaystyle \frac{\partial}{\partial   {{p}}^{\omega_{\mu}\nu}_{IJ} }  =  (\frac{\partial e_{\sigma}^{L}  }{\partial   {{p}}^{\omega_{\mu}\nu}_{IJ}    }  )   \frac{\partial  }{\partial e_{\sigma}^{L} }  = \left(  -  (1/6)  \epsilon^{IJKL} \epsilon_{\mu\nu\rho\sigma}     e_{K}^{\rho}   \right)   \frac{\partial  }{\partial e_{\sigma}^{L} }  $, 
so that 
\bee
\Xi (     \dttQ_{\omega , \psi}  )   =  -   {\big{(}} \omega^{IJ}_\mu \frac{\partial \psi^{\mu\nu}}{\partial x^{\nu}}     {\big{)}}    \frac{\partial}{\partial \varkappa}         -        (1/6)  \psi^{\mu\nu} (x)   \epsilon^{IJKL} \epsilon_{\mu\nu\rho\sigma}    e_{K}^{\rho}   \frac{\partial  }{\partial e_{\sigma}^{L} }  . 
\eee 
Finally, we obtain the expression  
 \[
\left.
\begin{array}{lll}
\displaystyle     \Xi (      \dttQ_{\omega , \psi}   )    \big |_{{{\cal C}}}      \iN {\pmb{\omega}}^{\hbox{\tiny{\sffamily Palatini}}}  & = & \displaystyle  \Xi (     \dttQ_{\omega , \psi} )   \iN {\big{(}}  \hbox{d} \varkappa \wedge {\beta} - (1/2)  \epsilon_{IJKL} \epsilon^{\mu\nu\rho\sigma} e^{K}_{\rho} \hbox{d} e^{L}_{\sigma} \wedge
\hbox{d} \omega_{\mu}^{IJ} \wedge  {\beta}_{\nu}  {\big{)}},
\\ 
\displaystyle   & = & \displaystyle      -   \omega^{IJ}_\mu \left(    {\partial_{\nu} \psi^{\mu\nu}_{IJ}}    \right)   \vol  \underbrace{ -   (1/2) \epsilon_{IJKL} \epsilon^{\mu\nu\rho\sigma} e^{K}_{\rho}     \hbox{d} e^{L}_{\sigma} (\Xi (     \dttQ_{\omega , \psi} ))  \hbox{d} \omega_{\mu}^{IJ} \wedge  {\beta}_{\nu}
}_{ ({\mathfrak{3}})  }
   \end{array}
\right.
\]
In the appendix \ref{app:lpkojihu88}, we explicitly prove that  $ ({\mathfrak{3}})    =  \psi^{\mu\nu}_{IJ}    \hbox{d} \omega_{\mu}^{IJ} \wedge  {\beta}_{\nu} $.
 
Finally,    we also consider the Hamiltonian $(n-1)$-form $  \dttQ_{\varkappa , \tau } =   \tau^{\nu}_{\mu} (x) X^{\mu}     \vol_{\nu}  $ and $        \dttP_{ \varkappa}   =   \varkappa {  X}^{\alpha}  {\beta}_{\alpha}   - { {p}}^{\omega_\mu \nu}_{IJ} {  X}^{\alpha} \dd \omega_{\mu}^{IJ} \wedge {\beta}_{\alpha\nu}$. We will use them in section \ref{subsec:Lie-algebra-02}   to give an example of  an homotopy Lie  structure.  Working on  the constraint hypersurface  ${{{\cal C}}} $ defined in section \ref{subsec:2.3}, 
 \bee
 \left.
\begin{array}{lll}
  \displaystyle   \dttQ_{ \varkappa}  \big |_{{{\cal C}}}    & =  & \displaystyle    =   \iota^{\star}  \dttQ_{ \varkappa}  =   \tau^{\nu}_{\mu} (x) X^{\mu}     \vol_{\nu}  =  \dttQ_{e }  ,
  \\
    \displaystyle     \dttP_{ \varkappa}  \big |_{{{\cal C}}}    & =  & \displaystyle   \iota^{\star}       \dttP_{ \varkappa}  =   \varkappa {  X}^{\alpha}  {\beta}_{\alpha}  - e e^{[\mu}_I e^{\nu]}_J {  X}^{\alpha} \dd \omega_{\mu}^{IJ} \wedge {\beta}_{\alpha\nu} .
   \end{array}
\right.
\eee
\begin{lemm}\label{lem:extderivobservables-additional-momentum}
   The Hamiltonian vector field  related to the Hamiltonian $(n-1)$-form  
{\em $ \dttP_{\varkappa} $}    is 
 {\em $
    \Xi (   \dttP_{ \varkappa}    ) =       X^{\mu} (x )   {\partial}_\mu       - \left(   \varkappa   \left(   {\partial_{\mu}   X^{\mu}    }  \right)       \right)  {\partial}/{\partial \varkappa}     +   {p}^{{\omega}_\mu \sigma}_{IJ}      
  \left(     ( {\partial_\sigma  X^{\nu} }  )   -  \delta^{\nu}_{\sigma}     ( {\partial  X^{\lambda} }/{\partial x^{\lambda}} )       \right) 
    {\partial}/{ \partial {p}^{\omega_\mu \nu}_{IJ} } $.}
\end{lemm}
{\sffamily Proof}.  The interior product $ \Xi (   \dttP_{ \varkappa}    )   \iN {\pmb{\omega}}^{\tiny{\hbox{\sffamily DW}}}$ yields
  \[
\left.
\begin{array}{lll}
\displaystyle  \Xi (   \dttP_{ \varkappa}    )   \iN {\pmb{\omega}}^{\tiny{\hbox{\sffamily DW}}} & = &  
\displaystyle     \left(     X^{\rho} (x )   {\partial}/{\partial  x^{\rho} }      -    \varkappa   \left(   {\partial_{\rho}   X^{\rho}    }  \right)    {\partial}/{\partial \varkappa}    \right)  \iN {\big{(}}  \hbox{d} \varkappa \wedge {\beta}  +   \dd  {{p}}^{ \omega_{\mu}\nu}_{IJ} \wedge    \hbox{d} \omega_{\mu}^{IJ} \wedge  {\beta}_{\nu}  {\big{)}} ,
  
  \\ 
 \displaystyle   &   & \displaystyle 
   +  
   {p}^{{\omega}_\lambda \sigma}_{KL}      \left(   ( {\partial_\sigma  X^{\rho} }  )   -  \delta^{\rho}_{\sigma}     ({\partial_\kappa  X^{\kappa} }  )    \right)    
    {\partial}/{ \partial {p}^{\omega_\lambda \rho}_{KL} } \iN \left(   \dd  {{p}}^{ \omega_{\mu}\nu}_{IJ} \wedge    \hbox{d} \omega_{\mu}^{IJ} \wedge  {\beta}_{\nu}  \right),
\\ 
 \displaystyle   & = & \displaystyle      -    \varkappa   \left(   {\partial_{\mu}   X^{\mu}    }  \right)     \vol - X^{\mu} \dd \varkappa \wedge \vol_{\mu} + X^{\rho}  \dd
   {{p}}^{
\omega_{\mu}\nu}_{IJ} \wedge 
  \hbox{d} \omega_{\mu}^{IJ} \wedge  {\beta}_{\rho\nu} ,
  \\ 
 \displaystyle   &   & \displaystyle 
   +  
   {p}^{{\omega}_\mu \sigma}_{IJ}      \left(  ( {\partial_{\sigma}  X^{\nu} }  )   -  \delta^{\nu}_{\sigma}     ( {\partial_\lambda  X^{\lambda} }  )     \right)   
     \hbox{d} \omega_{\mu}^{IJ} \wedge  {\beta}_{\nu}  .
 \end{array}
\right.
\]
Note that $    \dd \dttP_{ \varkappa}   =  {  X}^{\alpha}      \dd \varkappa \wedge {\beta}_{\alpha}  + \varkappa (\partial_{\mu} X^{\mu}) \vol - {  X}^{\alpha} \dd { {p}}^{\omega_\mu \nu}_{IJ}  \wedge \dd \omega_{\mu}^{IJ} \wedge {\beta}_{\alpha\nu} -  { {p}}^{\omega_\mu \nu}_{IJ} \dd {  X}^{\alpha}  \wedge \dd \omega_{\mu}^{IJ} \wedge {\beta}_{\alpha\nu} $. The last term is equivalently written   as
$ -
\left( {p}^{{\omega}_\mu \rho}_{IJ}         ( {\partial_{\rho}  X^{\nu} }  )           -   {p}^{{\omega}_\mu \nu}_{IJ}      ( {\partial_\rho  X^{\rho} }  )       \right)
     \hbox{d} \omega_{\mu}^{IJ} \wedge  {\beta}_{\nu} 
     $,    where we have used $  \dd x^{\alpha} \wedge  {\beta}_{\rho\nu} = \delta^{\alpha}_{\rho} \vol_{\nu} - \delta^{\alpha}_{\nu} \vol_{\rho}  $.

  \subsection{{{Brackets of Hamiltonian   $(n-1)$-forms, Lie and  homotopy  Lie structures }}}\label{subsec:saison002}
  
In this section, we study    bracket operations   between Hamiltonian $(n-1)$-forms. In particular,   the exactness or the failure   of the Jacobi property is clarified  along with simple examples.  First,  in section  \ref{subsec:Lie-algebra-01} we give an example  of  an exact Lie  algebra  $\textswab{A}_{\mathfrak{1}} :=  {{{\{}} \textswab{a}_{\mathfrak{1}}   , {\{} , {\}}  {{\}}}}$, where $ \textswab{a}_{\mathfrak{1}} $ is the set of Hamiltonian $(n-1)$-forms $  {{\{}}      \dttQ_{e , \chi} ,  \dttQ_{\omega ,  \psi} ,      \dttP_{e, \zeta}  , \dttP_{\omega, \varphi}         {{\}}} $. Then,  in  section     \ref{subsec:Lie-algebra-02}  we present some aspects of  an        homotopy   Lie algebra    $\textswab{A}_{\mathfrak{2}} :=  {{{\{}} \textswab{a}_{\mathfrak{2}}   , {\{} , {\}}  {{\}}}    } $, where $\textswab{a}_{\mathfrak{2}}$ is the set of Hamiltonian $(n-1)$-forms $ {{\{}}      \dttQ_{e , \chi} ,  \dttQ_{\omega ,  \psi} ,    \dttP_{\varkappa} ,  \dttP_{e, \zeta}  , \dttP_{\omega, \varphi} {{\}}}$.  Finally, in     section \ref{subsec:Lie-algebra-03},  we present   a third algebraic structure  on the set of Hamiltonian $(n-1)$-forms $\textswab{a}_{\mathfrak{3}} := {{\{}}     {\pmb{\cal C}}_{e^{I}_{\mu} }  ,  {{\pmb{\cal C}}}_{\omega^{IJ}_{\mu} }     {{\}}}$. This one   reproduces some aspects of  the formulation  of     vielbein gravity in polymomentum variables  \cite{kannnnnnnaa,kannnnnnnaa1}.
  
  \subsubsection{{{Lie algebraic  structure }}}\label{subsec:Lie-algebra-01}

 We construct some   bracket relations with the Hamiltonian $(n-1)$-forms introduced in      section \ref{subsec:SAOFex}. Let us consider the Hamiltonian $(n-1)$-forms $  \dttQ_{\omega,\psi}   =  \psi^{\mu\nu}_{IJ}  (x) \omega^{IJ}_{\mu}     \vol_{\nu} $,  $   \dttQ_{\omega,{\overline{\psi}}}   =  {\overline{\psi}}{}^{\mu\nu}_{IJ}  (x) \omega^{IJ}_{\mu}     \vol_{\nu}  $,  $  \dttP_{\omega, \varphi} =     \varphi_{\mu}^{IJ} (x) {{p}}^{\omega_{\mu}\nu}_{IJ}  \vol_{\nu}$, and $  \dttP_{\omega, {\overline{\varphi}}} =   {\overline{\varphi}}{}_{\mu}^{IJ} (x) {{p}}^{\omega_{\mu}\nu}_{IJ}  \vol_{\nu}$. Note that $\phi_{\mu}^{IJ}(x) , {\overline{\phi}}{}^{IJ}_{\mu}(x), \psi_{IJ}^{\mu\nu}(x)$, and  ${{\overline{\psi}}}{}_{IJ}^{\mu\nu} (x)$ are smooth functions on the space-time manifold ${\cal X}$,    where $\psi^{\mu\nu}(x)  = - \psi^{\nu\mu}(x)$ and  ${\overline{\psi}}^{\mu\nu}(x)  = - {\overline{\psi}}^{\nu\mu}(x)$.
 \begin{prop}\label{prop:amqwv1}  
  On the multisymplectic manifold,  {\em ${ ( {{\cal M}}_{\hbox{\tiny{\sffamily DW}}}   , {\pmb{\omega}}^{\tiny{\hbox{\sffamily DW}}})}$}, the brackets   on the set of Hamiltonian $(n-1)$-forms {\em $    \dttQ_{\omega , \psi}   ,   \dttQ_{\omega , \overline{\psi}}      , \dttP_{\omega, \varphi}$, and $\dttP_{\omega, {\overline{\varphi}}} \in {\Omega}_{\tiny{\hbox{\sffamily Ham}}}^{{{n-1}}} (  {\cal M}_{{\hbox{\tiny\sffamily DW}}}  )$}  are given by     {\em \bee\label{qmdlzzo8888}
  {\big{\{}}   \dttQ_{\omega , \psi}   ,    \dttQ_{\omega , \overline{\psi}}      {  {\}}}  =  {  {\{}}    \dttP_{\omega, \varphi}    ,    \dttP_{\omega, {\overline{\varphi}}}     {  {\}}}  = 0,    
\quad \quad     
 \displaystyle      {  {\{}}   \dttQ_{\omega,\psi}   ,    \dttP_{\omega, \varphi}  {  {\}}}    =    
      - \psi^{\mu\nu}_{IJ} (x)    \varphi^{IJ}_{\mu} (x ) {\dd}\mathfrak{y}_{\nu}.     
 \eee}
\end{prop}
{\sffamily Proof}. The brackets are easily computed using lemma \ref{lem:observable12}
  \bee
\left.
\begin{array}{rcl}
\displaystyle   {  {\{}}     \dttQ_{\omega , \psi}   ,    \dttQ_{\omega,{\overline{\psi}}}   {  {\}}}   &  =  & \displaystyle  -     \Xi (   \dttQ_{\omega , \psi}    ) \iN \Xi (    \dttQ_{\omega,{\overline{\psi}}}    ) \iN {\pmb{\omega}}^{\tiny{\hbox{\sffamily DW}}}  =   \Xi (   \dttQ_{\omega , \psi}    ) \iN \dd  \dttQ_{\omega,{\overline{\psi}}}   ,
\\
 \displaystyle        
&  =  & \displaystyle      \Xi (   \dttQ_{\omega , \psi}  ) \iN \left(
   \omega^{IJ}_{\mu} \left(   {\partial_\nu  {\overline{\psi}}{}^{\mu\nu}_{IJ}   }    \right) \vol 
 +
{\overline{\psi}}{}^{\mu\nu}_{IJ} \dd  \omega^{IJ}_{\mu} \wedge     \vol_{\nu}  \right) = 0  ,
\\
\\
\displaystyle  
  {  {\{}}    \dttP_{\omega, \varphi}    ,    \dttP_{\omega, {\overline{\varphi}}}     {  {\}}} &  =  & \displaystyle   -    \Xi (  \dttP_{\omega, \varphi}     ) \iN \Xi (  \dttP_{\omega, {\overline{\varphi}}}     ) \iN {\pmb{\omega}}^{\tiny{\hbox{\sffamily DW}}}  =      \Xi (  \dttP_{\omega, \varphi}     ) \iN  \dd   \dttP_{\omega, {\overline{\varphi}}}  ,   \\
 \displaystyle     
&  =  & \displaystyle        \Xi (  \dttP_{\omega, \varphi}     ) \iN   \left(   { {\varphi}}{}_{\mu}^{IJ} (x)       \dd   {{p}}^{\omega_{\mu}\nu}_{IJ}   \wedge \vol_{\nu}    + {{p}}^{\omega_{\mu}\nu}_{IJ}  \partial_{\nu}     { {\varphi}}{}_{\mu}^{IJ}    (x)    \vol         \right)      = 0 ,
\\
\\
\displaystyle   {  {\{}}    \dttQ_{\omega , \psi}    ,      \dttP_{\omega, \varphi}     {  {\}}}   &  =  & \displaystyle       \Xi (    \dttQ_{\omega , \psi}    ) \iN   \left(   {{\varphi}}{}_{\mu}^{IJ} (x)       \dd   {{p}}^{\omega_{\mu}\nu}_{IJ}   \wedge \vol_{\nu}    + {{p}}^{\omega_{\mu}\nu}_{IJ}  \partial_{\nu}     {{\varphi}}{}_{\mu}^{IJ}    (x)    \vol         \right)   ,
\\
 \displaystyle  
&  =  & \displaystyle -    [  \delta^{I}_{K} \delta^{J}_{L}   \delta^{\rho}_{\mu} \delta^{\sigma}_{\nu}   ]  \psi^{\rho\sigma}_{KL} (x)    \varphi_{\mu}^{IJ} (x )   \vol_{\nu}
= -
    \psi^{\mu\nu}_{IJ} (x)    \phi_{\mu}^{IJ} (x )   \vol_{\nu}  .
   \\
   \end{array}
\right.
\nonumber
\eee
 \begin{prop}\label{proposition-Lie-algebra}  
 $\textswab{A}_{\mathfrak{1}}   $ is a Lie algebra. 
 \end{prop}  
{\sffamily Proof}. We consider the   Hamiltonian $(n-1)$-forms    $  \dttQ_{\omega ,  {\psi}_{{\mathfrak{1}}} }   ,  \dttQ_{\omega ,  {\psi}_{{\mathfrak{2}}} } ,  \dttQ_{\omega ,  {\psi}_{{\mathfrak{3}}} }  \in {\Omega}_{\tiny{\hbox{\sffamily Ham}}}^{{{n-1}}} (  {\cal M}_{{\hbox{\tiny\sffamily DW}}}  )   $  and  $  \dttP_{\omega ,  {\varphi}_{{\mathfrak{1}}} }   ,  \dttP_{\omega ,  {\varphi}_{{\mathfrak{2}}} } ,  \dttP_{\omega ,  {\varphi}_{{\mathfrak{3}}} }  \in {\Omega}_{\tiny{\hbox{\sffamily Ham}}}^{{{n-1}}} (  {\cal M}_{{\hbox{\tiny\sffamily DW}}}  )   $.    The     brackets   $  {  {\{}}      {  {\{}}     \dttQ_{\omega , \psi_{\mathfrak{1}}}    ,     \dttQ_{\omega , \psi_{\mathfrak{2}}}        {  {\}}}  ,    \dttQ_{\omega , \psi_{\mathfrak{3}}}        {  {\}}}  $, $ 
    {  {\{}}      {  {\{}}       \dttQ_{\omega , \psi_{\mathfrak{1}}}      ,       \dttQ_{\omega , \psi_{\mathfrak{3}}}         {  {\}}}  ,     \dttQ_{\omega , \psi_{\mathfrak{2}}}      {  {\}}}  $,   ${  {\{}}      {  {\{}}       \dttQ_{\omega , \psi_{\mathfrak{2}}}      ,       \dttQ_{\omega , \psi_{\mathfrak{3}}}         {  {\}}}  ,     \dttQ_{\omega , \psi_{\mathfrak{1}}}      {  {\}}}    $,    $ {  {\{}}      {  {\{}}     \dttP_{\omega , \varphi_{\mathfrak{1}}}    ,     \dttP_{\omega , \varphi_{\mathfrak{2}}}        {  {\}}}  ,    \dttP_{\omega , \varphi_{\mathfrak{3}}}        {  {\}}} $, 
 $   {  {\{}}      {  {\{}}       \dttP_{\omega , \varphi_{\mathfrak{1}}}      ,       \dttP_{\omega , \varphi_{\mathfrak{3}}}         {  {\}}}  ,     \dttP_{\omega , \varphi_{\mathfrak{2}}}      {  {\}}}       $, as well as the bracket $   {  {\{}}      {  {\{}}       \dttP_{\omega , \varphi_{\mathfrak{2}}}      ,       \dttP_{\omega , \varphi_{\mathfrak{3}}}         {  {\}}}  ,     \dttP_{\omega , \varphi_{\mathfrak{1}}}      {  {\}}}      $    are identically vanishing. 
  We also have 
   \[
\left.
\begin{array}{ccl}
\displaystyle   {  {\{}}      {  {\{}}     \dttQ_{\omega , \psi_{\mathfrak{1}}}    ,   \dttP_{\omega , \varphi }      {  {\}}}  ,    \dttQ_{\omega , \psi_{\mathfrak{2}}}       {  {\}}}   &  =  & \displaystyle  -   \Xi (    \dttQ_{\omega , \psi_{\mathfrak{2}}}       )  \iN \dd     {  {\{}}     \dttQ_{\omega , \psi_{\mathfrak{1}}}    ,   \dttP_{\omega , \varphi }      {  {\}}}   ,    \\
 \displaystyle        
&  =  & \displaystyle     \Xi (    \dttQ_{\omega , \psi_{\mathfrak{2}}}       )  \iN     \left(   
     \partial_{\nu}  \psi^{\mu\nu}_{IJ} (x) \varphi^{IJ}_{\mu} (x )   + \psi^{\mu\nu}_{IJ} (x)   \partial_{\nu}   \varphi^{IJ}_{\mu} (x )      \right)   {\dd}\mathfrak{y}      = 0  .
\end{array}
\right.
\]
Analogously, ${  {\{}}      {  {\{}}   \dttP_{\omega , \varphi }    ,  \dttQ_{\omega , \psi_{\mathfrak{2}}}   {  {\}}}  ,   \dttQ_{\omega , \psi_{\mathfrak{1}}}   {  {\}}} =  {  {\{}}      {  {\{}}  \dttQ_{\omega , \psi_{\mathfrak{2}}} ,  \dttQ_{\omega , \psi_{\mathfrak{1}}}        {  {\}}}  ,  \dttP_{\omega , \varphi }      {  {\}}} = 0 $. Finally, the  last   brackets  $   {  {\{}}       {  {\{}} \dttP_{\omega ,  {\varphi}_{{\mathfrak{1}}} }  ,    \dttQ_{\omega , \psi }    {  {\}}}    ,   \dttP_{\omega ,  {\varphi}_{{\mathfrak{2}}} }   {  {\}}}   
$, $ 
  {  {\{}}      {  {\{}}   \dttQ_{\omega , \psi }    ,   \dttP_{\omega ,  {\varphi}_{{\mathfrak{2}}} }   {  {\}}}  ,   \dttP_{\omega ,  {\varphi}_{{\mathfrak{1}}} }   {  {\}}}  $, and $    {  {\{}}      {  {\{}}   \dttP_{\omega ,  {\varphi}_{{\mathfrak{2}}} } ,  \dttP_{\omega ,  {\varphi}_{{\mathfrak{1}}} }        {  {\}}}  ,  \dttQ_{\omega , \psi }      {  {\}}}     $  are also identically vanishing.
The  Jacobi property is satisfied exactly,  {\em i.e.}  
\bee
\left.
\begin{array}{rcl}
\displaystyle   0   &  =  & \displaystyle    {  {\{}}       {  {\{}} \dttQ_{\omega , \psi_{\mathfrak{1}}}  ,    \dttP_{\omega , \varphi }    {  {\}}}    ,  \dttQ_{\omega , \psi_{\mathfrak{2}}}   {  {\}}}    +   {  {\{}}      {  {\{}}   \dttP_{\omega , \varphi }    ,  \dttQ_{\omega , \psi_{\mathfrak{2}}}   {  {\}}}  ,   \dttQ_{\omega , \psi_{\mathfrak{1}}}   {  {\}}}  +    {  {\{}}      {  {\{}}  \dttQ_{\omega , \psi_{\mathfrak{2}}} ,  \dttQ_{\omega , \psi_{\mathfrak{1}}}        {  {\}}}  ,  \dttP_{\omega , \varphi }      {  {\}}}  ,
\\
  \displaystyle     0   
&  =  & \displaystyle {  {\{}}       {  {\{}} \dttP_{\omega ,  {\varphi}_{{\mathfrak{1}}} }  ,    \dttQ_{\omega , \psi }    {  {\}}}    ,   \dttP_{\omega ,  {\varphi}_{{\mathfrak{2}}} }   {  {\}}}  + 
  {  {\{}}      {  {\{}}   \dttQ_{\omega , \psi }    ,   \dttP_{\omega ,  {\varphi}_{{\mathfrak{2}}} }   {  {\}}}  ,   \dttP_{\omega ,  {\varphi}_{{\mathfrak{1}}} }   {  {\}}}  +    {  {\{}}      {  {\{}}   \dttP_{\omega ,  {\varphi}_{{\mathfrak{2}}} } ,  \dttP_{\omega ,  {\varphi}_{{\mathfrak{1}}} }        {  {\}}}  ,  \dttQ_{\omega , \psi }      {  {\}}}      .
\end{array}
\right.
\eee
Note that  
$
[  \Xi ( \dttQ_{\omega ,   \psi }    ) ,  \Xi (  \dttP_{\omega , \varphi }  ) ] \iN  {\pmb{\omega}}^{\tiny{\hbox{\sffamily DW}}}  =  \Xi (      {  {\{}}     \dttQ_{\omega , \psi }    ,   \dttP_{\omega , \varphi }      {  {\}}}   ) \iN  {\pmb{\omega}}^{\tiny{\hbox{\sffamily DW}}} =  - \dd  (  {  {\{}}     \dttQ_{\omega , \psi_{\mathfrak{1}}}    ,   \dttP_{\omega , \varphi }      {  {\}}} ) 
$. In this case the Hamiltonian  vector field is $   \Xi (      {  {\{}}     \dttQ_{\omega , \psi_{\mathfrak{1}}}    ,   \dttP_{\omega , \varphi }      {  {\}}}   )  =    \left(   
     \partial_{\nu}  \psi^{\mu\nu}_{IJ} (x) \varphi^{IJ}_{\mu} (x )   + \psi^{\mu\nu}_{IJ} (x)   \partial_{\nu}   \varphi^{IJ}_{\mu} (x )      \right)     {\partial }/{\partial \varkappa}$.

  \subsubsection{{{Homotopy Lie Algebraic structure  }}}\label{subsec:Lie-algebra-02}

In this section, we     work with the   set of Hamiltonian $(n-1)$-forms  $\textswab{a}_{\mathfrak{2}} := {\{}   \textswab{a}_{\mathfrak{1}}  , \dttP_{\varkappa}    {\}} $.  We present the failure of the Jacobi identity, {\em i.e.} the homotopy type of the Lie algebraic structure.    Here we only focus on  the brackets between the $(n-1)$-forms  $\dttP_{\varkappa}  ,  \dttQ_{\omega , \psi}      ,  \dttP_{\omega, \varphi}  \in {\Omega}_{\tiny{\hbox{\sffamily Ham}}}^{{{n-1}}} (  {\cal M}_{{\hbox{\tiny\sffamily DW}}}  ) $.
 \begin{prop}\label{prop:amqwv2}  
  On the multisymplectic manifold   {\em ${ ( {{\cal M}}_{\hbox{\tiny{\sffamily DW}}}   , {\pmb{\omega}}^{\tiny{\hbox{\sffamily DW}}})}$}, the bracket operations between    the     Hamiltonian $(n-1)$-forms {\em $  \dttP_{\varkappa}  \in {\Omega}_{\tiny{\hbox{\sffamily Ham}}}^{{{n-1}}} (  {\cal M}_{{\hbox{\tiny\sffamily DW}}}  ) $} and    {\em $     \dttQ_{\omega , \psi}      ,  \dttP_{\omega, \varphi}  \in {\Omega}_{\tiny{\hbox{\sffamily Ham}}}^{{{n-1}}} (  {\cal M}_{{\hbox{\tiny\sffamily DW}}}  )$}  are given by
 {\em \bee\label{qmdlzzo8888}
  \left.
\begin{array}{rcl}
\displaystyle  {  {\{}}   \dttP_{\varkappa}   ,     \dttQ_{\omega , \psi}     {  {\}}}    &  =  & \displaystyle        X^{\rho} (x )    \left(  \omega^{IJ}_{\mu} \left(   {\partial_\nu  {{\psi}}{}^{\mu\nu}_{IJ}  (x )     }    \right) \vol_{\rho}   -
 { {\psi}}{}^{\mu\nu}_{IJ}  (x )   \dd  \omega^{IJ}_{\mu} \wedge     \vol_{\rho\nu}         
   \right) ,
\\
\displaystyle  {  {\{}}    \dttP_{\varkappa}    ,      \dttP_{\omega, \varphi}     {  {\}}}  &  =  & \displaystyle    X^{\rho} (x )   \left(         {{p}}^{\omega_{\mu}\nu}_{IJ}  \partial_{\nu}     {{\varphi}}{}_{\mu}^{IJ}    (x)    \vol_\rho -  {{\varphi}}{}_{\mu}^{IJ} (x)       \dd   {{p}}^{\omega_{\mu}\nu}_{IJ}   \wedge \vol_{\rho\nu}   \right)   ,
   \\ 
 \displaystyle &     & \displaystyle  
     +   {p}^{{\omega}_\mu \sigma}_{IJ}       ( {\partial_{\sigma}  X^{\nu} }  )     {{\varphi}}{}_{\mu}^{IJ} (x)   \vol_{\nu}      -  {p}^{{\omega}_\mu \nu}_{IJ}          ( {\partial_\lambda  X^{\lambda} }  )       {{\varphi}}{}_{\mu}^{IJ} (x)     \vol_{\nu}       .
 \end{array}
\right.
\nonumber
\eee
  }
 \end{prop}
{\sffamily Proof}. By a straightforward calculation,  using lemma \ref{lem:observable12}  and  lemma \ref{lem:observable1}, we obtain
  \[
\left.
\begin{array}{ccl}
\displaystyle   {\big{\{}}     \dttP_{\varkappa}   ,    \dttQ_{\omega,{{\psi}}}   {\big{\}}}   &  =  & \displaystyle      \Xi (   \dttP_{\varkappa}    ) \iN \dd  \dttQ_{\omega,{{\psi}}}  =     X^{\rho}   (x )    \partial_\rho    \iN \left(
   \omega^{IJ}_{\mu} \left(   {\partial_\nu  {{\psi}}{}^{\mu\nu}_{IJ}  (x )     }    \right) \vol 
+
{{\psi}}{}^{\mu\nu}_{IJ}  (x )   \dd  \omega^{IJ}_{\mu} \wedge     \vol_{\nu}  \right)  ,
\\
 \displaystyle        
&  =  & \displaystyle      -     X^{\rho}  (x )    
  \omega^{IJ}_{\mu}  \dd    {{\psi}}{}^{\mu\nu}_{IJ}    \wedge \vol_{\rho\nu}      -  X^{\rho}  (x )    
 {{\psi}}{}^{\mu\nu}_{IJ}  (x )  \dd  \omega^{IJ}_{\mu} \wedge     \vol_{\rho\nu} ,
    \end{array}
\right.
\]
\[ 
\left.
\begin{array}{ccl}\displaystyle   {\big{\{}}    \dttP_{\varkappa}    ,      \dttP_{\omega, \varphi}     {\big{\}}}   &  =  & \displaystyle     \Xi (   \dttP_{\varkappa}    ) \iN \dd  \dttP_{\omega,{{\psi}}}   =      X^{\rho} (x ) \partial_\rho   \iN  \left(   {{\varphi}}{}_{\mu}^{IJ} (x)       \dd   {{p}}^{\omega_{\mu}\nu}_{IJ}   \wedge \vol_{\nu}    + {{p}}^{\omega_{\mu}\nu}_{IJ}  \partial_{\nu}     {{\varphi}}{}_{\mu}^{IJ}    (x)    \vol         \right)  
   ,
   \\
\displaystyle       &    & \displaystyle     +    {p}^{{\omega}_\alpha \sigma}_{KL}      \left(  ( {\partial_{\sigma}  X^{\beta} }  )   -  \delta^{\beta}_{\sigma}     ( {\partial_\lambda  X^{\lambda} }  )     \right)  \partial / \partial {p}^{{\omega}_\alpha \beta}_{KL}      \iN  \left(   {{\varphi}}{}_{\mu}^{IJ} (x)       \dd   {{p}}^{\omega_{\mu}\nu}_{IJ}   \wedge \vol_{\nu}            \right)   ,
 \\
 \displaystyle        
&  =  & \displaystyle -      X^{\rho} (x)       {{p}}^{\omega_{\mu}\nu}_{IJ}  \dd    {{\varphi}}{}_{\mu}^{IJ}       \wedge  \vol_{\rho \nu}    -       X^{\rho} (x)    {{\varphi}}{}_{\mu}^{IJ} (x)       \dd   {{p}}^{\omega_{\mu}\nu}_{IJ}   \wedge \vol_{\rho\nu}       ,
\\
\displaystyle       &    & \displaystyle  +    {{\varphi}}{}_{\mu}^{IJ} (x)   \left( {p}^{{\omega}_\mu \rho}_{IJ}         ( {\partial_{\rho}  X^{\nu} }  )           -   {p}^{{\omega}_\mu \nu}_{IJ}      ( {\partial_\rho  X^{\rho} }  )       \right)   {\beta}_{\nu}  
  . 
   \\   \end{array}
\right.
\]
Then,  
 \bee\label{ext-derivative-01}
\left.
\begin{array}{ccl}
   \displaystyle    \dd \left(  {\big{\{}}     \dttP_{\varkappa}   ,    \dttQ_{\omega,{{\psi}}}   {\big{\}}}   \right)   &  =  & \displaystyle  
   -  \left(  X^{\rho}          (x)  \partial_\rho { {\psi}}{}^{\mu\nu}_{IJ}  (x)    +  { {\psi}}{}^{\mu\nu}_{IJ}  (x)    \partial_\rho  X^{\rho} (x)    \right) \dd  \omega^{IJ}_{\mu}    \wedge     \vol_{\nu}     ,
\\
 \displaystyle  &    & \displaystyle 
+  \left(  X^{\nu}          (x)  
\partial_\rho { {\psi}}{}^{\mu\rho}_{IJ}  (x)    +  { {\psi}}{}^{\mu\rho}_{IJ}  (x)    \partial_\rho  X^{\nu} (x)    \right)     \dd  \omega^{IJ}_{\mu}    \wedge   \vol_{\nu}  ,
  \\
 \displaystyle        
&     & \displaystyle   +  (  \partial_\nu    {{\psi}}{}^{\mu\nu}_{IJ}    )     X^{\rho}  (x )    
 \dd  \omega^{IJ}_{\mu}   \wedge   \vol_{\rho}   - (  \partial_\rho    {{\psi}}{}^{\mu\nu}_{IJ}     )      X^{\rho}  (x )    
 \dd  \omega^{IJ}_{\mu}  \wedge   \vol_{\nu}        ,
 \\
 \displaystyle        
&     & \displaystyle      +   \omega^{IJ}_{\mu}     \partial_\rho    X^{\rho}  \vol    -    \omega^{IJ}_{\mu}     \partial_\nu    X^{\rho}       \vol     , 
    \end{array}
\right.
\eee
   \bee\label{ext-derivative-02}
\left.
\begin{array}{ccl}
 \displaystyle  \dd   {\big{\{}}    \dttP_{\varkappa}    ,      \dttP_{\omega, \varphi}     {\big{\}}}   &  =  & \displaystyle  \displaystyle -    X^{\rho} (x)    \partial_\rho    {{\varphi}}{}_{\mu}^{IJ}     \dd     {{p}}^{\omega_{\mu}\nu}_{IJ}   \wedge    \vol_{\nu}     +    X^{\rho} (x)    \partial_\nu    {{\varphi}}{}_{\mu}^{IJ}     \dd     {{p}}^{\omega_{\mu}\nu}_{IJ}   \wedge    \vol_{\rho}  ,
          \\
 \displaystyle       &    & \displaystyle +     {{p}}^{\omega_{\mu}\nu}_{IJ}  (\partial_\rho  X^{\rho} (x))         \partial_{\nu}    {{\varphi}}{}_{\mu}^{IJ}            \vol , 
  -
    \left(   {{p}}^{\omega_{\mu}\nu}_{IJ}  \partial_\nu  X^{\rho} (x)    \right)    \partial_{\rho}    {{\varphi}}{}_{\mu}^{IJ}           \vol ,
 \\
 \displaystyle       &    & \displaystyle 
   +     \left(  X^{\rho} (x)  \partial_\rho     {{\varphi}}{}_{\mu}^{IJ} (x) +   {{\varphi}}{}_{\mu}^{IJ} (x)    \partial_\rho    X^{\rho} (x)     \right)   \dd   {{p}}^{\omega_{\mu}\nu}_{IJ}   \wedge    \vol_{\nu} ,         
\\
 \displaystyle       &    & \displaystyle 
-    \left(  X^{\rho} (x)  \partial_\nu     {{\varphi}}{}_{\mu}^{IJ} (x) +   {{\varphi}}{}_{\mu}^{IJ} (x)    \partial_\nu    X^{\rho} (x)     \right)   \dd   {{p}}^{\omega_{\mu}\nu}_{IJ}   \wedge   \vol_{\rho}  ,          
 \\
 \displaystyle       &    & \displaystyle  +      ( {\partial_{\rho}  X^{\nu} }  )        {{\varphi}}{}_{\mu}^{IJ} (x)       \dd   {p}^{{\omega}_\mu \rho}_{IJ}    \wedge    {\beta}_{\nu}   +      ( {\partial_{\rho}  X^{\nu} }  )            {p}^{{\omega}_\mu \rho}_{IJ}      \partial_\nu {{\varphi}}{}_{\mu}^{IJ} (x)       {\beta}        ,
       \\
 \displaystyle       &    & \displaystyle           -    ( {\partial_\rho  X^{\rho} }  )    {{\varphi}}{}_{\mu}^{IJ} (x)   \dd    {p}^{{\omega}_\mu \nu}_{IJ}      \wedge     
      {\beta}_{\nu}     -   ( {\partial_\rho  X^{\rho} }  )          {p}^{{\omega}_\mu \nu}_{IJ}   \partial_\nu   {{\varphi}}{}_{\mu}^{IJ} (x)           {\beta}    . 
   \end{array}
\right.
 \eee
We have  used  in \eqref{ext-derivative-01} and \eqref{ext-derivative-02} the definition $\vol_{\mu\nu} := \partial_\mu \iN \partial_\nu \iN \vol := \partial_\nu \wedge \partial_\mu \iN \vol $ and  the algebraic identity $ \dd x^{\alpha} \wedge  {\beta}_{\rho\nu} = \delta^{\alpha}_{\rho} \vol_{\nu} - \delta^{\alpha}_{\nu} \vol_{\rho}  $.  The   brackets obtained by cyclic permutations are given by ${{\{}} {{\{}} \dttP_{\varkappa}  ,   \dttQ_{\omega , \psi}    { {\}}}    ,   \dttP_{\omega, \varphi}   { {\}}} =  -   \Xi ( \dttP_{\omega, \varphi}     )  \iN     \dd \left(  { {\{}}     \dttP_{\varkappa}   ,    \dttQ_{\omega,{{\psi}}}   { {\}}}   \right)     
$, $     { {\{}}      { {\{}}   \dttQ_{\omega , \psi}    ,   \dttP_{\omega, \varphi}   { {\}}}  ,   \dttP_{\varkappa}   { {\}}}  =   -  \Xi (\dttP_{\varkappa}  )  \iN    \dd \left(       { {\{}}   \dttQ_{\omega , \psi}    ,   \dttP_{\omega, \varphi}   { {\}}} \right)      
$, and  $ { {\{}}      { {\{}}    \dttP_{\omega, \varphi}   ,   \dttP_{\varkappa}   { {\}}}  ,  \dttQ_{\omega , \psi}       { {\}}} =    \Xi ( \dttQ_{\omega , \psi} )     \iN    \dd    \left(        { {\{}}   \dttP_{\varkappa}   ,    \dttP_{\omega, \varphi}      { {\}}}    \right)$. Thus, we obtain   
 \bee\label{cyclica001}
\left.
\begin{array}{ccl}
\displaystyle   { {\{}}       { {\{}} \dttP_{\varkappa}  ,    \dttQ_{\omega , \psi}    { {\}}}    ,   \dttP_{\omega, \varphi}   { {\}}}   &  =  &   \displaystyle 
 -    \varphi^{IJ}_{\mu}      \left(  X^{\rho}          (x)  
\partial_\rho { {\psi}}{}^{\mu\nu}_{IJ}  (x)    +  { {\psi}}{}^{\mu\nu}_{IJ}  (x)    \partial_\rho  X^{\rho} (x)    \right)          \vol_{\nu}     ,
 \\
\displaystyle  &    & \displaystyle 
+    \varphi^{IJ}_{\mu}    \left(  X^{\nu}          (x)  
\partial_\rho { {\psi}}{}^{\mu\rho}_{IJ}  (x)    +  { {\psi}}{}^{\mu\rho}_{IJ}  (x)    \partial_\rho  X^{\nu} (x)    \right)      \vol_{\nu}  ,
 \\
\displaystyle  &    & \displaystyle 
 + \varphi^{IJ}_{\mu}     (  \partial_\nu    {{\psi}}{}^{\mu\nu}_{IJ}    )     X^{\rho}  (x )    
    \vol_{\rho}   - \varphi^{IJ}_{\mu}     (  \partial_\rho    {{\psi}}{}^{\mu\nu}_{IJ}     )      X^{\rho}  (x )    
    \vol_{\nu},
     \end{array}
\right.
\eee
\bee\label{cyclica002}
\left.
\begin{array}{ccl}
\displaystyle   { {\{}}      { {\{}}   \dttQ_{\omega , \psi}    ,   \dttP_{\omega, \varphi}   { {\}}}  ,   \dttP_{\varkappa}   { {\}}}   &  =  & \displaystyle      
    -        X^{\rho} (x )   \psi^{\mu\nu}_{IJ} (x)  \dd   \varphi^{IJ}_{\mu} \wedge {\dd}\mathfrak{y}_{\rho\nu}  -   X^{\rho} (x )  \varphi^{IJ}_{\mu} (x )  \dd   \psi^{\mu\nu}_{IJ} \wedge {\dd}\mathfrak{y}_{\rho\nu}  ,
   \\
      \displaystyle     &  =  & \displaystyle     -      X^{\rho} (x )   \psi^{\mu\nu}_{IJ} (x)  \partial_{\rho}   \varphi^{IJ}_{\mu}   \vol_{\nu}    +   X^{\rho} (x )   \psi^{\mu\nu}_{IJ} (x)  \partial_{\nu}   \varphi^{IJ}_{\mu}  \vol_{\rho}       ,
\\
      \displaystyle     &    & \displaystyle       -     X^{\rho} (x )  \varphi^{IJ}_{\mu} (x )  \partial_{\rho}    \psi^{\mu\nu}_{IJ}    \vol_{\nu}   +  X^{\rho} (x )  \varphi^{IJ}_{\mu} (x )  \partial_{\nu}    \psi^{\mu\nu}_{IJ}       \vol_{\rho}    ,
  \end{array}
\right.
\eee
 \bee\label{cyclica003}
\left.
\begin{array}{ccl}
\displaystyle   { {\{}}      { {\{}}    \dttP_{\omega, \varphi}   ,   \dttP_{\varkappa}   { {\}}}  ,  \dttQ_{\omega , \psi}       { {\}}}   &  =  &   \displaystyle 
      -   \psi^{\mu\nu}_{IJ}          \left(    -    X^{\rho} (x)    \partial_\rho    {{\varphi}}{}_{\mu}^{IJ}          \vol_{\nu}     +    X^{\rho} (x)    \partial_\nu    {{\varphi}}{}_{\mu}^{IJ}       \vol_{\rho}       \right)   ,
 \\
 \displaystyle     &     & \displaystyle        -   \psi^{\mu\nu}_{IJ}        \left(       \left(  X^{\rho} (x)  \partial_\rho     {{\varphi}}{}_{\mu}^{IJ} (x) +   {{\varphi}}{}_{\mu}^{IJ} (x)    \partial_\rho    X^{\rho} (x)     \right)       \vol_{\nu}        \right)   ,
 \\
 \displaystyle     &     & \displaystyle     +   \psi^{\mu\nu}_{IJ}     \left(      \left(  X^{\rho} (x)  \partial_\nu     {{\varphi}}{}_{\mu}^{IJ} (x) +   {{\varphi}}{}_{\mu}^{IJ} (x)    \partial_\nu    X^{\rho} (x)     \right)       \vol_{\rho}            
     \right)  ,
 \\
  \displaystyle     &     & \displaystyle          -   \psi^{\mu\nu}_{IJ}      \left(            ( {\partial_{\rho}  X^{\nu} }  )        {{\varphi}}{}_{\mu}^{IJ} (x)       \dd   {p}^{{\omega}_\mu \rho}_{IJ}    \wedge    {\beta}_{\nu}    -    ( {\partial_\rho  X^{\rho} }  )    {{\varphi}}{}_{\mu}^{IJ} (x)   
      {\beta}_{\nu}        \right)  = 0.      
       \end{array}
\right.
 \eee
 
   Let us denote    $({\hbox{\sffamily cyc}}) :=   { {\{}}  {{\{}} \dttP_{\varkappa}  ,    \dttQ_{\omega , \psi}    { {\}}}    ,   \dttP_{\omega, \varphi}   { {\}}}  +  {{\{}}      {{\{}}   \dttQ_{\omega , \psi}    ,   \dttP_{\omega, \varphi}   { {\}}}  ,   \dttP_{\varkappa}   {{\}}}    +  {{\{}}      {{\{}}    \dttP_{\omega, \varphi}   ,   \dttP_{\varkappa}   { {\}}}  ,  \dttQ_{\omega , \psi}       {{\}}}$,   the sum of cyclic permutations. Using \eqref{cyclica001} - \eqref{cyclica003}, we obtain
\bee\label{cyclic001434}
\left.
\begin{array}{ccl}
 \displaystyle  ({\hbox{\sffamily cyc}})  & =    & \displaystyle 
    -   \varphi^{IJ}_{\mu}       { {\psi}}{}^{\mu\nu}_{IJ}  (x)    \partial_\rho  X^{\rho} (x)             \vol_{\nu}     
 +   \varphi^{IJ}_{\mu}    { {\psi}}{}^{\mu\rho}_{IJ}  (x)    \partial_\rho  X^{\nu} (x)         \vol_{\nu}  
    -       X^{\rho} (x )   \psi^{\mu\nu}_{IJ} (x)  \partial_{\rho}   \varphi^{IJ}_{\mu}   \vol_{\nu} ,
      \\
   \displaystyle     &     & \displaystyle  
      -    X^{\nu} (x )   \psi^{\mu\rho}_{IJ} (x)  \partial_{\rho}   \varphi^{IJ}_{\mu}  \vol_{\nu}    ,
   -    \varphi^{IJ}_{\mu}     X^{\rho}          (x)  
\partial_\rho { {\psi}}{}^{\mu\nu}_{IJ}  (x)         \vol_{\nu}  
+    \varphi^{IJ}_{\mu} X^{\nu}          (x)  
\partial_\rho { {\psi}}{}^{\mu\rho}_{IJ}  (x)       \vol_{\nu}  .
 \\
        \end{array}
\right.
\nonumber
\eee 
We denote by $\textswab{S}$  the $(n-2)$-form  $   \Xi (    \dttP_{\varkappa}  )  \wedge  \Xi (    \dttQ_{\omega , \psi}     )  \wedge  \Xi (    \dttP_{\omega, \varphi}  )   \iN  {\pmb{\omega}}^{\tiny{\hbox{\sffamily DW}}}  \in \Omega^{n-2}( {\cal M}_{{\hbox{\tiny\sffamily DW}}} )  $. Then, 
 $ \textswab{S}  =       \Xi (    \dttP_{\omega, \varphi}  )    \iN  \Xi (    \dttQ_{\omega , \psi}     )  \iN \Xi (    \dttP_{\varkappa}  )  \iN  {\pmb{\omega}}^{\tiny{\hbox{\sffamily DW}}}  $. Then,  
       \[
\left.
\begin{array}{ccl}
\displaystyle  \textswab{S}   
 &  =  & \displaystyle      \Xi (    \dttP_{\omega, \varphi}  )    \iN  \Xi (    \dttQ_{\omega , \psi}     )  \iN   
\left(     -    \varkappa   \left(   {\partial_{\mu}   X^{\mu}    }  \right)     \vol - X^{\mu} \dd \varkappa \wedge \vol_{\mu} + X^{\rho}  \dd
   {{p}}^{
\omega_{\mu}\nu}_{IJ} \wedge 
  \hbox{d} \omega_{\mu}^{IJ} \wedge  {\beta}_{\rho\nu}  \right),
  \\  
 \displaystyle   &     & \displaystyle +  
             \Xi (    \dttP_{\omega, \varphi}  )    \iN  \Xi (    \dttQ_{\omega , \psi}     )  \iN    \left(  \left(        { {p}}^{\omega_\mu \rho}_{IJ}  \partial_\rho X^{\nu} (x) -  { {p}}^{\omega_\mu \nu}_{IJ}  \partial_\rho X^{\rho} (x)  \right)   \dd \omega_{\mu}^{IJ} \wedge     {\beta}_{\nu} \right),
     \\
 \displaystyle   &  =  & \displaystyle      \Xi (    \dttP_{\omega, \varphi}  )    \iN   \left(        -   \omega^{IJ}_\mu   \left(  {\partial_{\nu} \psi^{\mu\nu}_{IJ}}   \right)    {\partial}/{\partial \varkappa}    \right)    \iN   
\left(    - X^{\mu} \dd \varkappa \wedge \vol_{\mu}      \right),
  \\
 \displaystyle   &     & \displaystyle +      \Xi (    \dttP_{\omega, \varphi}  )    \iN   \left(   -    \psi^{\alpha\beta}_{KL}    {\partial}/{\partial {{p}}^{\omega_{\alpha}\beta}_{KL}} 
 \right)    \iN   
\left(      X^{\rho}  \dd
   {{p}}^{
\omega_{\mu}\nu}_{IJ} \wedge 
  \hbox{d} \omega_{\mu}^{IJ} \wedge  {\beta}_{\rho\nu}  \right),
    \\
     \displaystyle   &  =  & \displaystyle  
       \Xi (    \dttP_{\omega, \varphi}  )    \iN     \left(     X^{\mu}       \omega^{IJ}_\mu   \left(  {\partial_{\nu} \psi^{\mu\nu}_{IJ}}   \right)        \vol_{\mu} - X^{\rho}    \psi^{\mu\nu}_{IJ}       
  \hbox{d} \omega_{\mu}^{IJ} \wedge  {\beta}_{\rho\nu}    \right) ,
         \\
   \displaystyle   &  =  & \displaystyle   -
      \left(    \varphi_{\alpha}^{KL}   {\partial}/{\partial {\omega}_{\alpha}^{KL} }      \right)   \iN     \left(       X^{\rho}    \psi^{\mu\nu}_{IJ}       
  \hbox{d} \omega_{\mu}^{IJ} \wedge  {\beta}_{\rho\nu}    \right) 
     =       -   X^{\rho}    \psi^{\mu\nu}_{IJ}    \varphi_{\mu}^{IJ}       
    {\beta}_{\rho\nu} . 
    \end{array}
\right.
\]
Therefore, 
     \[
\left.
\begin{array}{ccl} 
 \dd  \textswab{S}&  =  & \displaystyle      -   \left(          \psi^{\mu\nu}_{IJ} (x)    \varphi_{\mu}^{IJ} (x ) \dd       X^{\rho} (x )         +        X^{\rho} (x )  \psi^{\mu\nu}_{IJ} (x)   \dd   \varphi_{\mu}^{IJ} (x ) +    \varphi_{\mu}^{IJ} (x )     X^{\rho} (x )  \dd \psi^{\mu\nu}_{IJ} (x)     \right)   \wedge    \vol_{\rho\nu} ,
\\
  \displaystyle   &  =  & \displaystyle      -     \psi^{\mu\nu}_{IJ} (x)    \varphi_{\mu}^{IJ} (x )         \partial_\rho       X^{\rho} (x )  \vol_\nu   +     \psi^{\mu \rho}_{IJ} (x)     \varphi_{\mu}^{IJ} (x )   \partial_\rho       X^{\nu} (x )            \vol_{\nu}       -            X^{\rho} (x )      \psi^{\mu\nu}_{IJ} (x)   
  \partial_\rho   \varphi_{\mu}^{IJ} (x )     \vol_{\nu}    ,
\\
\displaystyle   &     & \displaystyle  
   +    X^{\nu} (x )      \psi^{\mu\rho}_{IJ} (x)    \partial_\rho   \varphi_{\mu}^{IJ} (x )      \vol_{\nu}     
      -         \varphi_{\mu}^{IJ} (x )          X^{\rho} (x )   \partial_\rho   \psi^{\mu\nu}_{IJ} (x)   \vol_{\nu}    +        \varphi_{\mu}^{IJ} (x )  X^{\nu} (x )   \partial_\rho   \psi^{\mu \rho}_{IJ} (x)       \vol_{\nu} ,
 \end{array}
\right.
\]
 is identically equal to  the sum of cyclic permutations: $ \dd \mathfrak{S} = ({\hbox{\sffamily cyc}})   $.  Hence, we have proven the   Jacobi  property  up to coherent homotopy, {\em i.e.}
 \bee
 \left.
\begin{array}{ccl}
\displaystyle   \dd \mathfrak{S}  
&  =  & \displaystyle            {{\{}}       {{\{}} \dttP_{\varkappa}  ,    \dttQ_{\omega , \psi}    {{\}}}    ,   \dttP_{\omega, \varphi}   {{\}}}  +  {{\{}}      {{\{}}   \dttQ_{\omega , \psi}    ,   \dttP_{\omega, \varphi}   {{\}}}  ,   \dttP_{\varkappa}   {{\}}}  +  {{\{}}      {{\{}}    \dttP_{\omega, \varphi}   ,   \dttP_{\varkappa}   {{\}}}  ,  \dttQ_{\omega , \psi}       {{\}}} .
\\
       \end{array}
\right.
\nonumber
\eee
 Using the notation   ${\textswab{S}}_{[n]}   :=  (\Xi_{1} \wedge \cdots \wedge \Xi_{n} ) \iN  {\pmb{\omega}}^{\tiny{\hbox{\sffamily DW}}} $ (where  $\Xi_1 , \cdots , \Xi_n   \in {\Omega}_{\tiny{\hbox{\sffamily Ham}}}^{{{n-1}}} (  {\cal M}_{{\hbox{\tiny\sffamily DW}}}  ) $ are  Hamiltonian vectors fields),  the Jacobi identity, up to a coherent   homotopy, is equivalently contained in   the    formula
 \bee \dd
{\textswab{S}}_{[n]}    = (-1)^{n} \sum_{1 \leq i <  j \leq n}  \left( [\Xi_i , \Xi_{j} ]  \wedge \Xi_{1} \wedge \cdots \Xi_{i-1} \wedge \Xi_{i+1} \wedge \cdots \wedge \Xi_{j-1} \wedge \Xi_{j+1} \wedge \cdots \Xi_{n} \right) \iN   {\pmb{\omega}}^{\tiny{\hbox{\sffamily DW}}}.
\nonumber
\eee
For a detailed proof, we refer to \cite{Rogers}, page $25$. Applying it to our example with  $ {{{\textswab{S}}}}_{[n]} :=    {{{\textswab{S}}}} =      \Xi (    \dttP_{\omega, \varphi}  )    \iN  \Xi (    \dttQ_{\omega , \psi}     )  \iN \Xi (    \dttP_{\varkappa}  )  \iN  {\pmb{\omega}}^{\tiny{\hbox{\sffamily DW}}}$),  we obtain
 \[
\left.
\begin{array}{ccl}
\displaystyle \dd  {\textswab{S}} &  =  & \displaystyle  - \left(  [ \Xi (    \dttP_{\varkappa} ) ,  \Xi (    \dttQ_{\omega , \psi}     ) ]   \wedge  \Xi (    \dttP_{\omega, \varphi}  )   \iN  {\pmb{\omega}}^{\tiny{\hbox{\sffamily DW}}}    \right)     
 -
  \left(  [ \Xi (    \dttP_{\varkappa} ) ,  \Xi (    \dttP_{\omega, \varphi}      ) ]   \wedge  \Xi (    \dttQ_{\omega , \psi}    )   \iN  {\pmb{\omega}}^{\tiny{\hbox{\sffamily DW}}}   \right)     ,
\\
 \displaystyle   &     & \displaystyle   - \left(  [ \Xi (    \dttQ_{\omega , \psi}     )  ,   \Xi (    \dttP_{\omega, \varphi}  )   ]   \wedge  \Xi (    \dttP_{\varkappa}    )     \iN  {\pmb{\omega}}^{\tiny{\hbox{\sffamily DW}}}     \right)     ,
 \\
 \displaystyle   &  =  & \displaystyle  - \left(     \Xi (    \dttP_{\omega, \varphi}  )   \iN    [ \Xi (    \dttP_{\varkappa} ) ,  \Xi (    \dttQ_{\omega , \psi}     ) ]     \iN {\pmb{\omega}}^{\tiny{\hbox{\sffamily DW}}}    \right)     
 -
  \left(     \Xi (    \dttQ_{\omega , \psi}    )   \iN  [ \Xi (    \dttP_{\varkappa} ) ,  \Xi (    \dttP_{\omega, \varphi}      ) ]    \iN   {\pmb{\omega}}^{\tiny{\hbox{\sffamily DW}}}   \right)    , 
\\
 \displaystyle   &     & \displaystyle   - \left(     \Xi (    \dttP_{\varkappa}    )     \iN    [ \Xi (    \dttQ_{\omega , \psi}     )  ,   \Xi (    \dttP_{\omega, \varphi}  )   ]    \iN {\pmb{\omega}}^{\tiny{\hbox{\sffamily DW}}}     \right)     ,
 \end{array}
\right.
\]
which  is easily verified.

  \subsubsection{{{Algebraic structure  on ${\pmb{\cal C}}_{e^{I}_{\mu} }  ,  {{\pmb{\cal C}}}_{\omega^{IJ}_{\mu} }$}}}\label{subsec:Lie-algebra-03}
 
Let us  denote by $\textswab{a}_{\mathfrak{3}} $ the set of two $(n-1)$-forms  ${\pmb{\cal C}}_{e^{I}_{\mu} }  ,  {{\pmb{\cal C}}}_{\omega^{IJ}_{\mu} }$, where ${{\pmb{\cal C}}}_{e^{I}_{\mu} } :=     {{p}}^{e_\mu \nu}_{I} \vol_{\nu}$ and $ {{\pmb{\cal C}}}_{\omega^{IJ}_{\mu} }: =   
  {{p}}^{{\omega}_{\mu}\nu}_{IJ}      \vol_{\nu} +  \dt^{[\mu}_{I} e^{\nu]}_{J}  \vol_{\nu}$.
  {Note that $ 
\displaystyle  \dd  {{\pmb{\cal C}}}_{e^{I}_{\mu} } =  \dd    {{p}}^{e_\mu \nu}_{I} \wedge \vol_{\nu} $
 and $
\displaystyle  \dd  {{\pmb{\cal C}}}_{\omega^{IJ}_{\mu} } =   
  \dd {{p}}^{{\omega}_{\mu}\nu}_{IJ}   \wedge   \vol_{\nu} 
 +   (1/2)    \epsilon_{IJKL} \epsilon^{\mu\nu\rho\sigma} e^{K}_{\rho} \dd    e^{L}_{\sigma}  \wedge \vol_{\nu}
  $.}   The related  Hamiltonian vector fields $  \Xi (   {{\pmb{\cal C}}}_{e^{I}_{\mu} }  )   $  and $ \Xi ({{\pmb{\cal C}}}_{\omega^{IJ}_{\mu} })  $ are given by
 \bee\label{Ham-vector-fields}
     \Xi (   {{\pmb{\cal C}}}_{e^{I}_{\mu} }  )   =       {\partial}/{\partial {e}_{\mu}^{I} }  ,
\quad \quad \quad \quad
      \Xi ({{\pmb{\cal C}}}_{\omega^{IJ}_{\mu} })  = 
 {\partial }/{\partial \omega^{IJ}_{\mu} }
   -  (1/2)   \epsilon_{IJKL} \epsilon^{\mu\nu\rho\sigma} e^{K}_{\rho}   {\partial}/{\partial {{p}}^{e_{\sigma}\nu}_{L}}.
 \eee
The   interior products of the Hamiltonian vector fields $  \Xi (   {{\pmb{\cal C}}}_{e^{I}_{\mu} }  )$ and $    \Xi ({{\pmb{\cal C}}}_{\omega^{IJ}_{\mu} })$  with the multisymplectic  form give  $ \Xi (   {{\pmb{\cal C}}}_{e^{I}_{\mu} }  ) \iN {\pmb{\omega}}^{\tiny{\hbox{\sffamily DW}}}   =
- \dd {{p}}^{e_{\mu}\nu}_I
 \wedge   {\beta}_{\nu} =
 -
  \dd  {{\pmb{\cal C}}}_{e^{I}_{\mu} }  $ and 
$ \Xi (   {{\pmb{\cal C}}}_{\omega^{IJ}_{\mu} }   ) \iN {\pmb{\omega}}^{\tiny{\hbox{\sffamily DW}}}  =
       -
  \dd {{p}}^{{\omega}_{\mu}\nu}_{IJ}   \wedge   \vol_{\nu} 
  -  (1/2)    \epsilon_{IJKL} \epsilon^{\mu\nu\rho\sigma} e^{K}_{\rho} \dd    e^{L}_{\sigma}  \wedge \vol_{\nu}
=  -  
  \dd  {{\pmb{\cal C}}}_{\omega^{IJ}_{\mu} }  
$, respectively. Note that, by definition,  ${{\pmb{\cal C}}}_{e^{I}_{\mu} }    |_{{\pmb{\cal C}}}   =  {{\pmb{\cal C}}}_{\omega^{IJ}_{\mu} }    |_{{\pmb{\cal C}}}       = 0 $.  We now calculate the bracket operations between the Hamiltonian $(n-1)$-forms $   {\pmb{\cal C}}_{e^{I}_{\mu}} \in {\Omega}_{\tiny{\hbox{\sffamily Ham}}}^{{{n-1}}} (  {\cal M}_{{\hbox{\tiny\sffamily DW}}}  ) $  and ${{\pmb{\cal C}}}_{\omega^{IJ}_{\mu} }      \in {\Omega}_{\tiny{\hbox{\sffamily Ham}}}^{{{n-1}}} (  {\cal M}_{{\hbox{\tiny\sffamily DW}}}  )  $: 
  \begin{equation}\label{09DD00z}
\left.
\begin{array}{rcl}
\displaystyle  {\big{\{}}  {{\pmb{\cal C}}}_{e^{I}_{\mu} }   ,   {{\pmb{\cal C}}}_{\overline{e}^{I}_{\mu} }    {\big{\}}}     &  =  & \displaystyle   - 
  \Xi (   {{\pmb{\cal C}}}_{e^{I}_{\mu} }  )
\iN  \Xi (   {{\pmb{\cal C}}}_{{\overline{e}}^{I}_{\mu} }  )     
\iN
\left(  \dd {{p}}^{e_{\mu}\nu}_I  \wedge \hbox{d} e_{\mu}^I \wedge   {\beta}_{\nu}  + \dd   {{p}}^{\omega_{\mu}\nu}_{IJ} \wedge 
  \hbox{d} \omega_{\mu}^{IJ} \wedge  {\beta}_{\nu} \right) ,
\\
 \displaystyle      &  =  & \displaystyle   - 
  \Xi (   {{\pmb{\cal C}}}_{e^{I}_{\mu} }  )
\iN   
 {\big{(}}    \dd {{p}}^{e_{\mu}\nu}_I   \wedge   {\beta}_{\nu} {\big{)}}  = 0 ,
\\
\\
\displaystyle  {\big{\{}}  {{\pmb{\cal C}}}_{\omega^{IJ}_{\mu} }   ,   {{\pmb{\cal C}}}_{\overline{\omega}^{IJ}_{\mu} }    {\big{\}}}     &  =  & \displaystyle   - 
  \Xi (   {{\pmb{\cal C}}}_{\omega^{IJ}_{\mu} }  )
\iN  \Xi (   {{\pmb{\cal C}}}_{{\overline{\omega}}^{IJ}_{\mu} }  )     
\iN
\left(  \dd {{p}}^{e_{\mu}\nu}_I  \wedge \hbox{d} e_{\mu}^I \wedge   {\beta}_{\nu}  + \dd   {{p}}^{\omega_{\mu}\nu}_{IJ} \wedge 
  \hbox{d} \omega_{\mu}^{IJ} \wedge  {\beta}_{\nu} \right),
\\
 \displaystyle      &  =  & \displaystyle  
    - 
 \Xi (   {{\pmb{\cal C}}}_{\omega^{IJ}_{\mu} }  ) \iN   
 {\big{(}}   \dd {{p}}^{{\omega}_{\mu}\nu}_{IJ}   \wedge   \vol_{\nu} 
 +  (1/2)   \epsilon_{IJKL} \epsilon^{\mu\nu\rho\sigma} e^{K}_{\rho} \dd    e^{L}_{\sigma}  \wedge \vol_{\nu} {\big{)}} = 0,
\\
\\
\displaystyle  {\big{\{}}    {{\pmb{\cal C}}}_{e^{L}_{\sigma} }    ,   {{\pmb{\cal C}}}_{{\omega}^{IJ}_{\mu} }    {\big{\}}}     &  =  & \displaystyle   -
  \Xi (  {{\pmb{\cal C}}}_{e^{L}_{\sigma} }    )
\iN  \Xi (   {{\pmb{\cal C}}}_{{{\omega}}^{IJ}_{\mu} }  )     
\iN
 {\big{(}}   \dd {{p}}^{e_{\mu}\nu}_I
 \wedge
  \hbox{d} e_{\mu}^I \wedge   {\beta}_{\nu}  + \dd
   {{p}}^{
\omega_{\mu}\nu}_{IJ} \wedge 
  \hbox{d} \omega_{\mu}^{IJ} \wedge  {\beta}_{\nu} {\big{)}} ,
\\
 \displaystyle      &  =  & \displaystyle  
 -  \Xi (  {{\pmb{\cal C}}}_{e^{L}_{\sigma} }     )
\iN   
\left(   
   \dd {{p}}^{{\omega}_{\mu}\nu}_{IJ}   \wedge   \vol_{\nu} 
 +   (1/2)   \epsilon_{IJKL} \epsilon^{\mu\nu\rho\sigma} e^{K}_{\rho} \dd    e^{L}_{\sigma}  \wedge \vol_{\nu} \right),
    \\
    \displaystyle      &  =  & \displaystyle    - (1/2)    \epsilon_{IJKL} \epsilon^{\mu\nu\rho\sigma} e^{K}_{\rho}    \vol_{\nu} .
\end{array}
\right.
\nonumber
\end{equation}
 
Note that 
$
 {\partial }/{\partial e^{L}_{\sigma} }  {\big{(}}   \dtt^{[\mu}_{I} e^{\nu]}_{J}   {\big{)}}
=  - 
{ {\partial}/{\partial e^{L}_{\sigma}}}
 {\big{(}}  (1/4) \epsilon_{IJKL} \epsilon^{\mu\nu\rho\sigma} e^{K}_{\rho} e^{L}_{\sigma}  {\big{)}}
= 
 -   (1/2)    \epsilon_{IJKL} \epsilon^{\mu\nu\rho\sigma} e^{K}_{\rho} 
$. We reproduce the result of Kanatchikov \cite{kannnnnnnaa,kannnnnnnaa1}, which underlines his constraints analysis of {DW} formulation of vielbein gravity and its precanonical quantization. In  particular, we refer to   equations $(19)$ page $6$ in \cite{kannnnnnnaa1}.  The brackets   are   written  as 
\begin{equation}\label{fzhdifzhef8767} 
    {\big{\{}}  {{\pmb{\cal C}}}_{e^{I}_{\mu} }   ,   {{\pmb{\cal C}}}_{\overline{e}^{J}_{\nu} }    {\big{\}}}       =    {\big{\{}}  {{\pmb{\cal C}}}_{\omega^{IJ}_{\mu} }   ,   {{\pmb{\cal C}}}_{\overline{\omega}^{KL}_{\nu} }    {\big{\}}}     =   
0,
\quad \quad 
    {\big{\{}}    {{\pmb{\cal C}}}_{e^{L}_{\sigma} }    ,   {{\pmb{\cal C}}}_{{\omega}^{IJ}_{\mu} }    {\big{\}}}   =       - \frac{\partial }{\partial e^{L}_{\sigma} }  {\big{(}}   \dtt^{[\mu}_{I} e^{\nu]}_{J}   {\big{)}}   \vol_{\nu}  .
 \end{equation}
 \begin{prop}\label{proposition-Lie-algebra-3}  
$\textswab{A}_{\mathfrak{3}} := {{\{}}  \textswab{a}_{\mathfrak{3}}   \ ; \ {{\{}}  , {{\}}} {{\}}}$ is a   Lie algebra, where  $ \textswab{a}_{\mathfrak{3}}  $ is the set of forms $ {{\pmb{\cal C}}}_{e^{L}_{\sigma} }    ,   {{\pmb{\cal C}}}_{{\omega}^{IJ}_{\mu} }  $ and where  the bracket  operation is ${ {\{}}     ,  { {\}}} $. 
 \end{prop} 
 {\sffamily Proof}. We consider       the Hamiltonian $(n-1)$-forms $ {{\pmb{\cal C}}}_{e^{M}_{\lambda} }   ,     {{\pmb{\cal C}}}_{e^{L}_{\sigma} }    ,   {{\pmb{\cal C}}}_{{\omega}^{IJ}_{\mu} } $. The following bracket operations based on  the cyclic permutations are found:
  \bee\label{cyclic001}
\left.
\begin{array}{ccl}
\displaystyle   { {\{}}       { {\{}} {{\pmb{\cal C}}}_{e^{M}_{\lambda} }  ,    {{\pmb{\cal C}}}_{e^{L}_{\sigma} }    { {\}}}    ,   {{\pmb{\cal C}}}_{{\omega}^{IJ}_{\mu} }   { {\}}}   &  =  & \displaystyle   -    \Xi ( {{\pmb{\cal C}}}_{{\omega}^{IJ}_{\mu} }     )  \iN \dd      { {\{}} {{\pmb{\cal C}}}_{e^{M}_{\lambda} }  ,    {{\pmb{\cal C}}}_{e^{L}_{\sigma} }    { {\}}}  =  0 ,
\\
\displaystyle   { {\{}}      { {\{}}   {{\pmb{\cal C}}}_{e^{L}_{\sigma} }    ,   {{\pmb{\cal C}}}_{{\omega}^{IJ}_{\mu} }   { {\}}}  ,   {{\pmb{\cal C}}}_{e^{M}_{\lambda} }   { {\}}}   &  =  & \displaystyle  -   \Xi ({{\pmb{\cal C}}}_{e^{M}_{\lambda} }  )  \iN \dd    { {\{}}   {{\pmb{\cal C}}}_{e^{L}_{\sigma} }    ,   {{\pmb{\cal C}}}_{{\omega}^{IJ}_{\mu} }   { {\}}} =      (1/2)    \epsilon_{IJML} \epsilon^{\mu\nu\lambda\sigma}  
   \vol_{\nu}     ,
   \\ 
\displaystyle   { {\{}}      { {\{}}   {{\pmb{\cal C}}}_{{\omega}^{IJ}_{\mu} } ,  {{\pmb{\cal C}}}_{e^{M}_{\lambda} }        { {\}}}  ,  {{\pmb{\cal C}}}_{e^{L}_{\sigma} }      { {\}}}   &  =  & \displaystyle  -   \Xi ( {{\pmb{\cal C}}}_{e^{L}_{\sigma} }      )  \iN \dd        { {\{}}   {{\pmb{\cal C}}}_{{\omega}^{IJ}_{\mu} } ,  {{\pmb{\cal C}}}_{e^{M}_{\lambda} }        { {\}}}      =    -   (1/2)    \epsilon_{IJML} \epsilon^{\mu\nu\lambda\sigma}   
   \vol_{\nu}         .
\end{array}
\right.
\eee
Also, let us   consider the   Hamiltonian $(n-1)$-forms   $ {{\pmb{\cal C}}}_{e^{M}_{\lambda} }   ,        {{\pmb{\cal C}}}_{{\omega}^{IJ}_{\mu} }   ,   {{\pmb{\cal C}}}_{{\omega}^{KL}_{\nu} }   \in  {\Omega}_{\tiny{\hbox{\sffamily Ham}}}^{{{n-1}}} (  {\cal M}_{{\hbox{\tiny\sffamily DW}}}  )   $. The    brackets based on the cyclic permutations of the Jacobi identity are 
  \bee\label{cyclic002}
\left.
\begin{array}{ccl}
\displaystyle    { {\{}}       { {\{}} {{\pmb{\cal C}}}_{e^{M}_{\lambda} }  ,     {{\pmb{\cal C}}}_{{\omega}^{KL}_{\nu} }      { {\}}}    ,   {{\pmb{\cal C}}}_{{\omega}^{IJ}_{\mu} }   { {\}}}   &  =  & \displaystyle  -    \Xi ( {{\pmb{\cal C}}}_{{\omega}^{IJ}_{\mu} }     )  \iN \dd      { {\{}} {{\pmb{\cal C}}}_{e^{M}_{\lambda} }  ,    {{\pmb{\cal C}}}_{{\omega}^{KL}_{\nu} }       { {\}}}  =  0 ,

\\
\displaystyle   { {\{}}      { {\{}}    {{\pmb{\cal C}}}_{{\omega}^{KL}_{\nu} }     ,   {{\pmb{\cal C}}}_{{\omega}^{IJ}_{\mu} }   { {\}}}  ,   {{\pmb{\cal C}}}_{e^{M}_{\lambda} }   { {\}}}   &  =  & \displaystyle  -   \Xi ({{\pmb{\cal C}}}_{e^{M}_{\lambda} }  )  \iN \dd    { {\{}}  {{\pmb{\cal C}}}_{{\omega}^{KL}_{\nu} }     ,   {{\pmb{\cal C}}}_{{\omega}^{IJ}_{\mu} }   { {\}}}  =   0,

\\
 \displaystyle   { {\{}}      { {\{}}   {{\pmb{\cal C}}}_{{\omega}^{IJ}_{\mu} } ,  {{\pmb{\cal C}}}_{e^{M}_{\lambda} }        { {\}}}  ,  {{\pmb{\cal C}}}_{{\omega}^{KL}_{\nu} }       { {\}}}        
&  =  & \displaystyle  \displaystyle  -   \Xi (  {{\pmb{\cal C}}}_{{\omega}^{KL}_{\nu} }     )  \iN \dd     { {\{}}   {{\pmb{\cal C}}}_{{\omega}^{IJ}_{\mu} } ,  {{\pmb{\cal C}}}_{e^{M}_{\lambda} }        { {\}}}    = 0     .
\end{array}
\right.
\eee
  Then, using \eqref{cyclic001} and \eqref{cyclic002}, we obtain the Jacobi identity
  \bee
\left.
\begin{array}{ccl}
\displaystyle   0 &  =  & \displaystyle { {\{}}       { {\{}} {{\pmb{\cal C}}}_{e^{M}_{\lambda} }  ,    {{\pmb{\cal C}}}_{e^{L}_{\sigma} }    { {\}}}    ,   {{\pmb{\cal C}}}_{{\omega}^{IJ}_{\mu} }   { {\}}}    +  { {\{}}      { {\{}}   {{\pmb{\cal C}}}_{e^{L}_{\sigma} }    ,   {{\pmb{\cal C}}}_{{\omega}^{IJ}_{\mu} }   { {\}}}  ,   {{\pmb{\cal C}}}_{e^{M}_{\lambda} }   { {\}}}  +    { {\{}}      { {\{}}   {{\pmb{\cal C}}}_{{\omega}^{IJ}_{\mu} } ,  {{\pmb{\cal C}}}_{e^{M}_{\lambda} }        { {\}}}  ,  {{\pmb{\cal C}}}_{e^{L}_{\sigma} }      { {\}}}  ,

\\
 \displaystyle     0     
&  =  & \displaystyle
 { {\{}}       { {\{}} {{\pmb{\cal C}}}_{e^{M}_{\lambda} }  ,    {{\pmb{\cal C}}}_{{\omega}^{KL}_{\nu} }    { {\}}}    ,   {{\pmb{\cal C}}}_{{\omega}^{IJ}_{\mu} }   { {\}}}    +  { {\{}}      { {\{}}   {{\pmb{\cal C}}}_{{\omega}^{KL}_{\nu} }    ,   {{\pmb{\cal C}}}_{{\omega}^{IJ}_{\mu} }   { {\}}}  ,   {{\pmb{\cal C}}}_{e^{M}_{\lambda} }   { {\}}}  +    { {\{}}      { {\{}}   {{\pmb{\cal C}}}_{{\omega}^{IJ}_{\mu} } ,  {{\pmb{\cal C}}}_{e^{M}_{\lambda} }        { {\}}}  ,  {{\pmb{\cal C}}}_{{\omega}^{KL}_{\nu} }      { {\}}}      .
\end{array}
\right.
\eee

   \subsection{{{Towards the canonical  forms for vielbein gravity}}}\label{subsec:saison003}

 The   quantization of  gravity  within  the {MG} formulation   is still in its infancy.  
However, some progress have been made by  Kanatchikov  within his  precanonical quantization based on his polysymplectic approach.  The description of  fundamental brackets, using the graded structure presented in \ref{subsec:saison001},   between   Hamiltonian $(n-1)$-forms and Hamiltonian $0$-forms is found in  \cite{kannnnnnnaa,kannnnnnnaa1}.   In particular,  the constraints analysis involves   a  generalization of the Dirac bracket to the polysymplectic context, see   \cite{KanaDirac}.   

Another example of canonical Poisson bracket, {\em i.e.} a  bracket between    canonically conjugate  forms, is  obtained by  using the copolarization of algebraic observable forms developed in the work of H\'{e}lein and Kouneiher \cite{HK-03}.   We present briefly  the formulation of a Poisson bracket on observable functionals for  vierbein gravity.  The functionals are built on the pair $(\omega , {\pmb{\varpi}})$   of   canonically conjugate  forms,  {\em i.e} ${\{ {\pmb{\varpi}} , {{\omega}}} \}  = 1$, where    $\omega := \omega^{IJ} \otimes \Delta_{IJ}  \in  {\Omega}_{\tiny{\hbox{\sffamily Ham}}}^{{{1}}} (  {\cal M}_{{\hbox{\tiny\sffamily DW}}}  ) \otimes {\textswab{g}}  $ and $  {\pmb{\varpi}} :=  {\pmb{\varpi}}_{IJ}  \otimes \Delta^{IJ}    \in  {\Omega}_{\tiny{\hbox{\sffamily Ham}}}^{{{n-2}}} (  {\cal M}_{{\hbox{\tiny\sffamily DW}}}  )  \otimes {\textswab{g}}^{\star} $.  We denote  $    
    {\pmb{\varpi}}_{I}   :=           (1/2) \sum_{\mu,\nu} {{p}}^{e_{\mu}\nu}_{I} {\beta}_{\mu\nu}   \in  {\Omega}_{\tiny{\hbox{\sffamily Ham}}}^{{{n-2}}} (  {\cal M}_{{\hbox{\tiny\sffamily DW}}}  )  $ and   
$
    {\pmb{\varpi}}_{IJ}  :=       (1/2) \sum_{\mu,\nu} {{p}}^{\omega_{\mu}\nu}_{IJ} {\beta}_{\mu\nu}       \in  {\Omega}_{\tiny{\hbox{\sffamily Ham}}}^{{{n-2}}} (  {\cal M}_{{\hbox{\tiny\sffamily DW}}}  )   
$.   When   restricted   to the constraint hypersurface ${\pmb{\cal C}}$, the $(n-2)$-forms are denoted $  {\pmb{\varpi}}_{I}   |_{{\pmb{\cal C}}}  := \iota^{\star} {\pmb{\varpi}}_{I}  = 0 $ and $ {\pmb{\varpi}}_{IJ}     |_{{\pmb{\cal C}}} :=  \iota^{\star} {\pmb{\varpi}}_{IJ} =   - ({{1}/{2}}) \sum_{\mu,\nu}  e e^{[\mu}_I e^{\nu]}_J  {\beta}_{\mu\nu} =   - (1/8)  \epsilon^{\mu\nu\sigma\rho} \epsilon_{IJKL} e^{K}_{\sigma} e^{L}_{\rho} 
\beta_{\mu\nu}$. 
Since 
 \[
\left.
\begin{array}{rcl}
\displaystyle \dd  e^{I} \wedge  \dd {\pmb{\varpi}}_{I}   & =  & \displaystyle
   (1/2)  \dd {{p}}^{e_{\mu}\nu}_{I}  \wedge   \dd  e^{I}_{\rho} \wedge  {\big{(}}  \delta^{\rho}_{\mu} \vol_{\nu} - \delta^{\rho}_{\nu} \vol_{\mu}  {\big{)}} =  \dd {{p}}^{e_{\mu}\nu}_I \wedge \hbox{d} e_{\mu}^I \wedge   {\beta}_{\nu} ,
\\
 \displaystyle  \dd  \omega^{IJ} \wedge \dd {\pmb{\varpi}}_{IJ}   & =  & \displaystyle  (1/2)
 \dd {{p}}^{\omega_{\mu}\nu}_{IJ} \wedge \dd  \omega^{IJ}_{\rho}     
  \wedge  {\big{(}}  \delta^{\rho}_{\mu} \vol_{\nu} - \delta^{\rho}_{\nu} \vol_{\mu}  {\big{)}}  =
  \dd
   {{p}}^{
\omega_{\mu}\nu}_{IJ} \wedge 
  \hbox{d} \omega_{\mu}^{IJ} \wedge  {\beta}_{\nu} ,
\end{array}
\right.
\]
the multisymplectic form is written as $   {\pmb{\omega}}^{\tiny{\hbox{\sffamily DW}}} = 
\dd \varkappa \wedge \vol + \dd  e^{I} \wedge \dd {\pmb{\varpi}}_{I} + \dd  \omega^{IJ} \wedge \dd {\pmb{\varpi}}_{IJ}$. Following the method found in  \cite{HK-02,HK-03}, we   construct a bracket between   the observable functionals $ {F} [{\omega}  , \Sigma \cap {\pmb{\gamma}}_{\pmb{\omega}}   ]  := \int_{\Sigma \cap {\pmb{\gamma}}_{\pmb{\omega}} } {\omega}   $   and  $ {F} [{\pmb{\varpi}}  , \Sigma \cap {\pmb{\gamma}}_{{\pmb{\varpi}}}   ]  := \int_{\Sigma \cap {\pmb{\gamma}}_{{\pmb{\varpi}}} } {\pmb{\varpi}}   $, where $\Sigma$ is a $1$-codimensional slice \cite{HK-02}, and $\Sigma \cap {\pmb{\gamma}}_{\pmb{\omega}}  $ and $\Sigma \cap {\pmb{\gamma}}_{{\pmb{\varpi}}}  $ are submanifolds of codimension $n-2$ and $n-3$, respectively.  We construct the Poisson bracket  $ {\Big{\{}} \int_{\Sigma \cap {\pmb{\gamma}}_{{\pmb{\varpi}}} } {\pmb{\varpi}}  , \int_{\Sigma \cap {\pmb{\gamma}}_{\pmb{\omega}} } \omega 
{\Big{\}}} (\pmb{\Gamma}) = \sum_{ m \in \Sigma \cap {\pmb{\gamma}}_{{\pmb{\varpi}}}  \cap {\pmb{\gamma}}_{\pmb{\omega}}  \cap \pmb{\Gamma} } {\pmb{c}} (m), $ where  ${\pmb{c}}(m) $ is a counting function and ${{\Gamma}}$ is a Hamiltonian $n$-curve. We refer to  a forthcoming paper \cite{Veygrav3} for an analysis of canonically conjugate  forms and Poisson brackets in the {DW} Hamiltonian formulation of vielbein gravity.
 
 \section{{{Conclusion}}}  

In this paper, we have presented several geometrical frameworks for the {DW} Hamiltonian formulation of   vielbein gravity. We have chosen to  work in a local trivialization of the principal fiber bundle $({{{\cal{P}}}} ,  {{\cal{X}}} , \pi , {{SO}}(1,3) )$.      The  covariant  configuration space is   the fiber  bundle ${\cal Y} :=   \mathfrak{iso}(1,3)   \otimes T^{\star} {\cal X}  $  over ${\cal X}$, see   section \ref{sec:section2}.     We have described the {DW} Hamilton equations in geometrical form     in sections \ref{sec:section2} and \ref{sec:section17}.  In section \ref{sec:section2}   we studied 
 the  Hamilton equations in the     multimomentum phase space ${{\cal M}}_{\hbox{\tiny{\sffamily DW}}}  := \Lambda^{n}_{\mathfrak{1}}T^{\star} {\cal{{{Y}}}}  $, which is described by the set of local coordinates $(x^{\mu},e_{\mu}^{I},\omega_{\mu}^{IJ},  \varkappa, p_I^{e_\mu \nu} ,  {{p}}^{\omega_{\mu}\nu}_{IJ})$.   Working with    $({\pmb{\cal C}}       ,  \iota^{\star} {\Omega}^{\tiny{\hbox{\sffamily DW}}} )$, the  {\sffamily DW} Hamilton equations   $X^{\footnotesize{{\pmb{\cal C}}}}   \iN  ( \iota^{\star}   {\pmb{\omega}}^{{\tiny\hbox{\sffamily  DW}}}) =  (-1)^n \hbox{d}  (  \iota^{\star}   {{{{\cal{H}}}}}^{{\tiny\hbox{\sffamily  DW}}} ) $, reproduce  the Einstein system of equations. In section   \ref{sec:section17} we consider the $n$-phase space formulation of dreibein and vierbein gravity, following the formalism developed by  Kijowski and Szczyrba    \cite{JK-01,KS0,KS1,KS2},  and  H\'elein \cite{H-02}. We present the {DW} Hamilton equations on the pre-multisymplectic phase space ${{(}} {\pmb{\cal C}}_\circ , {\pmb{\omega}}^\circ   {{)}} $. Then,  in the multisymplectic case, when   working  on the constraint hypersurface   $  {\pmb{\cal C}}$,  the     {DW} Hamilton equations  are given  by \eqref{fdoxx65} and \eqref{qlskdkq000}   for dreibein and vierbein gravity, respectively.  In the  pre-multisymplectic case, and working on $({\pmb{\cal C}}_{\circ})$, the equations are given by  \eqref{HE3Dpre} and \eqref{HE3D}.  This fact is related to the first order nature of the Einstein-Palatini gravity. We have  reproduced   in the context of the  {DW} Hamiltonian formulation developed in \cite{FH-01,HK-01,HK-02,HK-03} some of the   results found in \cite{bruno,bruno1,Espo,Rovelli002,Carlo1}.       In     section \ref{sec:section18} we give some     examples of Hamiltonian $(n-1)$-forms, their related Hamiltonian vectors fields,   and some    Poisson    brackets, which lead to 
   the  Lie   or   homotopy Lie algebra. 
    
    One of the interesting  questions  beyond the scope of the {DW} formulation  is to find a   multisymplectic manifold $ ( {\cal M}_{\tiny{\hbox{\sffamily Lepage}}}  ,  {\iota}_{\mathfrak{2}}^{\star} \Omega  ) $ contained in the following inclusion of spaces:
$   {{\cal M}}_{\hbox{\tiny{\sffamily DW}}} \hookrightarrow  {\cal M}_{\tiny{\hbox{\sffamily Lepage}}} \hookrightarrow {\cal M} $, 
 such that a more general Lepagean  Legendre  correspondence    \cite{HK-01,HK-02,HK-03} is non singular.  Note that  ${\iota}_{\mathfrak{2}} : {{{\cal M}}}_{\tiny{\hbox{\sffamily Lepage}}}   \hookrightarrow {{{\cal M}}}$  is the canonical inclusion. The idea is to use a formulation based on a higher Lepagean equivalent of   the Poincaré-Cartan $n$-form, denoted by $ \theta^{\hbox{\sffamily\tiny Lepage}} := \iota_{\mathfrak{2}}^{\star} {\theta}$. In such a context we   use the   multimomentum phase space  ${\cal M}_{\tiny{\hbox{\sffamily Lepage}}} := \Lambda^{n}_{\mathfrak{2}} T^{\star} (\textswab{p} \otimes T^{\star}{\cal X}) $. Then, for any point $(q,p)$ in ${\cal M}_{\tiny{\hbox{\sffamily Lepage}}} $,    
 \bee\label{fmdlq88}
\left.
\begin{array}{rcl}
\displaystyle    \theta^{\hbox{\sffamily\tiny Lepage}}_{(q,p)}     & :=  &  \displaystyle {{\theta}}^{{\tiny{\hbox{\sffamily DW}}}}_{(q,p)}  
  + {p}^{e_\rho^{I} \omega_\sigma^{JK} \mu \nu}  \hbox{d} e_{\rho}^I  \wedge  \hbox{d} \omega_{\sigma}^{JK} \wedge  {\beta}_{\mu\nu} 
  +   {p}^{e_\rho^{i} e_\sigma^{J} \mu \nu}  \hbox{d} e_{\rho}^I  \wedge  \hbox{d} e_{\sigma}^{J} \wedge  {\beta}_{\mu\nu} ,
    \\
\displaystyle    &    &  \displaystyle    
  +   {p}^{\omega_\rho^{IJ} \omega_\sigma^{KL} \mu \nu}  \hbox{d} \omega_{\rho}^{IJ}  \wedge  \hbox{d} \omega_{\sigma}^{KL} \wedge  {\beta}_{\mu\nu}  ,
\end{array}
\right.
\eee
where we have introduced additional multimomenta ${p}^{e_\alpha^{K} \omega_\beta^{IJ} \mu \nu}  , {p}^{e_\alpha^{I} e_\beta^{J} \mu \nu}$, and ${p}^{\omega_\alpha^{IJ} \omega_\beta^{KL} \mu \nu} $. Within this geometrical formulation  we could be able to construct  an isomorphism between a subset of the multimomenta   and  the field derivatives $\displaystyle \partial_{\mu} e_{\nu}^{i} $  and $ \partial_{\mu} \omega_\nu^{IJ}$. This viewpoint might allows  us to  avoid the   primary constraints at all,  and eventually  shed new light  on the        problem of quantization.  Another problem for further research, already mentioned in section \ref{subsec:confplace},  is to describe a {fully} covariant setting for vielbein gravity and to establish  connections with the work of Bruno {\em et al.}  \cite{bruno,bruno1,bruno2,bruno3}  and  H\'{e}lein \cite{pataym}.   
 
The most interesting problem    related on the quantization of vielbein gravity would      include    the   classification  of the full set of algebraic and dynamical observable forms and   the   search of    good conjugate  forms. 
  We hope to present elsewhere \cite{Veygrav3}     results   on the construction of      canonical forms  $({\pmb{\varpi}}_{IJ}  , \omega^{IJ} )$,  canonical brackets and a pre-quantum theory, in the sense of geometric quantization,  for vielbein gravity. {The canonically conjugate  forms are  the  connection $1$-form $\omega^{IJ} = \omega_{\mu}^{IJ} \dd x^{\mu}$ and the $2$-form   $   {\pmb{\varpi}}_{IJ} =  (1/2) \sum_{\mu,\nu}  e e^{[\nu}_I e^{\mu]}_J  {\beta}_{\mu\nu}$}.  Note that interesting results have been obtained   by  Kanatchikov  within his precanonical quantization scheme  for  vielbein gravity   \cite{kannnnnnnaa,kannnnnnnaa1}.   
 
\
 
\noindent ${{\textsf{Acknowledgments}}}$.  {I am   grateful to Fr\'ed\'eric H\'elein  and Joseph Kouneiher   for discussions   about the topic of multisymplectic geometry and   vielbein gravity.}  I also thank the referees of {CQG} for helpful suggestions.  
   
\appendix

 \section{{{First order Palatini action   of vielbein gravity}}}\label{app:lagrang}
  
First, we consider   the first order Palatini action functional of vierbein gravity 
\begin{equation}\label{izeuue987}
{{{\cal S}}}_{\tiny{\hbox{\sffamily Palatini}}} [e,\omega] =  {\kappa}   \int_{{\cal{X}}} {\hbox{\sffamily  vol}}  (e) e^\mu_I e^\nu_J {F}^{IJ}_{\mu\nu} [\omega] ,
\end{equation}
   also called the {\guillemotleft {\em Hilbert-Palatini}\guillemotright}  action functional    in Peldan's review  \cite{PP}, and which  corresponds to the  {\guillemotleft {\em frame-affine}\guillemotright}  framework  in     \cite{Fatibene}. The  functionals  $  {{{\cal S}}}_{\tiny{\hbox{\sffamily EH}}} [e]  := \kappa \int_{{{{{\cal X}}}}} {\hbox{vol}}  (e) e^\mu_I e^\nu_J {F}^{IJ}_{\mu\nu} [\omega (e)]$ and  $  {{{\cal S}}}_{\tiny{\hbox{\sffamily EP}}} [{\textsf{g}},\Gamma ]  :=  \kappa  \int_{{{{{{\cal X}}}}}}  \sqrt{-{\textsf{g}}}{R} [\Gamma] \vol :=  \kappa  \int_{{{{{{\cal X}}}}}}   {R}  \hbox{vol}({\textsf{g}})   $    are termed the   {\guillemotleft {\em Einstein-Hilbert}\guillemotright}   and the {{\guillemotleft}{\em Einstein-Palatini}{\guillemotright}} action functional   in Peldan's review  \cite{PP}. They correspond, in the framework developed by Fatibene  and Francaviglia     \cite{Fatibene}, to    the  {\guillemotleft{\em purely-frame}\guillemotright} and  the {\guillemotleft{\em metric-affine}\guillemotright}    formulations,  respectively.   Let us sketch the passage   from  $ {{{\cal S}}}_{\tiny{\hbox{\sffamily EP}}} [{\textsf{g}},\Gamma ]   $   to   ${{{\cal S}}}_{\tiny{\hbox{\sffamily Palatini}}} [e,\omega]   $,  using 
some vielbein algebraic relations.
    \begin{lemm}\label{cur2}
The  Palatini action functional {\em ${{{\cal S}}}_{\tiny{\hbox{\sffamily Palatini}}} [e,\omega]$} is written as 
{\em \bee \label{detail001}
{{{\cal S}}}_{\tiny{\hbox{\sffamily Palatini}}} [e,\omega] =  {{\kappa}\over{4}} \int_{{\cal{X}}}  \epsilon_{IJKL}{\epsilon}^{\mu\nu\rho\sigma} e^I_\mu  e^J_\nu {{F}}_{\rho\sigma}^{KL} [\omega] \beta 
= \frac{1}{64\pi G}   \int_{{\cal{X}}}  \epsilon_{IJKL}{\epsilon}^{\mu\nu\rho\sigma} e^I_\mu  e^J_\nu {{F}}_{\rho\sigma}^{KL} [\omega] \beta .
\eee}
\end{lemm}
{\sffamily Proof}.  Note that $  {\hbox{\sffamily  vol}}  (e)   e^\mu_I e^\nu_J {F}^{IJ}_{\mu\nu} =  \vol  e e^\mu_I e^\nu_J {F}^{IJ}_{\mu\nu} =  \vol \sqrt{-{\textsf{g}}} {R}^{\mu\nu}_{\mu\nu} = \vol \sqrt{-{\textsf{g}}} {R} =  \hbox{\sffamily{vol}} ({\textsf{g}}) {R}$. Alternatively,  we have the straightforward calculation:
\begin{equation}\label{EP001}
\left.
\begin{array}{rcl}
\displaystyle  \sqrt{-{\textsf{g}}}{R} \vol  & = &  \displaystyle     \sqrt{-{\textsf{g}}} \delta_{[\alpha}^{\rho}\delta_{\beta]}^{\sigma}  {{R}^{\alpha \beta}}_{\rho \sigma}  \vol =   (1/4)     \sqrt{-{\textsf{g}}}   {\epsilon}_{\mu\nu\alpha\beta} {\epsilon}^{\mu\nu\rho\sigma}   {{R}^{\alpha \beta}}_{\rho \sigma}  \vol ,
  \\
  \displaystyle & = &  \displaystyle     (1/4)    {\boldsymbol{\epsilon}}_{\mu\nu\alpha\beta} {\epsilon}^{\mu\nu\rho\sigma}   {{R}^{\alpha \beta}}_{\rho \sigma} \vol =   (1/4)    \epsilon_{IJKL}  e^I_\mu e^J_\nu e^K_\alpha e^L_\beta  {\epsilon}^{\mu\nu\rho\sigma}   {{R}^{\alpha \beta}}_{\rho \sigma} \vol    ,
  \\
  \displaystyle & = &  \displaystyle 
      (1/4)    \epsilon_{IJKL}  e^I_\mu e^J_\nu  {\epsilon}^{\mu\nu\rho\sigma}   [ e^K_\alpha e^L_\beta {{R}^{\alpha \beta}}_{\rho \sigma}  ]   \vol  
    =
      (1/4)   \epsilon_{IJKL}{\epsilon}^{\mu\nu\rho\sigma} e^I_\mu  e^J_\nu {F}_{\rho\sigma}^{KL} [\omega]  \vol,
\end{array}
\right.
\eee
where we have used $  \delta_{[\alpha}^{\rho}\delta_{\beta]}^{\sigma} = (1/2) [ \delta^\rho_\alpha  \delta^\sigma_\beta - \delta^\rho_\beta \delta^\sigma_\alpha ] =    (1/4) {\epsilon}_{\mu\nu\alpha\beta} {\epsilon}^{\mu\nu\rho\sigma}   $.   In the first line of   \eqref{EP001}  the  Levi-Civita tensor is written as $ {\boldsymbol{\epsilon}}_{\mu\nu\alpha\beta} =  \sqrt{-{\textsf{g}}} {\epsilon}_{\mu\nu\alpha\beta} $. We have used   
 ${\boldsymbol{\epsilon}}_{\mu\nu\alpha\beta} =     {\epsilon}_{IJKL}  e^I_\mu e^J_\nu e^K_\alpha e^L_\beta
 $  and $  e^K_\alpha e^L_\beta {{R}^{\alpha \beta}}_{\rho \sigma}  = {F}^{KL}_{\rho\sigma} $  in the second and the last line of  \eqref{EP001}, respectively. 
 We pass from the  Einstein-Palatini    action functional 
$   
{{{\cal S}}}_{\tiny{\hbox{\sffamily EP}}}[{\textsf{g}},\Gamma] =   \kappa \int_{{\cal{X}}} {{{L}}}_{\tiny{\hbox{\sffamily EP}}}[{\textsf{g}},\Gamma]  \vol $   to the functional
\bee   
{{{\cal S}}}_{\tiny{\hbox{\sffamily Palatini}}}[e,\omega]  = \frac{\kappa}{2} \int  \epsilon_{IJKL} e^I \wedge e^J \wedge {F}^{KL} = \frac{1}{32\pi G}    \epsilon_{IJKL} e^I \wedge e^J \wedge {F}^{KL} ,
\eee
  written in terms of differential forms.
   
   {\sffamily Proof}.  Let us evaluate $ \hbox{vol}({\textsf{g}})   {R} = \vol \sqrt{-{\textsf{g}}}{R} $, the integrand of the Einstein-Hilbert action.   Contracting the Riemannn curvature tensor we have the following equality ${R} = {{R}^{\alpha \beta}}_{\rho \sigma}\delta_{[\alpha}^{\rho}\delta_{\beta]}^{\sigma} $. Therefore, 
\[
   {{{L}}}_{\tiny{\hbox{\sffamily EH}}}[{\textsf{g}}]   \hbox{vol}({\textsf{g}})  =  \kappa     \hbox{vol}({\textsf{g}})  {R} = \kappa    \hbox{vol}({\textsf{g}}) \delta_{[\alpha}^{\rho}\delta_{\beta]}^{\sigma}  {{R}^{\alpha \beta}}_{\rho \sigma} =   {{\kappa}\over{4}}    \hbox{vol}({\textsf{g}})  (-1)^s  {\boldsymbol{\epsilon}}_{\mu\nu\alpha\beta} Ý {\boldsymbol{\epsilon}}^{\mu\nu\rho\sigma} {{R}^{\alpha \beta}}_{\rho \sigma} ,
\]
where we   use  the   relation $\delta_{[\alpha}^{\rho}\delta_{\beta]}^{\sigma}  p! (n-p)! (-1)^s =  {\boldsymbol{\epsilon}}_{\mu\nu\alpha\beta} Ý {\boldsymbol{\epsilon}}^{\mu\nu\rho\sigma}$  (see the algebraic identity \eqref{algebraicvvveilbe} in  appendix \ref{app:generalized-Kronecker-symbols}, with $n=4$ and p = 2). 
Then, in a  integrable moving  co-frame ${\bf e}^{\mu} :=  \dd x^\mu $, the volume form $\hbox{vol}({\textsf{g}})  = \sqrt{-{\textsf{g}}} \dd x^0 \wedge \dd x^1  \wedge \dd x^2 \wedge \dd x^3 $  is written as  
\[
\hbox{vol}({\textsf{g}})   = {{ \sqrt{-{\textsf{g}}} }\over{4!}} {\epsilon}_{\lambda\kappa\tau\gamma } \dd x^\lambda \wedge \dd x^\kappa  \wedge \dd x^\tau \wedge \dd x^\gamma =   {{1}\over{4!}}  {\boldsymbol{\epsilon}}_{\lambda\kappa\tau\gamma } \dd x^\lambda \wedge \dd x^\kappa  \wedge \dd x^\tau \wedge \dd x^\gamma  . 
\]
We refer to appendix \ref{app:volume-levi} for   details on the relation  between the volume form and  the Levi-Civita symbols.
Since, see the formula  \eqref{algebraicvvveilbe},  $ \displaystyle  {\boldsymbol{\epsilon}}^{\mu\nu\rho\sigma}  {\boldsymbol{\epsilon}}_{\lambda\kappa\tau\gamma }  = (-1)^s 4! \delta^{[\mu}_{\lambda}\delta^{\nu}_{\kappa}\delta^{\rho}_{\tau}\delta^{\sigma]}_{\gamma}$ the Einstein-Palatini functional is written as
 \[
\left.
\begin{array}{rcl}
\displaystyle {{\cal{S}}}_{\tiny{\hbox{\sffamily EP}}}[{\textsf{g}},\Gamma]    & = &  \displaystyle 
{{\kappa}\over{4}}   \int_{{\cal{X}}}     {{ (-1)^s}\over{4!}}  {\boldsymbol{\epsilon}}_{\mu\nu\alpha\beta}  {\boldsymbol{\epsilon}}^{\mu\nu\rho\sigma}  {\boldsymbol{\epsilon}}_{\lambda\kappa\tau\gamma }  {{R}^{\alpha \beta}}_{\rho \sigma} \dd x^\lambda \wedge \dd x^\kappa  \wedge \dd x^\tau \wedge \dd x^\gamma ,
\\
 \displaystyle     & = &  \displaystyle  {{\kappa}\over{4}}   \int_{{\cal{X}}}    {{ (-1)^s (-1)^s 4! }\over{4!}}  {\boldsymbol{\epsilon}}_{\mu\nu\alpha\beta}  \delta^{[\mu}_{\lambda}\delta^{\nu}_{\kappa}\delta^{\rho}_{\tau}\delta^{\sigma]}_{\gamma} {{R}^{\alpha \beta}}_{\rho \sigma} \dd x^\lambda \wedge \dd x^\kappa  \wedge \dd x^\tau \wedge \dd x^\gamma ,
 \\
 \displaystyle     & = &  \displaystyle
   ({{\kappa}/{4}})    \int_{{\cal{X}}}  {\boldsymbol{\epsilon}}_{\mu\nu\alpha\beta}  {{R}^{\alpha \beta}}_{\rho \sigma} \dd x^\mu \wedge \dd x^\nu  \wedge \dd x^\rho \wedge \dd x^\sigma  
=  ({{\kappa}/{2}})    \int_{{\cal{X}}}   {\boldsymbol{\epsilon}}_{\mu\nu\alpha\beta} \dd x^\mu \wedge \dd x^\nu  \wedge  {{R}^{\alpha \beta}} ,
\end{array}
\right.
\]
where in the last equality we use the curvature $2$-form   $\displaystyle {R}^{\alpha\beta} = (1/2) {{R}^{\alpha \beta}}_{\rho \sigma} \dd x^\rho \wedge \dd x^\sigma 
 $. Finally, using the relation $  {\boldsymbol{\epsilon}}_{\mu\nu\alpha\beta}  = e^I_\mu e^J_\nu e^K_\alpha e^L_\beta { {\epsilon}}_{IJKL} $   (between the volume element $ {\pmb{\epsilon}}_{\mu\nu\alpha\beta} $ of ${\textsf{g}}_{\mu\nu} = e^I_\mu e^J_\nu {\textsf{h}}_{IJ}$ and  the volume element ${{\epsilon}}_{IJKL}$ of the Minkowski metric ${\textsf{h}}_{IJ}$),  the Palatini functional action is written as   
  \[
\left.
\begin{array}{rcl}
\displaystyle   {{{\cal S}}}_{\tiny{\hbox{\sffamily Palatini}}}  [e,\omega]     & = &  \displaystyle  {{\kappa}\over{2}} \int_{{\cal{X}}}  e^I_\mu e^J_\nu e^K_\alpha e^L_\beta \epsilon_{IJKL} \dd x^\mu \wedge \dd x^\nu  \wedge  {{R}^{\alpha \beta}} ,
\\
 \displaystyle     & = &  \displaystyle  {{\kappa}\over{2}} \int_{{\cal{X}}} \epsilon_{IJKL}  e^I_\mu \dd x^\mu \wedge e^J_\nu  \dd x^\nu  \wedge  e^K_\alpha e^L_\beta  {{R}^{\alpha \beta}} = {{\kappa}\over{2}} \int  \epsilon_{IJKL} e^I \wedge e^J \wedge {F}^{KL}.   
  \end{array}
\right.
\]
Analogously, in the formulation of dreibein gravity,  the  Einstein-Hilbert action functional $   {{{\cal S}}}_{\tiny{\hbox{\sffamily EH}}} [{\textsf{g}}_{\mu \nu}] =     \int_{{\cal{X}}} \sqrt{-{\textsf{g}}} {R} \vol $ 
is equivalent   to the action functional $   {{{\cal S}}}_{\tiny{\hbox{\sffamily Palatini}}}  = \int  \epsilon_{IJK} e^I  \wedge {R}^{JK}$.
   
   {\sffamily Proof}. Let us evaluate $  \hbox{vol}  ({\textsf{g}})  {R} = \vol \sqrt{-{\textsf{g}}}{R} $, the integrand of the Einstein-Hilbert action. Contracting the Riemann curvature tensor, we have   ${R} = {{R}^{\alpha \beta}}_{\rho \sigma}\delta_{[\alpha}^{\rho}\delta_{\beta]}^{\sigma} $. Then,  
\[
{{{\cal S}}}_{\tiny{\hbox{\sffamily EP}}} [{\textsf{g}},\Gamma]  =  \int_{ {{\cal{X}}}}   \hbox{vol} ({\textsf{g}})  {R} = \int_{ {{\cal{X}}}}   \hbox{vol} ({\textsf{g}}) \delta_{[\alpha}^{\rho}\delta_{\beta]}^{\sigma}  {{R}^{\alpha \beta}}_{\rho \sigma}  .
\]
We also have the   relation $\delta_{[\alpha}^{\rho}\delta_{\beta]}^{\sigma}  1! 2! (-1)^s =  {\boldsymbol{\epsilon}}_{\mu\alpha\beta} Ý {\boldsymbol{\epsilon}}^{\mu\rho\sigma}$, see the algebraic identity \eqref{algebraicvvveilbe} in  appendix \ref{app:generalized-Kronecker-symbols}, with $n=3$ and $p = 1$. Thus, 
 \[
{{{\cal S}}}_{\tiny{\hbox{\sffamily EP}}} [{\textsf{g}},\Gamma] = {{1}\over{2}} \int_{ {{\cal{X}}}}   \hbox{vol} ({\textsf{g}})  (-1)^s  {\boldsymbol{\epsilon}}_{\mu\alpha\beta} Ý {\boldsymbol{\epsilon}}^{\mu\rho\sigma} {{R}^{\alpha \beta}}_{\rho \sigma}  .
\]
The volume form is written: 
$ \displaystyle
 \hbox{vol} ({\textsf{g}})  = \sqrt{-{\textsf{g}}}  \dd x^1  \wedge \dd x^2 \wedge \dd x^3 = ({{ \sqrt{-{\textsf{g}}} }/{3!}}) {\epsilon}_{\lambda\kappa\tau} \dd x^\lambda \wedge \dd x^\kappa  \wedge \dd x^\tau  =   ({{1}/{3!}})  {\boldsymbol{\epsilon}}_{\lambda\kappa\tau } \dd x^\lambda \wedge \dd x^\kappa  \wedge \dd x^\tau  . $
Then, we have 
\[
\left.
\begin{array}{rcl}
\displaystyle {{{\cal S}}}_{\tiny{\hbox{\sffamily EP}}} [{\textsf{g}},\Gamma]   & = &  \displaystyle 
  {{1}\over{2}} \int_{{\cal{X}}}    {{ (-1)^s}\over{3!}}  {\boldsymbol{\epsilon}}_{\mu\alpha\beta}  {\boldsymbol{\epsilon}}^{\mu\rho\sigma}  {\boldsymbol{\epsilon}}_{\lambda\kappa\tau }  {{R}^{\alpha \beta}}_{\rho \sigma} \dd x^\lambda \wedge \dd x^\kappa  \wedge \dd x^\tau   ,
   \\
  \displaystyle & = &  \displaystyle 
   {{1}\over{2}} \int_{{\cal{X}}}   {\boldsymbol{\epsilon}}_{\mu\alpha\beta}  {{R}^{\alpha \beta}}_{\rho \sigma} \dd x^\mu   \wedge \dd x^\rho \wedge \dd x^\sigma 
=
   \int_{{\cal{X}}}   {\boldsymbol{\epsilon}}_{\mu\alpha\beta} \dd x^\mu \wedge    {{R}^{\alpha \beta}} , 
\end{array}
\right.
\]
where we used $ {\boldsymbol{\epsilon}}^{\mu\rho\sigma}  {\boldsymbol{\epsilon}}_{\lambda\kappa\tau }  = (-1)^s 3! \delta^{[\mu}_{\lambda} \delta^{\rho}_{\kappa}\delta^{\sigma]}_{\tau}$ and  since  the curvature $2$-form is   written as ${R}^{\alpha\beta} =  (1/2) {{R}^{\alpha \beta}}_{\rho \sigma} \dd x^\rho \wedge \dd x^\sigma 
 $. Using the identity $  {\boldsymbol{\epsilon}}_{\mu\alpha\beta}  = e^I_\mu   e^J_\alpha e^K_\beta \epsilon_{IJK} $, we finally obtain
\[
{{{\cal S}}}_{\tiny{\hbox{\sffamily Palatini}}} [e,\omega]   =  \int_{{\cal{X}}}  e^I_\mu   e^J_\alpha e^K_\beta \epsilon_{IJK} \dd x^\mu \wedge   {{R}^{\alpha \beta}} = \int_{{\cal{X}}} \epsilon_{IJK}  e^I_\mu \dd x^\mu   e^J_\alpha e^K_\beta  {{R}^{\alpha \beta}}   =  \int  \epsilon_{IJK} e^I  \wedge {F}^{JK}.  
\]

 \section{{Algebraic relations, volume form and vielbein}}\label{app:algebraicvielbein}

In this section we present  the basic algebraic properties of the Levi-Civita symbols, generalized Kronecker symbols, Levi-Civita tensors, and    densities constructed on the vielbein field.   

\subsection{{{{Levi-Civita symbols}}} }

We denote by  ${{\epsilon}}_{\mu_1 , \cdots , \mu_n} $ the Levi-Civita symbol   and  by  ${\pmb{\epsilon}}_{\mu_1 , \cdots , \mu_n} $   the Levi-Civita tensor.   Let ${\cal{S}}_n$ be the set of all permutations of $n$ elements. The signature  of the permutation $  \sigma \in {\cal{S}}_n$ is denoted  by $\hbox{\sffamily sgn} (\sigma) $ with value $1$ and $-1$,  when the permutation is even or odd, respectively.  By definition, $ \epsilon_{\mu_1, \cdots ,\mu_n}  = +1$   if  $(\mu_1 ,\cdots ,\mu_n)$   is an even permutation of  $ (1, \cdots , n)  $,  $ \epsilon_{\mu_1 , \cdots , \mu_n}  = -1 $   if  $(\mu_1 , \cdots , \mu_n)$   is an odd permutation of $(1, \cdots , n) $,   and  $ \epsilon_{\mu_1 , \cdots , \mu_n}    = 0$ otherwise. 

 The determinant $ \hbox{det} (\textsc{m})   $ of a matrix $ \textsc{m} = \{{\textsc{m}^{\mu}}_{\nu}\}_{1 \leq \mu , \nu \leq n} $ is given by the Leibniz formula
\bee\label{determinasha}
\left.
\begin{array}{rcl}
\displaystyle \hbox{det} (\textsc{m})     & = &  \displaystyle \sum_{\sigma \in  {\cal S}_n}  \hbox{\sffamily sgn} (\sigma)    {\textsc{m}^{\sigma(1)}}_{1}  \cdots   {\textsc{m}^{\sigma(n)}}_{n}   = \sum_{\sigma \in  {\cal S}_n}  \hbox{\sffamily sgn} (\sigma)   {\textsc{m}^{1}}_{\sigma(1)}  \cdots   {\textsc{m}^{n}}_{\sigma (n)}    ,
 \\
\end{array}
\right.
\eee
and is  equivalently written as  $ \hbox{det} (\textsc{m})   =  \sum_{1 \leq \mu_1  \cdots  \mu_n \leq n}    \epsilon_{\mu_1 , \cdots , \mu_n}  {\textsc{m}^{\mu_1}}_{1}  \cdots   {\textsc{m}^{\mu_n}}_{n}  $. 

\subsection{{{{Generalized Kronecker symbols}}}}\label{app:generalized-Kronecker-symbols}

We  introduce the generalized Kronecker symbols $\delta^{\mu_1 , \cdots , \mu_n}_{\nu_1, \cdots, \nu_n}$. By definition    $ \delta^{\mu_1 , \cdots , \mu_n}_{\nu_1, \cdots, \nu_n} = +1$   if  $(\mu_1 , \cdots , \mu_n)$   is an even permutation of  $ (\nu_1, \cdots, \nu_n) $,  $ \delta^{\mu_1 , \cdots , \mu_n}_{\nu_1, \cdots, \nu_n} =   -1 $   if  $(\mu_1 , \cdots , \mu_n)$   is an an odd permutation of $(\nu_1, \cdots, \nu_n) $,   and  $ \epsilon_{\mu_1 , \cdots , \mu_n}    = 0$ otherwise.  The generalized Kronecker symbol provides a way to write the anti-symmetric Levi-Civita symbols   $ \displaystyle \epsilon_{\mu_1 , \cdots , \mu_n} = \delta^{1 \cdots n}_{\mu_1 , \cdots , \mu_n } 
$ and $ \displaystyle
 \epsilon^{\mu_1 , \cdots , \mu_n} = \delta_{1 \cdots n}^{\mu_1 , \cdots , \mu_n }  
 $.    We adopt the anti-symmetry conventions of  Wald \cite{Wald} {\em i.e.}
\bee
\delta^{\mu_1}_{[\nu_{1}} \cdots \delta^{\mu_n}_{\nu_{n}]} = \frac{1}{n!} \sum_{\sigma \in {\cal S}_{n} } \hbox{\sffamily sgn} (\sigma) \delta^{\mu_1}_{\sigma(\nu_1)} \cdots \delta^{\mu_n}_{\sigma(\nu_n)} = \frac{1}{n!}  \epsilon^{\nu_1, \cdots, \nu_n} \delta^{\mu_1}_{\nu_1} \cdots \delta^{\mu_n}_{\nu_n},
\eee
then, 
$
\delta^{\mu_1 , \cdots , \mu_n}_{\nu_1, \cdots, \nu_n} = n! \delta^{\mu_1}_{[\nu_{1}} \cdots \delta^{\mu_n}_{\nu_{n}]}
=  \sum_{\sigma \in {\cal S}_{n} }  \hbox{\sffamily sgn} (\sigma)  \delta^{\mu_1}_{\sigma(\nu_1)} \cdots \delta^{\mu_n}_{\sigma(\nu_n)}  
=    \epsilon^{\nu_1, \cdots, \nu_n} \delta^{\mu_1}_{\nu_1} \cdots \delta^{\mu_n}_{\nu_n}    
$.
 For any $1 \leq p \leq n $,  we also have the identity 
  \bee\label{identity001}
 ({1}/{p!}) \epsilon^{\mu_1 \cdots \mu_{n-p} \rho_{1} \cdots \rho_{p} } \epsilon_{\nu_1 \cdots \nu_{n-p} \rho_{1} \cdots \rho_{p} } = \delta^{\mu_1 \cdots \mu_{n-p}}_{\nu_1 \cdots \nu_{n-p}}.
 \eee
 The identity \eqref{identity001} is very useful and give
 \begin{equation}\label{algebraicvvveilbe}
\left.
\begin{array}{rcl}
\displaystyle   {{\epsilon}}^{\mu_1 ... \mu_p\alpha_1 ... \alpha_{n-p}} {{\epsilon}}_{\mu_1 ... \mu_p\beta_1 ... \beta_{n-p}}   & =   &     p ! (n-p) ! \delta_{\beta_1}^{[\alpha_1} ... \delta_{\beta_{n-p}}^{\alpha_{n-p}]}  ,
\\ 
\displaystyle     {{\epsilon}}^{\mu_1 ... \mu_n} {{\epsilon}}_{\nu_1 ... \nu_n}   & = &       n !  \delta_{\nu_1}^{[\mu_1} ... \delta_{\nu_{n}}^{\mu_{n}]}  ,
\\ 
\displaystyle      {{\epsilon}}^{\mu_1 ... \mu_n} {{\epsilon}}_{\mu_1 ... \mu_n}  &  =     &     n !    .
\end{array}
\right.
\end{equation}
   Finally, using the generalized Kronecker symbol,  the general formula for the determinant of a matrix $\textsc{m} \in \hbox{\sffamily Mat}_{n} (\R)$ is written as
$
\hbox{det} (\textsc{m}) = ({1}/{n!})
  \sum_{  \mu_1  \cdots  \mu_n \nu_1  \cdots  \nu_n }     \delta^{\nu_1, \cdots, \nu_n}_{\mu_1 , \cdots , \mu_n} 
 {\textsc{m}^{\mu_1}}_{\nu_1}  \cdots   {\textsc{m}^{\mu_n}}_{\nu_n} $.

\subsection{{{Volume form, Levi-Civita tensor, Levi-Civita tensor density}}}\label{app:volume-levi}

Let   $({{\cal{X}}},{\textsf{g}})$ be  a Riemannian manifold.  The canonical volume form, a nowhere vanishing  $n$-form on ${{\cal{X}}}$ is denoted by $\hbox{vol}({\textsf{g}}) \in \Lambda^nT^{\star}{{{\cal{X}}}}$ is related to the metric    ${\textsf{g}}_{\mu\nu}$ by    $\hbox{vol}({\textsf{g}}) = \sqrt{{\textsf{g}}} \dd x^1 \wedge .... \wedge \dd x^n = \sqrt{{\textsf{g}}} \vol$, where $ {\textsf{g}} := |  {\textsf{g}} | := | \hbox{det} ({\textsf{g}}_{\mu\nu}) |$. The Levi-Civita tensor is connected   to the volume form $ \hbox{vol}({\textsf{g}})$ by the following  formulae: 
 \bee\label{vol001}
{\pmb{\epsilon}}_{\mu_1 ... \mu_n} =  {\sqrt{|{\textsf{g}}|}} {\epsilon}_{\mu_1 ... \mu_n},
\quad \quad \quad \quad 
{\pmb{\epsilon}}^{\mu_1 ... \mu_n} =   (-1)^{\pmb{\sigma}}  ({{1}/{\sqrt{|{\textsf{g}}|}}}) {\epsilon}^{\mu_1 ... \mu_n} ,
\eee
where $ {\pmb{\sigma}}$ is the number of negative values in the signature  of the metric  {\em i.e.} $  (-1)^{\pmb{\sigma}} =1 $   and $ (-1)^{\pmb{\sigma}}  = - 1$  in the Riemannian  and Lorentzian cases,  respectively.    We construct the tensorial invariant volume $n$-form    $ \hbox{vol}({\textsf{g}}) =   \sqrt{ | {\textsf{g}} |} \vol$,    where   $  \vol  = 
\dd x^1 \wedge ... \wedge \dd x^{n} = {({1}/{n!})}  {\epsilon}_{\mu_1 ... \mu_n} \dd x^{\mu_1} \wedge ... \wedge \dd x^{\mu_n}   $.  We have $\left.
\begin{array}{lll}
  \displaystyle  \hbox{vol} ({\textsf{g}})   &  = &    \displaystyle   (1/n!)  {\pmb{\epsilon}}_{\mu_1 ... \mu_n} \dd x^{\mu_1} \wedge ... \wedge \dd x^{\mu_n} =  (1/n!) \sqrt{ | {\textsf{g}} |}   {\epsilon}_{\mu_1 ... \mu_n} \dd x^{\mu_1} \wedge ... \wedge \dd x^{\mu_n}   .
  \end{array}
\right.$
 
Finally,  the important formula ${\pmb{\epsilon}}^{\mu_1 ... \mu_p\alpha_1 ... \alpha_{n-p}} {\pmb{\epsilon}}_{\mu_1 ... \mu_p\beta_1 ... \beta_{n-p}} =   (-1)^{\pmb{\sigma}} p ! (n-p) ! \delta_{\beta_1}^{[\alpha_1} ... \delta_{\beta_{n-p}}^{\alpha_{n-p}]} $  specializes to   $  {\pmb{\epsilon}}^{\mu_1 ... \mu_n} {\pmb{\epsilon}}_{\nu_1 ... \nu_n}  =   (-1)^{\pmb{\sigma}} n !  \delta_{\nu_1}^{[\mu_1} ... \delta_{\nu_{n}}^{\mu_{n}]}  $  and  $    {\pmb{\epsilon}}^{\mu_1 ... \mu_n} {\pmb{\epsilon}}_{\mu_1 ... \mu_n} =   (-1)^{\pmb{\sigma}} n !   $.  
 
\subsection{{{Volume form and vielbein}}}

We introduce the covariant volume form  $\hbox{vol} ({\textsf{g}}) $, from the vielbein viewpoint. We denote $\hbox{vol}({e}) = e \vol = e^{I_1} \wedge \cdots \wedge e^{I_n}  $, where    
$ \displaystyle \vol = 
\dd x^1 \wedge  \cdots \wedge \dd x^n = {({1}/{n!})} {\epsilon}_{\mu_1 , \cdots , \mu_n} \dd x^{\mu_1} \wedge  \cdots  \wedge \dd x^{\mu_n}  $.   The space-time Levi-Civita symbols  $\epsilon^{\mu_1 , \cdots , \mu_n}$ and  $\epsilon_{\mu_1 , \cdots , \mu_n}$ have a counterpart in the vielbein setting. They correspond to the alternating symbols  with   tangent space indices $\epsilon^{I_1 \cdots I_n}$ and $\epsilon_{I_1 \cdots I_n}$, respectively. For any $  1 \leq j \leq n$ we have  $e^{I_j} := e_{\mu_j}^{I_j} \dd x^{\mu_j} $, thus $\hbox{vol} (e)$ is written as
\bee\label{popododo01}
\hbox{vol} (e) =  ({1}/{n!}) \epsilon_{I_1 \cdots I_n} e^{I_1} \wedge \cdots \wedge e^{I_n} =  ({1}/{n!})  \epsilon_{I_1 \cdots I_n} e^{I_1}_{\mu_1}   \cdots   e^{I_n}_{\mu_n} \dd x^{\mu_1} \wedge \cdots \wedge \dd x^{\mu_n}  .
\eee
Using  
$  
  e = \hbox{det} (e^{I}_{\mu}) =  {({1}/{n!})}  \epsilon_{I_1 \cdots I_n}  \epsilon^{\mu_1 , \cdots , \mu_n}   e^{I_1}_{\mu_1}   \cdots   e^{I_n}_{\mu_n} 
$ (the formula for the determinant of the vielbein  matrix),     \eqref{popododo01} is  now  written as
$  {\hbox{vol}} (e) = (1/n!)   \epsilon_{I_1 \cdots I_n} e^{I_1}_{\mu_1}   \cdots   e^{I_n}_{\mu_n} (n!)  \epsilon^{\mu_1 , \cdots , \mu_n} \vol =   e \vol$. Then, the determinant  $e_{[\mathfrak{3}]}   := \hbox{det} (e^{I}_{\mu})$     of the dreibein and    the determinant       $e_{[\mathfrak{4}]}   := \hbox{det} (e^{I}_{\mu})$ of the vierbein   are given by
\bee\label{Tiopodl01}
 \left.
\begin{array}{lcl}
\displaystyle e_{[\mathfrak{3}]}    & = & \displaystyle     ({1}/{3!}) \epsilon_{IJK}  \epsilon^{\mu\nu\rho}   e^{I}_{\mu}   e^{J}_{\nu}   e^{K}_{\rho} ,
 \end{array}
\right.
\quad      
 \left.
\begin{array}{lcl}
\displaystyle   e_{[\mathfrak{4}]}    & = & \displaystyle      ({1}/{4!}) \epsilon_{IJKL}  \epsilon^{\mu\nu\rho\sigma}   e^{I}_{\mu}   e^{J}_{\nu}   e^{K}_{\rho}   e^{L}_{\sigma} ,
 \end{array}
\right.
\eee
respectively. The determinant   of the inverse vielbein matrix  $\hbox{det} {\big{(}}  {{(}} e^{I}_{\mu} {{)}}^{-1}  {\big{)}} $  is given by 
\bee\label{Tiopodl05}
e^{-1} =\hbox{det} {\big{(}}  {\big{(}} e^{I}_{\mu} {\big{)}}^{-1}  {\big{)}} = {\big{(}}\hbox{det} {\big{(}}   e^{\mu}_{I}  {\big{)}}  {\big{)}}^{-1} =\hbox{det} (e_{I}^{\mu}) = ({1}/{n!}) \epsilon^{I_1 \cdots I_n}  \epsilon_{\mu_1 , \cdots , \mu_n}   e_{I_1}^{\mu_1}   \cdots   e_{I_n}^{\mu_n}   .
\eee
Let us note that the formula in \eqref{vol001} are equivalently, in the vielbein formalism, written as
  \bee\label{vol002}
 {\pmb{\epsilon}}_{\mu_1 ... \mu_n} =   e {\epsilon}_{\mu_1 ... \mu_n},
\quad \quad     
{\pmb{\epsilon}}^{\mu_1 ... \mu_n} =   (-1)^{\pmb{\sigma}} e^{-1} {\epsilon}^{\mu_1 ... \mu_n} .
 \eee
We have 
$   {\boldsymbol{\epsilon}}_{\mu\nu\alpha}  =  {\epsilon}_{IJK}  e^I_\mu e^J_\nu e^K_\alpha$, and $
 {\boldsymbol{\epsilon}}_{\mu\nu\alpha\beta}  =        {\epsilon}_{IJKL}  e^I_\mu e^J_\nu e^K_\alpha e^L_\beta $,
  where $  {\boldsymbol{\epsilon}}_{\mu\nu\alpha}         =  e {{\epsilon}}_{\mu\nu\alpha}   $ and $  {\boldsymbol{\epsilon}}_{\mu\nu\alpha\beta}     =  e {{\epsilon}}_{\mu\nu\alpha\beta}    $, for the dreibein   and vierbein formulation, respectively.  Note that we have also the relations $     {\epsilon}_{IJK} =      {\boldsymbol{\epsilon}}_{\mu\nu\alpha}     e_I^\mu e_J^\nu e_K^\alpha    
 $ and $      {\epsilon}_{IJKL}  =     {\boldsymbol{\epsilon}}_{\mu\nu\alpha\beta}    e_I^\mu e_J^\nu e_K^\alpha e_L^\beta 
 $.  

\subsection{{{Vielbein   densities}}}\label{app:generalizedVD}

We introduce the    vielbein   densities,   denoted by $\displaystyle \dtt^{\mu_{1} \cdots \mu_{p} }_{I_{1} \cdots I_{p}}  $, with $1 \leq p \leq n$. They are  constructed on the determinant of the vielbein  $\hbox{det}(e)$ and $p$ vielbeins $e^{\mu_1}_{I_1} \cdots e^{\mu_p}_{I_p} $ such that
$
 \dtt^{\mu_{1} \cdots \mu_{p} }_{I_{1} \cdots I_{p}}  = \hbox{det}(e)      \prod_{  j  } e^{\mu_j}_{I_j}  =   e e^{\mu_1}_{I_1} \cdots e^{\mu_p}_{I_p}  
$. We consider the anti-symmetrized object,  {\em i.e.}
 \bee\label{dzefzefezfe}
  \displaystyle \left.
\begin{array}{rcl}
 \displaystyle
\dtt^{[\mu_1 \cdots \mu_p]}_{I_{1} \cdots I_{p} }   & = &    \displaystyle
     \frac{1}{p!} \sum_{\sigma \in {\cal S}_{n}}  \dtt^{ \mu_{\sigma(1)} \cdots \mu_{\sigma(p)}  }_{I_{1} \cdots I_{p} }        = \frac{1}{p!}  \delta^{\mu_1 \cdots \mu_{p}}_{\nu_1 \cdots \nu_{p}}   \dtt^{\nu_1 \cdots \nu_p}_{I_{1} \cdots I_{p} }   =  \frac{1}{p!}  \delta^{\mu_1 \cdots \mu_{p}}_{\nu_1 \cdots \nu_{p}}    \dtt^{[\nu_1 \cdots \nu_p]}_{I_{1} \cdots I_{p} }
 \\
 \displaystyle
  & = &    \displaystyle     \frac{1}{(n-p)!} \epsilon_{\mu_1 \cdots \mu_p \rho_{1} \cdots \rho_{n-p} }  \frac{1}{p!}  \epsilon^{\nu_1 \cdots \nu_p \rho_{1} \cdots \rho_{n-p} }   
\dtt^{\mu_1 \cdots \mu_p}_{I_{1} \cdots I_{p} } 
\end{array}
\right.
\eee
First, we are interested by the density $\dtt^{\mu}_{I} = e e^{\mu}_{I}  =  \hbox{det} ({e^{\mu}_{I}})     e^{\mu}_{I}   $.   
        We have, for $p := 1$ ($n$ is the dimension of the space-time manifold), $\dtt^{\mu}_{I}      =  {\big{(}}  ({1}/{n!}) \epsilon_{I_1 \cdots I_n}  \epsilon^{\mu_1 , \cdots , \mu_n}   e^{I_1}_{\mu_1}   \cdots   e^{I_n}_{\mu_n}  {\big{)}}e^{\mu}_{I} $ or equivalently $ \dtt^{\mu}_{I}     =
 ({1}/{(n-1)!}) \epsilon^{\mu \mu_1 ... \mu_{n-1} } \epsilon_{I I_{1} ... I_{n-1} } 
e^{I_1}_{\mu_1} \cdots e^{I_{n-1}}_{\mu_{n-1}} $. This relation  is straightforwardly  derived. Let us denote $(\mathfrak{1}) :=  \epsilon^{\mu \mu_1 ... \mu_{n-1} } \epsilon_{I I_{1} ... I_{n-1} }    e^{I_1}_{\mu_1} \cdots e^{I_{n-1}}_{\mu_{n-1}} $. Using the algebraic relation  $\epsilon_{I I_{1} ... I_{n-1} }  =  {\pmb{\epsilon}}_{\nu \nu_1 ... \nu_{n-1} } e_{I}^{\nu} e_{I_1}^{\nu_1} \cdots e_{I_{n-1}}^{\nu_{n-1}}  $,   we obtain 
\[
\left.
\begin{array}{rcl}
 \displaystyle  (\mathfrak{1})  & = &  \displaystyle    \epsilon^{\mu \mu_1 ... \mu_{n-1} }  {\big{(}}   {\pmb{\epsilon}}_{\nu \nu_1 ... \nu_{n-1} } e_{I}^{\nu} e_{I_1}^{\nu_1} \cdots e_{I_{n-1}}^{\nu_{n-1}}     {\big{)}}    e^{I_1}_{\mu_1} \cdots e^{I_{n-1}}_{\mu_{n-1}} 
=
   e \delta^{\mu \mu_1 ... \mu_{n-1} }_{\nu \nu_1 ... \nu_{n-1} }e_{I}^{\nu} e_{I_1}^{\nu_1} \cdots e_{I_{n-1}}^{\nu_{n-1}}       e^{I_1}_{\mu_1} \cdots e^{I_{n-1}}_{\mu_{n-1}} ,
  \\
\displaystyle     & = &  \displaystyle  e \delta^{\mu \mu_1 ... \mu_{n-1} }_{\nu \nu_1 ... \nu_{n-1} }e_{I}^{\nu}  {\big{(}} \delta^{\nu_1}_{\mu_1} \cdots \delta^{\nu_{n-1}}_{\mu_{n-1}}  {\big{)}}  
=  
e \delta^{\mu \mu_1 ... \mu_{n-1} }_{\nu \mu_1 ... \mu_{n-1} }e_{I}^{\nu}   
=   \displaystyle   
  e (n-1)! \delta^{\mu}_{\nu} e^{\nu}_{I}  =  \displaystyle     e (n-1)! e^{\mu}_{I}  .
\end{array}
\right.
\]
 Now, we are interested  in  the density $ \dtt^{[\mu\nu]}_{IJ} = e e^{[\mu}_{I} e^{\nu]}_{J} = (1/2) e (e^{\mu}_{I} e^{\nu}_{J}  -  e^{\nu}_{I} e^{\mu}_{J} )$, which is written as
  $   \dtt^{[\mu\nu]}_{IJ} = \left(  \frac{1}{n!} \epsilon_{I_1 \cdots I_n}  \epsilon^{\mu_1 , \cdots , \mu_n}   e^{I_1}_{\mu_1}   \cdots   e^{I_n}_{\mu_n} \right) e^{[\mu}_{I}e^{\nu]}_{J} 
=    \frac{1}{2!(n-2)!} \epsilon^{\mu \nu \mu_1 ... \mu_{n-2} } \epsilon_{I JI_{1} ... I_{n-2} } 
e^{I_1}_{\mu_1} \cdots e^{I_{n-2}}_{\mu_{n-2}}  $. This relation is obtained as follows. Let us denote  $(\mathfrak{2}):=    \epsilon^{\mu \nu \mu_1 ... \mu_{n-2} } \epsilon_{I JI_{1} ... I_{n-2} } 
e^{I_1}_{\mu_1} \cdots e^{I_{n-2}}_{\mu_{n-2}} $. By the straightforward calculation 
\[
\left.
\begin{array}{rcl}
 \displaystyle
(\mathfrak{2}) & = &  \displaystyle    \epsilon^{\mu\nu \mu_1 ... \mu_{n-2} }  {\big{(}}   {\pmb{\epsilon}}_{\rho\sigma \nu_1 ... \nu_{n-2} } e_{I}^{\rho} e_{J}^{\sigma}  e_{I_1}^{\nu_1} \cdots e_{I_{n-2}}^{\nu_{n-2}}     {\big{)}}    e^{I_1}_{\mu_1}  \cdots e^{I_{n-2}}_{\mu_{n-2}}  ,
  \\
 \displaystyle     & = &  \displaystyle     e \delta^{\mu\nu \mu_1 ... \mu_{n-2} }_{\rho\sigma \nu_1 ... \nu_{n-2} }  e_{I}^{\rho} e_{J}^{\sigma}  e_{I_1}^{\nu_1} \cdots e_{I_{n-2}}^{\nu_{n-2}}       e^{I_1}_{\mu_1} \cdots e^{I_{n-2}}_{\mu_{n-2}} 
  =
   e \delta^{\mu \nu \mu_1 ... \mu_{n-2} }_{\rho\sigma \nu_1 ... \nu_{n-2} } e_{I}^{\rho}  e_{J}^{\sigma}  {\big{(}} \delta^{\nu_1}_{\mu_1} \cdots \delta^{\nu_{n-2}}_{\mu_{n-2}}  {\big{)}} ,
 \\
   & = &  \displaystyle  
e   \delta^{\mu \nu \mu_1 ... \mu_{n-2} }_{\rho\sigma \mu_1 ... \mu_{n-2} }  e_{I}^{\rho}  e_{J}^{\sigma}   
=
e  e_{I}^{\rho}  e_{J}^{\sigma}    (n-2)!  \delta^{\mu\nu}_{\rho\sigma} =
     e  e_{I}^{\rho}  e_{J}^{\sigma}    (n-2)! 2! \delta^{[\mu}_{\rho} \delta^{\nu]}_{\sigma} =   (n-2)!  2!    \dtt^{[\mu\nu]}_{IJ} ,
 \end{array}
\right.
\]
where we use  the formula   $   \delta^{\mu \nu \mu_1 ... \mu_{n-2} }_{\rho\sigma \mu_1 ... \mu_{n-2} }  =   \frac{(n-2)!}{(n-2- (n-2) )!}    \delta^{\mu\nu}_{\rho\sigma} =  (n-2)!  \delta^{\mu\nu}_{\rho\sigma} $    to pass from the second to the third line.
    \begin{lemm}\label{lem:cur2222}
Let us consider the   vielbein density  \eqref{dzefzefezfe}, with $p = 2$, {\em i.e.} {\em $\displaystyle \dtt^{[\mu\nu]}_{IJ}  = e e^{[\mu}_{I} e^{\nu]}_{J} $.} Then, {\em
 $\dtt^{[\mu\nu]}_{IJ} = 
\frac{1}{2!(n-2)!} \epsilon_{IJ I_1 \cdots I_{n-2}}    e^{I_1}_{\rho_1}    \cdots   e^{I_{n-2}}_{\rho_{n-2}}    \epsilon^{ \mu \nu \rho_{1} \cdots \rho_{n-2} }$    .
 }
\end{lemm}
    {\sffamily Proof}. By    the straightforward calculation  
 \[
\left.
\begin{array}{rcl}
\displaystyle  \dtt^{[\mu\nu]}_{IJ}    & = & 
\displaystyle    \left(   ({1}/{n!}) \epsilon_{I_1 \cdots I_n}  \epsilon^{\mu_1 , \cdots , \mu_n}   e^{I_1}_{\mu_1}   \cdots   e^{I_n}_{\mu_n}  \right) \left(
\delta^{[\mu}_{\rho} \delta^{\nu]}_{\sigma}  \right) e^{\rho}_{I} e^{\sigma}_{J},
  \\
\displaystyle    &  = & \displaystyle   \left(   ({1}/{n!}) \epsilon_{I_1 \cdots I_n}  \epsilon^{\mu_1 , \cdots , \mu_n}   e^{I_1}_{\mu_1}   \cdots   e^{I_n}_{\mu_n}  \right)(1/(2!(n-2)!))^{-1}  \left( \epsilon^{\nu_1 \cdots \nu_{n-2} \mu \nu } \epsilon_{\nu_1 \cdots \nu_{n-2}\rho\sigma} \right) e^{\rho}_{I} e^{\sigma}_{J},
\\
\displaystyle    &  = & \displaystyle  (1/(2!(n-2)!))^{-1}   \epsilon_{I_1 \cdots I_n}   \epsilon^{\nu_1 \cdots \nu_{n-2} \mu \nu }  \delta^{[\mu_1}_{\nu_1}  \cdots  \delta^{\mu_{n-2} }_{\nu_{n-2}}   \delta^{\mu_{n-1}}_{\rho}   \delta^{\mu_n]}_{\sigma}       \left(   e^{I_1}_{\mu_1}   \cdots   e^{I_n}_{\mu_n}  e^{\rho}_{I} e^{\sigma}_{J}  \right),
 \\
 \displaystyle    &  = & \displaystyle (1/(2!(n-2)!))^{-1}   \epsilon_{I_1 \cdots I_n}   \epsilon^{\mu_1 \cdots \mu_{n-2} \mu \nu }       \left(   e^{I_1}_{\mu_1}   \cdots   e^{I_{n-2}}_{\mu_{n-2}}  e^{I_{n-1}}_{\rho}   e^{I_n}_{\sigma}  e^{\rho}_{I} e^{\sigma}_{J}  \right),
  \\
  \displaystyle    & =& \displaystyle (1/(2!(n-2)!))^{-1}  \epsilon_{IJ I_1 \cdots I_{n-2}}    e^{I_1}_{\rho_1}    \cdots   e^{I_{n-2}}_{\rho_{n-2}}    \epsilon^{ \mu \nu \rho_{1} \cdots \rho_{n-2} }  .  
\end{array}
\right.
\]
In particular when $n=3$ and $n=4$ we have:
   \begin{lemm}\label{lem:cur222233}
The densities {\em $\displaystyle \dtt^{[\mu\nu]}_{IJ}  = e e^{[\mu}_{I} e^{\nu]}_{J} $}, which are  constructed with two   dreibeins and  vierbeins are  given by  
{\em$
 \dtt^{[\mu\nu]}_{IJ}  =   
 ({1}/{2}) \epsilon_{IJ K}    e^{K}_{\rho}     \epsilon^{ \mu \nu \rho}   ,
$ and $    
  \dtt^{[\mu\nu]}_{IJ}  =   
  ({1}/{4}) \epsilon_{IJ KL}    e^{K}_{\rho}      e^{L}_{\sigma}    \epsilon^{ \mu \nu \rho\sigma}$}, in the case where the dimension of the space-time manifold is $n=3$ and $n=4$, respectively.\footnote{In   Peldan's review \cite{PP}  we found the relation $ e e^{[\mu}_{I} e^{\nu]}_{J} = (1/2)  \epsilon_{IJ KL}    e^{K}_{\rho}      e^{L}_{\sigma}    \epsilon^{ \mu \nu \rho\sigma}$ since there the terms     in    antisymmetric sums are  weighted with $1$,   {\em e.g.} $   e^{[\mu}_{I} e^{\nu]}_{J}  =    e^{\mu}_{I} e^{\nu}_{J} -   e^{\nu}_{I} e^{\mu}_{J} $.  In our conventions, $   e^{[\mu}_{I} e^{\nu]}_{J}  =   (1/2)( e^{\mu}_{I} e^{\nu}_{J} -   e^{\nu}_{I} e^{\mu}_{J}) $, thus  $ e e^{[\mu}_{I} e^{\nu]}_{J} = (1/4)  \epsilon_{IJ KL}    e^{K}_{\rho}      e^{L}_{\sigma}    \epsilon^{ \mu \nu \rho\sigma}$.}
\end{lemm}

  \section{{Calculation of $   \Xi (  \dttQ_{\omega , \psi} )   \iN {\pmb{\omega}}^{\hbox{\tiny{\sffamily Palatini}}} $}}\label{app:lpkojihu88}
  The interior product $   \Xi (  \dttQ_{\omega , \psi} )   \iN {\pmb{\omega}}^{\hbox{\tiny{\sffamily Palatini}}} $ is given by the straightforward computation:
  \bee\label{gldpsodkfj7676}
\left.
\begin{array}{lll}
\displaystyle   ({\mathfrak{3}})   & = & \displaystyle - (1/2)
\epsilon_{IJKL} \epsilon^{\mu\nu\rho\sigma} e^{K}_{\rho}     \hbox{d} e^{L}_{\sigma} 
\left(
  -       (1/6)    \psi^{\alpha\beta} (x)   \epsilon^{OPQL} \epsilon_{\alpha\beta\eta\sigma}    e_{Q}^{\eta}    {\partial  }/{\partial e_{\sigma}^{L} } 
\right)\hbox{d} \omega_{\mu}^{IJ} \wedge  {\beta}_{\nu} ,
    \\
\displaystyle   &=   & \displaystyle  
(1/12) \epsilon^{OPQL}  \epsilon_{IJKL} \epsilon^{\mu\nu\rho\sigma}  \epsilon_{\alpha\beta\eta\sigma}   e^{K}_{\rho}         e_{Q}^{\eta}        \psi^{\alpha\beta} (x) 
  \hbox{d} \omega_{\mu}^{IJ} \wedge  {\beta}_{\nu} ,
    \\
   \displaystyle   & =  & \displaystyle  
  (1/12)  (3!) \delta_{I}^{[O} \delta_{J}^{P} \delta_{K}^{Q]}   (3!) \delta_{\alpha}^{[\mu} \delta_{\beta}^{\nu} \delta_{\eta}^{\rho]}     e^{K}_{\rho}         e_{Q}^{\eta}        \psi^{\alpha\beta} (x) 
  \hbox{d} \omega_{\mu}^{IJ} \wedge  {\beta}_{\nu} ,
    \end{array}
\right.
\eee
 Since
$
   \epsilon^{\mu\nu\rho\sigma}  \epsilon_{\alpha\beta\eta\sigma}   =
(3!)  \delta_{\alpha}^{[\mu} \delta_{\beta}^{\nu} \delta_{\eta}^{\rho]}  
= 
   ({(3!)}/{3})  {\big{(}} \delta^{[\mu}_\alpha \delta^{\nu]}_\beta \delta^{\rho}_\eta    
  +\delta^{[\rho}_\alpha   
\delta^{\mu]}_\beta  \delta^{\nu}_\eta    + \delta^{[\nu}_\alpha  \delta^{\rho]}_\beta   \delta^{\mu}_\eta {\big{)}} 
$,
we obtain 
{\footnotesize{
 \[
\left.
\begin{array}{lll}
\displaystyle    ({\mathfrak{3}})  & = & \displaystyle      (1/12) 
    \delta_{I}^{[O} \delta_{J}^{P} \delta_{K}^{Q]}   
    {\Big{(}} 
  \delta^{\mu}_\alpha \delta^{\nu}_\beta \delta^{\rho}_\eta  - \delta^{\mu}_\alpha \delta^{\rho}_\beta \delta^{\nu}_\eta  -  (\delta^{\nu}_\alpha  \delta^{\mu}_\beta   \delta^{\rho}_\eta - \delta^{\nu}_\alpha  \delta^{\rho}_\beta   \delta^{\mu}_\eta) +  \delta^{\rho}_\alpha   
\delta^{\mu}_\beta  \delta^{\nu}_\eta  - \delta^{\rho}_\alpha   
\delta^{\nu}_\beta  \delta^{\mu}_\eta  
   {\Big{)}}            e^{K}_{\rho}         e_{Q}^{\eta}        \psi^{\alpha\beta}  
  \hbox{d} \omega_{\mu}^{IJ} \wedge  {\beta}_{\nu} ,
   \\ 
 \displaystyle   &   =  & \displaystyle  
 (1/12)   \delta_{I}^{[O} \delta_{J}^{P} \delta_{K}^{Q]}  
   {\Big{(}}   
  \delta^{\mu}_\alpha \delta^{\nu}_\beta \delta^{\rho}_\eta    e^{K}_{\rho}         e_{Q}^{\eta}        \psi^{\alpha\beta}   
  \hbox{d} \omega_{\mu}^{IJ} \wedge  {\beta}_{\nu}  
           - \delta^{\mu}_\alpha \delta^{\rho}_\beta \delta^{\nu}_\eta    e^{K}_{\rho}         e_{Q}^{\eta}        \psi^{\alpha\beta}  
  \hbox{d} \omega_{\mu}^{IJ} \wedge  {\beta}_{\nu} 
                        -  \delta^{\nu}_\alpha  \delta^{\mu}_\beta   \delta^{\rho}_\eta     e^{K}_{\rho}         e_{Q}^{\eta}        \psi^{\alpha\beta}  
  \hbox{d} \omega_{\mu}^{IJ} \wedge  {\beta}_{\nu} ,
              \\ 
          \displaystyle   &   & \displaystyle
         +  \delta^{\nu}_\alpha  \delta^{\rho}_\beta   \delta^{\mu}_\eta    e^{K}_{\rho}         e_{Q}^{\eta}        \psi^{\alpha\beta}  
  \hbox{d} \omega_{\mu}^{IJ} \wedge  {\beta}_{\nu} 
        +  \delta^{\rho}_\alpha   
\delta^{\mu}_\beta  \delta^{\nu}_\beta   e^{K}_{\rho}         e_{Q}^{\eta}        \psi^{\alpha\beta}  
  \hbox{d} \omega_{\mu}^{IJ} \wedge  {\beta}_{\nu} 
                    - \delta^{\rho}_\alpha   
\delta^{\nu}_\beta  \delta^{\mu}_\beta 
 e^{K}_{\rho}         e_{Q}^{\eta}        \psi^{\alpha\beta}   
  \hbox{d} \omega_{\mu}^{IJ} \wedge  {\beta}_{\nu}     {\Big{)}}   ,
   \\ 
    \displaystyle   &   = & \displaystyle   (1/12)    \delta_{I}^{[O} \delta_{J}^{P} \delta_{K}^{Q]}  
   {\Big{(}}   
      e^{K}_{\rho}         e_{Q}^{\rho}        \psi^{\mu\nu}  
                 -       e^{K}_{\rho}         e_{Q}^{\nu}        \psi^{\mu\rho}  
           -          e^{K}_{\rho}         e_{Q}^{\rho}        \psi^{\nu \mu} 
         +         e^{K}_{\rho}         e_{Q}^{\mu}        \psi^{\nu \rho}  
        +     
   e^{K}_{\rho}         e_{Q}^{\nu}        \psi^{\rho\mu} 
                     -     
 e^{K}_{\rho}         e_{Q}^{\mu}        \psi^{\rho\nu} 
  {\Big{)}}      \hbox{d} \omega_{\mu}^{IJ} \wedge  {\beta}_{\nu}   ,
   \\ 
       \displaystyle   &   = & \displaystyle     (1/12)      \delta_{I}^{[O} \delta_{J}^{P} \delta_{K}^{Q]}  
   {\Big{(}}   
      \delta^{K}_{Q}              \psi^{\mu\nu}  
                 -       e^{K}_{\rho}         e_{Q}^{\nu}        \psi^{\mu\rho}  
         -  \delta^{K}_{Q}        \psi^{\nu \mu} 
         +         e^{K}_{\rho}         e_{Q}^{\mu}        \psi^{\nu \rho}  
        +     
   e^{K}_{\rho}         e_{Q}^{\nu}        \psi^{\rho\mu} 
                     -     
 e^{K}_{\rho}         e_{Q}^{\mu}        \psi^{\rho\nu} 
  {\Big{)}}      \hbox{d} \omega_{\mu}^{IJ} \wedge  {\beta}_{\nu}  , 
   \\ 
       \displaystyle   &   = & \displaystyle   (1/6)     \delta_{I}^{[O} \delta_{J}^{P} \delta_{K}^{Q]}  
   {\Big{(}}   
       \delta^{K}_{Q}              \psi^{\mu\nu}  
                 -        e^{K}_{\rho}         e_{Q}^{\nu}        \psi^{\mu\rho}   
                          +          e^{K}_{\rho}         e_{Q}^{\mu}        \psi^{\nu \rho}  
        {\Big{)}}      \hbox{d} \omega_{\mu}^{IJ} \wedge  {\beta}_{\nu}   ,
   \end{array}
\right.
\]
and since  
 $ \displaystyle
 \epsilon^{OPQ}  \epsilon_{IJK}      =   \delta^{O}_I \delta^{P}_J \delta^{Q}_K  - \delta^{O}_I \delta^{Q}_J \delta^{P}_K  -  (\delta^{P}_I  \delta^{O}_J   \delta^{Q}_K - \delta^{P}_I  \delta^{Q}_J   \delta^{O}_K) +  \delta^{Q}_I   
\delta^{O}_J  \delta^{P}_K  - \delta^{Q}_I   
\delta^{P}_J  \delta^{O}_K 
$,
then  $ ({\mathfrak{3}}) $ is written as 

\bee\label{developping}
\left.
\begin{array}{lll}
\displaystyle     ({\mathfrak{3}})   & = &     \displaystyle   (1/6)     {\big{(}} \delta^{O}_I \delta^{P}_J \delta^{Q}_K  - \delta^{O}_I \delta^{Q}_J \delta^{P}_K  -  (\delta^{P}_I  \delta^{O}_J   \delta^{Q}_K - \delta^{P}_I  \delta^{Q}_J   \delta^{O}_K) +  \delta^{Q}_I   
\delta^{O}_J  \delta^{P}_K  - \delta^{Q}_I   
\delta^{P}_J  \delta^{O}_K   {\big{)}} 
\\
 \displaystyle    &  &     \displaystyle   
     {\big{(}}   
      \delta^{K}_{Q}              \psi^{\mu\nu}  
                 -       e^{K}_{\rho}         e_{Q}^{\nu}        \psi^{\mu\rho}   
                          +       e^{K}_{\rho}         e_{Q}^{\mu}        \psi^{\nu \rho}  
        {\big{)}}        \hbox{d} \omega_{\mu}^{IJ} \wedge  {\beta}_{\nu}   ,
        \\
        \displaystyle     & = & \displaystyle     (1/6)      \BL
\delta^{O}_I \delta^{P}_J \delta^{Q}_K       {\Big{(}}    \delta^{K}_{Q}     \psi^{\mu\nu}    \hbox{d} \omega_{\mu}^{IJ} \wedge  {\beta}_{\nu}      {\Big{)}}  
- \delta^{O}_I \delta^{Q}_J \delta^{P}_K   {\Big{(}}    \delta^{K}_{Q}     \psi^{\mu\nu}    \hbox{d} \omega_{\mu}^{IJ} \wedge  {\beta}_{\nu}      {\Big{)}}  
-  \delta^{P}_I  \delta^{O}_J   \delta^{Q}_K  {\Big{(}}    \delta^{K}_{Q}     \psi^{\mu\nu}    \hbox{d} \omega_{\mu}^{IJ} \wedge  {\beta}_{\nu}      {\Big{)}}  ,
    \\ 
      \displaystyle   &   & \displaystyle
+ \delta^{P}_I  \delta^{Q}_J   \delta^{O}_K {\Big{(}}    \delta^{K}_{Q}     \psi^{\mu\nu}    \hbox{d} \omega_{\mu}^{IJ} \wedge  {\beta}_{\nu}      {\Big{)}}  
 +  \delta^{Q}_I   \delta^{O}_J  \delta^{P}_K  {\Big{(}}    \delta^{K}_{Q}     \psi^{\mu\nu}    \hbox{d} \omega_{\mu}^{IJ} \wedge  {\beta}_{\nu}      {\Big{)}}  
 - \delta^{Q}_I   \delta^{P}_J  \delta^{O}_K    {\Big{(}}    \delta^{K}_{Q}     \psi^{\mu\nu}    \hbox{d} \omega_{\mu}^{IJ} \wedge  {\beta}_{\nu}      {\Big{)}}  ,
    \\ 
          \displaystyle   &   & \displaystyle  +
          \delta^{O}_I \delta^{P}_J \delta^{Q}_K    {\Big{(}}       -       e^{K}_{\rho}         e_{Q}^{\nu}        \psi^{\mu\rho}   
                       \hbox{d} \omega_{\mu}^{IJ} \wedge  {\beta}_{\nu}      {\Big{)}} 
          - \delta^{O}_I \delta^{Q}_J \delta^{P}_K    {\Big{(}}       -       e^{K}_{\rho}         e_{Q}^{\nu}        \psi^{\mu\rho}   
                       \hbox{d} \omega_{\mu}^{IJ} \wedge  {\beta}_{\nu}      {\Big{)}} 
           -   \delta^{P}_I  \delta^{O}_J   \delta^{Q}_K   {\Big{(}}       -       e^{K}_{\rho}         e_{Q}^{\nu}        \psi^{\mu\rho}   
                       \hbox{d} \omega_{\mu}^{IJ} \wedge  {\beta}_{\nu}      {\Big{)}} ,
                           \\ 
          \displaystyle   &   & \displaystyle
          + \delta^{P}_I  \delta^{Q}_J   \delta^{O}_K   {\Big{(}}       -       e^{K}_{\rho}         e_{Q}^{\nu}        \psi^{\mu\rho}   
                       \hbox{d} \omega_{\mu}^{IJ} \wedge  {\beta}_{\nu}      {\Big{)}} 
           +  \delta^{Q}_I \delta^{O}_J  \delta^{P}_K   {\Big{(}}       -       e^{K}_{\rho}         e_{Q}^{\nu}        \psi^{\mu\rho}   
                       \hbox{d} \omega_{\mu}^{IJ} \wedge  {\beta}_{\nu}      {\Big{)}} 
            - \delta^{Q}_I   \delta^{P}_J  \delta^{O}_K    {\Big{(}}       -       e^{K}_{\rho}         e_{Q}^{\nu}        \psi^{\mu\rho}   
                       \hbox{d} \omega_{\mu}^{IJ} \wedge  {\beta}_{\nu}      {\Big{)}}   , 
    \\ 
          \displaystyle   &   & \displaystyle   + \delta^{O}_I \delta^{P}_J \delta^{Q}_K      {\Big{(}}       +       e^{K}_{\rho}         e_{Q}^{\mu}        \psi^{\nu \rho}    \hbox{d} \omega_{\mu}^{IJ} \wedge  {\beta}_{\nu}  
        {\Big{)}} 
           - \delta^{O}_I \delta^{Q}_J \delta^{P}_K       {\Big{(}}       +       e^{K}_{\rho}         e_{Q}^{\mu}        \psi^{\nu \rho}    \hbox{d} \omega_{\mu}^{IJ} \wedge  {\beta}_{\nu}  
        {\Big{)}} 
          -   \delta^{P}_I  \delta^{O}_J   \delta^{Q}_K      {\Big{(}}       +       e^{K}_{\rho}         e_{Q}^{\mu}        \psi^{\nu \rho}    \hbox{d} \omega_{\mu}^{IJ} \wedge  {\beta}_{\nu}  
        {\Big{)}} ,
            \\ 
          \displaystyle   &   & \displaystyle
          + \delta^{P}_I  \delta^{Q}_J   \delta^{O}_K      {\Big{(}}       +       e^{K}_{\rho}         e_{Q}^{\mu}        \psi^{\nu \rho}    \hbox{d} \omega_{\mu}^{IJ} \wedge  {\beta}_{\nu}  
        {\Big{)}} 
           +  \delta^{Q}_I   \delta^{O}_J  \delta^{P}_K      {\Big{(}}       +       e^{K}_{\rho}         e_{Q}^{\mu}        \psi^{\nu \rho}    \hbox{d} \omega_{\mu}^{IJ} \wedge  {\beta}_{\nu}  
        {\Big{)}}  
           - \delta^{Q}_I   \delta^{P}_J  \delta^{O}_K       {\Big{(}}       +       e^{K}_{\rho}         e_{Q}^{\mu}        \psi^{\nu \rho}    \hbox{d} \omega_{\mu}^{IJ} \wedge  {\beta}_{\nu}  
        {\Big{)}}  \BR  ,
  \\
\displaystyle      & = &         \displaystyle   (1/6)     \BL  {\Big{(}} 
 +     4 \psi^{\mu\nu}    \hbox{d} \omega_{\mu}^{PO} \wedge  {\beta}_{\nu}     
   {\Big{)}}    
   +
          {\Big{(}}       -       \delta^{\nu}_{\rho}             \psi^{\mu\rho}   
                       \hbox{d} \omega_{\mu}^{OP} \wedge  {\beta}_{\nu}      {\Big{)}} 
          +
               {\Big{(}}           e^{P}_{\rho}         e_{Q}^{\nu}        \psi^{\mu\rho}   
                       \hbox{d} \omega_{\mu}^{OQ} \wedge  {\beta}_{\nu}      {\Big{)}} 
      +      {\Big{(}}          \delta^{\nu}_{\rho}            \psi^{\mu\rho}   
                       \hbox{d} \omega_{\mu}^{PO} \wedge  {\beta}_{\nu}      {\Big{)}} ,
                  \\
   &   & \displaystyle
          +     {\Big{(}}       -       e^{O}_{\rho}         e_{Q}^{\nu}        \psi^{\mu\rho}   
                       \hbox{d} \omega_{\mu}^{PQ} \wedge  {\beta}_{\nu}      {\Big{)}} 
           +         {\Big{(}}       -       e^{P}_{\rho}         e_{Q}^{\nu}        \psi^{\mu\rho}   
                       \hbox{d} \omega_{\mu}^{QO} \wedge  {\beta}_{\nu}      {\Big{)}} 
            -    {\Big{(}}       -       e^{O}_{\rho}         e_{Q}^{\nu}        \psi^{\mu\rho}   
                       \hbox{d} \omega_{\mu}^{QP} \wedge  {\beta}_{\nu}      {\Big{)}}  ,   
    \\ 
          \displaystyle   &   & \displaystyle    +
            {\Big{(}}       + \delta^{\mu}_{\rho}        \psi^{\nu \rho}    \hbox{d} \omega_{\mu}^{OP} \wedge  {\beta}_{\nu}  
        {\Big{)}} 
           -        {\Big{(}}       +       e^{P}_{\rho}         e_{Q}^{\mu}        \psi^{\nu \rho}    \hbox{d} \omega_{\mu}^{OQ} \wedge  {\beta}_{\nu}  
        {\Big{)}} 
          -        {\Big{(}}       +       \delta^{\mu}_{\rho}            \psi^{\nu \rho}    \hbox{d} \omega_{\mu}^{PO} \wedge  {\beta}_{\nu}  
        {\Big{)}} ,
            \\ 
          \displaystyle   &   & \displaystyle
          +        {\Big{(}}       +       e^{O}_{\rho}         e_{Q}^{\mu}        \psi^{\nu \rho}    \hbox{d} \omega_{\mu}^{PQ} \wedge  {\beta}_{\nu}  
        {\Big{)}} 
           +         {\Big{(}}       +       e^{P}_{\rho}         e_{Q}^{\mu}        \psi^{\nu \rho}    \hbox{d} \omega_{\mu}^{QO} \wedge  {\beta}_{\nu}  
        {\Big{)}}  
           -       {\Big{(}}       +       e^{O}_{\rho}         e_{Q}^{\mu}        \psi^{\nu \rho}    \hbox{d} \omega_{\mu}^{QP} \wedge  {\beta}_{\nu}  
        {\Big{)}} \BR   ,
 \\
\displaystyle       & = & \displaystyle   (1/6)      \BL   {\Big{(}} 
 +     6 \psi^{\mu\nu}    \hbox{d} \omega_{\mu}^{PO} \wedge  {\beta}_{\nu}     
   {\Big{)}}  
+
    {\Big{(}} 
               {\Big{(}}           e^{P}_{\rho}         e_{Q}^{\nu}        \psi^{\mu\rho}   
                       \hbox{d} \omega_{\mu}^{OQ} \wedge  {\beta}_{\nu}      {\Big{)}} 
                       -   {\Big{(}}              e^{O}_{\rho}         e_{Q}^{\nu}        \psi^{\mu\rho}   
                       \hbox{d} \omega_{\mu}^{PQ} \wedge  {\beta}_{\nu}      {\Big{)}} ,
                           \\ 
          \displaystyle   &   & \displaystyle
           +         {\Big{(}}       -       e^{P}_{\rho}         e_{Q}^{\nu}        \psi^{\mu\rho}   
                       \hbox{d} \omega_{\mu}^{QO} \wedge  {\beta}_{\nu}      {\Big{)}} 
            -    {\Big{(}}       -       e^{O}_{\rho}         e_{Q}^{\nu}        \psi^{\mu\rho}   
                       \hbox{d} \omega_{\mu}^{QP} \wedge  {\beta}_{\nu}      {\Big{)}}      
  +
         {\Big{(}}               e^{O}_{\rho}         e_{Q}^{\mu}        \psi^{\nu \rho}    \hbox{d} \omega_{\mu}^{PQ} \wedge  {\beta}_{\nu}  
        {\Big{)}} ,
            \\ 
          \displaystyle   &   & \displaystyle     -        {\Big{(}}             e^{P}_{\rho}         e_{Q}^{\mu}        \psi^{\nu \rho}    \hbox{d} \omega_{\mu}^{OQ} \wedge  {\beta}_{\nu}  
        {\Big{)}} 
          +       
              {\Big{(}}            e^{P}_{\rho}         e_{Q}^{\mu}        \psi^{\nu \rho}    \hbox{d} \omega_{\mu}^{QO} \wedge  {\beta}_{\nu}  
        {\Big{)}}  
           -       {\Big{(}}             e^{O}_{\rho}         e_{Q}^{\mu}        \psi^{\nu \rho}    \hbox{d} \omega_{\mu}^{QP} \wedge  {\beta}_{\nu}  
        {\Big{)}} \BR   ,
\\
\displaystyle     & = & \displaystyle   \BL   {\Big{(}} 
 +     6    (1/6)     \psi^{\mu\nu}    \hbox{d} \omega_{\mu}^{PO} \wedge  {\beta}_{\nu}     
   {\Big{)}}  
+     (1/6)     
    {\Big{(}}  2
               {\Big{(}}           e^{P}_{\rho}         e_{Q}^{\nu}        \psi^{\mu\rho}   
                       \hbox{d} \omega_{\mu}^{OQ} \wedge  {\beta}_{\nu}      {\Big{)}} 
                       -      (1/6)     {\Big{(}}      2        e^{O}_{\rho}         e_{Q}^{\nu}        \psi^{\mu\rho}   
                       \hbox{d} \omega_{\mu}^{PQ} \wedge  {\beta}_{\nu}      {\Big{)}} ,
                           \\ 
          \displaystyle   &   & \displaystyle
            +     (1/6)  
         {\Big{(}}       2        e^{O}_{\rho}         e_{Q}^{\mu}        \psi^{\nu \rho}    \hbox{d} \omega_{\mu}^{PQ} \wedge  {\beta}_{\nu}  
        {\Big{)}}  
             -        (1/6)    {\Big{(}}    2         e^{P}_{\rho}         e_{Q}^{\mu}        \psi^{\nu \rho}    \hbox{d} \omega_{\mu}^{OQ} \wedge  {\beta}_{\nu}  
        {\Big{)}} 
       \BR  .
   \end{array}
\right.
\nonumber
\eee
Therefore, we conclude that 
$ \displaystyle
 ({\mathfrak{3}})     = 
 6   (1/6)  
  \psi^{\mu\nu}    \hbox{d} \omega_{\mu}^{IJ} \wedge  {\beta}_{\nu}   
  =      (1/6)   ( (1/6) )^{-1}
  \psi^{\mu\nu}    \hbox{d} \omega_{\mu}^{IJ} \wedge  {\beta}_{\nu}       
=  \psi^{\mu\nu}    \hbox{d} \omega_{\mu}^{IJ} \wedge  {\beta}_{\nu}.$

}}

     \protect\label{conclusion}

 \protect\label{mvf}
 
\pdfbookmark[1]{References}{ref}

\LastPageEnding

\end{document}